\documentclass[a4paper,11pt]{article}
\pdfoutput=1 

\usepackage{jhep_like} 

\newcommand{\HL}[1]{\textcolor{magenta}{#1}}

\newcommand{\SV}[1]{\textcolor{red}{#1}}

\usepackage{tensor}
\usepackage{braket}
\usepackage{comment}

\usepackage{dsfont}
\usepackage{bbm}
\usepackage{tikz}
\usepackage{upgreek}

\usepackage[normalem]{ulem}
\usepackage{amsmath}
\usepackage{enumerate}
\usepackage{epsfig}
\usepackage{mathbbol}

\newcommand\half{{\ensuremath{\frac{1}{2}}}}

\newcommand\field[1]{{\ensuremath{\mathbb{{#1}}}}}

\newcommand\vev[1]{{\ensuremath{\left\langle{#1}\right\rangle}}}

\newcommand{\RR}{\field{R}}

\newcommand{\Zpt}{\sZ_n^{(\text{PT})}}

\newcommand{\be}{\begin{equation}}
\newcommand{\ee}{\end{equation}}
\newcommand{\bea}{\begin{eqnarray}}
\newcommand{\eea}{\end{eqnarray}}
\newcommand{\bega}{\begin{gather}}
\newcommand{\eega}{\end{gather}}

\newcommand{\bi}{\begin{itemize}}
\newcommand{\ei}{\end{itemize}}
\newcommand{\ben}{\begin{enumerate}}
\newcommand{\een}{\end{enumerate}}
\newcommand{\bca}{\begin{cases}}
\newcommand{\eca}{\end{cases}}
\newcommand{\bln}{\begin{align}}
\newcommand{\eln}{\end{align}}
\newcommand{\bst}{\begin{split}}
\newcommand{\est}{\end{split}}
\def\ie{\begin{equation}\begin{aligned}}
\def\fe{\end{aligned}\end{equation}}
\newcommand{\bma}{\le(\begin{matrix}}
\newcommand{\ema}{\end{matrix}\ri)}
\newcommand{\bwt}{\begin{widetext}}
\newcommand{\ewt}{\end{widetext}}

\newcommand\al{{\alpha}}
\def\b{{\beta}}
\newcommand\ep{\epsilon}
\newcommand\sig{\sigma}

\newcommand\lam{\lambda}
\newcommand\Lam{\Lambda}

\newcommand\da{{\dagger}}

\newcommand\ov{\over}
\newcommand\ha{{\half}}

\def\le{\left}
\def\ri{\right}

\newcommand\sA{{\ensuremath{{\mathcal A}}}}
\newcommand\sB{{\ensuremath{{\mathcal B}}}}
\newcommand\sC{{\ensuremath{{\mathcal C}}}}
\newcommand\sD{{\ensuremath{{\mathcal D}}}}
\newcommand\sE{{\ensuremath{{\mathcal E}}}}

\newcommand\sI{{\ensuremath{{\mathcal I}}}}

\newcommand\sH{{\ensuremath{{\mathcal H}}}}

\newcommand\sL{{\ensuremath{{\mathcal L}}}}

\newcommand\sN{{\ensuremath{{\mathcal N}}}}
\newcommand\sO{{\ensuremath{{\mathcal O}}}}
\newcommand\sP{{\ensuremath{{\mathcal P}}}}

\newcommand\sR{{\mathcal R}}
\newcommand\sS{{\mathcal S}}

\newcommand\sY{{\mathcal Y}}
\newcommand\sZ{{\mathcal Z}}

\newcommand{\Tr}{\text{Tr}}

\title{\boldmath Mixed-state entanglement and information recovery in thermalized states and evaporating black holes}

\author[a, *]{Shreya Vardhan,} 
\author[b, *]{Jonah Kudler-Flam,}
\author[c]{Hassan Shapourian,}
\author[a]{and Hong Liu}
\affiliation[a]{Center for Theoretical Physics, 
Massachusetts Institute of Technology, Cambridge, MA 02139 }
\affiliation[b]{Kadanoff Center for Theoretical Physics,
University of Chicago, Chicago, IL 60637 }
\affiliation[c]{Microsoft Station Q, 
Santa Barbara, CA 93109
 }
 \note[*]{These authors contributed equally to the work. }

\emailAdd{vardhan@mit.edu}
\emailAdd{jkudlerflam@uchicago.edu}
\emailAdd{hassan.shapp@gmail.com}
\emailAdd{hong\_liu@mit.edu}

\preprint{MIT-CTP/5362}

\abstract{We study the universal behavior of quantum information-theoretic quantities in thermalized isolated quantum many-body systems and evaporating black holes. In particular, we study a genuine mixed-state entanglement measure called the logarithmic negativity, other correlation measures including the Renyi negativities and the mutual information, and a signature of multipartite entanglement called the reflected entropy. We also probe the feasibility of recovering quantum information from subsystems of a thermalized quantum many-body system or from the radiation of an evaporating black hole, using quantities such as relative entropy and Petz map fidelity. A recently developed technique called the equilibrium approximation allows us to probe these quantities at finite temperature. We find striking qualitative differences from the infinite temperature case, which has been the topic of previous studies using Haar-random states.
In particular, we find regimes where the logarithmic negativity is extensive but the mutual information is sub-extensive, indicating a large amount of undistillable, bound entanglement in thermalized states. For evaporating black holes at finite temperature, both the logarithmic negativity and the Petz map fidelity reveal an important new time scale $t_b$, which is earlier than the Page time $t_p$ by a finite fraction of the total evaporation time. We find that $t_b$, as opposed to $t_p$, is the time scale at which quantum entanglement between different parts of the radiation becomes extensive, and the fidelity of information recovery for a large diary thrown into the black hole starts to grow.
}

\begin{document}

\maketitle


\newpage

\section{Introduction}

A chaotic quantum many-body system initially in a far-from-equilibrium pure state should eventually 
approach a macroscopic equilibrium. In equilibrium, despite the fact that the system is in a pure state, we can use 
an equilibrium density operator $\rho^{\text{(eq)}}$ to characterize its macroscopic properties using quantities such as 
 temperature, thermal entropy, and free energy. Furthermore, expectation values and correlation functions of generic few-body observables can also be reliably calculated using $\rho^{\text{(eq)}}$.

 Surprisingly, it has been recognized recently that fine-grained quantum-informational quantities such as the Renyi and von Neumann entropies of various subsystems can also be calculated using $\rho^{\text{(eq)}}$, in a way which is compatible with unitarity~\cite{2020arXiv200801089L}. The method, called the equilibrium approximation, provides a powerful tool for extracting universal quantum-informational properties of a chaotic quantum many-body system. For example, it enables one to make predictions for the entanglement structure of the system at a finite temperature, and can be used to explain certain semi-classical gravity calculations of Renyi and von Neumann entropies for evaporating black holes~\cite{2020JHEP...09..002P, 2019JHEP...12..063A, 2019arXiv191111977P, 2020JHEP...05..013A}.

In this paper, we generalize the equilibrium approximation to a number of other quantum-informational measures, including 
 Renyi and logarithmic negativities, relative entropy, the fidelity of the Petz map, and reflected entropy. These generalizations 
 enable us to probe and make predictions both for the mixed-state entanglement structure and for information recovery in a system at a {\it finite} temperature. Besides their implications for general quantum chaotic systems, these issues are also of much interest for probing the quantum nature of black hole evaporation. The results of this paper can be used both to make predictions for the entanglement structure hidden in the Hawking radiation emitted by a black hole, and for understanding how and when quantum information is transferred from a black hole to its radiation. We find various surprising phenomena at finite temperature that have no infinite temperature analog, underscoring the importance of energy conservation.

 \begin{figure}[] 
\centering 
\includegraphics[width=\textwidth]{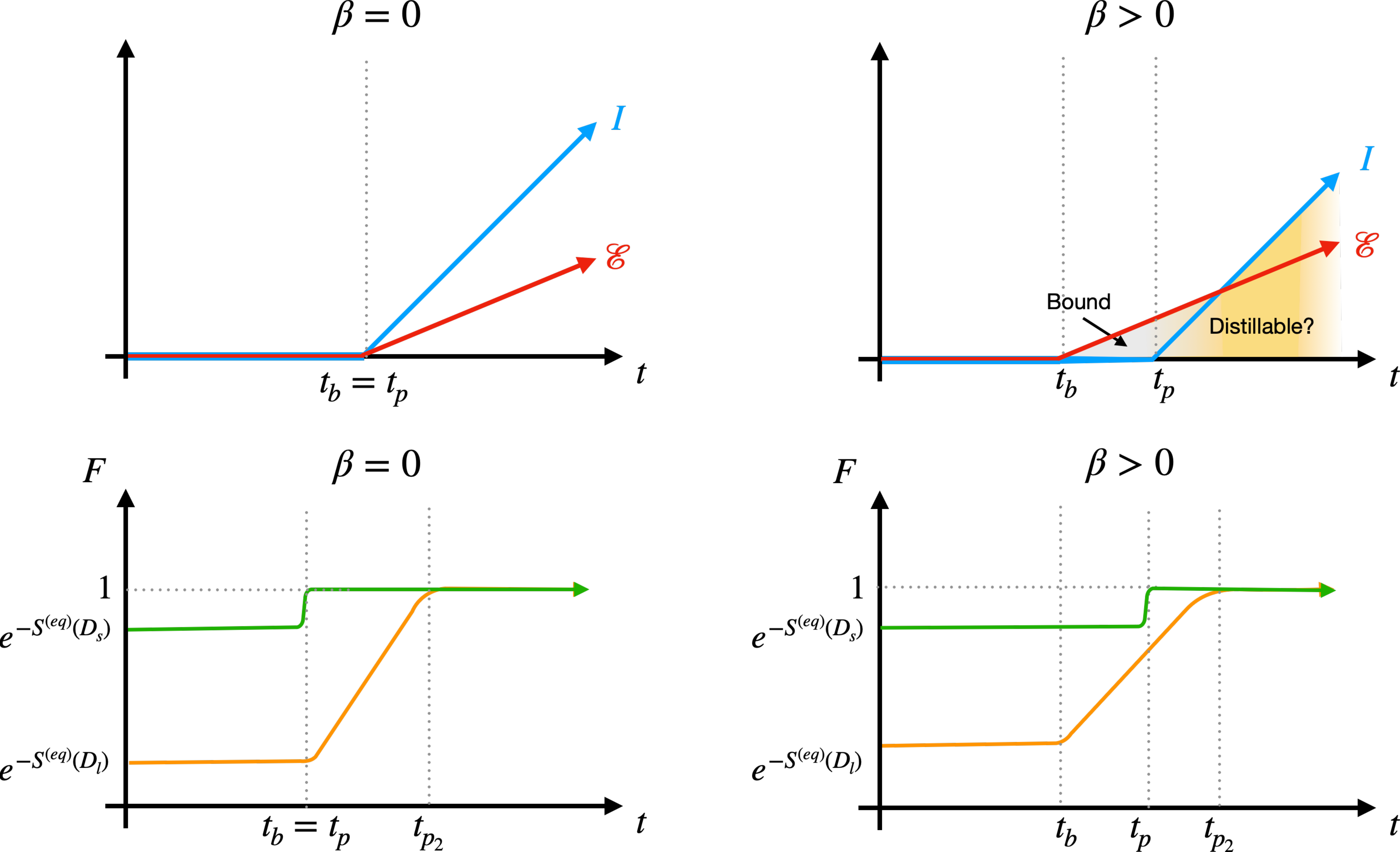}
\caption{Top row: The general behavior of logarithmic negativity (red) and mutual information (blue) is shown for finite temperature equilibrated pure states, specifically evaporating black holes. While the mutual information within the radiation does not become extensive until the Page time $t_p$, the negativity starts to become extensive at the earlier time $t_b$, signaling the existence of quantum correlations in the radiation prior to the Page time that cannot be distilled into EPR pairs. After the Page time, we expect the entanglement to be distillable but do not have a rigorous proof. Bottom row: We also plot the fidelity of the Petz map in the Hayden-Preskill experiment for small (green) and large (orange) diaries. For small diaries, the fidelity increases rapidly from its initial value to 1 at $t_p$. For sufficiently large diaries, the fidelity begins to increase at the time scale $t_b< t_p$, and reaches a value close to 1 at $t_{p_2}> t_p$. $t_{p_2}$ is defined as the time when the entropy of the radiation is equal to the entropy of the black hole plus the equilibrium entropy of the diary. The cartoon plots in the bottom row are schematic, but should be thought to have a logarithmically scaled y-axis.}
\label{fig:bound_cartoon}
\end{figure}

Our main results can be summarized as follows:\footnote{Some of the results in the first point below have been summarized in an earlier paper~\cite{2021arXiv211002959V}.}
 
 \ben 
 
 \item Consider a system $AB$ in a macroscopically equilibrated pure state $\ket{\Psi}$, and a subsystem $A = A_1 \cup A_2$ whose reduced density operator is given by $\rho_A = \Tr_B \ket{\Psi} \bra{\Psi}$.
 We study the Renyi negativities $R_n (A_1, A_2)$ and the logarithmic negativity $\sE (A_1, A_2)$, which encode bi-partite entanglement between $A_1$ and $A_2$ in the mixed state $\rho_A$. 
 The logarithmic negativity $\sE$ is non-zero only if 
$\rho_A$ is not separable,
and can be used to lower-bound the PPT entanglement cost, $\mathcal{E} (A_1, A_2) \leq E_c^{\rm (ppt, exact)}(A_1, A_2)$ \cite{audenaert2003entanglement}. Depending on the sizes of $A$ and $A_1$, the behaviors of the negativities give rise to an intricate phase diagram, exhibiting a rich entanglement structure. In particular, finite temperature has a significant effect on the qualitative structure of the entanglement phase diagram. 

The behavior of the negativities can further be contrasted with that of the Renyi mutual informations $I_n (A_1, A_2)$ and the mutual information $I (A_1, A_2)$ between $A_1$ and $A_2$. Unlike the logarithmic negativity, the mutual information $I$ can contain information about both classical and quantum correlations in the state, but is nevertheless of importance for understanding the entanglement structure as it upper-bounds the distillable entanglement, $E_d (A_1, A_2) \leq \ha I (A_1, A_2)$ \cite{2004JMP....45..829C}. 
The phase structure indicated by the mutual information $I (A_1, A_2)$ appears to be insensitive to finite temperature, although the Renyi mutual informations can be sensitive to finite temperature. Comparison between (Renyi) mutual information and negativities indicate that these quantities likely capture complementary aspects of mixed state entanglement. 

A particularly surprising result is that in the thermodynamic limit, there can be a finite region in the entanglement phase diagram where the logarithmic negativity is extensive, but the mutual information is sub-extensive, implying that there is a large amount of \textit{bound entanglement}. This phenomenon does not take place in the infinite temperature case previously studied in \cite{2021PRXQ....2c0347S}. This observation also has important implications for black hole physics. It implies the existence of a new time scale $t_b$ before the Page time $t_p$, after which there is already a significant amount of entanglement correlation 
within the Hawking radiation. In the following point, we give an operational interpretation of $t_b$. See Fig.~\ref{fig:bound_cartoon}. 

These results on the behavior of the negativity and mutual information, together with the behavior of a quantity called reflected entropy, also suggest the existence of nontrivial multi-partite entanglement among different parts of the radiation and the black hole.

 \item We consider the Hayden-Preskill thought experiment \cite{2007JHEP...09..120H}\footnote{Throw a diary into an evaporating black hole, and see when the information of the diary is recoverable from the radiation.} for information transfer from a black hole to its Hawking radiation at a finite temperature. We take two perspectives on this process of information transfer, viewing it as~(i) growth of the mutual information between an auxiliary reference system and the radiation subsystem, as in~\cite{2007JHEP...09..120H}; (ii) growth of the fidelity of a recovery map for extracting the information from the radiation. The recovery map is in turn studied from two approaches: from a lower bound on its fidelity provided by relative entropy, and from a direct calculation of the fidelity of an explicit map called the Petz map. We show that in both approaches, the results at finite temperature can be expressed in terms of natural quantum information-theoretic measures of the thermal state. 
In particular, for a ``small'' diary whose size does not scale with the total volume of the system, the fidelity rapidly grows from its initial small value to its maximum value of 1 at the standard Page time $t_p$. For a ``large'' diary whose size does scale with the volume, the fidelity becomes exponentially close to 1 at a time $t_{p_2} > t_p$, which was previously identified in \cite{2019JHEP...12..007H}. But intriguingly, the fidelity of the large diary first starts to grow from its small initial value at the same time scale $t_b < t_p$ when the logarithmic negativity between parts of the radiation starts to grow.\footnote{As we vary the size of the diary from $O(1)$ to $O(V)$ where $V$ is the volume of the system, the time scale for the beginning of fidelity growth gradually changes from $t_p$ to $t_b$.} This result is in contrast to the prior literature based mostly on infinite temperature calculations, which always took $t_p$ and $t_{p_2}$ to be the only relevant time scales for information recovery. These statements can be viewed as giving operational definitions of the time scales $t_b$, $t_p$, and $t_{p_2}$, which apply both at infinite temperature and finite temperature. See Fig.~\ref{fig:bound_cartoon}.

 \een

The plan of the paper is as follows. In Section \ref{review_sec}, we review some relevant concepts about mixed-state entanglement, the Renyi and logarithmic negativities, and the Petz recovery map, as well as the equilibrium approximation developed in \cite{2020arXiv200801089L}. In Section \ref{eq_approx_sec}, we explain how to generalize the equilibrium approximation to the various information-theoretic quantities studied in this paper. In Section \ref{phase_diag_sec}, we use these methods to find entanglement phase diagrams for a variety of universality classes of thermalized states. Section \ref{sec:mixen} discusses the operational implications of these entanglement phase diagrams, and the consequences for black hole dynamics. We study the question of information recovery from the radiation of an evaporating black hole at finite temperature in Section \ref{transfer_sec}, and conclude with some discussion of future directions in Section \ref{conclusion}. Many details are relegated to appendices.

\section{Review of background materials} 
\label{review_sec}

In this section we review topics and technical tools that will be used in subsequent discussions. 

\subsection{Entanglement in mixed states}

Consider a state $\rho$ in a bipartite system $\sH_A = \sH_{A_1} \otimes \sH_{A_2}$. $\rho$ is said to be a separable state if it can be written as a convex combination of product states,
\be 
\rho = \sum_{i=1}^q p_i \, (\rho_i)_{A_1}\otimes (\tilde{\rho}_i)_{A_2} \, , \quad 0 \leq p_a \leq 1 , \quad \sum_{a=1}^q p_a = 1\, . \label{sep}
\ee
Such a state has no quantum entanglement, as the correlations in it can be given a classical hidden-variable description \cite{werner1989quantum}, and it can be prepared using only local operations and classical communications (LOCC) without any need for EPR pairs between $A_1$ and $A_2$. Any state that is not separable is said to be entangled. 

If we know that $\rho$ is a pure state, then its entanglement entropy or von Neumann entropy in $A_1$, defined as 
\be 
S_{A_1}(\rho) := - \text{Tr}[\rho_{A_1} \log \rho_{A_1}], \quad \rho_{A_1} := \text{Tr}_{A_2}[\rho] \label{vn}
\ee
is sufficient to determine whether it is separable or entangled. $\rho$ is an entangled state if $S_{A_1}(\rho)$ is non-zero, and a separable state with $q=1$ otherwise. 

If $\rho$ is not a pure state, then no easily calculable quantity is known which is zero when the state is unentangled and non-zero when it is entangled. In fact, it is an NP-hard problem to determine whether an arbitrary state is entangled \cite{gurvits2004classical}. 

One familiar quantity that captures correlations between regions $A_1$ and $A_2$ in the state $\rho$ is the mutual information, 
\be 
I(A_1, A_2) := S_{A_1}(\rho) +S_{A_2}(\rho) - S_{A}(\rho), \label{mut} 
\ee 
where $S_{A_i}(\rho)$ is the von Neumann entropy defined in \eqref{vn}. While the mutual information is non-zero for any entangled state, it is also nonzero for a separable state as in \eqref{sep} with $q> 1$, and can hence reflect both classical correlations and quantum entanglement. 

Another useful quantity for studying entanglement in mixed states is the logarithmic negativity. Suppose the density matrix $\rho$ has matrix elements 
$\rho_{a_1 a_2, b_1 b_2}$ in some basis $\ket{a_1 a_2}$ for $\sH_{A_1}\otimes \sH_{A_2}$. We define the partial transpose $\rho^{T_2}$ of $\rho$ with respect to $\sH_{A_2}$ as 
\be 
\rho^{T_2}_{a_1 a_2, b_1 b_2} := \rho_{a_1 b_2, b_1 a_2} \ . \label{pt}
\ee
The eigenvalues of $\rho^{T_2}$ can now in principle be negative. The presence of negative eigenvalues in $\rho^{T_2}$ reflects that subsystems $A_1$ and $A_2$ are entangled; for a separable state \eqref{sep}, $\rho^{T_2} = \sum_{i=1}^q p_i \, (\rho_i)_{A_1}\otimes (\tilde{\rho}_i)^{T}_{A_2}$ remains a legitimate density operator with non-negative eigenvalues, as the transpose of a positive operator is still positive. 
This motivates one to define the logarithmic negativity as\footnote{$\rho^{T_1}$ has the same eigenvalues as $\rho^{T_2}$, and thus the definition of $\sE$ is symmetric between $A_1$ and $A_2$.} \cite{1996PhLA..223....1H,1998PhRvA..58..883Z, 1996PhRvL..77.1413P, eisert1999comparison, 2000PhRvL..84.2726S, vidal2002computable, plenio2005logarithmic}
\be
\sE (A_1, A_2) 
 := \log \sum_{k} |\lam_k|,
\ee
where $\lam_k$ are the eigenvalues of $\rho^{T_2}$. By definition, $\rho^{T_2}$ has unit trace, and thus $\sum_{k} |\lam_k| \geq 1$, 
which implies that $\sE \geq 0$. States with $\sE=0$ are referred to as positive partial transpose (PPT) states, and could still be entangled. Separable states form a proper subset of PPT states. Hence, unlike non-zero mutual information, a non-zero value of the logarithmic negativity implies that $\rho$ is entangled.

Another natural way to quantify entanglement is through hypothetical protocols for interconversion between the state $\rho$ and some number of EPR pairs shared between $A_1$ and $A_2$, using only LOCC. This leads to two operationally motivated measures of entanglement, known as the entanglement cost $E_c$ and the distillable entanglement $E_d$ \cite{bennett1996mixed}. 

To define both of the above quantities, suppose we take $n$ copies of the original system, $A_1^{\otimes n} \otimes A_2^{\otimes n}$. We allow only local operations and classical communication between $A_1^{\otimes n}$ and $A_2^{\otimes n}$, and consider conversions between $\rho^{\otimes n}$ and $(\ket{\rm EPR}\bra{\rm EPR})^{\otimes m}$, 
where 
\be 
\ket{\rm EPR} = \frac{1}{\sqrt{2}}(\ket{0}_{x_1}\ket{0}_{x_2}+ \ket{1}_{x_1}\ket{1}_{x_2}), \quad x_1, x_2 \text{ are qubits in $A_1^{\otimes n}, A_2^{\otimes n}$.} 
\ee
First consider the conversion from $(\ket{\rm EPR}\bra{\rm EPR})^{\otimes m}$ to $\rho^{\otimes n}$ under different choices $\sL$ of LOCC operations, with vanishing error in the limit $n\rightarrow \infty$. The entanglement cost $E_c$ is defined as the minimum ratio $\frac{m}{n}$ over all choices of $\sL$ \cite{hayden2001asymptotic}. Next, consider the conversion from $\rho^{\otimes n}$ to $(\ket{\rm EPR}\bra{\rm EPR})^{\otimes m}$ under LOCC operations $\sL$, with vanishing error in the limit $n\rightarrow \infty$. Now the maximum ratio $\frac{m}{n}$ over all choices of $\sL$ is defined as the distillable entanglement $E_d$. 

For a pure state $\rho$, 
$E_c$ and $E_d$ are both equal to the entanglement entropy $S(\rho_{A_1})$ \cite{1996PhRvA..53.2046B}. This is another justification for seeing the entanglement entropy as a natural measure of entanglement for pure states. 
For mixed states, in general 
$E_c \geq E_d$, 
and neither of these quantities must be equal to $S^{(A_1)}(\rho)$ \cite{bennett1996mixed}. In fact, there can be ``bound-entangled'' states, for which $E_c$ is non-zero while $E_d=0$ \cite{horodecki1998mixed}. 

We can also consider replacing the LOCC operations of the above definitions with a larger set of operations called PPT-preserving transformations, which send any state $\sigma$ with $\sigma^{T_2}\geq 0$ to another state $\sigma'$ with $(\sigma')^{T_2}\geq 0$. The entanglement cost and distillable entanglement under such operations, $E_c^{(\rm ppt)}$ and $E_d^{(\rm ppt)}$, are then natural generalizations of the definitions for LOCC given above. 

Since LOCC operations are a proper subset of PPT-preserving operations, it is clear that $E_c^{(\rm ppt)}\leq E_c$, and $E_d^{(\rm ppt)}\geq E_d$. It turns out that 
it is not possible to find states where $E_c^{\rm (ppt)}$ is non-zero while $E_d^{\rm (ppt)}=0$ \cite{2001PhRvL..87y7902E}. However, there are known examples of states for which the preparation by PPT operations is irreversible, i.e.~$E_c^{\rm (ppt)}> E_d^{\rm (ppt)}$ \cite{2017PhRvL.119r0506W}. 

In the discussion below, we will sometimes refer to ``exact'' versions of each of the above entanglement costs and distillable entanglements. For these quantities, we modify the above definitions by requiring that the error in the conversion vanishes before taking the $n\rightarrow \infty$ limit. For example, $E_c^{\rm (exact)}$
refers to the minimum ratio $\frac{m}{n}$ in the exact conversion from $(\ket{\rm EPR}\bra{\rm EPR})^{\otimes m}$ to $\rho^{\otimes n}$ by LOCC in the $n\rightarrow \infty$ limit. $E_d^{\rm (exact)}$, $E_c^{(\rm ppt, exact)}$, and $E_d^{(\rm ppt, exact)}$ are defined similarly. It is then clear, for instance, that $E_c^{\rm (exact)} \geq E_c$, and $E_d^{\rm (ppt, exact)} \leq E_d^{\rm (ppt)}$.

While the entanglement cost and the distillable entanglement for mixed states are natural generalizations of the entanglement entropy for pure states from an operational perspective, these measures are difficult to compute in practice even for few-qubit systems. However, they can be related to computable measures such as logarithmic negativity and mutual information through upper and lower bounds. Two inequalities which will be useful in our subsequent discussion are a relation between the mutual information and the distillable entanglement~\cite{2004JMP....45..829C}, 
\be 
E_d (A_1, A_2) \leq \ha I (A_1, A_2)\, , 
\ee
and a relation between the logarithmic negativity and the exact PPT entanglement cost~\cite{audenaert2003entanglement}, 
\be 
\mathcal{E} (A_1, A_2) \leq E_c^{\rm (ppt, exact)}(A_1, A_2) \leq E_c^{\rm (exact)}(A_1, A_2) \, . 
\ee

\subsection{Renyi entropies and negativities} \label{sec:revneg}
 
Recall that in order to calculate the von Neumann entropy, it is often useful to introduce higher moments of the reduced density matrix, which we will refer to as Renyi partition functions, 
 \be 
 \sZ_{n, A} = \text{Tr}[(\rho_{A})^n].
 \ee
 These partition functions can be used to define the $n$-th Renyi entropies 
 \be 
 S_{n, A}(\rho)= -\frac{1}{n-1}\log \sZ_{n, A}, 
 \ee
 which can in turn be used to define the $n$-th Renyi mutual informations
 \be 
I_n(A_1, A_2) = S_{n, A_1} +S_{n, A_2} - S_{n,A} \, . \label{ren_m}
\ee 
The Renyi entropies and mutual informations provide further information about the entanglement structure in addition to the von Neumann entropy and mutual information; however, note that in some cases the Renyi mutual information can be negative, unlike the mutual information, which is one reason why the Renyi mutual information should only be considered as proxies for correlation measures. The von Neumann entropy can be written as a limit of the Renyi entropies in the index $n$, 
\be 
S_{A} = \lim_{n\rightarrow 1} S_{n, A} \, . 
\ee
In some cases, such as when the $S_{n, A}$ cannot be written as an analytic function of $n$, we may not be able to use analytic continuation to find the von Neumann entropy. Another method for obtaining the von Neumann entropy using the Renyi partition functions that turns out to be more generally applicable is through a quantity known as the resolvent, 
\be 
\sR(\lambda) =\text{Tr}\left(\frac{1}{\lambda \mathbf{1} -\rho_A}\right) = \frac{1}{\lambda}\sum_{n=0}^{\infty} \frac{1}{\lambda^{n}} \sZ_{n,A}, 
\label{Rdef}
\ee 
where we take $\lam$ to be a general complex parameter. Since the spectrum of $\rho_A$ is bounded, the power series in the second equality is convergent for sufficiently large $|\lam|$. We can then compute $\sR(\lam)$ for such $\lam$'s and then analytically continue to obtain other values of $\lam$.
The density of eigenvalues of $\rho_A$, known as the entanglement spectrum, can be obtained from the discontinuities of $\sR(\lam)$ across the real axis, 
\be
D (\lam) = \frac{1}{\pi } {\rm Im} \, \sR(\lambda- i\epsilon), \quad \lam \in \RR \ .
\ee
We can then use $D(\lam)$ to calculate the von Neumann entropy, 
\be 
S_A = -\int d \lam \, D(\lam) \, \lam \log \lam \ .
\label{Sdef}
\ee
 
It is similarly useful to define the higher moments of the partially transposed density matrix $\rho_A^{T_2}$ introduced in \eqref{pt}, which we refer to as partial transpose partition functions, 
\be \label{repar}
\sZ_n^{(\rm PT)} = \Tr \left(\rho_A^{T_{2}}\right)^{n} \ .
\ee
By definition $\sZ_1^{(\rm PT)} =1$, and it can be readily checked that $\sZ_2^{(\rm PT)} = \Tr \rho^2$, so nontrivial moments start with $n=3$. There are qualitative differences between even and odd $n$'s, as 
\be
\sZ_{2m}^{(\rm PT)}=\sum_{\lambda_{i}>0}\left|\lambda_{i}\right|^{2m}+\sum_{\lambda_{i}<0}\left|\lambda_{i}\right|^{2m} , \qquad
\sZ_{2m+1}^{(\rm PT)}=\sum_{\lambda_{i}>0}\left|\lambda_{i}\right|^{2m+1}-\sum_{\lambda_{i}<0}\left|\lambda_{i}\right|^{2m+1} \ .
\ee
The logarithmic negativity $\sE$ can be obtained by analytically continuing $\sZ_{2m}^{(\rm PT)}$ as 
\be
\mathcal{E}=\lim _{m \rightarrow \ha} \log \sZ_{2m}^{(\rm PT)} \ . \label{limit_1}
\ee
It can be shown that a PPT state satisfies~\cite{elben2020mixed} 
\be \label{ehn}
\sZ_{3}^{(\rm PT)} \geq \le(\sZ_{2}^{(\rm PT)}\ri)^2 \ .
\ee
Thus if $\sZ_{3}^{(\rm PT)} < \le(\sZ_{2}^{(\rm PT)}\ri)^2$, then it must be that $\sE > 0$, which provides a quick diagnostic. 
The condition~\eqref{ehn} is weaker than the PPT condition; there can be states that satisfy~\eqref{ehn} but still have 
$\sE > 0$. 

We can also use the partial transpose partition functions to define quantities called the $n$-th Renyi negativities, 
 \be 
\label{rn}
R_n(A_1, A_2) = b_n \log \le[{\sZ_n^{(\rm PT)} \ov \sZ_{n,A}} \ri] , \quad b_n = \bca {1 \ov 1-n} & n \, {\rm odd} \cr {1 \ov 2-n} & n \, {\rm even} \eca \ .
\ee 
$n=1,2$ should be understood as being defined through limits. More explicitly, 
\bega 
R_1 = -\sum_{\lambda_{i}>0}|\lambda_{i}| \log |\lam_i| + \sum_{\lambda_{i}<0}|\lambda_{i}| \log |\lam_i| - S_{A}(\rho) , \\
R_2 ={1 \ov \sZ_2} \le( - \sum_i |\lambda_{i}|^2 \log |\lam_i| + \sum_{i}|\tilde \lambda_{i}|^2 \log \tilde \lam_i \ri),
\end{gather}
where $\tilde \lam_i$ are eigenvalues of $\rho$. 
The logarithmic negativity can also be obtained from $R_{2m}$ by analytic continuation 
\be 
\sE = \lim_{m \to \ha} R_{2m} \ . \label{limit_2}
\ee

For a product state $\rho = \rho_{A_1} \otimes \rho_{A_2}$, we have 
$\sZ_n^{(\rm PT)} = \sZ_n$ and thus $R_n =0$. But for a general separable state, $R_n$ are in general nonzero. 
 Thus from nonzero $R_n$, we cannot conclude for certain that there is nonzero entanglement. 
Even so, there are indications that the Renyi negativities often have the same general behavior as the logarithmic negativity, and these quantities have proven useful in a variety of problems \cite{2014PhRvB..90f4401C,2018PhRvL.121o0503G,2020PhRvL.125n0603W,2020PhRvL.125k6801L,2020PhRvB.102f4304W,2020arXiv200706305E,2020arXiv200811727L,2021arXiv211007384W}.

It is often useful to compare the Renyi negativities with the Renyi mutual informations defined in \eqref{ren_m}. 
In fact, when $\rho$ is a pure state, the Renyi negativities and Renyi mutual informations are related in a simple way. More explicitly, in this case we have 
\begin{align} 
& R_{2m}(A_1, A_2) = S_{m, A_1} = \ha I_{m}(A_1, A_2) , \\
& R_{2m+1}(A_1, A_2) = S_{2m+1, A_1} =\ha I_{2m+1}(A_1, A_2) \ . \label{rjj}
\end{align} 
We also have 
\be \label{ejj}
\sE(A_1, A_2) = 
S_{\ha, A_1} = \ha I_{\ha}(A_1, A_2) \ .
\ee
Equation~\eqref{ejj} gives a rough intuition on the {relative normalization} between the logarithmic negativity and mutual information. 

In some cases, we will find that the $\sZ_n^{\rm (PT)}$ and $R_n$ cannot be written simply as analytic functions of $n$, so that we will not be able to use \eqref{limit_1} or \eqref{limit_2} to find $\sE$. Again, a more generally applicable method is to use the resolvent for $\rho_A^{T_2}$: 
\be 
R_N(\lambda) =\text{Tr}\left(\frac{1}{\lambda \mathbf{1} -\rho_A^{T_2}}\right) = \frac{1}{\lambda}\sum_{n=0}^{\infty} \frac{1}{\lambda^{n}} \sZ_n^{\rm (PT)} \ . 
\label{Rndef}
\ee 
The density of eigenvalues of $\rho_A^{T_2}$, known as the negativity spectrum, can be obtained from the discontinuities of $R_N(\lam)$ across the real axis, 
\be
D_N (\lam) = \frac{1}{\pi } {\rm Im} \, R_N(\lambda- i\epsilon), \quad \lam \in \RR \ , \label{Dndef}
\ee
and we can use $D_N(\lam)$ to calculate the logarithmic negativity, 
\be 
\sE(A_1, A_2) = \log\left(\int d \lam \, D_N(\lam) \, |\lambda|\right) \ .
\label{erdef}
\ee


\subsection{Brief review of the equilibrium approximation} \label{sec:revea}

In this subsection, we review the equilibrium approximation introduced in~\cite{2020arXiv200801089L}, which will be the main tool we use in this paper. We first discuss the formulation for a pure state, and then for a mixed state. 

\subsubsection{Pure states} 

We consider a system evolving from a far-from-equilibrium pure state $\rho_0 = \ket{\Psi_0} \bra{\Psi_0}$ to 
a state $\rho = \ket{\Psi} \bra{\Psi}$ with $\ket{\Psi} = U \ket{\Psi_0}$, which is in equilibrium at a macroscopic level. We assume that macroscopic physical properties of equilibrated pure state $\rho$ can be approximated by an equilibrium density operator 
\begin{align} \label{eqd} 
 \rho^{(\rm eq)} = \frac{\mathcal{I}_{\alpha}}{Z(\alpha)}, \quad Z(\alpha) = \Tr\, \mathcal{I}_{\alpha} ,
\end{align}
where $\al$ collectively denotes macroscopic parameters for the equilibrium state. 

Consider the $n$-th Renyi entropy of the equilibrated pure state with respect to a subsystem $A$ 
\be \label{rnyi}
\sZ_{n,A} = e^{- (n-1) S_{n,A}} = {\rm Tr}_A \rho_A^n = {\rm Tr}_A \le({\rm Tr}_{\bar A} U \rho_0 U^\da \ri)^n 
= \braket{\eta_A \otimes e_{\bar A}| (U \otimes U^{\dagger})^n| \rho_0,e},
\ee
where in the last equality we have written it as an amplitude in the replica space $(\sH \otimes \sH)^n$, with various notations defined as follows. 
For any operator $\sO$ acting on $\sH$, the state $\ket{\sO, \sig} \in (\sH \otimes \sH)^n$, where $\sig$ is an element of the permutation group $\sS_n$ of $n$ objects, is defined as 
\bega \label{aa1}
\vev{i_1 \bar i_1' i_2 \bar i_2' \cdots i_n \bar i_n'| \sO, \sig} = \sO_{i_1 i'_{\sig (1)}} \sO_{i_2 i'_{\sig (2)}} \cdots 
\sO_{i_n i'_{\sig (n)}} , \quad \sO_{ij} = \vev{i |\sO|j} \ .
\end{gather} 
Here $\{\ket{i_1 \bar i_1' i_2 \bar i_2' \cdots i_n \bar i_n'}\}$ is a 
basis for $(\sH \otimes \sH)^{n}$ and $\sig (i)$ denotes the image of $i$ under $\sig$. 
When $\sO$ is given by the identity operator, we will denote the states obtained this way simply as $\ket{\sig}$. 
When the system is divided into subsystems, we can similarly define states by associating different permutations to different subsystems. For example, suppose $\sH = \sH_A \otimes \sH_{\bar A}$, $\ket{\sO, \tau_A \otimes \sig_{\bar A}}$ with $\tau, \sig \in \sS_n$ is defined as 
\bega 
 \vev{i_{1_a} i_{1_b} \bar i_{1_a}' \bar i_{1_b}' \cdots i_{n_a} i_{n_b} \bar i_{n_a}' \bar i_{n_b}' |\sO, \tau_A \otimes \sig_{\bar A}} = 
 \sO_{i_{1_a} i_{1_b}, i_{\tau(1)_a}' i_{\sig(1)_b}'} \cdots \sO_{i_{n_a} i_{n_b}, i_{\tau(n)_a}' i_{\sig(n)_b}'} 
\label{zn_t}
\end{gather}
where $\ket{i_{k_a}}, \ket{\bar i'_{k_a}}$, $\ket{i_{k_b}}, \ket{\bar i'_{k_b}}$ denote respectively basis vectors for subsystem $A$ and $\bar A$ in the $k$-th replica of $\sH \otimes \sH$.
In~\eqref{rnyi}, $\ket{\eta_A \otimes e_{\bar A}}$ is a state associated with the identity operator, with $e$ representing the identity permutation and $\eta$ the cyclic permutation $(n, n-1, \cdots 1)$.

We can decompose the identity on the replica Hilbert space as
\begin{align} \label{idf}
 \mathbbm{1} = P_{\alpha} + Q, \quad P_{\alpha }Q = QP_{\alpha } = 0,\quad Q^2 = Q,
\end{align}
where $P_\al$ is the projector
\begin{align}
 P_{\alpha} = \frac{1}{Z_2^n}\sum_{\sigma,\tau} g^{\sigma \tau}\ket{\mathcal{I}_{\alpha}, \sigma}\bra{\mathcal{I}_{\alpha},\tau}, \quad
 g_{\tau \sigma} = \frac{\bra{\mathcal{I}_{\alpha},\tau}\mathcal{I}_{\alpha},\sigma\rangle}{\sqrt{\bra{\mathcal{I}_{\alpha},\tau}\mathcal{I}_{\alpha},\tau\rangle\bra{\mathcal{I}_{\alpha},\sigma}\mathcal{I}_{\alpha},\sigma\rangle}}, \quad Z_n := {\rm Tr} \sI_\al^n \ . \label{palpha}
\end{align}

We will be interested in systems with a large number of degrees of freedom, i.e.~$Z_1 \gg 1$. 
For such systems, inserting the identity twice in the last expression of~\eqref{rnyi} and neglecting the term involving $Q$, 
we find $\sZ_{n}^{(A)}$ can be approximated as 
\begin{align} 
\sZ_{n, A} & \approx [\sZ_{n, A}]_{\rm eq \; approx} := {1 \ov Z_2^n} \sum_{\sig, \tau} g^{\tau\sig} \vev{\eta_A \otimes e_{\bar A} | \sI_\al , \tau} \vev{\sI_\al, \sig|\rho_0,e} \label{fen_line1} \\
& = {1 \ov Z_1^n} \sum_{\sig, \tau} g^{\tau\sig} \vev{\eta_A \otimes e_{\bar A} | \sI_\al , \tau} \label{fen_line2}\\
& \approx {1 \ov Z_1^n} \sum_{\tau \in \sS_n} \vev{\eta_A \otimes e_{\bar A} | \sI_\al , \tau} \ . \label{fen}
\end{align} 
In going from \eqref{fen_line1} to \eqref{fen_line2}, we use the fact that for~\eqref{fen_line1} to be compatible with $\Tr \rho= 1$, $\sI_\al$ should satisfy 
a consistency requirement
\be \label{conw}
\Tr (\sI_ \al \rho_0 )= {Z_2 \ov Z_1} \ .
\ee
In going from \eqref{fen_line2} to \eqref{fen}, we use the fact that $g^{\tau \sigma}$ is approximately equal to the identity when $Z_1$ is large. 
 
 $\sZ_{n, A}$, as given in~\eqref{fen}, only depends on the equilibrium density operator $\sI_\al$, but satisfies the unitarity constraint 
\be \label{unit1}
 \sZ_{n, A} = \sZ_{n, \bar A} \ .
\ee

 The size of the terms we neglected in reaching~\eqref{fen} 
can be estimated from $\Delta$, defined by 
\be 
\label{Renyi_q} 
\Delta^2 := [(\sZ_{n,A})^2]_{\text{eq approx}}- \left([\sZ_{n, A}]_{\text{eq approx}}\right)^2 \ .
\ee
It was shown in Appendix B of \cite{2020arXiv200801089L} that
\be 
\frac{\Delta}{[\sZ_{n, A}]_{\text{eq approx}}} \sim Z_1^{-1/2} \ll 1\ .
\label{delta}
\ee

\begin{figure}[]
 \centering
 \includegraphics[width=5cm]{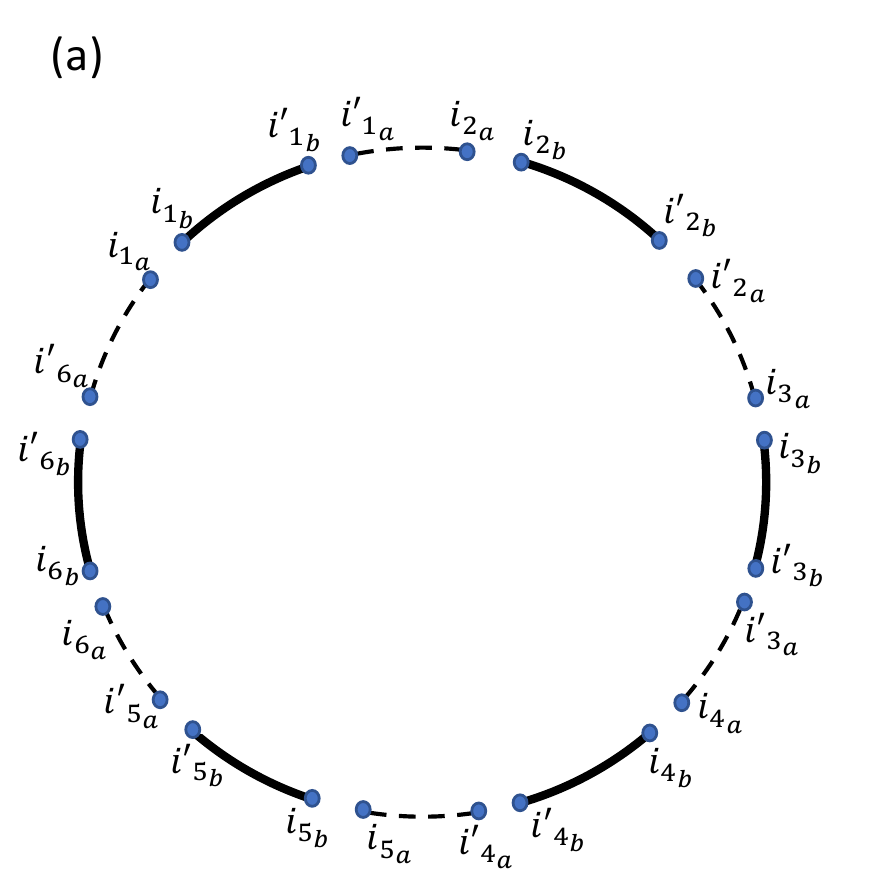} \includegraphics[width=5cm]{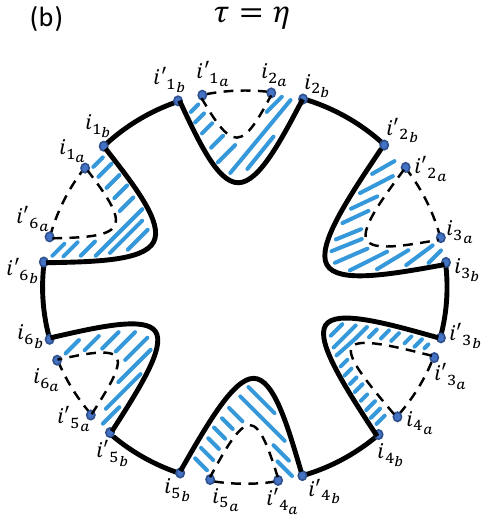}\includegraphics[width=5cm]{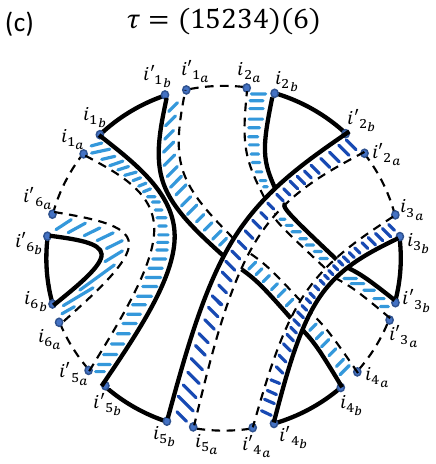}
 \caption{(a) shows the common exterior lines of all diagrams for different terms in \eqref{fen}, and (b) and (c) show examples of diagrams for two choices of $\tau$, for the case $n=6$.}
 \label{fig:Renyi_diagrams}
\end{figure}

Each term in the final expression in \eqref{fen} can be given a diagrammatic representation, as shown in Fig.~\ref{fig:Renyi_diagrams}. We can insert the identity to write 
\be 
\begin{gathered} 
\vev{\eta_A \otimes e_{\bar A} | \sI_\al , \tau} = \sum_{i_1, i'_1, ... i_n, i'_n} \braket{\eta_A \otimes e_{\bar{A}}|i_1 \bar{i'}_1 ... i_n \bar{i'}_n} \braket{i_1 \bar{i'}_1 ... i_n \bar{i'}_n| \sI_{\al}, \tau}, \\
\ket{i_m} = \ket{i_{m_a}}_{A} \ket{i_{m_b}}_{\bar{A}}, \quad \ket{\bar{i'}_m} = \ket{\bar{i'}_{m_a}}_{A} \ket{\bar{i'}_{m_b}}_{\bar{A}} \,.
\end{gathered} 
\ee
The exterior of the diagram, which is the same for all $\tau$, represents $\braket{\eta_A \otimes e_{\bar{A}}|i_1 \bar{i'}_1 ... i_n \bar{i'}_n}$ by connecting $i_{m_a}$ with $i'_{\eta(m)_a}$ using dashed lines, and $i_{m_b}$ with $i'_{m_b}$ using solid lines, as shown in Fig.~\ref{fig:Renyi_diagrams}(a). The interior of the diagram represents 
$\braket{i_1 \bar{i'}_1 ... i_n \bar{i'}_n| \sI_{\al}, \tau}$, by connecting $i_m$ with $i'_{\tau(m)}$, as shown for two examples in Fig.~\ref{fig:Renyi_diagrams}(b) and (c). Roughly, each solid loop in the resulting diagram gives a power of $d_{\bar{A}}$ and each dashed loop gives a power of $d_{A}$, where $d_A$ and $d_{\bar{A}}$ are respectively the effective Hilbert space dimensions of $A$ and $\bar{A}$.\footnote{When $\sI_\al$ can be factorized between $A$ and $\bar A$, i.e.~$\sI_\al = \sI_A \otimes \sI_{\bar A}$, we can define the effective dimensions as $d_A = \Tr_A \sI_A$ and $d_{\bar A} = \Tr_{\bar A} \sI_{\bar A}$. When $\sI_\al$ cannot be factorized, they can be estimated by counting the numbers of degrees of freedom of the subsystems.} The number of solid and dashed loops in a diagram is respectively equal to $k(\tau)$ (the number of cycles in permutation $\tau$) and $k(\eta^{-1}\tau)$. We therefore find 
\be 
\label{sat}
\braket{\eta_A \otimes e_{\bar{A}}| \sI_{\al}, \tau}\sim d_A^{k(\eta^{-1}\tau)} d_{\bar{A}}^{k(\tau)}.
\ee
For any permutation $\tau$, we have the inequality 
\be 
k(\tau)+ k(\eta^{-1}\tau) \leq n+1,
\ee
and the permutations for which this inequality is saturated are the ones associated with planar diagrams such as Fig.~\ref{fig:Renyi_diagrams}(b). 

From~\eqref{sat}, when $A$ is much smaller than $\bar A$, i.e.~$d_A \ll d_{\bar A}$, the permutation $\tau = e$, which maximizes $k (\tau)$, dominates, and we have 
\be
 \mathcal{Z}_{n, A} = \mathcal{Z}_{n, A}^{({\rm eq})},
 \ee
 where $ \mathcal{Z}_{n, A}^{({\rm eq})}$ is the Renyi partition function for $A$ in the state $\rho^{(\rm eq)}$. When $A$ is much larger than $\bar A$, i.e.~$d_A \gg d_{\bar A}$, $\tau = \eta$, which maximizes $k(\eta^{-1} \tau)$, dominates
 \be
 \mathcal{Z}_{n, A} = \mathcal{Z}_{n, \bar A}^{({\rm eq})} \ .
 \ee
 Except for a crossover region around $d_A \sim d_{\bar A}$ where the behavior may be more complicated, we
 then have 
 \
 \be \label{erem}
 S_{n,A} = {\rm min} \le(S_{n,A}^{\rm (eq)}, S_{n, \bar A}^{\rm (eq)} \ri), \qquad n \geq 2 
 \ee
where $S_{n,A}^{(\rm eq)}$ denotes the $n$-th Renyi entropy for subsystem $A$ in the equilibrium density operator $\rho^{(\rm eq)}$. 
 
 When $S_A$ can be obtained from $S_{n, A}$ by analytic continuation to $n=1$, \eqref{erem} implies 
 \be \label{endm}
 S_A = {\rm min} (S_A^{\rm (eq)}, S_{\bar A}^{\rm (eq)}) ,\, 
 \ee
where $S_A^{(\rm eq)}$ is the entanglement entropy for subsystem $A$ in $\rho^{(\rm eq)}$. In cases where the system $A\bar{A}$ is inhomogeneous, in general \eqref{endm} cannot be deduced by analytic continuation. We show in two such examples in Appendix \ref{app:page_finite} that \eqref{endm} still holds, and will assume in the later discussion in this paper that it holds in general.

When $\sI_\al$ can be factorized
\be \label{gdn}
\sI_\al \approx \sI_{A} \otimes \sI_{\bar A} ,
\ee
equations~\eqref{fen} can be written more explicitly in terms of partition functions of $A$ and $\bar A$ 
\be 
Z_{m, A} = {\rm Tr}_A \sI_{A}^{m} , \qquad Z_{m, \bar A }= {\rm Tr}_{\bar A} \sI_{\bar A}^{m} , 
\quad 
\ee 
 as 
\ie \label{fen2}
 \sZ_{n, A} & \approx {1 \ov Z_1^n} \sum_\tau 
 \le( Z_{m_1, A} \cdots Z_{m_l, A} \ri) \le( Z_{n_1, \bar A} \cdots Z_{n_k, \bar A} \ri) \ ,
\fe
where $k$ is the number of cycles of $\tau$ with $n_1, \cdots n_k$ the lengths of the corresponding cycles,
and $l$ is the number of cycles of $\tau \eta^{-1}$ with $m_1, \cdots m_l$ the lengths of the corresponding cycles.

\subsubsection{Mixed states} \label{sec:mixed}

The equilibrium approximation can also be applied to a system $A = A_1 \cup A_2$ in a mixed state $\rho_A$ that is in macroscopic equilibrium, but is far from the thermal density operators~\cite{2020arXiv200801089L}. 


Suppose the system starts with a far-from equilibrium mixed state $\rho_{0, A}$, which evolves under unitary evolution operator $U_A$ to $\rho_A = U_A \rho_{0, A} U^\da_A $. 
All the moments of $\rho_{0,A}$ are also preserved by the time evolution 
\be \label{ienm}
z_{n, A} = \Tr \rho_{0,A}^n = \Tr \rho^n_A = \Tr (U_A \rho_{0,A} U_A^\da)^n = \vev{\eta|(U_A \otimes U^{\da}_A)^n| \rho_{0,A},e}, \quad n=2, \cdots \ .
\ee
The statement that $\rho_{0, A} $ is far-from-equilibrium is imposed by requiring that the $n-$th Renyi entropy of $\rho_{0, A}$ is smaller than the equilibrium entropy of $A$. For example, this condition is satisfied if we have
\be 
z_{n,A} \sim Z_A^{-(n-1)f}\, , \quad 0 \leq f < 1 \, . \label{out_eq}
\ee
Assuming that $\rho_A$ can be approximated by an equilibrium density operator $\rho^{(\rm eq)}_A = {1 \ov Z_A} \sI_A $ 
and applying the equilibrium approximation by inserting the projector \eqref{palpha} in \eqref{ienm} and ignoring terms with $\sigma \neq \tau$, we have 
\be 
z_{n,A} \approx {1 \ov Z_{2,A}^n} \sum_\tau \vev{\eta|\sI_A , \tau} \vev{\sI_A, \tau| \rho_{0,A}, e} , 
\ee 
which can be further simplified to the following constraints on $\sI_A$ under the out-of-equilibrium assumption \eqref{out_eq}
\be \label{elk}
\Tr (\sI_A \rho_{0,A})^n \approx z_{n,A} {Z_{2,A}^n \ov Z_A^n} , \quad Z_{n, A} = \Tr_A \sI_A^n , \quad Z_A = Z_{1,A} \ .
\ee
The Renyi partition functions for $A_1$ can then be approximated as 
 \be\label{for11}
\sZ_{n, A_1} \approx [\sZ_{n, A_1} ]_{\rm eq \, approx} 
= {1 \ov Z_A^n} \sum_{\tau} \vev{\eta_{A_1} \otimes e_{A_2} | \sI_A , \tau} \prod_{i=1}^{k (\tau)} \Tr_A \rho_{0,A}^{n_i} 
, 
\quad n=2,3, \cdots 
\ee
where $k (\tau)$ is the number of cycles of $\tau$ with $n_1, \cdots n_{k (\tau)}$ the lengths of the cycles. 

The above discussion can be further generalized by embedding $A$ in a larger 
system $S = A \cup B$, with the total system $S$ in an initial pure state $\ket{\Psi_0}$ evolved to $\ket{\Psi}
= U \ket{\Psi_0}$ in macroscopic equilibrium\footnote{$A$ and $B$ in principle do not have to be in equilibrium with each other.}. 
Suppose $\ket{\Psi}$ can be approximated macroscopically by $\rho^{(\rm eq)} = {1 \ov Z_1} \sI$. 
We then have 
\be \label{manw}
\sZ_{n, A_1} \approx {1 \ov Z_1^n} \sum_{\tau} \vev{\eta_{A_1} \otimes e_{A_2 B} | \sI , \tau} \ .
\ee

This generalizes~\eqref{for11} as under the evolution of $U$ for the full system $S$, the evolution from the initial density operator $\rho_{0,A}$ to $\rho_A$ is in general not unitary. To recover~\eqref{for11} we take 
$ U = U_A \otimes U_B$ to be factorized between $A$ and $B$, in which case 
the equilibrium density operator $\sI = \sI_A \otimes \sI_B$ should also factorize, and~\eqref{manw} can be written as 
\be \label{for12} 
\sZ_{n, A_1} = {1 \ov Z_A^n} \sum_{\tau} \vev{\eta_{A_1} \otimes e_{A_2} | \sI_A , \tau} \prod_{i=1}^{k (\tau)} \hat Z_{n_i, B} , \quad 
\hat Z_{n, B} := {1 \ov Z_B^n} {\rm \Tr}_B \sI_B^n \ .
\ee
Equations~\eqref{for11} and~\eqref{for12} are equal provided that we choose the initial state $\rho_{0,B}$ such that 
\be \label{hhe1}
 {\rm Tr}_B \rho_{0, B}^m = {1 \ov Z_B^m} {\rm Tr}_B \sI_B^m \, . 
 \ee
Then since the initial state is pure, $z_{n, A}$ is also given by \eqref{hhe1}. The requirement that $\rho_{0, A}$ is out-of-equilibrium is then equivalent to the requirement that $Z_{B} \gg Z_{A}$.

The relation between~\eqref{manw} and~\eqref{for11} also gives a way to estimate which permutation dominates in~\eqref{for11}. 
From~\eqref{manw}, when $A_1$ is smaller (larger) than $A_2 B$, the dominant contribution is $\tau = e$ ($\tau = \eta$). Translating these statements to the notation of~\eqref{for11}, we conclude that 
\be \label{ejjj}
S_{n, A_1} = \bca S_{n, A_1}^{({\rm eq})} & S_{n, A_1}^{(\rm eq)} < S_{n, A_2}^{(\rm eq)} + S_{n,A} \cr
 S_{n, A_2}^{({\rm eq})} + S_{n,A} & S_{n, A_1}^{(\rm eq)} > S_{n, A_2}^{(\rm eq)} + S_{n,A} 
 \eca \ .
\ee 

In applying~\eqref{manw} to explicit calculations, we will need to make assumptions regarding $B$ in the equilibrium density operator $\rho^{(\rm eq)}$ for the full system, which may be considered as specifying different universality classes for $\rho_A$.

\subsection{The Petz recovery map}\label{sec:petz}

An important question concerning a quantum channel $\sN$ is whether or not it is reversible i.e.~whether there exists a recovery channel $\sN$ such that for any state $\rho$, $\sR \circ \sN (\rho) = \rho$. 
This question plays a central role, for example, in the theory of quantum error correction as well as quantum thermalization.
If the quantum channel is unitary, the initial state is perfectly recovered by acting on the output with the adjoint of the unitary. In the other extreme, a quantum channel could replace all states with the maximally mixed state, in which case an initial state is unrecoverable as the information about it is completely lost. A recovery map that is independent of the initial state $\rho$ is called \textit{universal}. 

It follows from a theorem by Petz~\cite{cmp/1104115260,10.1093/qmath/39.1.97,2003RvMaP..15...79P} that a quantum channel $\sN$ is reversible if and only if 
the data processing inequality is saturated
\begin{align}
 D(\rho\lvert \rvert \sigma) = D(\mathcal{N}(\rho)\lvert \rvert \mathcal{N}(\sigma))
\end{align}
for all $\rho, \sig$. Here, $ D(\rho\lvert \rvert \sigma) $ is the relative entropy. 
Furthermore, suppose we fix some reference state $\sigma$. Then, for any state $X$ in the support of $\sN(\sigma)$, there exists a recovery channel $\sR$. This channel is given explicitly by the Petz map, 
\begin{align}
 \mathcal{P}_{\sigma,\mathcal{N}}(X) = \sigma^{\frac{1}{2}} \mathcal{N}^{\dagger}\left((\mathcal{N}(\sigma))^{-\frac{1}{2}} X (\mathcal{N}(\sigma))^{-\frac{1}{2}}\right)\sigma^{\frac{1}{2}},
 \label{petz}
\end{align}
where $\mathcal{N}^{\dagger}$ is the adjoint map of $\sN$. As a basic check, relative entropy is invariant under unitary channels $\rho \to U \rho U^\da$, and we find from~\eqref{petz}, $ \mathcal{P}_{\sigma,\mathcal{N}}(X) = U^\da X U$.

Interestingly, it has recently been understood that the change of relative entropy under quantum channels places strict bounds on how well a state can be recovered. The basic idea is intuitive; if two states that were initially easily distinguishable become nearly indistinguishable under a channel, then it should be impossible to identify what the initial states were using only information about the indistinguishable output states. In particular, it can be shown~\cite{2015arXiv150907127J} that there exists a recovery map $\sR_{\sig, \sN}$ with $\mathcal{R}_{\sigma,\mathcal{N}}\circ \mathcal{N} (\sigma) = \sigma$ satisfying\footnote{This was proven for Type I von Neumann algebras but was recently generalized to the Type III algebras relevant to quantum field theory that we will model \cite{2020arXiv200608002F}. This technicality will not play an important role for us.}, 
\begin{align}
 F(\rho, [\mathcal{R}_{\sigma,\mathcal{N}}\circ \mathcal{N}](\rho)) \geq \exp (D(\mathcal{N}(\rho) || 
 \mathcal{N}(\sigma) ) - D(\rho|| \sigma) ),
 \label{petz_fidelity}
\end{align}
where $F$ is the fidelity, defined as 
\begin{align}
 F(\rho, \sig) := \left(\Tr \left[ \sqrt{\sqrt{\rho} \sigma \sqrt{\rho}}\right]\right)^2 \ .
\end{align}
For example, the bound \eqref{petz_fidelity} holds for an explicit but complicated recovery channel called the twirled Petz map \cite{2015arXiv150907127J}.

For the quantum channel that we are interested in this paper, consider a system $D \cup B = R \cup B'$ in the initial state $\rho_D \otimes \rho_B$, evolve it for a while, and then trace out a portion $B'$ of the full system. If we take $\rho_B$ to be some fixed state, this gives a quantum channel from $D$ to $R$,
\begin{align}
 \mathcal{N}(\rho_D)= \Tr_{B'}\left[ U\left(\rho_D \otimes \rho_B \right)U^{\dagger} \right] \ . 
\end{align}
The corresponding $\sN^\da$ from $R$ to $D$ is then given by 
\begin{align}
 \mathcal{N}^{\dagger}(\phi_R) = \Tr_B \le(\rho_B U^{\dagger}\left(\phi_R \otimes \mathbbm{1}_{B'}\right)U \ri) \ .
\end{align}
Applying~\eqref{petz} to this case and imposing a replica trick, we have 
\bega \label{ren1}
 \mathcal{P}_{\sigma_D, \mathcal{N}}\circ \mathcal{N} (\rho_D)
 = \lim_{m \to -\ha} \sig_D^\ha \Tr_B \le[\rho_B U^{\dagger} \le(\sN (\sig_D)^m \mathcal{N}(\rho_D) \sN (\sig_D)^m \otimes \mathbbm{1}_{B'} \ri) U \ri] \sig_D^\ha \ .
 \end{gather} 
 In some of our discussion below, an alternative analytic continuation will also be useful: 
 \bega \label{ren_alt}
 \mathcal{P}_{\sigma_D, \mathcal{N}}\circ \mathcal{N} (\rho_D)
 = \lim_{\substack{n_1 \to -\ha, \\ \, n_2 \to -\ha}} \sig_D^\ha \Tr_B \le[\rho_B U^{\dagger} \le(\sN (\sig_D)^{n_1} \mathcal{N}(\rho_D) \sN (\sig_D)^{n_2} \otimes \mathbbm{1}_{B'} \ri) U \ri] \sig_D^\ha \ .
 \end{gather} 
When we take $B'$ to be empty, then $\sN$ is unitary, and $\sN^\da$ is the inverse evolution, for which the above equation gives $\rho_D$. In the opposite limit, with $B'$ being the full system, we have $\mathcal{P}_{\sigma_D, \mathcal{N}}\circ \mathcal{N} (\rho_D) = \sig_D$ for any $\rho_D$. 

In the case where $\rho_D$ is a pure state, the fidelity of the Petz map can be written as an overlap 
\bega \label{ren2}
 F(\rho_D, \mathcal{P}_{\sigma_D, \mathcal{N}}\circ \mathcal{N}(\rho_D)) = \Tr_D \le( \rho_D \mathcal{P}_{\sigma_D, \mathcal{N}}\circ \mathcal{N}(\rho_D)\ri) \cr
 = \lim_{m \to -\ha} \Tr \le[ U \le(\sig_D^\ha \rho_D \sig_D^\ha \otimes \rho_B \ri) U^\da \sN (\sig_D)^m \mathcal{N}(\rho_D) \sN (\sig_D)^m \ri] \ .
 \end{gather} 
 or a similar overlap using \eqref{ren_alt}. 
\eqref{ren1},~\eqref{ren_alt}, and~\eqref{ren2} can all be evaluated by using the equilibrium approximation. 
If the Petz map works perfectly, without error, the fidelity will be one. At worst, the Petz map should output a random answer in which case the fidelity would be exponentially small in the entropy.

\section{Equilibrium approximation for quantum-informational quantities} 
\label{eq_approx_sec}

In addition to the Renyi and entanglement entropies, the equilibrium approximation can in principle be used to calculate any quantum informational quantities which can be defined using replicas. In this section, we consider a few quantities that will be used in the later discussion of the paper: Renyi and logarithmic negativities, relative entropy, reflected entropy, and the Petz recovery map. 

\subsection{Renyi and logarithmic negativities} \label{sec:negas}

We now describe the calculation of Renyi negativities between $A_1$ and $A_2$ by adding an auxiliary system $B$ to $A$, and applying the equilibrium approximation to an equilibrated pure state $\ket{\Psi}$ describing the full system $S = A \cup B$. 
There is also an alternative way to to calculate the negativities between $A_1$ and $A_2$ by directly applying the equilibrium approximation to the mixed state $\rho_A$, analogous to the discussion for the Renyi entropies in Section \ref{sec:mixed}.


Suppose the system is partitioned to $A \cup B$ with $\sH = \sH_A \otimes \sH_B$ and 
the subsystem $A$ is in turn partitioned to $A = A_1 \cup A_2$ with $\sH_A = \sH_1 \otimes \sH_2$. 
We are interested in
\begin{align}\label{repa}
 \mathcal{Z}_n^{(\rm PT)} := \Tr \left( \rho_{A}^{T_2} \right)^n = \Tr_{A} \left( \Tr_{B}\left( U\rho_0U^{\dagger}\right)^{T_2}\right)^n
\end{align}
and the corresponding logarithmic negativity $\mathcal{E}(A_1, A_2) = \lim_{n \to \ha} \log \mathcal{Z}_{2n}^{(\rm PT)}$. 

Using the replica space $(\sH \otimes \sH)^n$ we can write~\eqref{repa} as 
\begin{align} \label{repa1}
 \mathcal{Z}_n^{(\rm PT)} =\bra{\eta_{A_1} \otimes \eta^{-1}_{A_2} \otimes e_{B}}\left( U\otimes U^\da \right)^n\ket{\rho_0, e},
\end{align}
where $\eta = (n, n-1, \cdots 1)$ is the cyclic permutation and its inverse $\eta^{-1}$ is the anti-cyclic permutation.

We can insert the identity~\eqref{idf} twice in~\eqref{repa1} to arrive at
\begin{align}
\label{PQ}
 \mathcal{Z}_n^{\rm (PT)} = \bra{\eta_{A_1} \otimes \eta^{-1}_{A_2} \otimes e_{B}}P_{\alpha}\ket{\rho_0, e} + \bra{\eta_{A_1} \otimes \eta^{-1}_{A_2} \otimes e_{B}}Q\left( U\otimes U^\da \right)^nQ\ket{\rho_0, e} \ .
\end{align}
In the equilibrium approximation, we drop the second term in the above equation, which leads to a time-independent expression, 
\begin{align}
 \mathcal{Z}_n^{(\rm PT)} &\approx [\mathcal{Z}_{n}^{(\rm PT)}]_{\text{eq approx}} := \frac{1}{Z_2^n} \sum_{\sigma, \tau} g^{\tau \sigma}\bra{\eta_{A_1} \otimes \eta^{-1}_{A_2} \otimes e_{B}}\mathcal{I}_{\alpha},\tau\rangle\bra{\mathcal{I}_{\alpha},\sigma}\rho_0,e\rangle
 \nonumber
 \\
 &=\frac{1}{Z_1^n} \sum_{\sigma, \tau} g^{\tau \sigma}\bra{\eta_{A_1} \otimes \eta^{-1}_{A_2} \otimes e_{B}}\mathcal{I}_{\alpha},\tau\rangle
 \nonumber
 \\
 &\approx\frac{1}{Z_1^n} \sum_{ \tau}\bra{\eta_{A_1} \otimes \eta^{-1}_{A_2} \otimes e_{B}}\mathcal{I}_{\alpha},\tau\rangle \ .
 \label{mastereq}
\end{align}
In the second line above we used~\eqref{conw} and in the third line, we again used the fact that in the large $Z_1$ limit $g_{\sig \tau}$ can be approximated by the identity matrix. 
This final expression is independent of the initial state and only depends on the equilibrium density operator. 

\begin{figure}[]
 \centering
 \includegraphics[width=10cm]{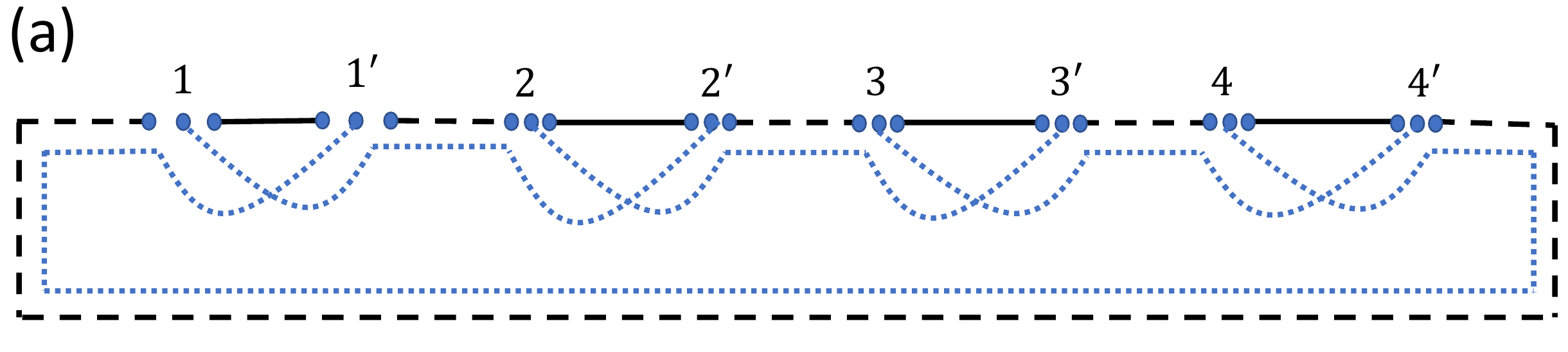}
 \includegraphics[width=10cm]{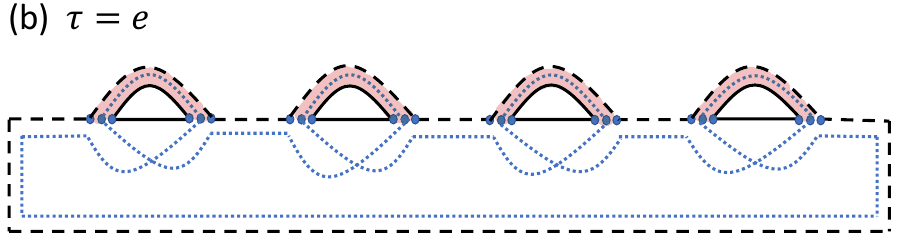}
 
 \vspace{0.5cm}
 
 \includegraphics[width=10cm]{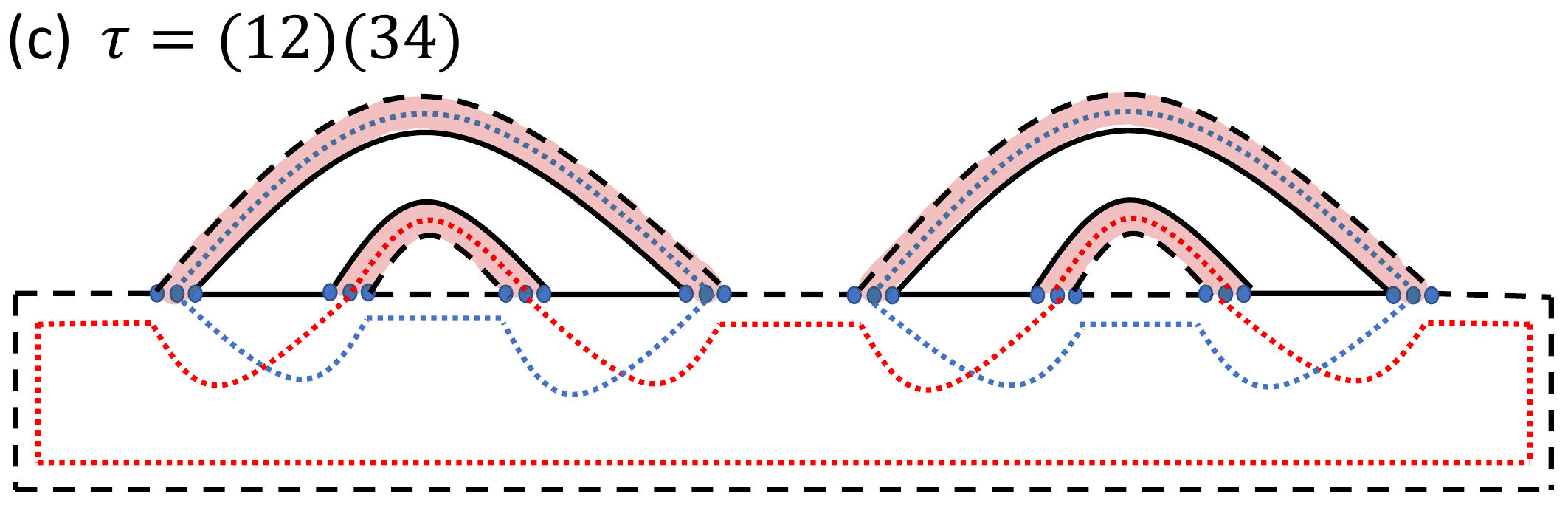}
 \caption{(a) shows the common lines of all diagrams for different terms in \eqref{mastereq}, and (b) and (c) show examples for two choices of $\tau$, for the case $n=4$. The dotted loops are shown in two different colors in (c) to make the distinction between different loops clear.}
 \label{fig:negativity_figs}
\end{figure}

Each term in the final expression in \eqref{mastereq} can be given a diagrammatic representation, as shown in Fig.~\ref{fig:negativity_figs}. We can insert the identity to write 
\be 
\label{negativity_diag}
\begin{gathered} 
\vev{\eta_{A_1} \otimes \eta^{-1}_{A_2} \otimes e_{B} | \sI_\al , \tau} = \sum_{i_1, i'_1, ... i_n, i'_n} \braket{\eta_{A_1} \otimes \eta^{-1}_{A_2} \otimes e_{B}|i_1 \bar{i'}_1 ... i_n \bar{i'}_n} \braket{i_1 \bar{i'}_1 ... i_n \bar{i'}_n| \sI_{\al}, \tau} \, , \\
\ket{i_m} = \ket{i_{m_a}}_{A_1} \ket{i_{m_{\bar{a}}}}_{A_2} \ket{i_{m_b}}_B, \quad \ket{\bar{i'}_m} = \ket{\bar{i'}_{m_a}}_{A_1} \ket{\bar{i'}_{m_{\bar{a}}}}_{A_2} \ket{\bar{i'}_{m_b}}_B \,.
\end{gathered} 
\ee
 The lower half of each diagram represents $\braket{\eta_{A_1} \otimes \eta^{-1}_{A_2} \otimes e_{B}|i_1 \bar{i'}_1 ... i_n \bar{i'}_n}$ by connecting $i_{m_a}$ with $i'_{\eta(m)_a}$ using dashed lines, $i_{m_{\bar{a}}}$ with $i'_{\eta^{-1}(m)_{\bar{a}}}$ using dotted lines, and $i_{m_b}$ with $i'_{m_b}$ using solid lines, as shown in Fig.~\ref{fig:negativity_figs}(a). The upper half of the diagram represents 
$\braket{i_1 \bar{i'}_1 ... i_n \bar{i'}_n| \sI_{\al}, \tau}$, by connecting $i_m$ with $i'_{\tau(m)}$, as shown for two examples in Fig.~\ref{fig:negativity_figs}(b) and (c). In the resulting diagram, roughly each solid loop gives a power of $d_{B}$, each dashed loop gives a power of $d_{A_1}$, and each dotted loop gives a power of $d_{A_2}$, where $d_P$ is the effective Hilbert space dimension of subsystem $P$. The number of solid, dashed, and dotted loops in a diagram is respectively equal to $k(\tau)$, $k(\eta^{-1}\tau)$, and $k(\eta \tau)$. 
 We therefore find 
\be 
\braket{\eta_{A_1}\otimes \eta^{-1}_{A_2}\otimes e_{B}| \sI_{\al}, \tau}\sim d_{A_1}^{k(\eta^{-1}\tau)}d_{A_2}^{k(\eta\tau)} d_{B}^{k(\tau)}.
\label{sat2}
\ee
The sum of these powers satisfies the inequality 
\be 
k(\eta^{-1}\tau)+ k(\eta\tau)+k(\tau) \leq \begin{cases} 3n/2+2 & n\text{ even}\\
(3n+3)/2 & n\text{ odd}
\end{cases} \ .
\ee
The permutations for which this inequality is saturated are described in Appendix \ref{app:perm}. 

We note 
\be \label{hty}
\bra{\eta_{A_1} \otimes \eta^{-1}_{A_2} \otimes e_{B}}\mathcal{I}_{\alpha},\tau\rangle
= \bra{(\mu_1 \eta)_{A_1} (\mu_2 \eta^{-1})_{A_2} (\mu_3)_{B}}\mathcal{I}_{\alpha},(\mu_1\tau)_{A_1} 
(\mu_2\tau)_{A_2} (\mu_3\tau)_{B} \rangle
\ee
for arbitrary $\mu_{1,2,3} \in \sS_n$, and 
\be \label{hty1}
\bra{\eta_{A_1} \otimes \eta^{-1}_{A_2} \otimes e_{B}}\mathcal{I}_{\alpha},\tau\rangle =
 \vev{(\eta \mu)_{A_1} (\eta^{-1} \mu)_{A_2} \mu_{B} | \sI_\al , (\tau \mu)_{A_1} (\tau \mu)_{A_2} (\tau \mu)_{B} } , \quad \mu \in \sS_n \ .
\ee

The expression~\eqref{mastereq} passes various consistency checks. Firstly, we have 
\be \label{syu}
 \mathcal{Z}_n^{(\rm PT)} (A_1, A_2) = \mathcal{Z}_n^{(\rm PT)} (A_2, A_1) ,
\ee
which can be shown as follows. Since $\eta$ and $\eta^{-1}$ have the same cycle structure, there exists some element $\sig \in S_n$ such that $ \eta^{-1} = \sig \eta \sig^{-1}$ and $\eta = \sig \eta^{-1} \sig^{-1}$. We then have from~\eqref{hty}--\eqref{hty1} 
\be 
\bra{\eta_{A_1} \otimes \eta^{-1}_{A_2} \otimes e_{B}}\mathcal{I}_{\alpha},\tau\rangle =
\bra{\eta^{-1}_{A_1} \otimes \eta_{A_2} \otimes e_{B}}\mathcal{I}_{\alpha}, \sig^{-1} \tau \sig \rangle 
\ee
and thus~\eqref{syu} results upon summing over $\tau$. 

Next, consider $n=2$, where $\mathcal{Z}_2^{(\rm PT)}$ should give the same answer as $\mathcal{Z}^{(A)}_2$, the second moment of the density matrix that has not been partially transposed. For $n = 2$, $\eta = \eta^{-1}$ because both are swap operators, so this check immediately passes.

Finally, consider the case where $B$ is not present, i.e.~$\rho_A$ is a pure state
\begin{align}
 \mathcal{Z}_n^{\rm (PT)}\simeq\frac{1}{Z_1^n} \sum_{ \tau
 \in \sS_n}\bra{\eta_{A_1} \otimes \eta^{-1}_{A_2}}\mathcal{I}_{\alpha},\tau\rangle \ .
\end{align}
From~\eqref{hty1} with $\mu = \eta$ we have 
\begin{align}
 \mathcal{Z}_n^{\rm (PT)}\simeq\frac{1}{Z_1^n} \sum_{ \tau
 \in \sS_n}\bra{\eta^2_{A_1} \otimes e_{A_2}}\mathcal{I}_{\alpha},\tau\rangle.
\end{align}
For even $n = 2m$, we have $\eta^2 = (2m-1, 2m-3, \cdots , 3 ,1) (2m, 2m-2, \cdots 4, 2)$. We can separate the sum over $\tau \in \sS_{2m}$ into that over the elements of the subgroup $\sS_m \times \sS_m$ which permutes separately odd and even numbers, and the rest. We then find 
\begin{align}
 \mathcal{Z}_{2m}^{\rm (PT)}\simeq \left(\mathcal{Z}_{m, A_1}\right)^2 + \sum_{ \tau
 \in \sS_{2m} - (\sS_{m} \times \sS_{m})}\bra{\eta^2_{A_1} \otimes e_{A_2}}\mathcal{I}_{\alpha},\tau\rangle
\end{align}
In the replica limit, the first term gives the $n=1/2$ Renyi entropy as expected for the pure state limit of negativity. The second term is
suppressed by additional factors of $Z_1$ as they correspond to non-planar diagrams.

{The size of the} second term in \eqref{PQ}, which we ignored in \eqref{mastereq}, can be estimated as $\Delta_N$, where 
\be 
\label{neg_q} 
\Delta_N^2 := [(\sZ_n^{\text{(PT)}})^2]_{\text{eq approx}}- \left([\sZ_n^{\text{(PT)}}]_{\text{eq approx}}\right)^2 
\ee
In Appendix \ref{app:fluctuations}, we show that 
\be 
\frac{\Delta_N}{[\sZ_n^{\text{(PT)}}]_{\text{eq approx}}} \sim Z_1^{-1/2}. 
\label{Delta}
\ee
Hence, the fluctuations around the equilibrium value of $\sZ_n^{\text{(PT)}}$ are suppressed. 

We can obtain the equilibrium approximation for the logarithmic negativity using analytic continuation in cases where $\sZ_n^{\rm (PT)}$ can be written as an analytic function of $n$, or, more reliably, by using the resolvent \eqref{Rndef}.

\subsection{Relative entropy}

Consider a system evolved from two possible initial states specified respectively by density operators $\rho_0, \sig_0$. We will assume that the support of $\rho_0$ lies inside that of $\sig_0$ so that the relative entropy 
\be 
D (\rho_0||\sig_0) = \Tr \rho_0 \log \rho_0 - \Tr \rho_0 \log \sig_0 
\ee
is finite. 
Suppose $\rho = U \rho_0 U^\da$ and $\sig= U \sig_0 U^\da$ can be approximated at the macroscopic level by the same equilibrium density operator $\rho^{(\rm eq)}$. We are interested in calculating the relative entropy 
\bea 
D (\rho_{A} || \sigma_{A}) &= & \Tr_{A} \rho_{A} \log \rho_{A} - \Tr_{A} \rho_{A} \log \sig_{A} \\
& = & \lim_{n\rightarrow 1} \frac{1}{n-1} \left(\log \Tr_{A} \left[\rho_{A}^n\right]- \log \Tr_{A} \left[\rho_{A} \sigma_{A}^{n-1}\right] \right)
 \label{relative_replica}
\eea
between the reduced density operators $\rho_{A}, \sig_{A}$ of some subsystem $A$. 
The calculation of the first term was already reviewed in Sec.~\ref{sec:revea}, see e.g~\eqref{for11}. Here we discuss how to use the equilibrium approximation to compute the second term, which depends on two distinct density matrices.

As before, we rewrite the second term in~\eqref{relative_replica} as a transition amplitude in the replica Hilbert space
\begin{align}
 \mathcal{D}_{n,A} = \Tr_{A} \left[\rho_{A} \sigma_{A}^{n-1}\right] = 
 \bra{\eta_{A} \otimes e_{\bar A}} (U\otimes U^{\dagger})^{\otimes n} \ket{\rho_0 \otimes \sigma_0^{\otimes (n-1)}, e} ,
\end{align}
where the state $\ket{\rho \otimes \sig^{\otimes (n-1)}, \tau}$ for a permutation $\tau$ is defined as 
\be 
\vev{i_1 \bar i_1' i_2 \bar i_2' \cdots i_n \bar i_n'|\rho \otimes \sig^{\otimes (n-1)}, \tau}
= \rho_{i_1 i'_{\tau (1)}} \sig_{i_2 i'_{\tau (2)}} \cdots 
\sig_{i_n i'_{\tau (n)}} 
\ee
which is inhomogeneous in the replicas. 
Applying the equilibrium approximation, we find 
\bea 
 \sD_{n,A} & \approx &
 \frac{1}{Z_2^n}\sum_{\tau \in \mathcal{S}_n} \bra{\eta_{A} \otimes e_{\bar A}}\mathcal{I}_\al, \tau \rangle \langle \mathcal{I}_\al, \tau \ket{\rho_0 \otimes \sigma_0^{\otimes (n-1)}, e} \\
 & =& \frac{1}{Z_2^n}\sum_{\tau \in \mathcal{S}_n} \bra{\eta_{A} \otimes e_{\bar A}}\mathcal{I}_\al, \tau \rangle \nonumber \\
 && \quad \times \Tr\left[\mathcal{I} \rho_0 \left( \mathcal{I}_\al \sigma_0 \right)^{m_1-1}\right]\Tr\left[\left( \mathcal{I}_\al \sigma_0 \right)^{m_2}\right]
 \cdots \Tr\left[\left( \mathcal{I} \sigma_0 \right)^{m_{k (\tau)}}\right] \\
 & = & {1 \ov Z_1^n} \sum_{\tau \in \mathcal{S}_n} \bra{\eta_{A} \otimes e_{\bar A}}\mathcal{I}_\al, \tau \rangle 
 {\Tr (\rho_0 \sig_0^{m_1-1}) \ov \Tr (\sig_0^{m_1})} \prod_{i=1}^{k (\tau)} \Tr (\sig_0^{m_i} )
 \label{relative_ent_perm_sum}
 \eea
where $m_i$ is the number of elements in $i$-th cycle of $\tau$, the $i=1$ cycle is taken to be the one containing the first copy of the Hilbert space, and we have used self-consistency conditions (derived similarly as~\eqref{elk})
\be 
 \Tr\left[\mathcal{I} \rho_0 \left( \mathcal{I}\sigma_0 \right)^{n-1}\right] \approx {Z_2^n \ov Z_1^n} \Tr (\rho_0 \sig_0^{n-1}), \quad
 \Tr\left[ \le(\mathcal{I}\sigma_0 \right)^{n}\right] \approx {Z_2^n \ov Z_1^n} \Tr (\sig_0^{n}) \ .
 \ee

We now make some general comments on the structure of~\eqref{relative_ent_perm_sum}. 
In~\eqref{relative_ent_perm_sum}, we can divide $\tau$'s into those with $m_1 =1$, and those with $m_1 > 1$. Denoting the two sets respectively as $\sS_{n,1}$ and $\sS_{n,2}$, we have 
\be \label{eons}
\sD_{n,A} = 
 \sum_{\tau \in \mathcal{S}_{n,1}} \sZ_{n, A} (\tau) + \sum_{\tau \in \sS_{n,2}} {\Tr (\rho_0 \sig_0^{m_1-1}) \ov \Tr (\sig_0^{m_1})} \sZ_{n, A} (\tau)
 \ee
where 
\be 
\sZ_{n, A} (\tau)= {1 \ov Z_1^n} \bra{\eta_{A} \otimes e_{\bar A}}\mathcal{I}_\al, \tau \rangle 
 \prod_{i=1}^{k (\tau)} \Tr (\sig_0^{m_i} )
\ee
is the contribution of $\tau$ to the Renyi partition function for $A$ with initial state $\sig_0$. 

From the discussion around~\eqref{ejjj}, we then conclude that when $A$ is small $\tau =e$ dominates\footnote{Note $\Tr (\rho_0 \sig_0^{m-1}) < \Tr (\sig_0^m)$.}, giving
\be 
\sD_{n,A} = \mathcal{Z}_{n, A}^{({\rm eq})}, \quad \text{for} \; \;
 S_{n, A}^{(\rm eq)} \ll S_{n, \bar A}^{(\rm eq)} + S_n (\sig_0) \ .
\ee
For $A$ to be sufficiently large we expect the first term is dominated by $\eta_{n-1}$ which cyclicly permutes $2, \cdots n$ 
and gives a contribution $ \sZ_{n, A} (\eta_{n-1})$, while the dominant permutation for the second term is $\eta$, giving a contribution 
$ \Tr (\rho_0 \sig_0^{n-1}) \mathcal{Z}_{n, \bar A}^{({\rm eq})}$. From our general discussion $ \sZ_{n, A} (\eta_{n-1})$ is smaller than $\mathcal{Z}_{n, \bar A}^{({\rm eq})}$ by at least a factor $Z_1^{-1}$. Thus for the case $\Tr (\rho_0 \sig_0^{n-1}) $ is not too small (i.e.~much larger than $Z_1^{-1}$) we then have 
\be 
\sD_{n,A} 
\approx 
 \Tr (\rho_0 \sig_0^{n-1}) \mathcal{Z}_{n, \bar A}^{({\rm eq})}
, \quad \text{for} \; \;
 S_{n, A}^{(\rm eq)} \gg S_{n, \bar A}^{(\rm eq)} + S_n (\sig_0) \ .
\ee

Now combining the above discussion with~\eqref{ejjj}, and assuming that we can analytically continue to $n=1$, we find~\eqref{relative_replica} can be written as 
\begin{align}
 D(\rho_A \lvert \rvert \sigma_A) \simeq \begin{cases}
 0 ,& S_{{A}}^{\rm (eq)} \ll S_{\bar{A}}^{\rm (eq)} + S(\rho_0) 
 \\
 S_{A}^{\rm (eq)} - S_{\bar{A}}^{\rm (eq)}-S(\rho_0), & S_{\bar{A}}^{\rm (eq)} + S(\rho_0) \ll S_{{A}}^{\rm (eq)} \ll S_{\bar{A}}^{\rm (eq)} + S(\sig_0) 
 \\
 D(\rho_0 \lvert \rvert \sigma_0), & S_{{A}}^{\rm (eq)}\gg S_{\bar{A}}^{\rm (eq)} + S(\sig_0) 
 \end{cases}.
 \label{relative_entropy_equilibrium}
\end{align}
where we have assumed $S(\sig_0) \gg S(\rho_0)$. The above expressions are intuitively reasonable. 
When subregion $A$ is sufficiently small, the two density matrices are entirely indistinguishable, a manifestation of thermalization in an isolated quantum system. 
Once we move beyond the first regime of \eqref{relative_entropy_equilibrium}, the state $\sigma_0$ becomes important. In particular, the relative entropy rises from $0$ to $D(\rho_0 || \sigma_0)$ where it plateaus; as we gain information, the density matrices become more and more distinguishable. The relative entropy never increases beyond $D(\rho_0 || \sigma_0)$ due to the monotonicity of relative entropy under quantum channels.

Less conservatively, if we trust the analytic continuations of dominant permutations, we find the following sharper version of \eqref{relative_entropy_equilibrium}: 
\begin{align}
 D(\rho_A \lvert \rvert \sigma_A) \simeq \begin{cases}
 0 ,& S_{{A}}^{\rm (eq)} < S_{\bar{A}}^{\rm (eq)} + S(\rho_0) 
 \\
 S_{A}^{\rm (eq)} - S_{\bar{A}}^{\rm (eq)}-S(\rho_0), & S_{\bar{A}}^{\rm (eq)} + S(\rho_0) < S_{{A}}^{\rm (eq)} < S_{\bar{A}}^{\rm (eq)} + S(\rho_0) + D(\rho_0|| \sigma_0) 
 \\
 D(\rho_0 \lvert \rvert \sigma_0), & S_{{A}}^{\rm (eq)}> S_{\bar{A}}^{\rm (eq)} + S(\rho_0) + D(\rho_0|| \sigma_0) 
 \end{cases}.
 \label{relative_entropy_finite_T}
\end{align}
We also found numerical evidence for this equation in small spin chains and it is consistent with an infinite temperature result we derive exactly in Section \ref{transfer_sec}. It would be interesting to test this equation more generally.

\subsection{The fidelity of the Petz map} \label{sec:petzA}

We now turn to the calculation of the fidelity~\eqref{ren2} of the Petz map using the equilibrium approximation. Recall that in \eqref{ren2}, we assume $\rho_D$ is pure. 

We can write~\eqref{ren2} as a transition amplitude in the replica space as
\begin{align}
 F(\rho_D, \mathcal{P}_{\sigma_D, \mathcal{N}}\circ \mathcal{N}(\rho_D)) &= \lim_{m\rightarrow -\frac{1}{2}} \bra{\eta_{{R}} \otimes e_{B'}} (U\otimes U^{\dagger})^{\otimes (2m+2)}\ket{\chi,e} := \lim_{m\rightarrow -\frac{1}{2}}F_m , \label{f_limit}
\end{align}
where $\chi$ has the form 
\be \label{defci}
\chi =\chi_1 \otimes \chi_2^{\otimes m} \otimes \chi_3 \otimes \chi_2^{\otimes m} , 
\quad \chi_1 = \sigma_D^{\frac{1}{2}}\rho_D\sigma_D^{\frac{1}{2}} \otimes \rho_B, 
\quad \chi_2 = \sig_D \otimes \rho_B, \quad \chi_3 = \rho_D \otimes \rho_B \ . 
\ee
Applying the equilibrium approximation (assuming the equilibrated states of $\chi_{1,2,3}$ can all be described by the same macroscopic state $\sI_\al$) 
\begin{align} 
 F_m = \frac{1}{Z_2^{2m+2}}\sum_{\tau \in \mathcal{S}_{2m +2}} \bra{\eta_{{R}} \otimes e_{B'}} \mathcal{I}_{\alpha},\tau\rangle \langle \mathcal{I}_{\alpha} ,\tau\ket{\chi,e} := \sum_{\tau \in \mathcal{S}_{2m +2}} F_m (\tau) \label{fm_tau}
\end{align}
where 
\bea \label{i_chi}
\langle \mathcal{I}_{\alpha} ,\tau\ket{\chi,e} &= & \prod_{i=1}^{k (\tau)}
\Tr \le[ (\sI_\al \chi_1)^{a_i} (\sI_\al \chi_2)^{b_i} (\sI_\al \chi_3)^{c_i} (\sI_\al \chi_2)^{d_i} \ri] \\
& = & {Z_2^{2m+2} \ov Z_1^{2m+2}} \prod_{i=1}^{k (\tau)}
\Tr \le[ \chi_1^{a_i} \chi_2^{b_i} \chi_3^{c_i} \chi_2^{d_i} \ri]
\eea
where $a_i, b_i, c_i, d_i$ denote the number of appearances of $\chi_{1,2,3}$ in~\eqref{defci} in the $i$-th cycle of $\tau$.\footnote{Note that in general, the factors of $\chi_1^{a_i}$, $\chi_2^{b_i}$, $\chi_3^{c_i}$, and $\chi_2^{d_i}$ can also appear in other orders.} Clearly $a_1 =1$ and $a_{i > 1} =0$. In the second line we again have used the consistency conditions as in~\eqref{elk}. 

The contribution from the identity permutation can be written 
 as 
 \be 
 F_m (e) = \sZ_{2m+2, R}^{(\rm eq)} \Tr (\sig_D \rho_D) = \sZ_{2m+2, R}^{(\rm eq)} F(\rho_D, \sig_D) , 
 \ee
 which we expect to dominate for $R$ much smaller than $B'$. Analytically continuing the above expression to $m =-\ha$ we then find 
 \be
 F(\rho_D, \mathcal{P}_{\sigma_D, \mathcal{N}}\circ \mathcal{N}(\rho_D)) = F(\rho_D, \sig_D) ,
 \ee
 which is consistent with statement below~\eqref{ren1} that in the limit where $R$ is small, the Petz map simply gives $\sig_D$ for any $\rho_D$. 
 
 For $R$ to be much larger than $B'$, we expect the contribution from $\tau = \eta$ dominates giving 
\be 
F_m (\eta) = \sZ_{2m+2, B'}^{(\rm eq)} \Tr (\rho_D \sig_D^{m+\ha} \rho_D \sig_D^{m+\ha}) \Tr \rho_B^{2m+2} \ . 
\ee
Analytically continuing to $m =-\ha$ we have 
\be 
 F(\rho_D, \mathcal{P}_{\sigma_D, \mathcal{N}}\circ \mathcal{N}(\rho_D)) = 1 \ .
 \ee
which is consistent with that in the limit $B'$ is empty, the Petz map becomes the identity map.

There is a crossover between the above two extremes where other permutations could become important. This crossover behaviour will be dependent on the choice of $\sI_{\al}$, and we explain how to calculate it for particular choices of $\sI_{\al}$ in Section \ref{transfer_sec}. 

Note that starting from \eqref{ren_alt}, we could alternatively use the equilibrium approximation to get expressions involving permutations in $\sS_{n_1 + n_2 +2}$ instead of $\sS_{2m +2}$ by a similar series of steps, which give the same results for $F(\rho_D, \mathcal{P}_{\sigma_D, \mathcal{N}}\circ \mathcal{N}(\rho_D))$ on analytic continuation to $n_1 \rightarrow - \ha,~ n_2 \rightarrow - \ha$. 

\subsection{Reflected entropy} \label{sec:refl}

Another interesting quantity that may be computed using the replica trick is the reflected entropy \cite{2019arXiv190500577D}. 
Consider $A = A_1 \cup A_2$ in a mixed state $\rho_{A}$ and its canonical purification $\rho_{A} \rightarrow \ket{\sqrt{\rho_{A}}}_{A A^*} \in \sH_A \otimes \sH_{A^*}$ with $A^* = A_1^* \cup A_2^*$. The reflected entropy is then defined to be the von Neumann entropy of $A_1 A_1^*$ in this pure state. 

To calculate the reflected entropy using replicas, we consider 
\be 
\sR_{n,m} = - {1 \ov n-1} \log \Tr_{A_1 A_1^*} \le(\Tr_{A_2 A_2^*} \ket{\rho_A^m} \bra{\rho_A^m} \ri)^n 
:= -{1 \ov n-1} \log \sY_{n,m} 
\ee
where $ \ket{\rho_A^m} \in \sH_A \otimes \sH_{A^*}$ is defined as
\be 
\vev{i \bar i' | \rho_A^m} = (\rho_A^m)_{i i'} 
\ee
where $\ket{i}$ is a basis for $\sH_A$. The reflected entropy is then given by 
\be 
S_\sR := \lim_{n \to 1} S_{\sR}^{(n)}= \lim_{n \to 1} \lim_{m \to \ha} \sR_{n,m} \label{SR_def}
\ee
where $S_{\sR}^{(n)}$ are the Renyi reflected entropies. 

We take $A$ to be embedded in a larger system $S = A \cup B$ with 
\be 
\rho_A = \Tr_B (U \rho_0 U^\da)
\ee
where $\rho_0$ is a pure state. We can then write $\sY_{n,m}$ as a transition amplitude in $\sH_S^{\otimes 2 mn}$ 
\be 
\sY_{n,m} = \vev{(\sig_1)_{A_1}\otimes (\sig_2)_{A_2} \otimes e_B| (U \otimes U^\da)^{2mn} |\rho_0, e} 
\ee
where $\sig_1$ and $\sig_2$ denote the following permutations in $\sS_{2mn}$
\bega
\sig_1 = \prod_{k = 1}^n (k, k+n, \dots, k+n(m-1),k+1+nm,\dots,k+1+n(2m-1)) , \\
\sig_2 = \prod_{k = 1}^n (k, k+n, \dots, k+n(2m-1)) \ .
\end{gather} 
In the equilibrium approximation we then have 
\be 
\sY_{n,m} \approx \frac{1}{Z_1^{2nm}} \sum_{ \tau \in S_{2nm}}\bra{(\sigma_{1})_{A_1} \otimes (\sigma_2)_{A_2} \otimes e_{B}}\mathcal{I}_{\alpha},\tau\rangle \ .
 \label{reflected_equil}
\ee

When $B$ is sufficiently larger than $A$, the identity element will dominate the sum, leading to
\begin{align}
 \sY_{n, m} = \frac{(Z_{2m,A})^n}{(Z_{1,A})^{2nm}}.
\end{align}
Plugging this into \eqref{SR_def}, we immediately find all Renyi reflected entropies to be zero.

When $A$ is sufficiently large, there will be correlations between $A_1$ and $A_2$. When $A_1$ is much larger than $A_2$, we can be confident that $\tau =\sigma_{{1}}^{-1} $ will dominate the sum. 
Note that 
\begin{align}
 {\sigma}_{{2}}\, \sigma_{{1}}^{-1}&= (1,2,\dots n)(n(m+1),n(m+1)-1,\dots, nm+1),
\end{align}
which has two cycles of length $n$ and $n(2m-2)$ trivial cycles. This leads to
\begin{align}
 \sY_{n,m} = \frac{(Z_{n,A_2})^2(Z_{2m,B})^n}{(Z_{1,A_2})^{2n}(Z_{1,B})^{2nm}}.
\end{align}
Taking $m\rightarrow \ha$, we find that the Renyi reflected entropy is given by twice the equilibrium Renyi entropy of $A_2$.
\begin{align}
 S_{\sR}^{(n)} = \frac{2}{1-n} \log \frac{Z_{n,A_2}}{(Z_{1,A_2})^{n}} = 2 S^{\rm (eq)}_{n, A_2}.
\end{align}
In the same way, when $\tau =\sigma_{\tilde{g}}^{-1} $ dominates, \begin{align}
 S_{\sR}^{(n)} = 2 S^{\rm (eq)}_{n,A_1}.
\end{align}
It is natural to ask whether there are additional important permutations interpolating between these limits. Indeed, such permutations are identified in Ref.~\cite{SR_RTN1, SR_RTN2} where the random tensor network (infinite temperature) result is studied in detail using the reflected entropy spectrum. However, in the $n\rightarrow 1$ limit, the naive analytic continuations of the previously discussed permutations always give the correct answer to leading order. Here, we simply assume that this remains to be the case at finite temperature when $n\rightarrow 1$ and leave a rigorous justification of this assumption to future work. 
To find the transition points between the phases, we maximize the partition functions at $n \sim 1$, giving
\be
 S_{\sR} 
 =\begin{cases}
 0, & S^{\rm (eq)}_A < S^{\rm (eq)}_B
 \\
 2S^{\rm (eq)}_{A_1}, & S^{\rm (eq)}_{A_1} < S^{\rm (eq)}_{A_2},~ S^{\rm (eq)}_A > S^{\rm (eq)}_B
 \\
 2S^{\rm (eq)}_{A_2}, & S^{\rm (eq)}_{A_2}< S^{\rm (eq)}_{A_1},~ S^{\rm (eq)}_A > S^{\rm (eq)}_B
 \end{cases}.
\ee
The interplay between the two replica numbers is delicate for the analytic continuation, and the order of limits in \eqref{SR_def} is an additional assumption in the computation. For a more reliable analysis, one must evaluate the reflected entropy spectrum.

\section{Entanglement phase diagram of an equilibrated mixed state} 
\label{phase_diag_sec}

\subsection{General setup} 

We would like to explore the entanglement structure of a system $A$ in a mixed state $\rho_A$, which is in a macroscopic equilibrium but can be far in trace distance from the usual equilibrium density operators describing thermal ensembles. 
For this purpose, we consider various bi-partite quantum entanglement measures between a subsystem $A_1$ and its complement $A_2$ in $A$, including the Renyi and logarithmic negativities, as well as the mutual information and Renyi mutual informations. As reviewed earlier, the logarithmic negativity and the mutual information are of particular interest: the former because it is sensitive only to quantum entanglement correlations and gives a lower bound on the PPT entanglement cost, $E_c^{\rm (ppt, exact)}\geq \sE$, and the latter because it gives an upper bound on the PPT distillable entanglement, $\frac{1}{2} I \geq E_d$. 

$\rho_A$ can be characterized by an infinite number of parameters: $z_n = \Tr \rho_A^n, \; n=2, 3, \cdots$, In principle, the entanglement structure between $A_1, A_2$ can depend on the relations among these infinite number of parameters in a complicated way. In other words, if we use an ``entanglement phase diagram'' to characterize different entanglement structures, the diagram is in principle drawn on an infinite dimensional space. For $\rho_A$ in macroscopic equilibrium, we expect that
the story should be much simpler in the thermodynamic limit, not depending on microscopic details of $\rho_A$. Our goal is to 
extract the universal behavior of the entanglement structure in this regime. 

A general setup for exploring entanglement correlations in a mixed state in $A$ is to imagine that $A$ is embedded in a larger 
system $S = A\cup B$, with the total system $S$ in a pure state $\ket{\Psi}$ in macroscopic equilibrium,
and $\rho_A = {\rm Tr}_B \ket{\Psi} \bra{\Psi}$. 
In many questions of interest, such a $B$ naturally exists. For example, consider the evaporation of a black hole formed from gravitational collapse of a star in a pure state, a system of central interest for the black hole information paradox. If we take $A$ to be the collection of the Hawking radiation and ask about the entanglement correlations between different parts of the radiation, the corresponding $B$ is the black hole emitting the Hawking radiation. If quantum gravity is compatible with the usual rules of quantum mechanics, the full system of the black hole plus the Hawking radiation would be in a pure state in macroscopic equilibrium. 
Alternatively, we may simply view $B$ as an auxiliary system used to purify $\rho_A$. 

In this setup, with $A_1 \cup A_2 \cup B$ in an equilibrated pure state $\ket{\Psi}$, 
the equilibrium approximation reviewed in Sec.~\ref{sec:revea} can be generalized to calculate the Renyi and logarithmic negativities between $A_1$ and $A_2$, as we discussed in Sec.~\ref{sec:negas}.

We will show that the entanglement structure, i.e.~the qualitative behaviors of the negativities and mutual informations between $A_1$ and $A_2$, can be characterized by the equilibrium density operator $\rho^{(\rm eq)}$~\eqref{eqd}, and two parameters describing the relative sizes of $A_1, A_2$ and of $A, B$. Suppose the system $A$ has a volume\footnote{For a lattice system $V_A$ correspond to the number of sites. For a system with no spatial extent, such as the SYK model, $V_A$ corresponds to the number of fermions in $A$.} 
$V_A$. We denote the von Neumann entropy of $\rho_A$ as $S_A$, with the entropy density given by 
\be 
s_A = {S_A \ov V_A} \ . 
\ee
We are interested in the thermodynamic limit $V_A\to \infty$ with $s_A$ finite. 
We will consider $\sE(A_1, A_2)$ and $I(A_1,A_2)$ at leading order in the thermodynamic limit. In this limit, \eqref{fen} and \eqref{mastereq} can both be approximated by terms from a subset of permutations $\tau$, which give the dominant contribution. These sets of permutations can change as we vary two parameters
\be \label{cde1}
\lam := {S_{A_1}^{(\rm eq)} \ov S_A^{(\rm eq)}} , \qquad c := {S_A^{(\rm eq)} \ov S_A^{(\rm eq)} + S_B^{(\rm eq)}},
\ee
where $S_{A_1, A, B}^{\rm (eq)}$ are respectively the von Neumann entropies for $A_1, A, B$ in the state $\rho^{(\rm eq)}$. These parameters can be seen as a way of measuring the relative sizes of the subsystems in the general case where the system $S = A \cup B$ is inhomogeneous; when the full system is homogeneous, $\lam$ and $c$ 
are simply the volume fractions of various subsystems, $\lam = {V_{A_1}/ V_A}$ and $c = {V_A /( V_A + V_B)}$. The change in the dominant contribution on varying $c$ and $\lambda$ leads to qualitative changes in the behaviors of $\sZ_n^{\rm (PT)}$ and $\sZ_{n, R}$, and correspondingly of $\sE(A_1, A_2)$ and $I(A_1,A_2)$. We refer to such changes as entanglement phase transitions.

\subsection{Entanglement structure at infinite temperature} \label{sec:inft} 



We now proceed to apply the techniques developed in Sec.~\ref{sec:negas} to calculate the negativities between $A_1$ and $A_2$ in various situations. We first consider the case where the system has a finite-dimensional Hilbert space 
and is sufficiently excited that it can be treated as being at infinite temperature. 
Here, we find a universal entanglement phase structure which is independent of any details of $A$ or $B$. 
The structure also coincides with that obtained from the Haar average of a random state~\cite{2021PRXQ....2c0347S}. 

The dimensions of the Hilbert spaces for $A, A_1, A_2$ and $B$ will be denoted respectively as $d_A, d_1, d_2, d_B$. 
Assuming the system is homogeneous, we then have $\log d_A = V_A \log q$ where $q$ is the dimension of the Hilbert space at a single site. The parameters $\lam$ and $c$~\eqref{cde1} can also be written in terms of dimensions of various Hilbert spaces as 
\be
S_A^{(\rm eq)} = \log d_A , \quad \lam := {\log d_1 \ov \log d_A} , \quad
c := {\log d_A \ov \log d_A + \log d_B} , \quad d_A = d_1 d_2 \ .
\ee
We will also denote $S_0 := \log d_A + \log d_B$. 
 Taking $\sI_{\al}= \mathbf{1}$ in \eqref{mastereq} and \eqref{fen}, we get the following approximations for the quantities $\sZ_n^{\text{(PT)}}(A_1, A_2)$ and $\sZ_n^{(A)}$:
\be
\sZ_n^{(\text{PT})}(A_1, A_2) \approx \frac{1}{e^{n S_0}} \sum_{\tau} e^{S_0 \sA(\tau)}, \quad \sZ_n^{(A)} \approx \frac{1}{e^{nS_0}} \sum_{\tau} e^{S_0 \sB(\tau)},
\ee
where 
\begin{align} 
\label{AB}
\sA(\tau) &= c\lambda~ k(\eta^{-1}\tau)+ c (1-\lambda)~ k(\eta\tau) + (1-c)~ k(\tau), \\ \sB(\tau) &= c~ k(\eta^{-1}\tau)+ (1-c)~ k(\tau).
\end{align}

For any choice of $c, \lambda$, there is some
set of permutations for which $\sA(\tau)$ is maximized, and some set for which $\sB(\tau)$ is maximized. Let $\tau_m, \tau'_{m}$ respectively refer to any elements of these sets. Then, in the thermodynamic limit, we have 
\be 
\log\Zpt \approx S_0 \sA(\tau_m) - n S_0, 
\ee
and 
\be 
 R_n(A_1, A_2) \approx b_n S_0 \left(\sA(\tau_m)-\sB(\tau'_m) \ri) \ . 
\ee

The set of permutations that maximize $B(\tau)$ was reviewed earlier in Sec.~\ref{sec:revea}, with $\tau'_m = e$ for $c<1/2$, and
$\tau'_m = \eta$ for $c>1/2$.

\begin{figure}[] 
\centering 
\includegraphics[width=8cm]{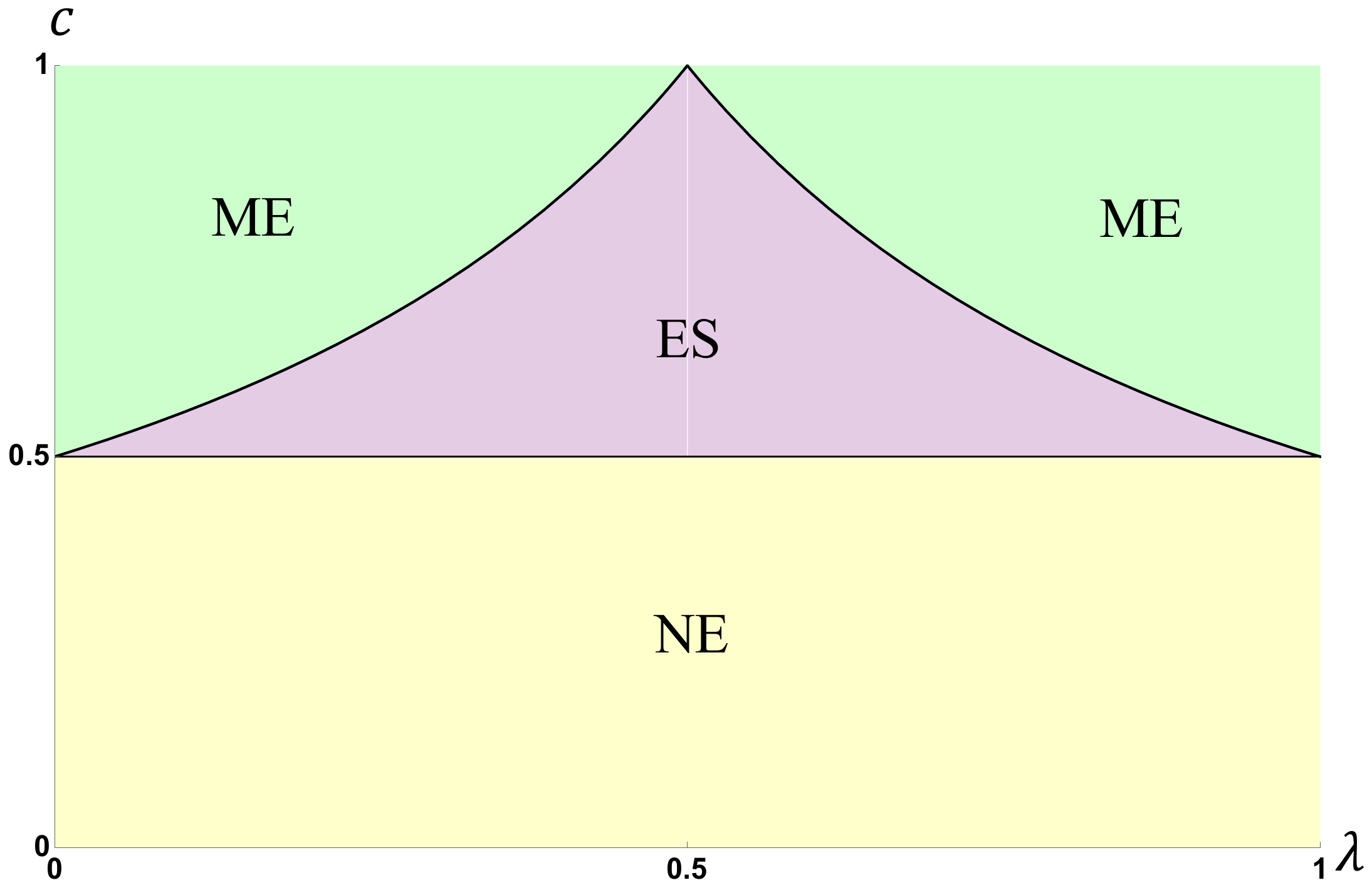}
\caption{Entanglement phase diagram for the infinite temperature equilibrated pure state.}
\label{fig:inf_phases}
\end{figure}

To maximize $\sA (\tau)$, it is convenient to write~\eqref{AB} in a few different ways, 
\begin{align} 
\sA(\tau) & = c \lambda \left(k(\eta^{-1}\tau)+k(\tau) \right) + c(1-\lambda) \left(k(\eta\tau)+k(\tau) \right) + (1-2c) k(\tau) \label{a1} \\
& = (2c \lambda-1) k(\eta^{-1}\tau) + c(1-\lambda)\left(k(\eta \tau)+ k(\eta^{-1}\tau)\right)+ (1-c) \left(k(\tau)+ k(\eta^{-1}\tau)\right) \label{a2} \\
& = c \lambda \left(k(\eta^{-1}\tau)+k(\eta\tau)\right) + (2c(1-\lambda)-1)k(\eta \tau) + (1-c)\left(k(\tau)+ k(\eta\tau)\right) \label{a3} \\
\begin{split} 
&=\left(\ha - c \lambda\right)\left(k(\eta\tau)+k(\tau)\right) + \left(\ha-c(1-\lambda)\right)\left( k(\eta^{-1}\tau) + k(\tau) \right) \\ & \quad + \left(c-\ha\right) \left(k(\eta\tau)+k(\eta^{-1}\tau)\right) \ .
\label{a4}
\end{split} 
\end{align}
Different expressions above are convenient for different ranges of parameters $c$ and $\lam$, from which 
we find three entanglement phases, which are shown in Fig.~\ref{fig:inf_phases}:



\begin{enumerate} 
\item \textbf{Phase of no entanglement}\footnote{``No entanglement'' here should be understood as no ``volume'' entanglement, i.e.~there is no contribution at the order of $O(\log d_A)$.} 

For $c<\ha$, i.e.~$S^{\rm (eq)}_A < S^{\rm (eq)}_B$, all coefficients in \eqref{a1} are positive and $\sA(\tau)$ is maximized by having $k(\eta^{-1}\tau)+ k(\tau)$, $k(\eta\tau)+ k(\tau)$ and 
$k(\tau)$ all reach their maximum values simultaneously. This happens for $\tau=e$, which gives 
\be
\begin{gathered} 
R_n(A_1, A_2) = 0, \quad n\geq 3, \quad \quad
\lim_{n\rightarrow 2} R_n(A_1, A_2) = 0, \\
\log \Zpt = S_A^{(\rm eq)} ~(1-n), \quad n \geq 2 \ .
\label{inf_un}
\end{gathered}
\ee
By analytically continuing either $R_{2m}$ or $\log \sZ_{2m}^{\text{(PT)}}$, we find that the logarithmic negativity is given by 
\be 
\sE(A_1, A_2) = \lim_{m\rightarrow \ha} R_{2m} = \lim_{m\rightarrow \ha}\log \sZ_{2m}^{\text{(PT)}} = 0 \ . \label{zeroinf}
\ee
Furthermore, we find that all Renyi mutual informations vanish
\be 
I_n(A_1, A_2)= 0 , \quad n\geq 2, \quad \quad I = \lim_{n\rightarrow 1} I_n(A_1, A_2) = 0 \ .
\ee
Since the equilibrium approximation calculates only the leading order contribution of order $O(\log d_A)$, in this phase there is no extensive entanglement. Both the negativities and mutual information may have nontrivial higher order sub-extensive contributions. 

It is quite intuitive that in this case there is no entanglement between any subsystems. 
In the language of purification, all degrees of freedom of $A$ are maximally entangled with those in $B$, and from the monogamy of entanglement, there is no entanglement within $A$. 


\item \textbf{Maximally entangled phase} 
 
For $c>\ha, \, \lambda>\frac{1}{2c}$, i.e.~$S_B^{\rm (eq)} < S_A^{(\rm eq)},\; S_{A_2}^{(\rm eq)} < \ha (S_A^{(\rm eq)} - S_B^{\rm (eq)})$, 
all coefficients in~\eqref{a2} are positive, and $\sA(\tau)$ is maximized by having $k(\eta^{-1}\tau)$, $k(\eta\tau)+ k(\eta^{-1}\tau)$ and 
$k(\tau) + k(\eta^{-1}\tau)$ reach maximum values simultaneously. This happens for $\tau = \eta$, which gives 
\be 
\begin{gathered} 
R_n(A_1, A_2) = \log d_2 = S_{A_2}^{(\rm eq)} 
\quad n\geq 3, \qquad 
\lim_{n\rightarrow 2} R_n(A_1, A_2) = \log d_2 = S_{A_2}^{(\rm eq)}
, \\
\log \Zpt = \begin{cases} 
- (n-2) S_{A_2}^{(\rm eq)} - (n-1) S_B^{\rm (eq)} 
& n \geq 2,~ n \text{ even} \\
- (n-1) (S_{A_2}^{(\rm eq)} + S_B^{\rm (eq)})
& n \geq 3,~ n \text{ odd}
\end{cases} 
\label{inf_me2}
\end{gathered} 
\end{equation} 
By analytic continuation, we get 
\be \label{legn}
\sE(A_1, A_2) = \lim_{m\rightarrow \ha} R_{2m} = \lim_{m\rightarrow \ha}\log \sZ_{2m}^{\text{(PT)}} =\log d_2 = S_{A_2}^{(\rm eq)} \ .
\ee
For this parameter range, $V_A > V_B, \; V_{A_2} < \ha (V_{A} - V_B) < V_{A_1}$,~\eqref{legn} is the maximal 
value of $\sE (A_1, A_2)$ can have, and implies that $A_2$ is maximally entangled with $A_1$. We also have 
$V_{A_2} + V_B < \ha (V_A + V_B) < V_{A_1}$, i.e.~the effective number of degrees of freedom in $A_2$ and $B$ together is smaller than that in $A_1$. Thus both $A_2$ and $B$ should be maximally entangled with $A_1$ and there should be no entanglement between $A_2$ and $B$.

Similarly for $c>\ha, ~~ \lambda<1-\frac{1}{2c}$, i.e.~$S_B^{\rm (eq)} < S_A^{(\rm eq)},\; S_{A_1}^{(\rm eq)} < \ha (S_A^{(\rm eq)} - S_B^{\rm (eq)})$, from \eqref{a3}, the dominant permutation for $\sA(\tau)$ is $\eta^{-1}$
and $A_1$ is maximally entangled with $A_2$, with $R_n$ and $\log \Zpt$ obtained by exchanging $A_1$ and $A_2$ in~\eqref{inf_me2} and~\eqref{legn}. 

From the equilibrium approximation for the mutual information, we find that for both range parameters 
\be
\begin{gathered} 
I_n(A_1, A_2) = 2 \, \text{min}( S^{\rm (eq)}_{A_1}, S^{\rm (eq)}_{A_2}) 
\quad n\geq 2, \\ I = \lim_{n\rightarrow 1} I_n(A_1, A_2) = 2 \, \text{min}( S^{\rm (eq)}_{A_1}, S^{\rm (eq)}_{A_2}) \, . 
 \end{gathered} 
\label{mutual_max}
\ee
which is also consistent with the maximally entangled interpretation given. 

\item \textbf{Entanglement saturation phase} 

For $c>\ha,\; 1-\frac{1}{2c}< \lambda< \frac{1}{2c}$, i.e.~$S_B^{\rm (eq)} < S_A^{(\rm eq)},\; \ha (S_A^{(\rm eq)} - S_B^{\rm (eq)}) < S_{A_1}^{(\rm eq)} < \ha (S_A^{(\rm eq)} + S_B^{\rm (eq)})$, 
 from \eqref{a4}, $\sA(\tau)$ is maximized by permutations which maximize simultaneously $k(\tau)+ k(\eta^{-1}\tau)$, $k(\tau)+ k(\eta\tau)$, and $k(\eta^{-1}\tau)+ k(\eta\tau)$. For even $n$, 
$\sA(\tau)$ is maximized by the set $\{\tau\}^{\ast}$ of non-crossing permutations which consist only of 2-cycles~\footnote{A $k$-cycle refers to a cycle with $k$ elements.}. For odd $n$, the dominant permutations for $\sA(\tau)$ are the set $\{\tau\}_{\text{odd}}^{\ast}$ of non-crossing permutations which consist of $\frac{n-1}{2}$ 2-cycles and one 1-cycle. See Appendix \ref{app:perm} for more details. We then find that~\footnote{Note that in the discussion here, we consider only the leading order $O(V)$ contribution to $R_n$, ignoring the $O(1)$ term coming from the degeneracy of different permutations.} 
\bega \label{egsm} 
R_n(A_1, A_2) = \ha (S_A^{(\rm eq)} - S_B^{\rm (eq)}) , 
\quad n\geq 2, 
\\
\log \Zpt = \begin{cases} 
- \le({n \ov 2} -1 \ri) S_A^{(\rm eq)} -{n \ov 2} S_B^{\rm (eq)},
& n \geq 2, ~n\text{~even} \\
- \frac{n-1}{2} (S_A^{(\rm eq)} + S_B^{\rm (eq)}) ,
& n \geq 3, ~n\text{~odd}
\end{cases} \ .
\label{inf_es}
\end{gather} 
From analytic continuation, the logarithmic negativity is given by 
\be 
\sE(A_1, A_2) = \lim_{m\rightarrow \ha} R_{2m} = \lim_{m\rightarrow \ha}\log\sZ_{2m}^{\text{(PT)}} =
\ha (S_A^{(\rm eq)} - S_B^{\rm (eq)}) \ .
\label{neg_inf_es}
\ee
The mutual information is given by 
\be 
I_n(A_1, A_2) = S_A^{(\rm eq)} - S_B^{\rm (eq)},\quad n\geq 2, \quad \quad I = \lim_{n\rightarrow 1} I_n(A_1, A_2) = S_A^{(\rm eq)} - S_B^{\rm (eq)} \ . 
\label{mutual_es}
\ee
Both the negativity and mutual information depend only on the difference $S_A^{(\rm eq)} - S_B^{\rm (eq)}$, and do not change as we vary the size of $A_1$ (as long as we stay in the aforementioned parameter range). For this reason, this is referred to as the entanglement saturation (ES) phase. Also notice that $R_n$ and $\sE$ are half of the corresponding values for the mutual information as in a pure state, even though here $\rho$ is mixed.

In terms of purification, in this range parameters, we have $V_A > V_B, \; \ha (V_A - V_B) < V_{A_1} , V_{A_2} < \ha (V_A + V_B)$, and the sum of any two $V_{A_1}, V_{A_2}, V_B$ is larger than the third. Thus, any two of the systems are entangled with each other. There is a simple intuitive interpretation of~\eqref{neg_inf_es} and~\eqref{mutual_es} in terms of mutual bi-partite entanglement: since $V_B < V_A$, $\log d_B = S_B^{\rm (eq)} $ degrees of freedom in $A$ are entanglement with $B$, and the remaining $\log d_A - \log d_B = S_A^{(\rm eq)} - S_B^{\rm (eq)}$ are entangled between $A_1$ and $A_2$, resulting in~\eqref{neg_inf_es} and~\eqref{mutual_es}. 

We should emphasize, however, that this ``mechanical'' way of assigning entanglement likely does not 
reflect the genuine entanglement structure of the system in this phase. 
A state of mostly bi-partite entanglement has been proposed in~\cite{2019CMaPh.376..609C} which satisfies the relation $S_R (A_1, A_2) - I (A_1, A_2) =0$ 
at leading order~\cite{2020JHEP...04..208A}, which in our current context means the absence of a volume law contribution. 
From the discussion of Sec.~\ref{sec:refl}, we find at leading order in the infinite temperature case 
\be 
S_R (A_1, A_2) - I (A_1, A_2) = \begin{cases} 
0 & \text{NE phase} \\
0 & \text{ME phase} \\
S_B^{\rm (eq)} - |S_{A_1}^{\rm (eq)} - S_{A_2}^{\rm (eq)}| & \text{ES phase}
\end{cases} 
 \ .
\ee
This gives some hint that the ES phase likely has significant multi-partite entanglement.


\end{enumerate} 
In the above discussion, we obtained the behavior of $\sE$ by analytic continuation in $n$, but it can checked that 
the same results follow from calculating the full negativity spectrum using the resolvent. See Appendix~\ref{app:mic_inf}.

\subsection{Entanglement phase structure at a finite temperature: general discussion} 

The infinite temperature case only applies to a system with a finite dimensional Hilbert space at sufficiently high energies or without energy conservation. When the energy of a system is not large enough, or if a system has an infinite dimensional Hilbert space such as in a field theory, energy constraints have to be imposed. 

Now there are many more possibilities for $\rho^{\rm (eq)}$, which depend on the ensembles we choose and properties of $B$. 
 We may view the different
possibilities as giving different universality classes. 
In this subsection, we summarize the general structure and the main results, leaving explicit discussions of specific choices of $\rho^{\rm (eq)}$
to subsequent subsections.

At infinite temperature, the Renyi negativities $R_n$, logarithmic negativity $\sE$, Renyi mutual information $I_n$, and mutual information $I$ all have the same $n$-independent phase structure. At finite temperature, in general each of these infinite number of quantities can give rise to a different phase diagram as a function of the parameters $c$ and $\lambda$ defined in \eqref{cde1}. This reveals intricate patterns of entanglement structure\footnote{Recall that other than the logarithmic negativities, all other quantities are sensitive to classical correlations.}, but also makes it all but impossible to draw a phase diagram which reflects the behavior of all these quantities. 
Of particular interest is the behavior of logarithmic negativity $\sE$, as it directly reflects quantum entanglement correlations, and as reviewed in Sec.~\ref{review_sec}, can be used to bound the PPT entanglement cost. Furthermore, its relation with mutual information $I$ can shed light on the irreversibility of the preparation of a mixed state by different types of operations. 
A significant technical complication at finite temperature is that the extraction of the logarithmic negativity using analytic continuation becomes a priori unreliable for various parameter ranges due to non-uniform dependence on $n$ of $R_n$. The resolvent is needed to calculate $\sE$, which fortunately can be done for some choices of $\rho^{(\rm eq)}$ and can be used to illustrate the general structure. Here is a summary of the main features we observe for finite temperature ensembles:

\ben

\item As we will discuss immediately below, the mutual information $I(A_1, A_2)$ has exactly the same phase structure as that at infinite temperature. Thus, the mutual information appears to be insensitive to finite temperature effects. 

The behavior of Renyi mutual informations in general become $n$-dependent, with $n$-dependent phase boundaries. See Fig.~\ref{fig:can_infa} for an example.

\item For negativities, each $n$ can have its own version of finite-temperature generalizations of the NE, ME, and ES phases, with phase boundaries between them, which in general are $n$-dependent. See Fig.~\ref{fig:n_disc} for an example.
 
\item In the phase diagram of the logarithmic negativity, there are finite-temperature generalizations of the NE, ME and ES phases. Between these phases, there can be new phases that have different $\sE$ behavior, and are separated from other phases by non-analytic dependences on the $c, \lam$ parameters. Such phases cannot be deduced by analytic continuation of $\sZ_n^{\rm (PT)}$ or $R_n$. See Fig.~\ref{fig:phase_diagram_a2_inf} for an example.

\item There are regimes where $\sE$ has a nonzero volume-law contribution, but the mutual information $I$ is sub-extensive. This is a surprising result, as it is generally believed that mutual information contains both quantum and classical correlations. That it is sub-extensive intuitively implies there cannot be any volume-like quantum entanglement correlations. Our results indicate this intuition cannot be correct. We will elaborate further on this relation between logarithmic negativity and mutual information in Sec.~\ref{sec:mixen}. 



\item For various choices of $\rho^{(\rm eq)}$, the Renyi and logarithmic negativities can often be expressed in terms of equilibrium Renyi entropies, but this appears to not always be the case.  

\item Due to the $n$-dependence of phase boundaries of $R_n$ and $I_n$, for a given $\lam, c$, there can be intricate patterns of entanglement. 
For example, it is possible that for all $n$, the negativities show the behavior of the ME phase, while for all $n$ the mutual information show the behavior of the ES phase. It can also happen that for different $n$, $R_n$ show different phases, i.e.~some $n$ can be in the ME phase while other $n$ are in the  ES or NE phases. See Fig.~\ref{fig:phaD} for some examples.

\een

To conclude this general discussion, let us now consider the mutual information. 
From our general discussion in Sec.~\ref{sec:revea}, we have 
\be \label{erem1}
S_{n, A}= {\rm min} \le(S_{n, A}^{({\rm eq})} , S_{n, \bar A}^{({\rm eq})}\ri) \ 
\ee
for $n \geq 2$, as well as 
\be 
S_{A}= {\rm min} \le(S_{A}^{({\rm eq})} , S_{\bar A}^{({\rm eq})}\ri) \, . 
\ee
Given that the von Neumann entropy of a thermal density operator is extensive, i.e.~
\be 
S_{A_1}^{( {\rm eq})} + S_{A_2}^{( {\rm eq})} = S_{A}^{({\rm eq})} , \quad
S_{A_1B}^{({\rm eq})} = S_{A_1}^{({\rm eq})} + S_{B}^{({\rm eq})} , 
\ee
we then have 
\be \label{nume}
I (A_1, A_2) = \bca 0 & c< \ha \; \; \text{i.e.} \; \; S_{A}^{( {\rm eq})} = S_B^{(\rm eq)} \cr
S_{A}^{({\rm eq})} - S_{B}^{( {\rm eq})} & c >\ha,\; 1-\frac{1}{2c}< \lambda< \frac{1}{2c} 
\cr
2 S_{A_2}^{({\rm eq})} & \lam > {1 \ov 2 c} \; \; \text{i.e.} \; \; S_{A_1}^{( {\rm eq})} > S_{A_2}^{( {\rm eq})} + S_B^{(\rm eq)} \cr
2 S_{A_1}^{({\rm eq})} & \lam < 1- {1 \ov 2 c} \; \; \text{i.e.} \; \; S_{A_2}^{( {\rm eq})} > S_{A_1}^{( {\rm eq})} + S_B^{(\rm eq)} 
\eca 
\ee
where $c$ and $\lam$ were defined in~\eqref{cde1}. Thus, we find exactly the same phase structure as that at infinite temperature and the same phase diagram Fig.~\ref{fig:inf_phases}. 

For the Renyi mutual information, we have a similar phase structure, but the transitions are in general at $n$-dependent values of $c, \lam$. More explicitly, suppose $\sI_\al = \sI_A \otimes \sI_B$ can be factorized between $A$ and $B$, then 
from~\eqref{erem1}, 
\be \label{nume1} 
I_n (A_1, A_2) = \bca S_{n, A_1}^{( {\rm eq})} + S_{n, A_2}^{( {\rm eq})} - S_{n, A}^{( {\rm eq})} & S^{\rm (eq)}_{n, A} < S^{\rm (eq)}_{n, B} \cr
S_{n, A_1}^{( {\rm eq})} + S_{n, A_2}^{( {\rm eq})} - S_{n, B}^{( {\rm eq})} & S^{\rm (eq)}_{n, A}>S^{\rm (eq)}_{n, B},\; S^{\rm (eq)}_{n, B} > | S^{\rm (eq)}_{n, A_1}- S^{\rm (eq)}_{n, A_2}| \cr
2 S_{n,A_2}^{({\rm eq})} & S^{\rm (eq)}_{n, A_1}> S^{\rm (eq)}_{n, A_2}+ S^{\rm (eq)}_{n, B} \cr
2 S_{n, A_1}^{({\rm eq})} & S^{\rm (eq)}_{n, A_2}> S^{\rm (eq)}_{n, A_1}+ S^{\rm (eq)}_{n, B}
\eca ,
\ee
Note that the Renyi entropies from subsystems may not be additive, i.e.~$S_{n, A_1}^{( {\rm eq})} + S_{n, A_2}^{( {\rm eq})} $
does not have to be equal to $S_{n, A}^{( {\rm eq})} $. 

The behavior of the Renyi and logarithmic negativities is technically much more complicated at finite temperature, and we turn to this next. 

\subsection{Canonical ensemble}

In this subsection, we consider more explicitly the entanglement structure in the canonical ensemble.

\subsubsection{General setup} \label{general_canonical}

In the canonical ensemble, we consider an effective identity operator of the form 
\be \label{canD} 
\sI = e^{-\b_A H_A} \otimes e^{-\b_B H_B} ,
\ee
which can be interpreted in several ways. One possible scenario is that after some brief interactions, subsystems $A$ and $B$ are separated and thus can achieve separate equilibria at different temperatures. When $\b_A = \b_B = \b$ it also applies to a system $AB$ with a local Hamiltonian $H = H_A + H_B + H_{AB}$ for which $e^{-\beta H} \approx e^{-\beta H_{A}} \otimes e^{-\beta H_{B}} $. We can drop the $H_{AB}$ term as it should give contributions proportional to the areas of the subsystems in the exponent, and thus can be neglected in the thermodynamic limit.

We assume that the system is homogeneous within $A$, so that $Z_{m,R}= \text{Tr}[e^{-m\beta_R H_R}]$ for different subsystems $R$ can be written as 
\be 
Z_{m,A_1} = e^{V_{A_1} f_A(m\beta_A)}, \quad Z_{m,A_2} = e^{V_{A_2} f_A(m\beta_A)}, \quad Z_{m,B} = e^{V_{B} f_B (m\beta_B)} \ .
\ee
The corresponding equilibrium Renyi entropy density can be written as 
\be \label{sn_eq}
s^{\rm (eq)}_{n, R}(\beta) := \frac{n f_R (\beta) - f_R (n\beta)}{n-1} , \qquad S^{\rm (eq)}_{n,R}(\beta) = V_R s^{\rm (eq)}_{n,R}(\beta) \ .
\ee
We allow $n$ to take any non-negative real value. The $n\rightarrow 1$ limit of \eqref{sn_eq} gives the density of the equilibrium von Neumann entropy (i.e.~the thermodynamic entropy density), 
\be \label{s1_eq}
s^{\rm (eq)}_R (\beta) := f_R (\beta) - \beta f_R '(\beta) , \qquad S^{\rm (eq)}_{R}(\beta) = V_R s^{\rm (eq)}_R (\beta) 
 \ .
\ee

The free energy densities $f_A(\beta)$ and $f_B(\beta)$ both satisfy the following properties: 
\begin{enumerate} 
\item Since $\braket{(H-\braket{H}_{\beta})^2}_{\beta} = N f''(\beta)$ is non-negative, 
\be 
f''(\beta)\geq 0. \label{in_1}
\ee
\item From the non-negativity of the thermodynamic entropy, 
\be 
f(\beta) - \beta f'(\beta) \geq 0. \label{in_2}
\ee
\item From the non-negativity of~\eqref{sn_eq},
\be 
nf(\beta)-f(n\beta) \geq 0, \quad n\geq 2. \label{in_3}
\ee
\item From the fact that $s_{n}^{\rm (eq)}$ decreases with the Renyi index $n$, we have 
\be 
\frac{d}{dn}\left(\frac{n f(\beta)- f(n\beta)}{n-1}\right) \leq 0 \Rightarrow f(\beta) - \beta f'(n\beta) \geq f(n\beta) - n \beta f'(n\beta) \label{in_4}
\ee
Note that \eqref{in_4}, together with \eqref{in_2} at inverse temperature $n\beta$, implies that 
\be 
f(\beta) - \beta f'(n\beta) \geq 0. \label{in_5}
\ee
\end{enumerate}

From \eqref{mastereq}, we find 
\be 
 \sZ_n^{\text{(PT)}}(A_1, A_2) = \frac{1}{e^{nF}} \sum_{\tau} e^{ \sA(\tau)}, \quad \sZ_n^{(A)} = \frac{1}{e^{n F}} \sum_{\tau} e^{\sB(\tau)}
 \label{eq_canonical}
\ee
where 
\begin{align} 
\label{defAs}
&\sA(\tau)= V_{A_1} \sum_{i=1}^{k(\eta^{-1}\tau)} f_A(a_i\beta_A)+ V_{A_2} \sum_{j=1}^{k(\eta\tau)} f_A(b_j\beta_A) + V_B \sum_{k=1}^{k(\tau)} f_B(c_k\beta_B), \\
\label{defBs} 
& \sB(\tau)= V_A \sum_{i=1}^{k(\eta^{-1}\tau)}f_A(a_i \beta_A) + V_B \sum_{k=1}^{k(\tau)} f_B(c_k \beta_B),\\
& F = V_A f_A (\b_A)+ V_B f_B (\b_B)
\end{align} 
and $\{a_i\}, \{b_j\}, \{c_k\}$ respectively are the numbers of elements of the cycles in $\eta^{-1}\tau$, $\eta\tau$ and $\tau$. 
Now if $\tau_m$ denotes one of the permutations that maximizes $\sA(\tau)$ and $\tau'_m$ denotes one of the permutations that maximize $\sB(\tau)$ for a particular choice of $V_{A_1}$, $V_{A_2}$, $V_B$, then 
\be 
\log \Zpt(A_1, A_2) \approx \sA(\tau_m) -n F 
\ee
and 
\be 
R_n(A_1, A_2) = b_n \left(\sA(\tau_m)- \sB(\tau'_m) \right)
= b_n \le(\log \sZ_n^{\rm (PT)} - \log \sZ_n^{(A)} \ri) \ .
\ee
Using the equilibrium Renyi entropies~\eqref{sn_eq} we can write the contribution from a permutation $\tau$ to~\eqref{eq_canonical} as 
\bega \label{canonical_gen}
 \log \sZ_n^{\rm (PT)}(\tau) = \sum_{i=1}^{k(\eta \tau^{-1})} -(a_i-1) S_{a_i, A_1}^{\rm (eq)} + \sum_{i=1}^{k(\eta \tau)} -(b_i-1) S_{b_i, A_2}^{\rm (eq)} + \sum_{i=1}^{k(\tau)} -(c_i -1) S_{c_i, B}^{\rm (eq)} \\
\log \sZ_{n,A} (\tau) = \sum_{i=1}^{k(\eta \tau^{-1})} -(a_i-1) S_{a_i, A}^{\rm (eq)} + \sum_{i=1}^{k(\tau)} -(c_i -1) S_{c_i, B}^{\rm (eq)} 
\label{renPn}
 \end{gather}

For the Renyi partition function~\eqref{renPn}, the dominant permutations were already reviewed in Sec.~\ref{sec:revea} and the end of last section, given by either $\tau = e$ or $\tau = \eta$, 
with 
\be 
\log \sZ_{n,A} (e) = - (n-1) S_{n, A}^{\rm (eq)} (\b_A) , \quad \log \sZ_{n,A} (\eta) = - (n-1) S_{n, B}^{\rm (eq)} (\b_B ) \ . 
\ee

To find the negativities we need to maximize~\eqref{canonical_gen} among different permutations $\tau$, which 
is a difficult task without knowing the explicit forms of functions $f_A, f_B$ and relative magnitude of $\b_A, \b_B$. We will give detailed discussion for some specific examples and here summarize some general features. In all our investigations we find that for a given $n$, only one of the following set of permutations could dominate: $\tau =e$, $\tau = \eta, \eta^{-1}$, $\tau = \tau_{ES}$, which may be considered as giving finite temperature generalizations of the NE, ME and ES phases. The set of $\tau_{ES}$ denotes, 
\be 
\tau_{ES} = \bca (12)(34)\cdots (n-1,n) \text{ and } (23)(45)\cdots (n1) ,& \text{even} \; n \cr
 \{(23)(45)...(n-1~n),$~ $(34)(56)...(n~1),~ \cdots~,(12)...(n-2~n-1)\} ,& \text{odd} \; n
 \eca,
\ee 
which is a subset of the permutations $\tau^{\ast}, \tau^{\ast}_{\rm odd}$ for the ES phase at the infinite temperature.\footnote{In some special finite temperature cases, all permutations in the $\tau^{\ast}$ or $\tau^{\ast}_{\rm odd}$ give degenerate contributions, but more generally the degeneracy is broken. We note that these permutations are also relevant to the moments of the partial transpose for tensor network states with non-maximally entangled bonds \cite{2021arXiv210111029D}.} The corresponding expressions of~\eqref{canonical_gen} for these permutations are given by 
\bea\label{er}
\log \sZ_n^{\rm(PT)}(e) &=& - (n-1) S_{n, A}^{\rm (eq)} (\b_A) \\
\label{er1}
\log \Zpt (\eta) &=& \begin{cases}
- (n-1) S_{n, B}^{\rm (eq)} (\b_B) - (n-2) S^{(\rm eq)}_{{n \ov 2}, A_2} (\b_A)
 & n \geq 2,~ n \text{ even} \\
- (n-1) (S_{n, A_2}^{(\rm eq)} (
\b_A) + S_{n, B}^{\rm (eq)} (\b_B)) & n \geq 3,~ n \text{ odd}
\end{cases} 
\\
\label{er2}
\log \Zpt (\eta^{-1}) &=& \begin{cases}
- (n-1) S_{n, B}^{\rm (eq)} (\b_B) - (n-2) S^{(\rm eq)}_{{n \ov 2}, A_1} (\b_A)
 & n \geq 2,~ n \text{ even} \\
- (n-1) (S_{n, A_1}^{(\rm eq)} (
\b_A) + S_{n, B}^{\rm (eq)} (\b_B)) & n \geq 3,~ n \text{ odd}
\end{cases} 
\\
 \log \Zpt (\tau_{ES}) &=& \begin{cases}
 -{n-2 \ov 2} S^{\rm (eq)}_{{n \ov 2}, A}(\beta_A) - {n \ov 2} S_{2, B}^{\rm (eq)}(\beta_B) & n \geq 2,~ n \text{ even} \\
 - {n-1 \ov 2} \le(S^{\rm (eq)}_{{n+1 \ov 2}, A}(\beta_A) + S_{2, B}^{\rm (eq)} (\b_B) \ri)
& n \geq 3,~ n \text{ odd}
\end{cases} \ .
\label{fin_es}
\eea 
The regions where the different permutations dominate can be found by comparing their magnitudes. For example, 
for even $n$, $\tau_{ES}$ dominates over $e$ when 
\be \label{echb} 
(n-1) S_{n, A}^{\rm (eq)} - {n-2 \ov 2} S_{\frac{n}{2}, A}^{\rm (eq)} \geq \frac{n}{2} S^{\rm (eq)}_{2, B}
\ee
and the equal sign defines the transition line for the change of dominance. These transitions lines are 
generally $n$-dependent.

If we can analytically continue the expressions~\eqref{er}--\eqref{fin_es} for even $n$ to $n =1$ we would obtain the logarithmic negativity for the corresponding finite temperature generalizations of NE, ME, and ES phases with 
\bega \label{ksn}
\sE_{NE} = 0 , \\
\label{ksn1}
\sE_{ME} = S_{\ha, A_2}^{\rm (eq)} (\b_A) \quad \text{or} \quad \sE_{ME} = S_{\ha, A_1}^{\rm (eq)} (\b_A) , \\
\sE_{ES} = \ha \le(S^{\rm (eq)}_{{1 \ov 2}, A}(\beta_A) - S_{2, B}^{\rm (eq)}(\beta_B) \ri) \ .
\label{ksn2} 
\end{gather} 
Comparing with the behavior of mutual information~\eqref{nume}, notice that in~\eqref{ksn1}, $S^{(\rm eq)}_{{1 \ov 2}, A_2} (\b) $ appears and is no longer related to the third line of~\eqref{nume} by a simple factor of $\ha$ as it was in the infinite temperature case. 
Similarly, in~\eqref{ksn2}, $S^{(\rm eq)}_{\ha , A}$ and $S^{(\rm eq)}_{2,B}$ appear, in contrast to the second line of~\eqref{nume}.
Analytically continuing~\eqref{echb} to $n=1$ gives 
\be 
S_{\frac{1}{2}, A}^{\rm (eq)} = S^{\rm (eq)}_{2, B} 
\label{n_es_e_can}
\ee 
which may be interpreted as the transition from NE to ES phase for $\sE$. 
As we emphasized before, the correctness of such analytic continuations is not warranted in particular in regions of parameters where the dependence on $n$ is not uniform. 
In Sec.~\ref{sec:Binf} we discuss a case where the resolvent can be calculated numerically, and we checked in the NE and ES regimes that the expressions~\eqref{ksn},~\eqref{ksn2} and \eqref{n_es_e_can} do apply. 

It is instructive to compare~\eqref{n_es_e_can} with the transition from NE to ES for the mutual information $I$ which from~\eqref{nume} is given by $c = \ha$, i.e.~
\be 
S_{A}^{\rm (eq)} = S^{\rm (eq)}_{B} \ . \label{mutual_ne_es}
\ee
Given that $S_n$ monotonically decreases with $n$, we have $S_{\frac{1}{2}, A}^{\rm (eq)} > S_{A}^{\rm (eq)} $
and $S^{\rm (eq)}_{2, B} < S^{\rm (eq)}_{B} $, so the transition~\eqref{n_es_e_can} must happen for
$S_{A}^{\rm (eq)} < S^{\rm (eq)}_{B} $, i.e.~somewhere with $c < \ha$. Therefore, \eqref{n_es_e_can} implies a region in the phase diagram where the logarithmic negativity is extensive, but the mutual information is sub-extensive. 

Below, we will consider the following representatives of two classes of examples:
\begin{enumerate} 
\item 
In discrete systems, $f(\beta)$ has a well-defined $\beta\rightarrow 0$ limit. An example is 
\be 
 \quad f(\beta) = \log 2 + \log\cosh(\beta J) \label{example_disc} 
\ee
which is observed in a variety of spin chain models. 
\item 
In continuum systems, $f(\beta)$ does not have a well-defined $\beta \rightarrow 0$ limit. For instance, one form that appears in quantum field theories is 
\be 
f(\beta) = a \beta^{-\alpha}, \quad \alpha \geq 0. \label{example_cont}
\ee
\end{enumerate}

\subsubsection{$A$ and $B$ are uniform}

As an explicit illustration, we consider the case where the Hamiltonian is translationally invariant and $\b_A = \b_B$ for which 
we have $f_A (\b)= f_B (\b) = f (\b)$. It will be useful to consider the explicit examples in both \eqref{example_disc} and \eqref{example_cont}. In this case, it is not clear how to perform a resolvent calculation, so we are not able to find the phase diagram for $\sE$ and will only present the structure for $R_n$. 

Since the system is homogeneous, for $\log \sZ_n^{(A)}$, $\tau =e$ dominates for $c < \ha$ and $\tau = \eta$ dominates for $c >\ha$. In this case, $c_n$ and $\lam_n$ coincide with $c$ and $\lam$ for all $n$, so the Renyi mutual information~\eqref{nume1} has 
the same phase structure as the mutual information~\eqref{nume}. 


To understand the structure of $\log \Zpt$, it is convenient to introduce $V = V_A + V_B$ and rewrite 
 $\sA(\tau)$ in four ways,
\begin{align}
\begin{split} 
{1 \ov V} \sA(\tau) =& c\lambda \left(G(\beta, \tau) + G(\beta, \eta^{-1}\tau)\right)+ c(1-\lambda) \left(G(\beta, \tau)+G( \beta, \eta\tau)\right) \\ &+ (1-2c) G(\beta, \tau)\end{split}\label{abeta1} \\
\begin{split} = & (2c\lambda-1)G(\beta, \eta^{-1}\tau) + c(1-\lambda) \left(G(\beta, \eta\tau)+G( \beta, \eta^{-1}\tau)\right) \\ &+ (1-c) \left(G(\beta, \tau)+ G(\beta, \eta^{-1}\tau)\right) \end{split} \label{abeta2}\\
\begin{split} =& c \lambda \left(G(\beta, \eta^{-1}\tau)+G(\beta, \eta \tau)\right) + (2c(1-\lambda)-1) G(\beta, \eta \tau) \\ &+ (1-c) \left(G(\beta, \tau)+ G(\beta, \eta \tau) \right) 
\end{split} \label{abeta3}\\
\begin{split} =& (\frac{1}{2}-c \lambda)\left(G(\beta, \tau)+G(\beta, \eta\tau)\right) + (\frac{1}{2}-c (1-\lambda))\left(G(\beta, \tau)+G(\beta, \eta^{-1}\tau)\right) \\ &+ (c-\frac{1}{2}) \left(G(\beta, \eta\tau)+ G(\beta, \eta^{-1}\tau)\right) 
\end{split} \label{abeta4} 
\end{align} 
where we have introduced 
\be 
G(\beta, \tau) := \sum_{i=1}^{k(\tau)} f(c_i \beta) \label{func_G}
\ee
with $\{c_i\}$ the lengths of the cycles in $\tau$. We show in Appendix~\ref{finitetemp_perms} that $G(\beta, \tau)$ and $G(\beta, \eta^{-1}\tau)$ are respectively maximized by $\tau=e$ and $\tau=\eta$, and 
$G(\beta, \tau)+G( \beta, \eta^{-1}\tau)$ is maximized by both $\tau=e$ and $\tau=\eta$. Similarly, $G(\beta, \eta\tau)$ is maximized by $\tau=\eta^{-1}$, and $G(\beta, \tau)+G( \beta, \eta\tau)$ is maximized by both $\tau=e$ and $\tau=\eta^{-1}$. Furthermore, we observe numerically that $G(\beta, \eta\tau)+ G(\beta, \eta^{-1}\tau)$ is maximized for even $n$ by four permutations: $\tau= \eta, \eta^{-1}$ and $\tau_{ES}= \{(12)(34)...(n-1~n),~ (23)(45)...(n1)\}$, and for odd $n$ by two permutations: $\tau= \eta, \eta^{-1}$. 

From the above, we then find: 
\begin{enumerate} 
\item For $c<\frac{1}{2}$, from \eqref{abeta1} $\tau=e$ is the dominant permutation for $\sA(\tau)$ for all $n\geq 3$, with $\log \Zpt$ given by~\eqref{er}. 
Since $\tau=e$ is also the dominant permutation for $\sB(\tau)$, we have 
\be\label{fin_un}
\begin{gathered} 
R_n(A_1, A_2) = 0, \quad n\geq 3, \quad \quad
\lim_{n\rightarrow 2} R_n(A_1, A_2) = 0, 
\end{gathered}
\ee
This is the natural generalization of the NE phase for the $R_n$. 

\item For $c>\ha, ~ \lambda>\frac{1}{2c}$, from \eqref{abeta2} the dominant permutation for $\sA(\tau)$ is $\eta$,
with $\log \Zpt$ given by~\eqref{er1} and
\begin{gather} 
\label{ehvs}
R_n(A_1, A_2) = \begin{cases} V_{A_2} \frac{n f(\beta)- f(n\beta)}{n-1} = S^{(\rm eq)}_{n, A_2} & n\geq 3, n \text{ odd}\\
V_{A_2} \frac{n f(\beta)- 2f(\frac{n}{2}\beta)}{n-2} = S^{(\rm eq)}_{{n \ov 2}, A_2} & n\geq 4, n \text{ even}
\end{cases} 
\\
\lim_{n\rightarrow 2} R_n(A_1, A_2) = S^{\rm (eq)}_{A_2} (\beta) ,
\end{gather} 
From analytic continuation, $\sE$ is given by~\eqref{ksn1}. 
This may naturally be interpreted as the finite temperature generalization of the maximally entangled (ME) phase. 
Notice in~\eqref{ehvs} for even $n$, $S^{(\rm eq)}_{{n \ov 2}, A_2} (\b) $ appears instead of $S^{(\rm eq)}_{n, A_2} (\b) $ as in the third line of~\eqref{nume1} for Renyi mutual information.

For $c>\ha, \lambda< 1- \frac{1}{2c}$, the dominant permutation is $\eta^{-1}$, and we get similar expressions with $A_2$ replaced by $A_1$. 

\item For $c>1/2$, $1-\frac{1}{2c}< \lambda < \frac{1}{2c}$, the dominant permutations now appear to depend on values of $c$ and $\lam$, and are model-dependent. 
We present the phase diagrams for examples~\eqref{example_disc} and~\eqref{example_cont} based on the behavior of even $n$ in Fig.~\ref{fig:phaD} (a) and (b) respectively.
In addition to the NE and ME phases discussed in items 1 and 2 above, there are also the following regions in the phase diagrams: 

 
\begin{figure}[] 
\includegraphics[width=8cm]{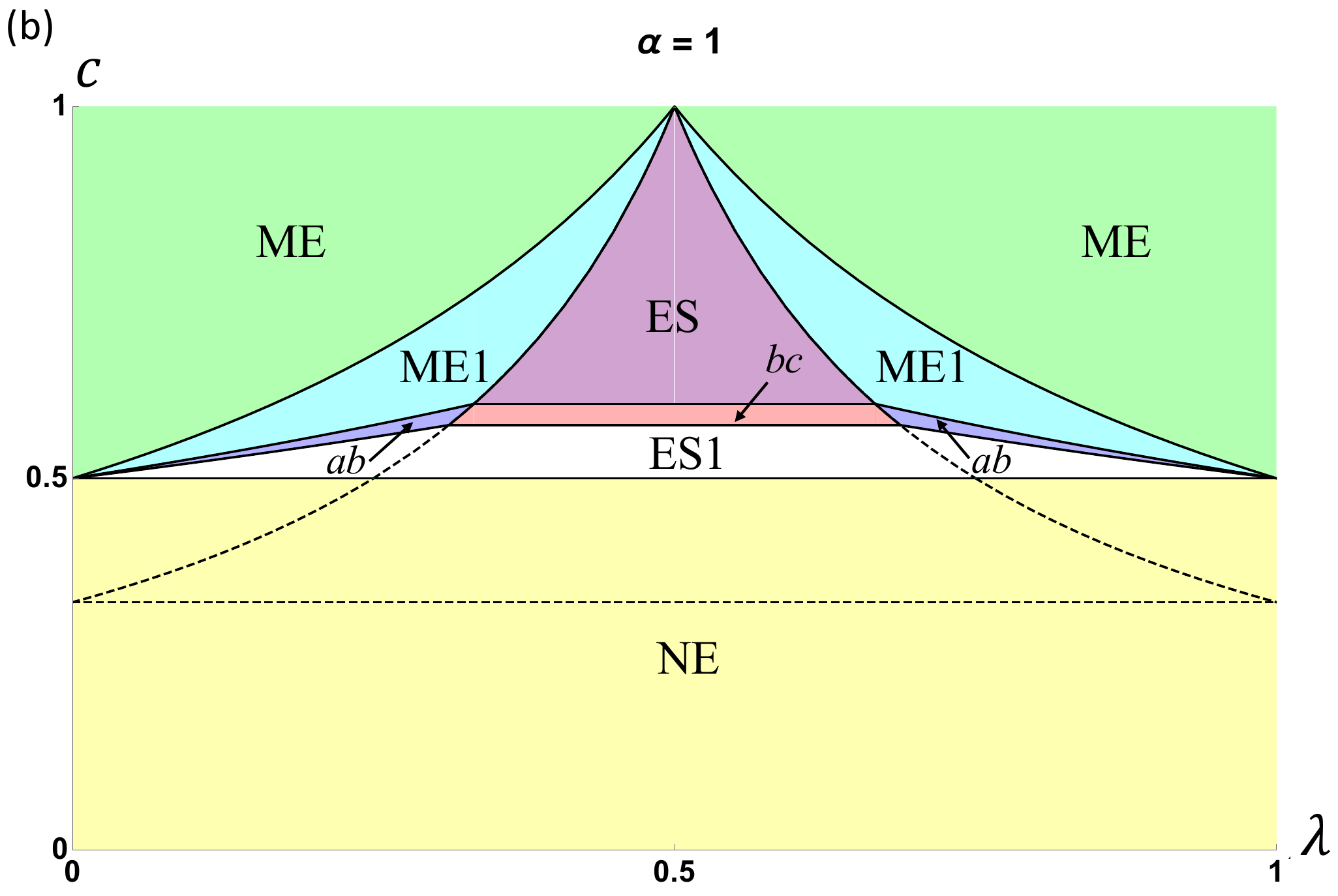} \includegraphics[width=8cm]{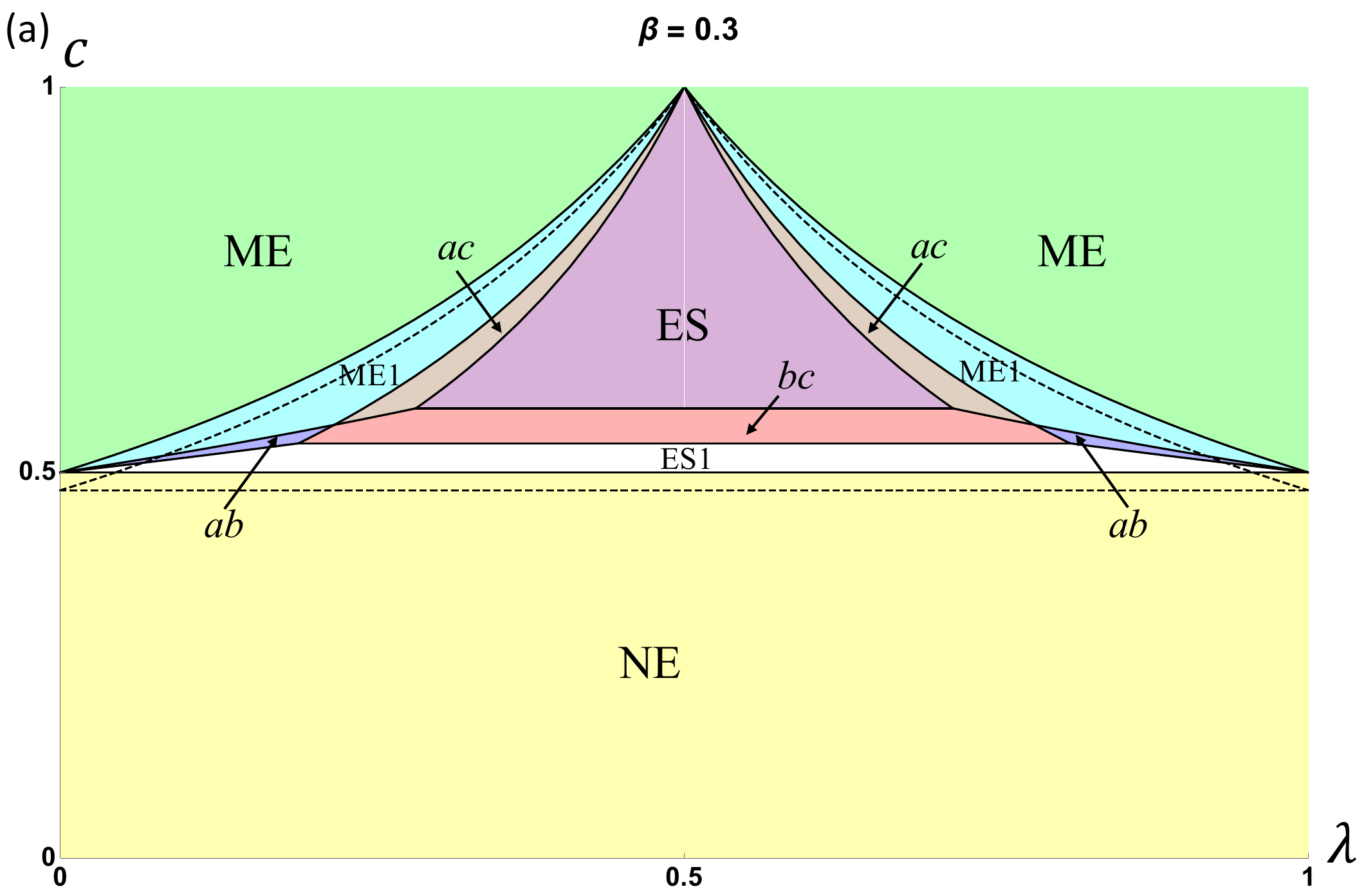} 
\caption{Phase diagrams for the canonical ensemble in a homogeneous system. (a) shows the case with $f(\beta)$ as in \eqref{example_disc}, and (b) shows the case with $f(\beta)$ in \eqref{example_cont}. Note that in (a), in addition to the $ab$, $bc$ and $ac$ regions, there is also a small $abc$ region near the intersections of these regions which is not shown explicitly in the figure. The dashed lines show the prediction for the phase transition in the logarithmic negativity from the naive analytic continuation prescription, where we set $\sE_{ES}=\sE_{NE}$ or $\sE_{ES}=\sE_{ME}$ in \eqref{ksn}-\eqref{ksn2}.} 
\label{fig:phaD}
\end{figure} 
 
\begin{enumerate} 
\item ME1 region in Fig.~\ref{fig:phaD} where $\tau_m = \eta$ or $\tau_m = \eta^{-1}$ dominates for all $n$. The behavior of Renyi and logarithmic negativities are the same as~\eqref{ehvs} and~\eqref{ksn1}. But note that the mutual information in this region are given by the second line of~\eqref{nume} rather than the third line or fourth line. So the mutual information and Renyi mutual information in this region differ from those of ME in item 2, and they should be considered as describing distinct phases. We refer to this region as the ME1 phase.
It is curious that in this region, the negativities behave like those in the maximally entangled phase, while the mutual informations behave like those for the entanglement saturation phase. In contrast to the relation \eqref{rjj} for pure states, we have for even $n$ 
\be 
R_n (A_1, A_2) > \frac{1}{2} I_{n/2}(A_1, A_2) \ .
 \label{egi}
\ee

\item ES1 region in Fig.~\ref{fig:phaD} where $\tau_m = e$ dominates for all $n\geq 3$. But since for $c>\ha$ the dominant permutation for $\sB(\tau)$ is $\tau=\eta$, the values of $R_n$ are different from those in~\eqref{fin_un}, 
\bega 
R_n(A_1, A_2) = \bca {n-1 \ov n-2} (S_{n,A}^{(\rm eq)} - S^{\rm (eq)}_{n, B}) & n \, {\rm even} \cr
S_{n,A}^{(\rm eq)} - S^{\rm (eq)}_{n, B} & n \, {\rm odd} 
\eca , \quad n\geq 3 \ 
\label{rn_ppt2}
\\
\log \sZ_n^{\rm (PT)} =\begin{cases} - (n-1) S_{n,A}^{(\rm eq)} 
 & n \geq 3 \\ 
- S^{\rm (eq)}_{2,B} & n = 2
\end{cases} \ .
\label{bwnq}
\end{gather} 
We cannot obtain $\sE(A_1, A_2)$ by analytic continuation from the above expressions, as the analytical continuation of~\eqref{bwnq} for even $n\geq 4$ to $n=2$ already gives the wrong expression, not to mention to $n=1$. 
The mutual information of this region is given by the second line of~\eqref{nume}, which is also nonzero. So this should describe a distinct phase from the NE phase of item 1. Since ~\eqref{rn_ppt2}--\eqref{bwnq} and the Renyi mutual informations in this phase only depend on $A$, and not on $A_1, A_2$, we will refer to it as the ES1 phase. 
For $n=2$, there is no ES1 region.

\item ES region in Fig.~\ref{fig:phaD} where $\tau_m = \tau_{ES}$ for all $n$ with $\log \Zpt$ given by~\eqref{fin_es}. We find 
\begin{gather} 
R_n(A_1, A_2) = \begin{cases} 
\ha S^{\rm (eq)}_{{n-1 \ov 2}, A}(\beta) - S^{\rm (eq)}_{n, B} (\b) + \ha S^{\rm (eq)}_{2, B} (\b) 
 & n\geq 3, n \text{ odd}\\
\frac{1}{2} (S^{\rm (eq)}_{{n \ov 2}, A}(\beta)- S^{\rm (eq)}_{{n \ov 2}, B}( 2\beta)) & n\geq 4, n \text{ even}
\end{cases} 
\\
\lim_{n\rightarrow 2} R_n(A_1, A_2) = \ha S_A^{(\rm eq)} (\b) -\ha S^{\rm (eq)}_B (2 \b) 
\end{gather} 
and the logarithmic negativity is given by~\eqref{ksn2}. 
Notice all quantities depend only on properties of $A$, not on $A_1$ or $A_2$. 
We can thus identity this as the finite temperature generalization of the ES phase. 
For even $n$ we also have \eqref{egi}. 

\item In the remaining regions of Fig.~\ref{fig:phaD}, the set of dominant permutations now depends on $n$. For example, in the region denoted by $ac$ there exists a critical value $n_c$ for $n \leq n_c$, the dominant permutations are that of item (a), while for $n > n_c$, the dominant permutations are those of item (c). Similarly with $bc$ and $ab$ regions. There is in general also an $abc$ region where all three options appear at a given $c, \lambda$ for different $n$. These regions also give potentially new phases. Clearly in these ``mixed'' regions, there is no way to perform analytic continuation in $n$ to obtain $\sE (A_1, A_2)$. 

In the above, we have assumed that for a region if a given set of permutations dominate for all $n$ and there is no apparent contradiction, we can analytically continue $n$ to obtain the logarithmic negativity. But we should caution that it is not warranted such a procedure always gives the correct answer.

\end{enumerate}

 \item As the temperature decreases, the ES phase region shrinks. 


\een

The $n$-dependent transition lines for $R_n$ are shown for a few examples of even and odd $n$ in Figs.~\ref{fig:n_disc} and \ref{fig:n_cont}. Fig.~\ref{fig:n_disc} shows the case of $f(\beta)$ as in \eqref{example_disc}, at two different values of $\beta$. Fig.~\ref{fig:n_cont} shows the case of $f(\beta)$ in \eqref{example_cont}, where the transition lines turn out to be independent of $\beta$, at two different values of the parameter $\alpha$. Note that there are qualitative differences between the diagrams for even and odd $n$. In particular, for sufficiently large $\beta$ or $\alpha$, the odd $n$ cases have no region where $\tau_{ES}$ dominates.

\begin{figure}[] 
\includegraphics[width=8cm]{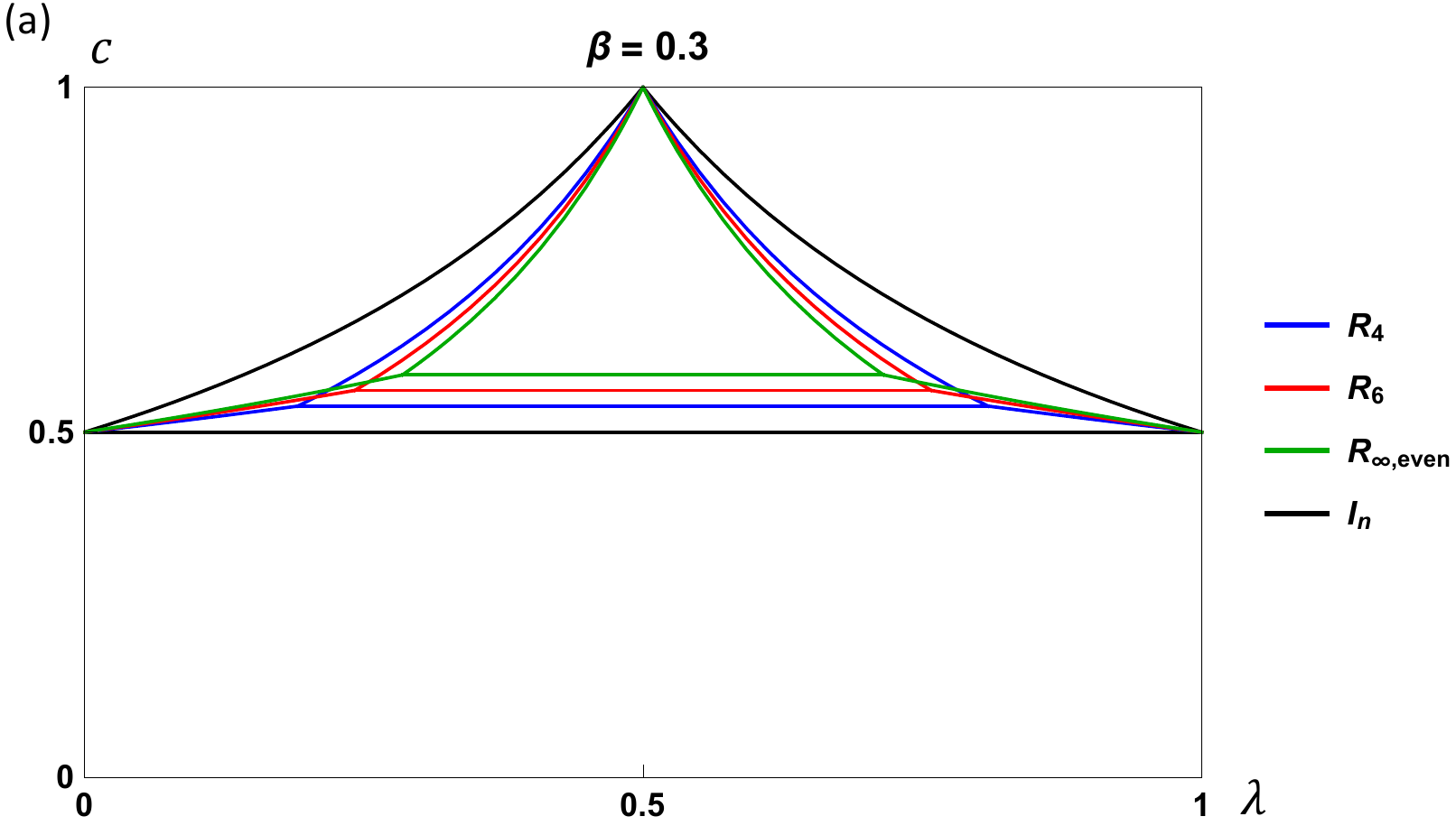} 
\includegraphics[width=8cm]{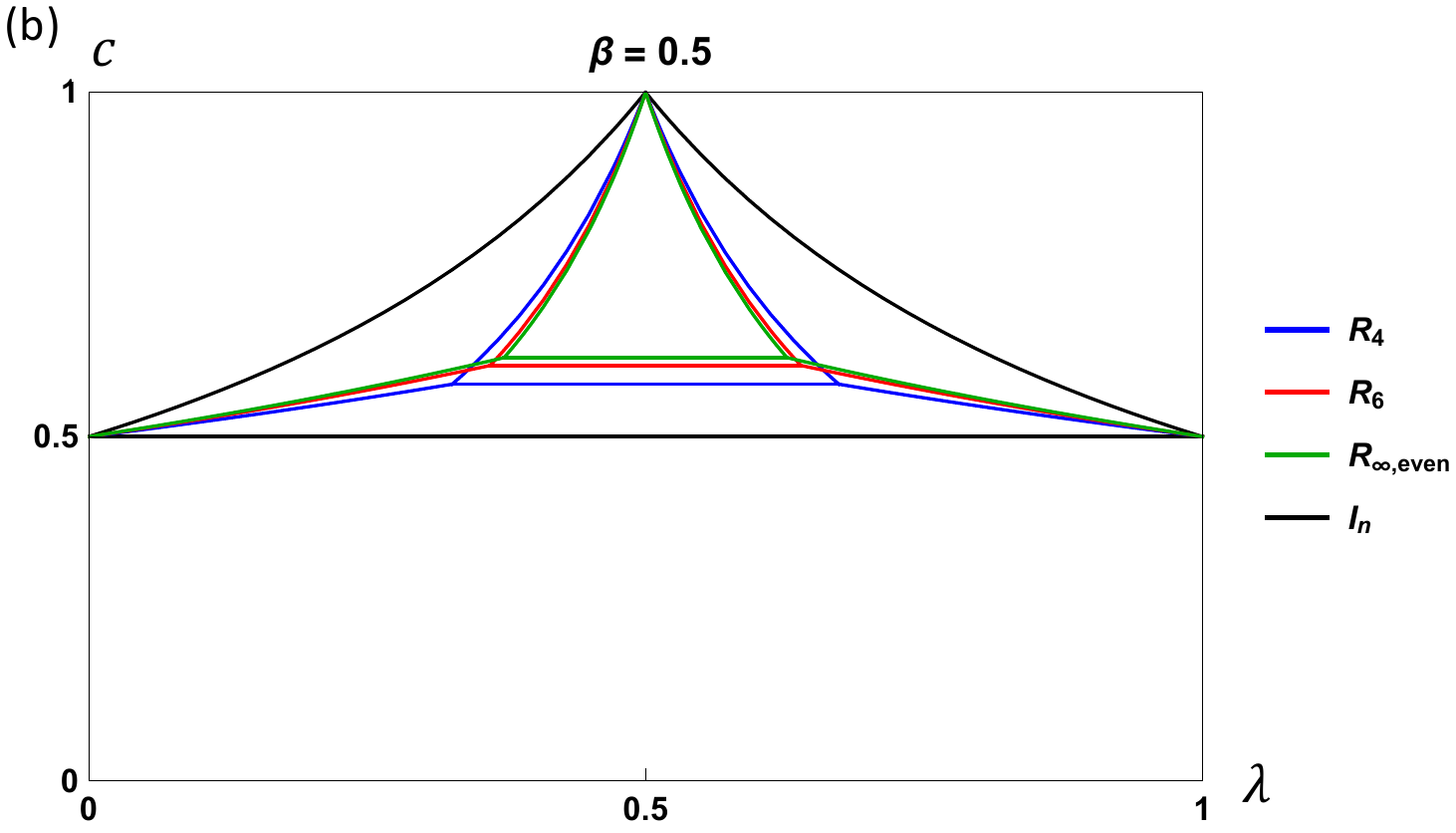} 
\includegraphics[width=8cm]{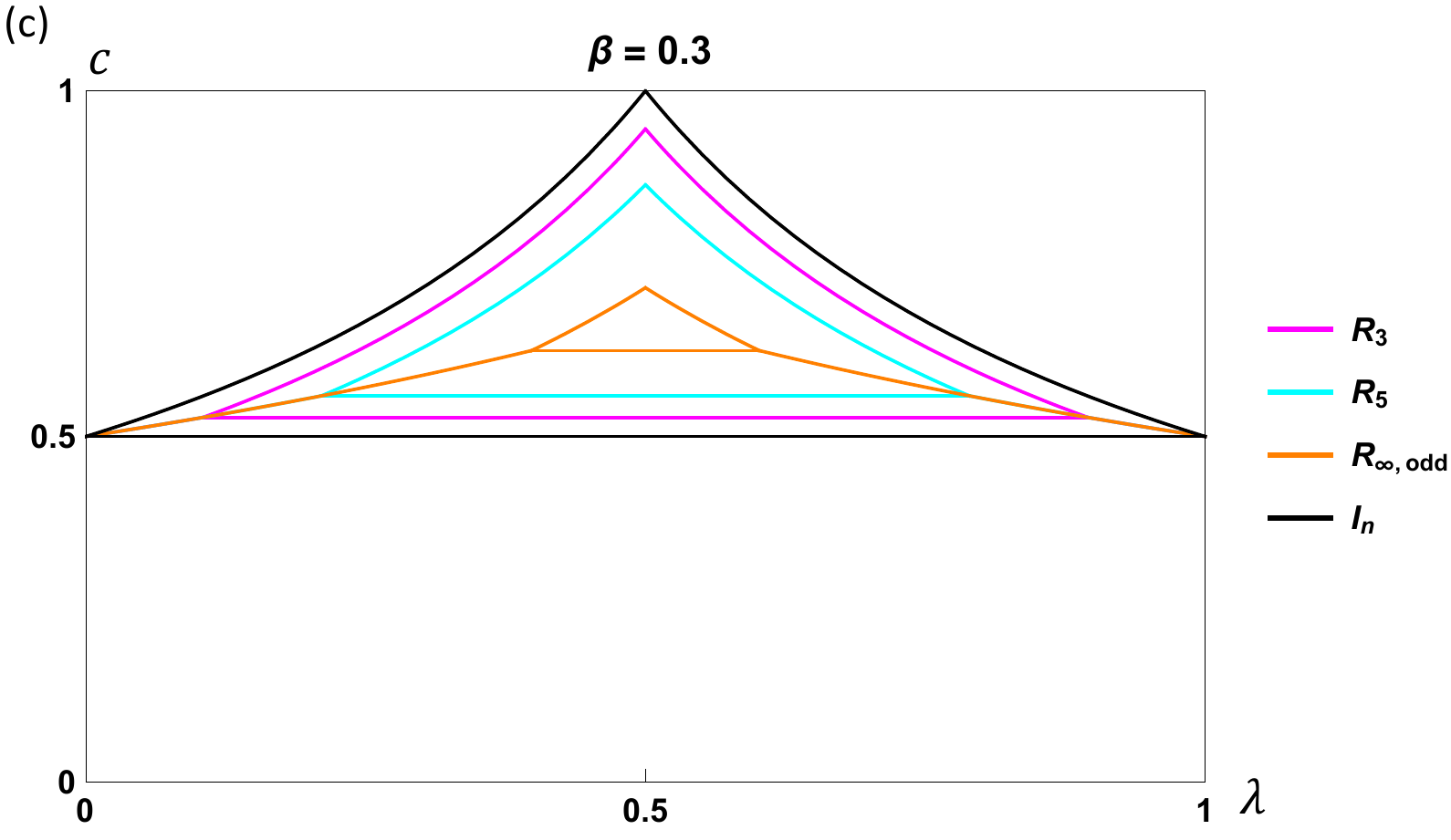} 
\includegraphics[width=8cm]{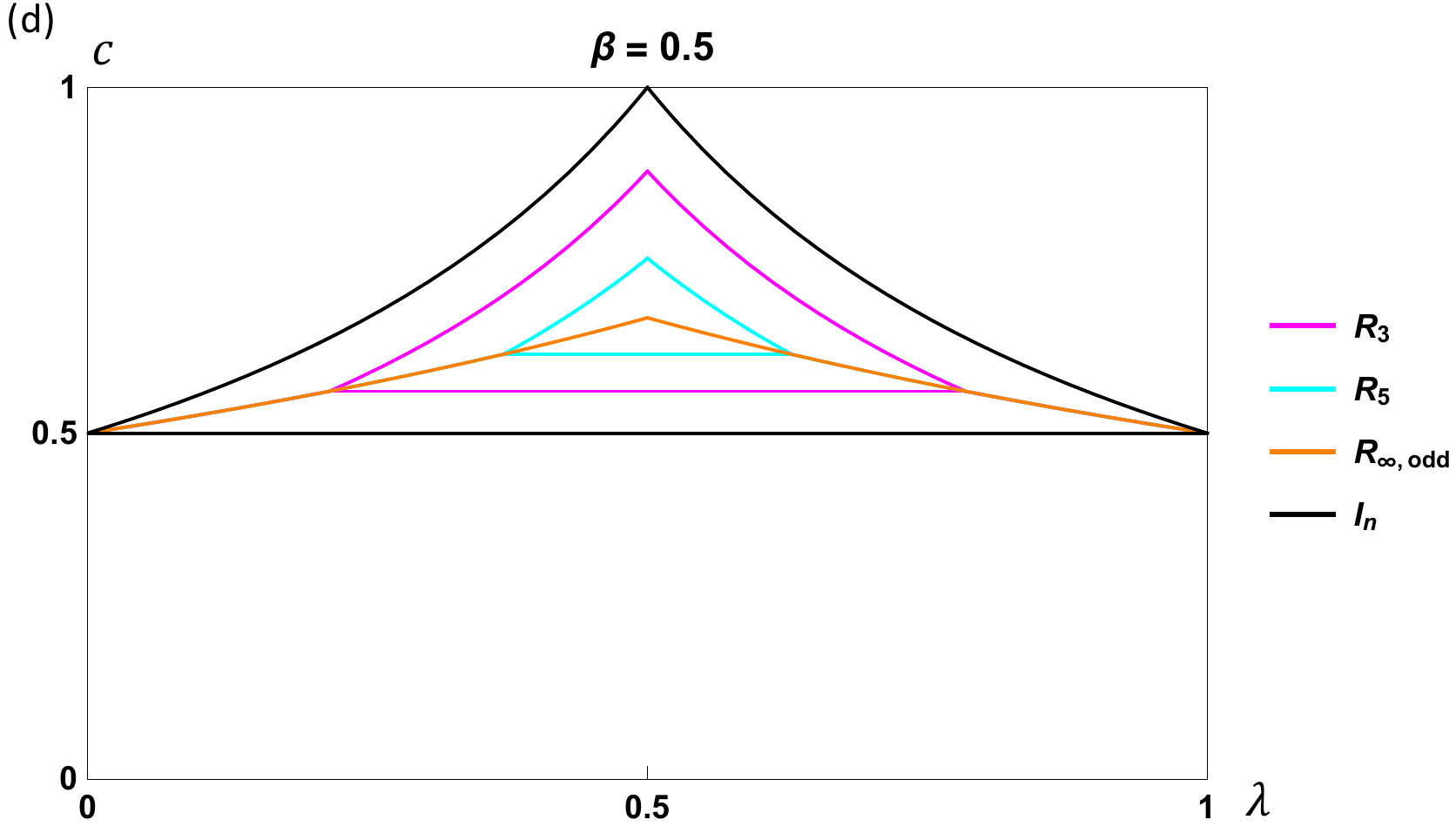} 
\caption{Transition lines for $R_n$ at two different values of $\beta$ for $f(\beta)$ in \eqref{example_disc}.} 
\label{fig:n_disc}
\end{figure} 

\begin{figure}[] 
\includegraphics[width=8cm]{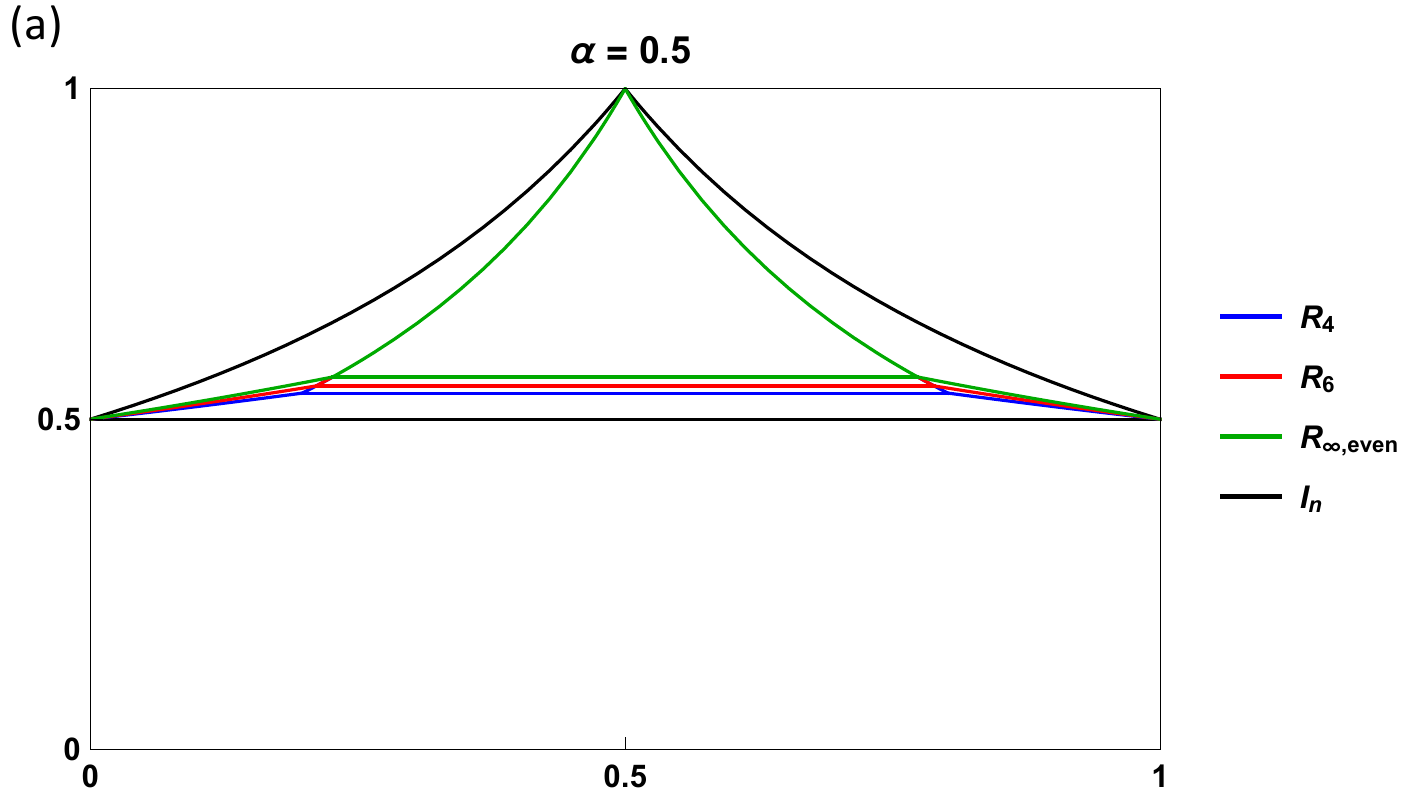} 
\includegraphics[width=8cm]{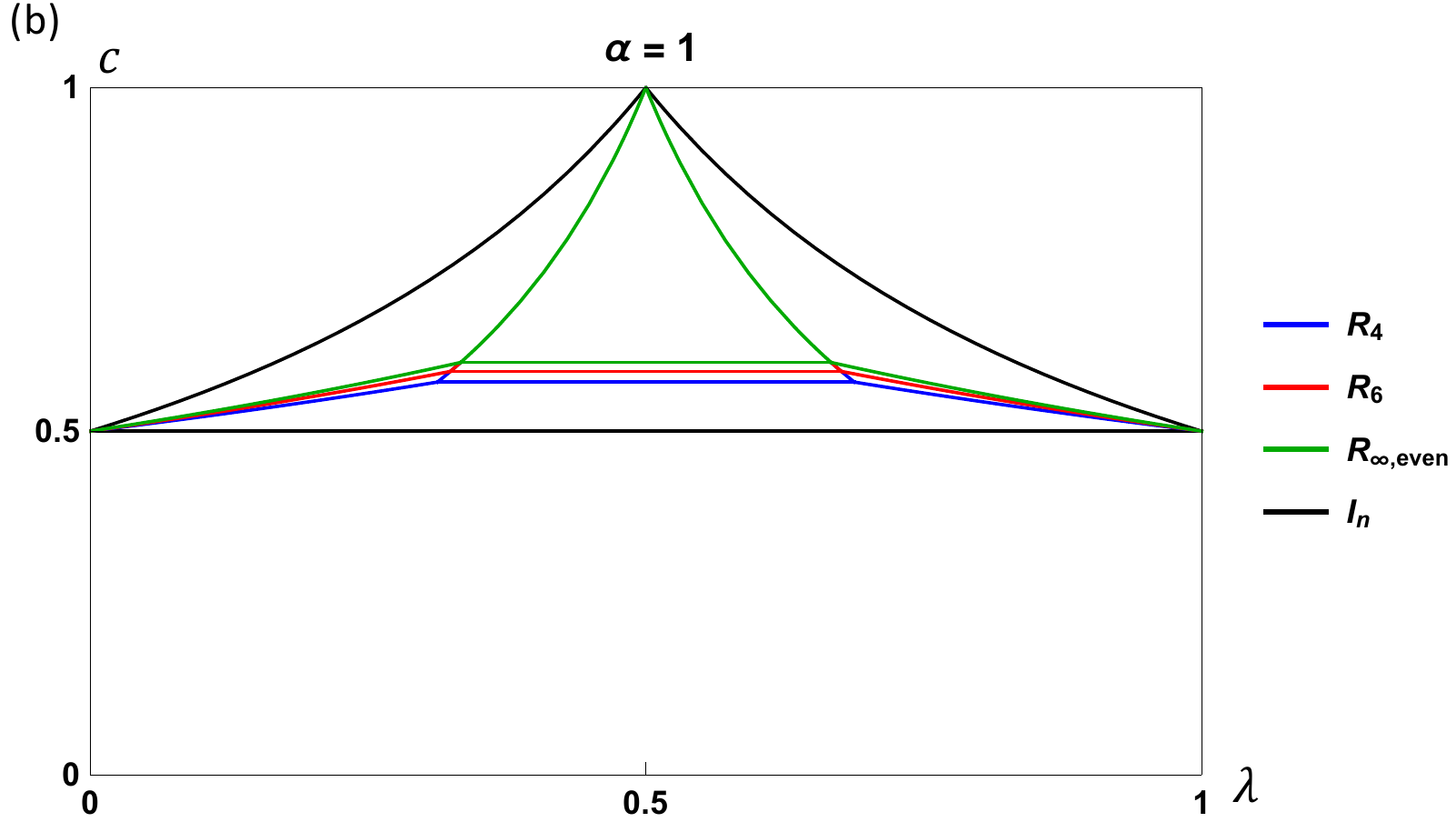} 
\includegraphics[width=8cm]{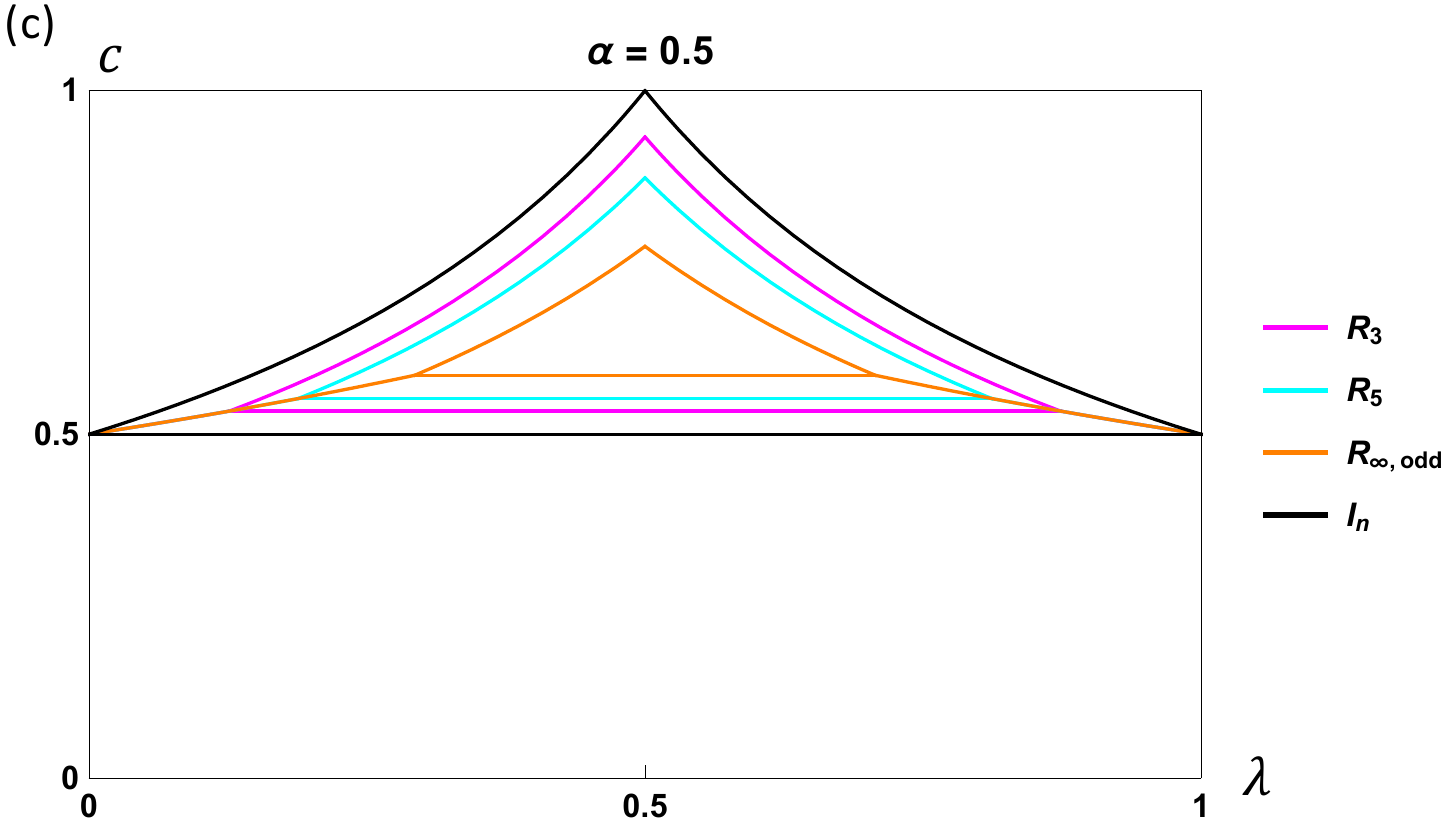} 
\includegraphics[width=8cm]{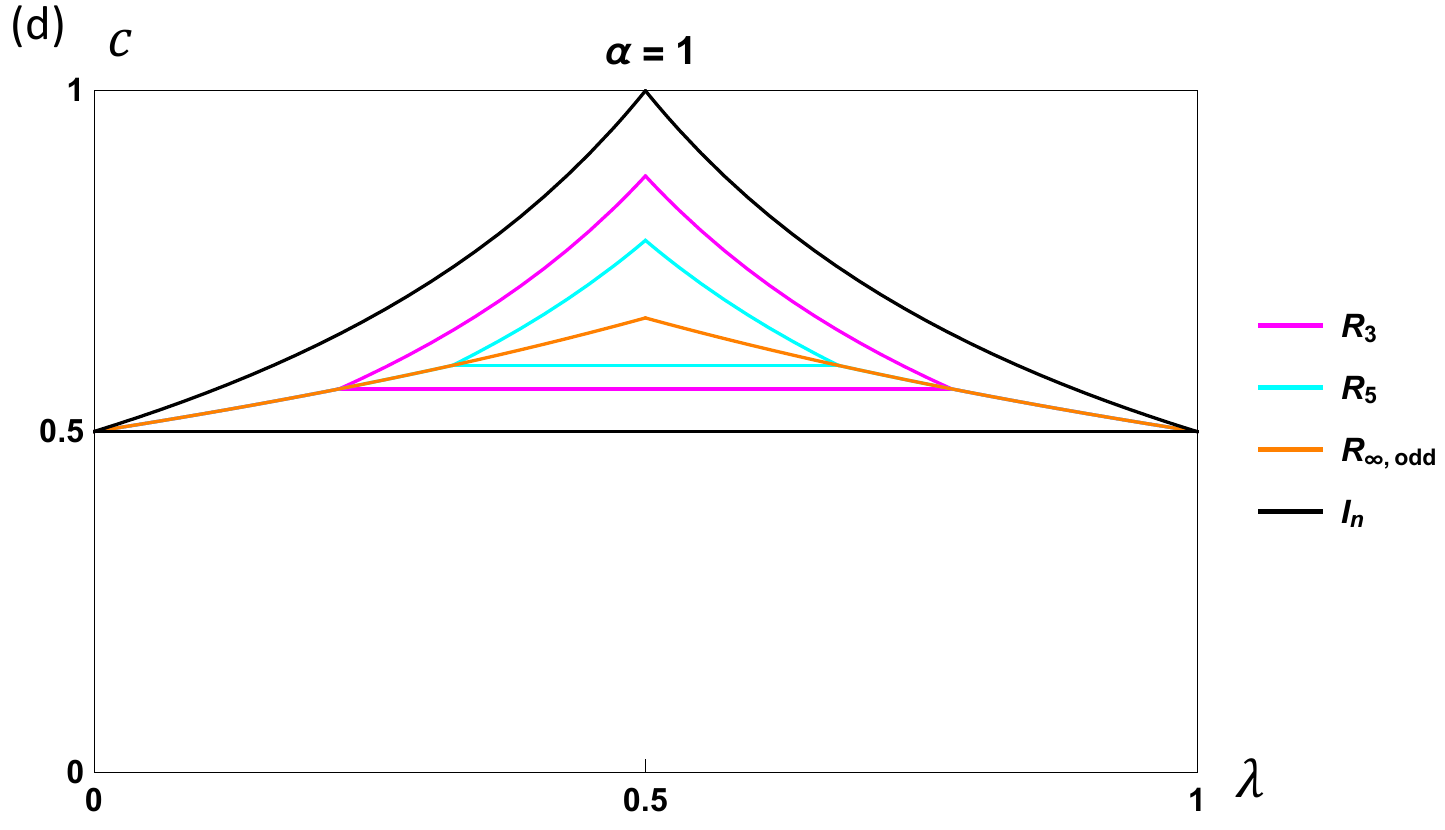} 
\caption{Transition lines for $R_n$ at two different values of $\alpha$ for $f(\beta)$ in \eqref{example_cont}.} 
\label{fig:n_cont}
\end{figure}

\subsubsection{$A$ at infinite temperature} \label{sec:Binf} 

As another illustration, here we consider the case where $A$ is at infinite temperature, while $B$ is at a finite temperature $\b$. 
We find numerically for $n=4, 6$ that the transition the dominant permutation contributing to $\sZ_n^{\rm (PT)}$ is always one out of $e$, $\tau^{\ast}$, and $\eta, \eta^{-1}$, and assume that this is true for all $n$. The corresponding expressions for $\sZ_n^{\rm(PT)}$ and analytically continued expressions for $\sE$ are given by \eqref{er}-\eqref{fin_es} and \eqref{ksn}-\eqref{ksn2}. The phase transitions for both the mutual information and the Renyi negativities are $n-$dependent in this case, and are shown in Fig.~\ref{fig:can_infa}. 

These phase diagrams have a similar structure for both discrete and continuum examples \eqref{example_disc} and \eqref{example_cont}. The phase diagrams for both $R_n$ and $I_n$ are $n$-dependent in this case. However, the transition from dominance of $\tau=e$ to $\tau=\tau_{ES}$ for all $R_n$ happens along the same line, indicated as the purple dashed line in the figure. This coincides with the transition in $I_2$ from NE to ES, and takes place at a smaller value of $c$ than the corresponding transition in $I$. As explained in Appendix \ref{app:mic_can}, the transition in $\sE$ obtained from the semicircle approximation in the resolvent also lies along the same line. More explicitly, in the semi-circle approximation, the logarithmic negativity is given by
\begin{align}
 {\cal E}(A_1,A_2) &=
\log \left(\frac{2}{\pi} \sin^{-1}\left[ \ha \sqrt{\frac{e^{S^{\rm (eq)}_{2,B}}}{d_A}} \, \right]+
    \frac{ \left(\sqrt{\frac{e^{S^{\rm (eq)}_{2,B}}}{d_A}}+8 \sqrt{\frac{d_A}{e^{S^{\rm (eq)}_{2,B}}}} \right) \sqrt{ 1- \frac{1}{4}{\frac{e^{S^{\rm (eq)}_{2,B}}}{d_A}} }}{3\pi }
    \right), 
    \\
    &\simeq
    \begin{cases}
      0 , & \log d_A < S^{\rm (eq)}_{2, B}
      \\
      \frac{1}{2}\left(\log d_A - S^{\rm (eq)}_{2,B}\right) + \log \frac{8}{3\pi}, & \log d_A > S^{\rm (eq)}_{2, B}
    \end{cases}
\label{eq:logneg}
\end{align}
Comparison between the mutual information and negativity is shown in Fig.~\ref{fig:canon_neg_vs_mi}. Here, we explicit see a regime before the Page time when the negativity is extensive and hence much larger than the mutual information.

\begin{figure}[!h] 
\begin{center}
\includegraphics[height=4.2cm]{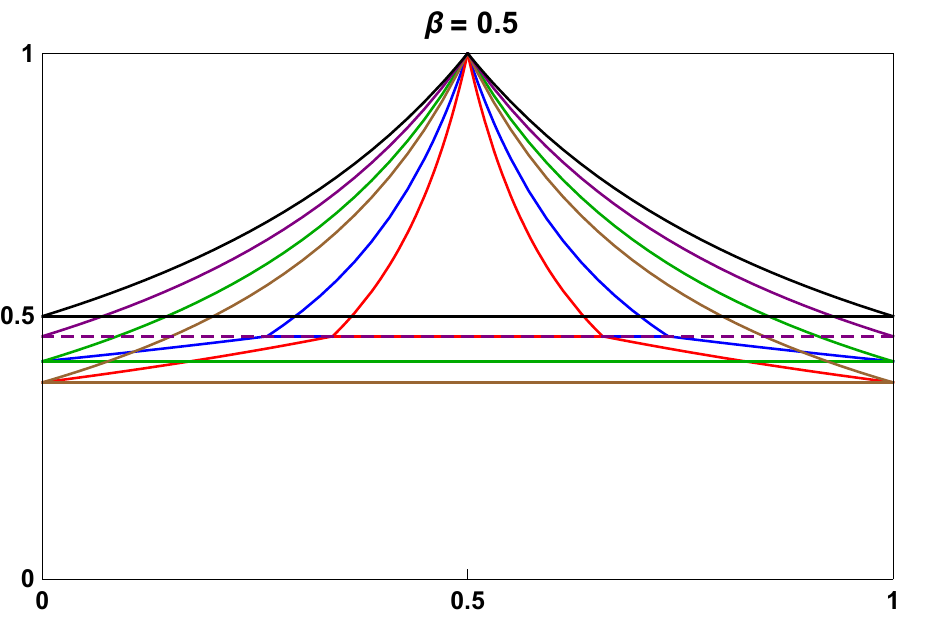}
\includegraphics[height=4.2cm]{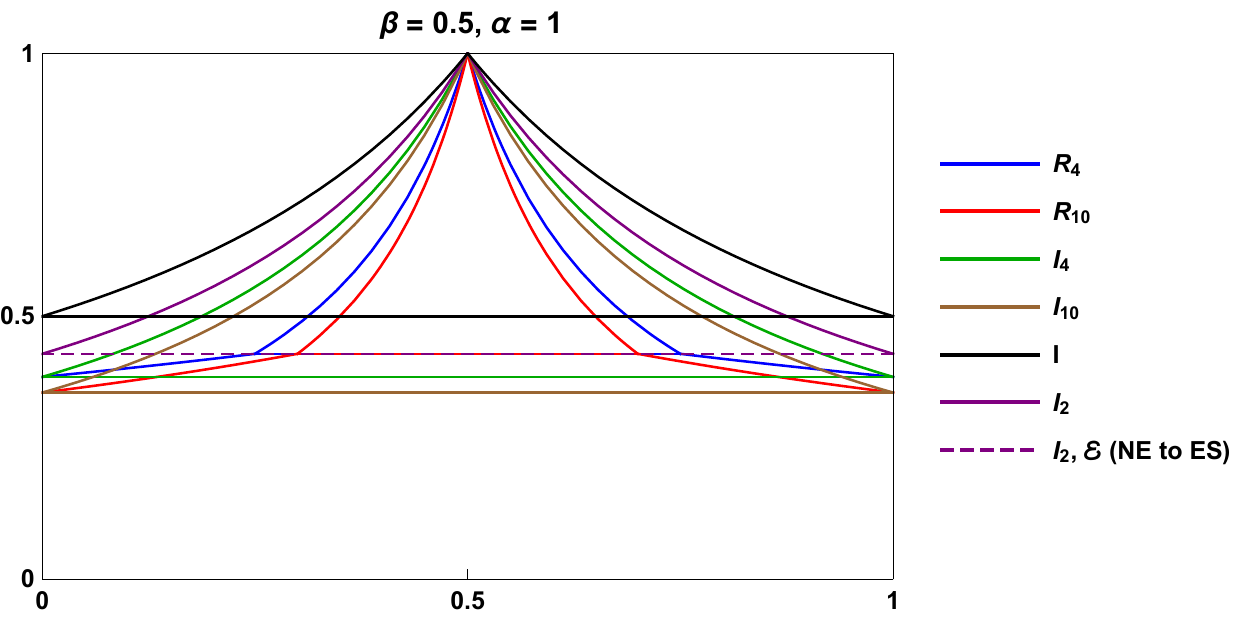}
\end{center}
\caption{Phase transition lines for $R_n$ and $I_n$ for a few different $n$ for $f_B$ as in \eqref{example_disc} and \eqref{example_cont} are shown on the left and right respectively.}
\label{fig:can_infa}
\end{figure}

\begin{figure}[!h] 
\begin{center}
\includegraphics[width = .6\textwidth]{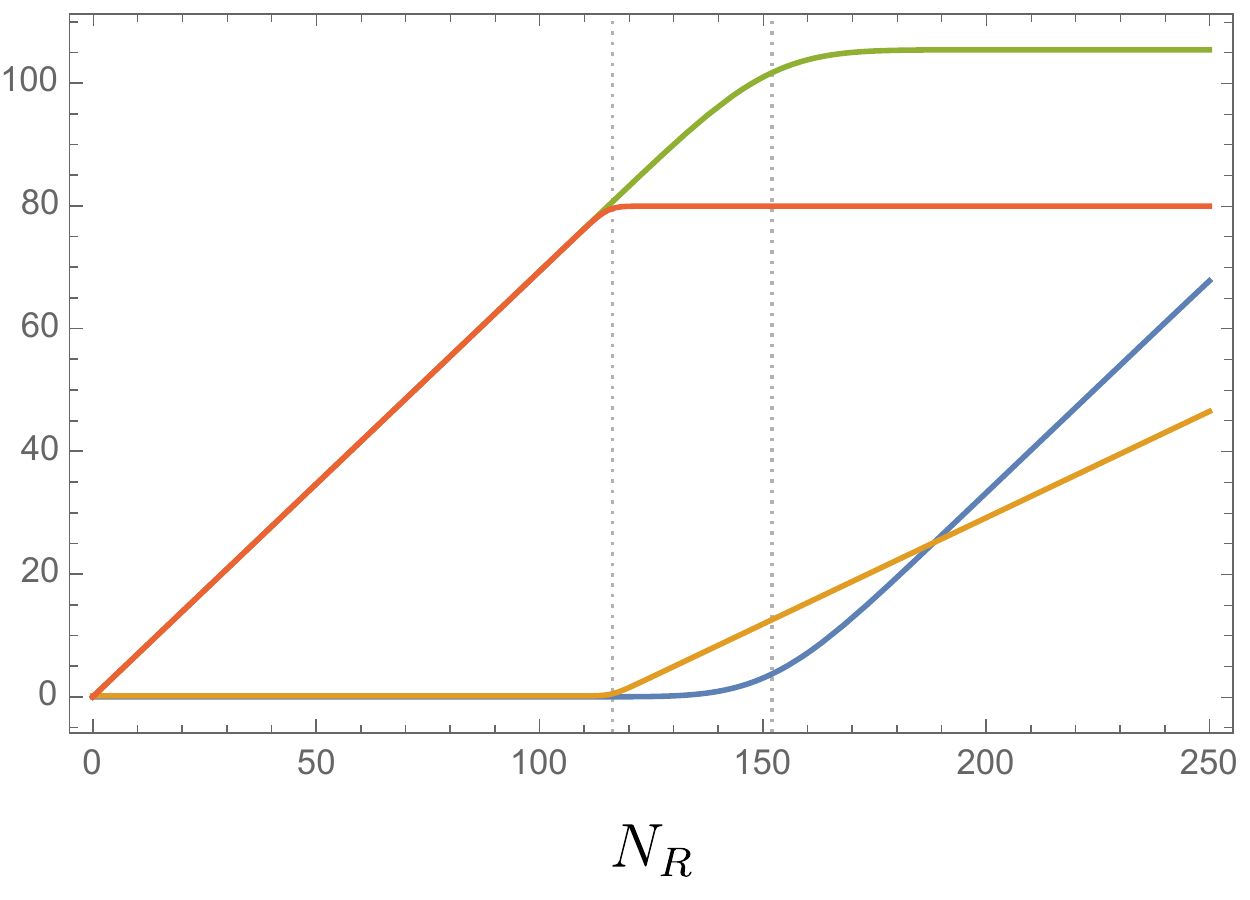}
\end{center}
\caption{The mutual information (blue), negativity (orange), von Neumann entropy of $R$ (green), and second R\'enyi entropy of $R$ (red) are shown when $B$ is in the canonical ensemble and $A$ at infinite temperature. While the mutual information within $A$ becomes volume-law when the von Neumann entropy of $A$ saturates, the negativity becomes volume-law when the second R\'enyi entropy saturates, a significantly earlier time. We also note that the finite temperature corrections around the transition point are significantly larger for the mutual information than they are for negativity, which has a sharper transition. We take $V_B = 100$ and $\beta = 1/2$ for a Cardy-like density of states for $B$, $\rho(E) = e^{V_B \sqrt{\frac{E}{V_B}}} $.}
\label{fig:canon_neg_vs_mi}
\end{figure}

\subsection{Microcanonical ensemble} \label{sec:micro}

\subsubsection{General description} 

In cases where the initial state is supported on a narrow interval $I_{E, \Delta} = [E- \Delta, E+\Delta]$ of energies, we can take the effective identity operator for the equilibrium approximation to be the projector onto the eigenstates in this microcanonical interval. Again assuming that the interaction terms between different subsystems $A_1, A_2$ and $B$ give small contributions to the energy compared to the terms involving only degrees of freedom within the subsystems, we can approximate this effective identity operator as 
\be 
\sI_{E} = \sum_{E_n \in I_{E, \Delta}} \ket{n}\bra{n} \approx \sum_{E_p^{A_1}+E_q^{A_2}+E_r^{B}\in I_{E, \Delta}} (\ket{p}\bra{p})_{A_1} \otimes (\ket{q}\bra{q})_{A_2} \otimes (\ket{r}\bra{r})_{B}. 
\label{full_mic}
\ee
where $\ket{p}, \ket{q}$ and $\ket{r}$ are respectively eigenstates of $H_{A_1}$, $H_{A_2}$ and $H_B$ with energies $E_p^{A_1}$, $E_q^{A_2}$ and $E_r^{B}$. Below we will denote $N_E := \text{Tr}[\sI_{E}]$. 
The contributions to the equilibrium approximation for $\sZ_n^{(\rm PT)}$ from general permutations $\tau$ with this choice of $\sI_{\alpha}$ in a general homogenous system are complicated to evaluate, and in particular do not have simple expressions in terms of the cycle numbers and lengths of various permutations like the canonical ensemble expressions \eqref{eq_canonical}. Some examples are shown in Appendix \ref{app:mic_hom}. We consider two simplifying cases below, with $B$ or $A_2$ at infinite temperature. 
These examples are useful as they allow a calculation of the logarithmic negativity through the resolvent, as explained in Appendices \ref{app:mic_b} and \ref{app:mic_a2}, revealing new features in their phase diagrams which do not have an analog at infinite temperature and cannot be deduced from the phase diagrams for the Renyi negativities. In both examples, we find that the logarithmic negativity becomes extensive in a regime where the mutual information is sub-extensive.

For concreteness, in both cases below, it will sometimes be useful to consider a specific example of the entropy density as a function of energy density in the part of the system with energy conservation, 
\be 
s(\epsilon) = g \sqrt{\epsilon}. 
\label{s_half}
\ee
where $g$ is a constant.

\subsubsection{$B$ at infinite temperature} 

Let us first assume that the total energy in $A$ is conserved and it equilibrates to the microcanonical ensemble, while $B$ equilibrates to infinite temperature. The effective identity operator is 
\be \label{mic_case1}
\sI_{\al} = \sum_{E^{A_1}_{a_1} + E^{A_2}_{a_2} \in I_{E, \Delta}} \ket{a_1}\bra{a_1} \otimes \ket{a_2}\bra{a_2} \otimes \mathbf{1}_B 
\ee
As we vary the volumes of the different subsystems, we fix the average energy density in $A$ to a value $\frac{E}{V_A} = \epsilon$, and the infinite temperature entropy density in $B$ to a value $s_0$. 

Note first that since $\sI_{\al}$ factorizes as $\sI_{A} \otimes \sI_{B}$ in this example, the Renyi mutual information is again given by the general expression \eqref{nume1}. Note that $S^{\rm (eq)}_{n, A}$ and $S^{\rm (eq)}_{n, B}$ for all $n$ in this example are equal to $S^{\rm (eq)}_{A}$ and $S^{\rm (eq)}_{B} = \log d_B$ respectively, so the transition from the first to the second line of \eqref{nume1} occurs at $S^{\rm (eq)}_{A} = \log d_B$ for all $n$. But note that in this example, the first line of \eqref{nume1} is not zero for $n\neq 1$: it is positive for $n<1$, and negative for $n>1$. 

The contributions to $\sZ_n^{\rm (PT)}$ for even $n\geq 2$ from different permutations are 
\bea\label{er_mic}
\log \sZ_n^{\rm(PT)}(e) &=& - (n-1) S_{A}^{\rm (eq)}\\
\label{er1_mic}
\log \Zpt (\eta) &=& 
- (n-1) \log d_B - (n-2) S^{(\rm eq)}_{{n \ov 2}, A_2} 
\\
\label{er2_mic}
\log \Zpt (\eta^{-1}) &=& 
- (n-1) \log d_B - (n-2) S^{(\rm eq)}_{{n \ov 2}, A_1} 
\\
 \log \Zpt (\tau_{ES}) &=& 
 -{n-2 \ov 2} S^{\rm (eq)}_{{n \ov 2}, A_1} -{n-2 \ov 2} S^{\rm (eq)}_{{n \ov 2}, A_2} - {n \ov 2} \log d_B\label{fin_es_mic}
\eea 
The derivation of these expressions is explained in Appendix \ref{app:mic_b}. On noticing that $S^{\rm (eq)}_{n, A}=S^{\rm (eq)}_{A}$ and $S^{\rm (eq)}_{n, B}=\log d_B$ for all $n$ in this example, and that the equilibrium Renyi entropies of different subsystems are additive in the canonical ensemble, we can see that these expressions in terms of the equilibrium entropies turn out to be the same as the contributions from the leading permutations in the canonical ensemble case, \eqref{er}-\eqref{fin_es}. From analytic continuation, the above expressions correspond to the following values of the logarithmic negativity: 
\bega \label{mksn}
\sE_{NE} = 0 , \\
\label{mksn1}
\sE_{ME} = S_{\ha, A_2}^{\rm (eq)} \quad \text{or} \quad \sE_{ME} = S_{\ha, A_1}^{\rm (eq)} , \\
\sE_{ES} = \ha \le(S^{\rm (eq)}_{{1 \ov 2}, A_1} + S^{\rm (eq)}_{{1 \ov 2}, A_2} - \log d_B \ri) \ .
\label{mksn2} 
\end{gather} 
Depending on the permutations $\tau_m$ and $\tau'_m$ that dominate in $\sZ_n^{\rm (PT)}$ and $\sZ_n^{(A)}$ respectively, the Renyi negativities for even $n$ can take three possible forms, 
\be 
R_n = \begin{cases} 
0 & \tau_m = e,~ \tau_m'= e \\
 \text{min}( S^{\rm eq}_{\frac{n}{2}, A_2} , S^{\rm eq}_{\frac{n}{2}, A_1}) & \tau_m = \eta^{-1},\eta,~ \tau_m'= \eta \\
\frac{1}{2}(S^{\rm eq}_{n/2, A_1} + S^{\rm eq}_{n/2, A_1} - \log d_B ) 
& \tau_m = \tau_{ES},~ \tau_m'= \eta 
\end{cases} 
\ee
Note that this example does not have an ES1 phase like \eqref{rn_ppt2} where $\tau_m = e,~ \tau_m'= \eta$ for $n>2$.
\begin{figure}[] 
\begin{center}
\includegraphics[width=10cm]{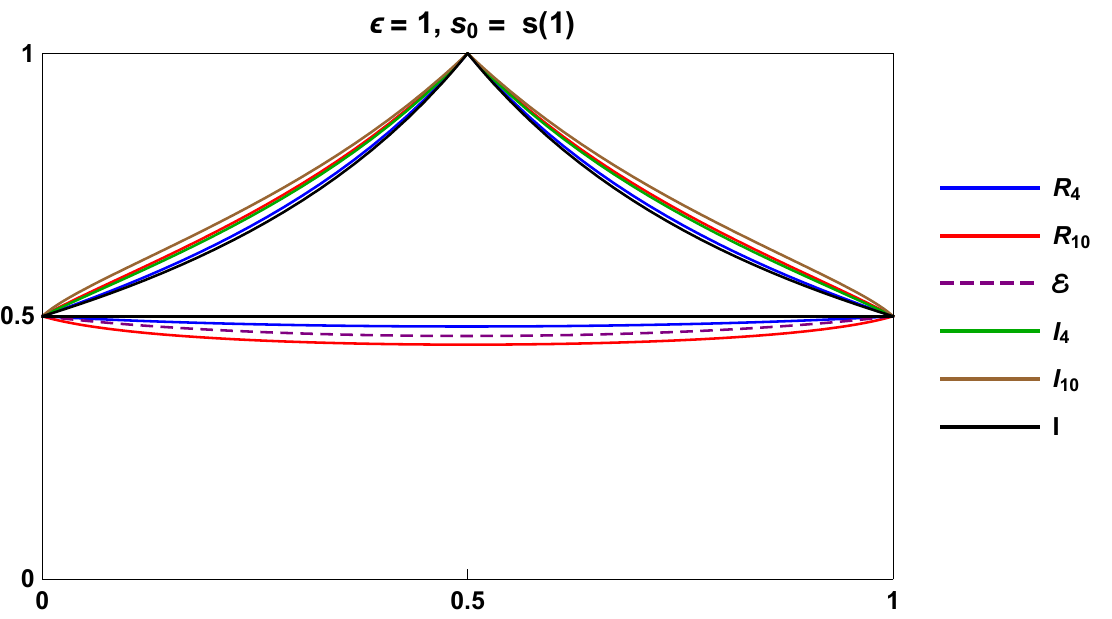}
\end{center}
\caption{Phase transition lines for $R_n$ and $I_n$ for $n=4, 10$. The transition from NE to ES for $\sE$ obtained from the semi-circle approximation is shown with the dashed purple line. The transition from the first to the second line of \eqref{nume1} for all Renyi mutual informations coincides with the black $n=1$ transition line in this case.}
\label{fig:mic_infb}
\end{figure} 
The $n$-dependent phase transition lines for $R_n$ and $I_n$ are shown in Fig.~\ref{fig:mic_infb}, taking the entropy density in $A$ to be of the form \eqref{s_half}. Analytic continuation of the condition $\Zpt(\tau_{ES})= \Zpt(e)$ suggests that the transition from the NE to the ES phase for the logarithmic negativity takes place when 
\be 
S^{\rm (eq)}_{1/2, A_1} + S^{\rm (eq)}_{1/2, A_2} = \log d_B \,.\label{inf_b_an}
\ee
This analytically continued phase transition line is indicated with the dashed purple line in Fig.~\ref{fig:mic_infb}. 
When this transition takes place, the mutual information is still sub-extensive from \eqref{nume}. As discussed in Appendix \ref{app:mic_b}, a computation of the resolvent in the ``semicircle approximation" for this case indicates that the naive analytic continuation of the phase transition \eqref{inf_b_an}, and the analytically continued value \eqref{mksn2} of the logarithmic negativity in the ES phase, are correct. However, our computation of the full resolvent in a different setup in the following subsection suggests that the semicircle approximation may in principle be insensitive to new phases that cannot be deduced by analytic continuation. Irrespective of whether such new phases also exist in this example, the conclusion that the negativity is extensive in a regime where the mutual information is sub-extensive holds. 

\subsubsection{$A_2$ at infinite temperature} 
 Consider an inhomogeneous example, where the system $A_2$ is taken to be at infinite temperature while $A_1 B$ is a homogenous system that satisfies energy conservation and equilibrates to the microcanonical ensemble. More explicitly, 
we take the effective identity operator to be 
\be 
\sI_{\alpha} = \mathbf{1}_{A_2} \otimes \sum_{E_p^{A_1} + E_r^B \in I_{E, \Delta}} (\ket{p}\bra{p})_{A_1} \otimes (\ket{r}\bra{r})_{B}. \label{mic_inf_temp}
\ee
As we vary the volumes of the different subsystems to find the phase diagram, we keep the average energy density in $A_1B$ fixed to some value $\frac{E}{V_{A_1}+V_B}=\epsilon$. We denote the infinite temperature equilibrium entropy density in $A_2$ by $s_0$. 

Using \eqref{erem}, we find that the Renyi mutual informations in this setup are given by 
\be 
I_n(A_1, A_2) = \label{in_mic2}
\begin{cases} 
 0 & S^{\rm (eq)}_{n, A_1} + \log d_{A_2} < S^{\rm (eq)}_{n, B} \\
 S^{\rm (eq)}_{n, A_1} +\log d_{A_2} - S^{\rm (eq)}_{n, B}, & S^{\rm (eq)}_{n, A_1} +\log d_{A_2} > S^{\rm (eq)}_{n, B}, ~ S^{\rm (eq)}_{n, A_1} < \log d_{A_2}+ S^{\rm (eq)}_{n, B}, \\
 &\log d_{A_2} < S^{\rm (eq)}_{n, A_1B} \\
 2 \log d_{A_2} & \log d_{A_2}> S^{\rm (eq)}_{n, A_1B} \\
 S_{n, A_1}^{\rm (eq)} + S_{n, A_1 B}^{\rm (eq)} - S_{n, B}^{\rm (eq)} & S^{\rm (eq)}_{n, A_1} > \log d_{A_2} + S^{\rm (eq)}_{n, B}
 \end{cases}
\ee
Note that this differs slightly from \eqref{nume1}, as the Renyi entropies for $n\neq 1$ in $A_1B$ are not additive in this example. For $I(A_1, A_2)$, we recover the general expression \eqref{nume}. The phase transitions for different $I_n$ are $n$-dependent in this example. 

We find that the dominant permutation in $\sZ_n^{\rm (PT)}$ for even $n$ is given by one of the choices $\tau= e$, $\tau=\eta, \eta^{-1}$ or $\tau=\tau_{ES}$ \footnote{We checked this numerically for $n=4$ and $n=6$ for the entropy density function in \eqref{s_half}.}, and the corresponding values are 
\begin{align} 
 \log Z_n^{\rm (PT)}(e) &= (1-n) \, (\log d_{A_2} + S_{n, A_1}^{\rm (eq)}) \label{mic2_e},\\
 \log Z_n^{\rm (PT)}(\eta^{-1}) &= S^{\rm (eq)}_{\ha, A_1} -(n-1) (S^{\rm (eq)}_B + S^{\rm (eq)}_{A_1} ) \label{mic2_n},\\
\log Z_n^{\rm (PT)}(\eta) &= (2-n)\log d_{A_2} +(1-n)S_{B, n}^{\rm (eq)} \label{mic2_n1},\\
 \log Z_n^{\rm (PT)}(\tau_{ES}) &= (1-\frac{n}{2}) \log d_{A_2} + S^{\rm (eq)}_{\frac{n}{n+2}, A_1}-\frac{n}{2}(S^{\rm (eq)}_{A_1}+S^{\rm (eq)}_{B}) .\label{mic2_es}
\end{align} 
From analytic continuation of the above expressions, we get the following forms of the logarithmic negativity: 
\bega \label{mbksn}
\sE_{NE} = 0 , \\
\label{mbksn1}
\sE_{ME} = \log d_{A_2}\quad \text{or} \quad \sE_{ME} = S_{\ha, A_1}^{\rm (eq)} , \\
\sE_{ES} = \ha \log d_{A_2} + S^{\rm (eq)}_{{1 \ov 3}, A_1} -\ha S^{\rm (eq)}_{A_1} - \ha S_{B}^{\rm (eq)} \ .
\label{mbksn2} 
\end{gather} 
We also find the following different forms of the Renyi negativities for even $n$ depending on the permutations $\tau_m$ and $\tau_m'$ that dominate in $\sZ_n^{\rm (PT)}$ and $\sZ_n^{(A)}$ respectively: 
\be 
R_n(A_1, A_2) = \begin{cases} 
\vspace{0.2cm}
0 & \tau_m = e,~ \tau_m'= e \\
\vspace{0.2cm}
\frac{n-1}{n-2}(S^{\rm (eq)}_{n, A} - S_{n, A}) & \tau_m = e,~ \tau_m'= \eta \\
\frac{1}{2} \log d_{A_2} + \frac{n}{2(n-2)}(S^{\rm (eq)}_{A_1}+S^{\rm (eq)}_{B}) & {} \\
\vspace{0.2cm}
- \frac{1}{n-2} S^{\rm (eq)}_{\frac{n}{n+2}, \, A_1} - \frac{n-1}{n-2} S^{\rm (eq)}_{n,B} & \tau_m = \eta^{-1},~ \tau_m'= \eta \\
\vspace{0.2cm}
S^{\rm (eq)}_{A_2} & \tau_m = \eta,~ \tau_m'= \eta \\
 \frac{(n-1)S^{\rm (eq)}_{1/n, A_1} - S^{\rm (eq)}_{1/2, A_1}}{n-2} & \tau_m = \tau_{ES},~ \tau_m'= \eta 
\end{cases} 
\ee
The phase transition lines for $R_n$ and $I_n$ for $n= 4, 6$ and $8$ are shown in Fig.~\ref{fig:phase_diagram_a2_inf}(a), where we have assumed that the entropy density in $A_1B$ is given by \eqref{s_half}. Qualitatively, this has a similar structure to the phase diagrams seen in previous cases, with analogs of each of the regions NE, ES1, ES, ME1, ME, $ab$, $bc$, $ac$ and $abc$ that we saw previously. 
\begin{figure}[]
\centering
 \includegraphics[height=4.4cm]{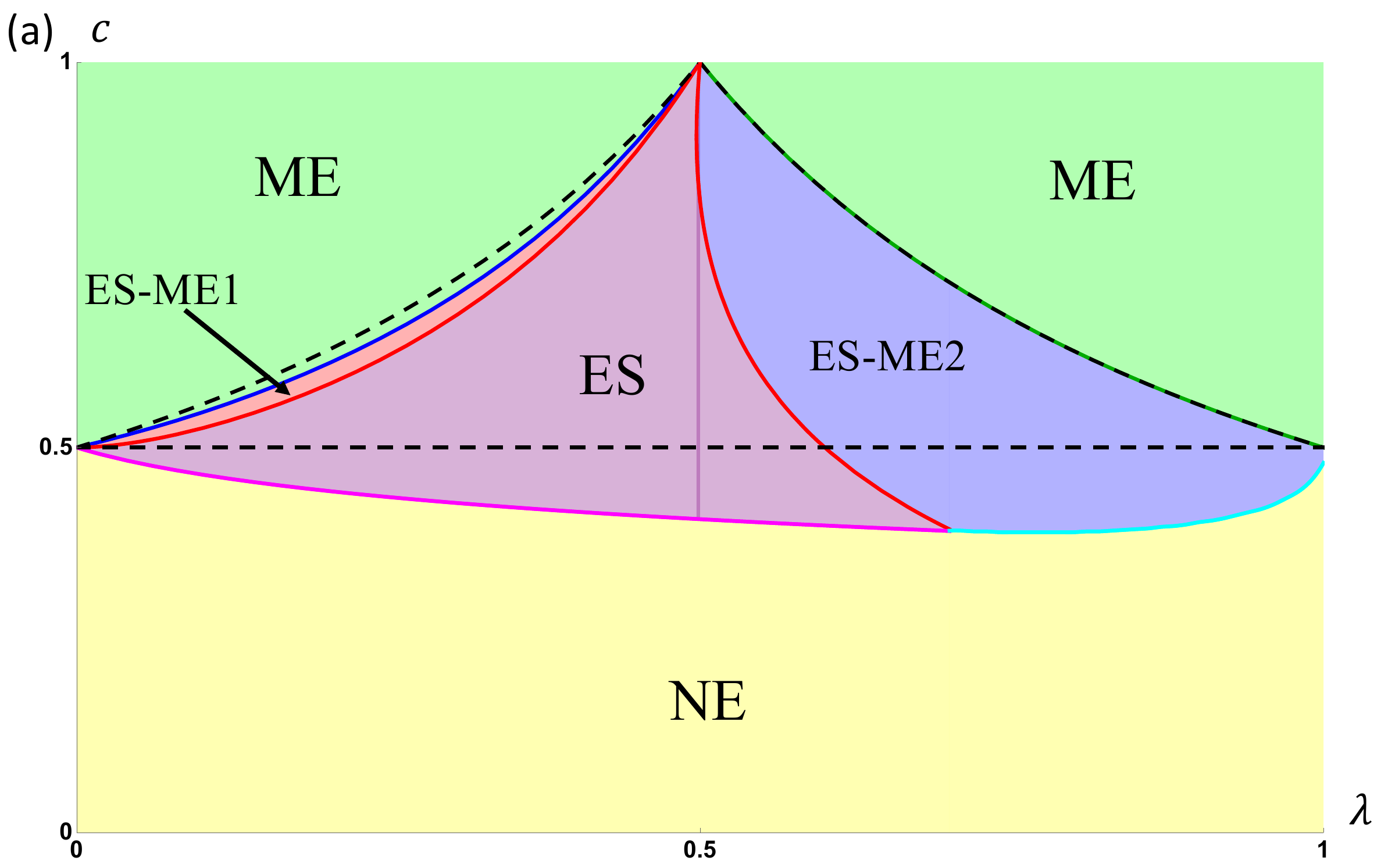}
 \includegraphics[height=4.4cm]{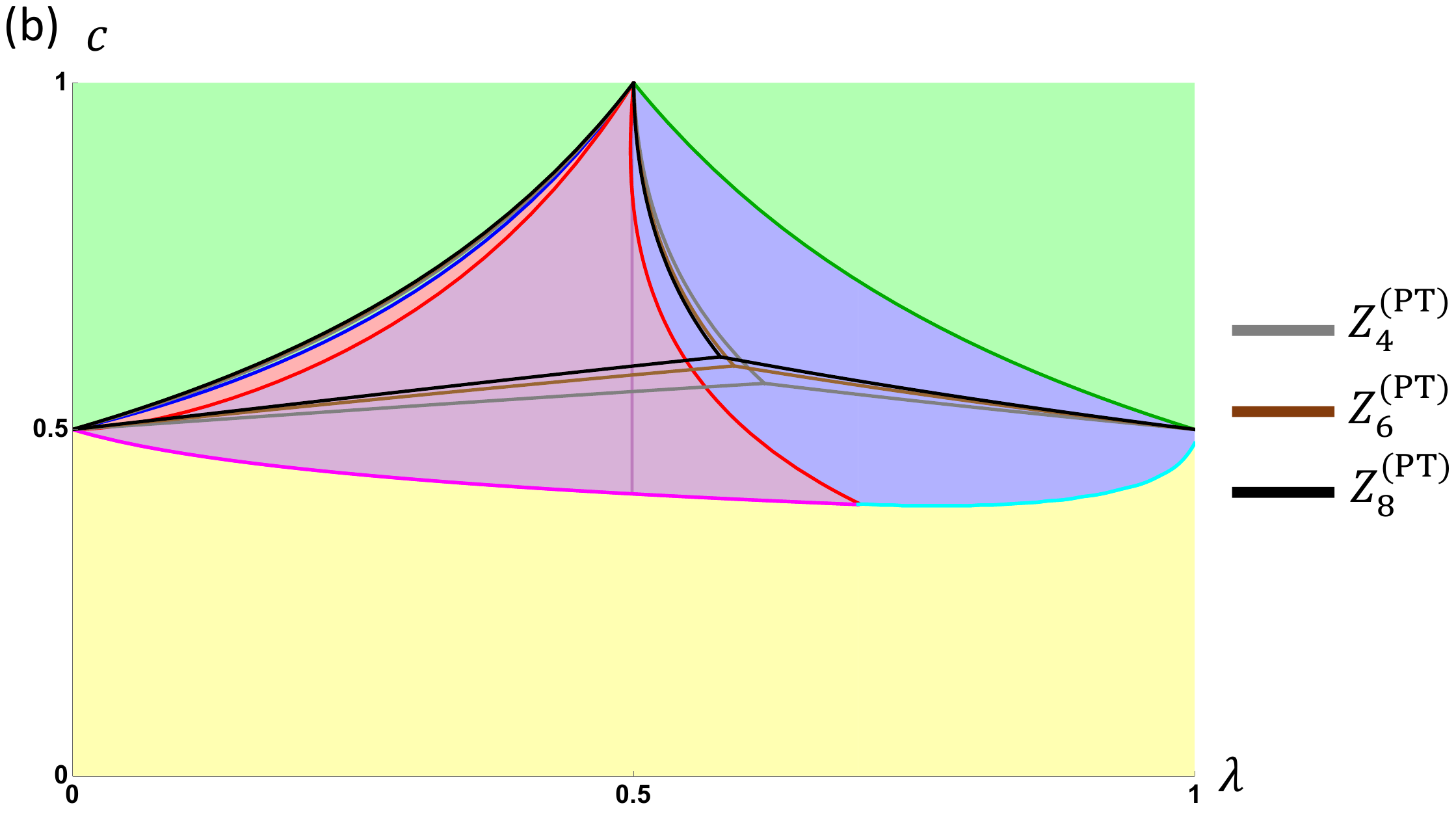}
\caption{(a) shows the phases of $\sE$ with $\sI_{\al}$ as in \eqref{mic_inf_temp}, $s(\epsilon) = \sqrt{\epsilon}$ in $A_1B$, and $\frac{E}{V}= 0.5$. The phase transitions of the mutual information are shown with dashed black lines. In (b), we show the same phase transition lines of $\sE$ alongside the phase transition lines of $\sZ_n^{\rm (PT)}$ for $n=4, 6, 8$. Note that the precise shape of all phase transition lines will depend on the choices of $s(\epsilon)$ and $\frac{E}{V}$. }
\label{fig:phase_diagram_a2_inf}
\end{figure} 
For this example, we are further able to use the resolvent to find the full phase diagram of the logarithmic negativity. We present the details of this calculation in Appendix \ref{app:mic_a2}. The resulting phase digram is shown in Fig.~\ref{fig:phase_diagram_a2_inf}(b). 
 Each of the five different shaded regions, labelled NE, ES, ME, ES-ME1 and ES-ME2, represents a distinct phase of the logarithmic negativity. The phase transition lines for the mutual information are also shown in the same figure with dashed lines, but the corresponding regions are not shaded.  In the top row of Fig.~\ref{fig:E_I_plots}, both $\sE$ and $I$ are shown as functions of the parameter $c=\frac{S^{\rm (eq)}_A}{S^{\rm (eq)}_A+S^{\rm (eq)}_B}$ along different vertical lines marked in the phase diagram of Fig.~\ref{fig:phase_diagram_a2_inf}(b), which have different fixed values of the ratio $\lambda=\frac{S^{\rm (eq)}_{A_1}}{S^{\rm (eq)}_{A}}$. The various phases shown in the diagram should be interpreted as follows: 
\begin{enumerate} 
\item In the phases labelled by NE, ES, and ME, $\sE$ is given by the analytically continued expressions \eqref{mbksn}-\eqref{mbksn2}.
\item In the phases labelled by ES-ME1 and ES-ME2, $\sE$ cannot be obtained by analytic continuation of $\sZ_n^{\rm (PT)}$ or $\sE$. The expressions for the negativity in these phases are given respectively by 
\be 
\sE_{ES-ME1} = S^{\rm (eq)}_{A_2} + V_{A_1} (s(\theta_3)- s(\epsilon)) - V_B s(\epsilon) \,. 
\label{ln_4_mt}
\ee
and 
\be 
\sE_{ES-ME2} = V_{A_1} \, (2 s(\theta_2)- s(\epsilon)) - V_B \, s(\epsilon) \, . 
\label{ln_5_mt}
\ee
where $\theta_2$ and $\theta_3$ are solutions to the equations 
\be 
\begin{gathered} 
V_{A_2}\, s_0 + V_B\, s\left(\frac{\epsilon\, (V_{A_1}+ V_B)- V_{A_1}\, \theta_2}{V_{B}} \right) = V_{A_1} s(\theta_2), \\
V_{A_1}\, s(\theta_3) + V_B\, s\left(\frac{\epsilon\, (V_{A_1}+ V_B)- V_{A_1}\, \theta_3}{V_{B}} \right) = V_{A_2} s_0\, . 
\end{gathered} 
\ee
\item Only one of the phase transition lines shown in the figure -- the pink line dividing the NE and ES phases -- can be correctly deduced by naive analytic continuation. Putting $n=1$ in the condition $Z_n^{\rm (PT)}(\tau_{ES})\geq Z_n^{\rm (PT)}(e)$, we would deduce that the transition line between the NE and ES phases for $\sE$ is given by 
\be 
\frac{1}{2}S_{A_2}^{\rm eq} + S_{1/3, A_1}^{\rm eq} - \frac{1}{2} S_{A_1}^{\rm eq} - \frac{1}{2} S^{\rm eq}_{B}=0 \label{n_1}
\ee
\eqref{n_1} agrees with the pink line in Fig \ref{fig:phase_diagram_a2_inf}(b), but it cannot be used to determine the point where the pink line ends and the cyan line begins, as the naive analytic continuation does not ``know about" the existence of the ES-ME2 phase. The semi-circle approximation for the resolvent in this example also correctly predicts the pink line for the transition from NE to ES, but is again insensitive to the ES-ME2 phase.
\item In the transition from NE to ES or ES-ME2, there is a discontinuity in the first derivative of $\sE$ with respect to $c$. The remaining phase transitions correspond to a discontinuity in the second derivative. We show $\frac{\partial \sE}{\partial c}$ along different vertical lines in the phase diagram in the bottom row of Fig.~\ref{fig:E_I_plots}. This is unlike the infinite temperature case, where all phase transitions involve discontinuities in the first derivative of $\sE$. 
\item In the region between the dashed black line and the pink and cyan lines, the mutual information is sub-extensive while the negativity is extensive. Comparing \eqref{n_1} with the NE to ES transition line for the mutual information in \eqref{mutual_ne_es}, we see quite generally that since $S^{\rm (eq)}_{1/3, A_1}> S^{\rm (eq)}_{A_1}$, such a regime should exist for any choice of $s(\epsilon)$, as long as the naive transition line for the logarithmic negativity in \eqref{n_1} is correct.

Note also that the NE-ES transition line for $I_{1/2}$ lies slightly below, but quite close to, that of the mutual information. In particular, it also lies above the lines where the logarithmic negativity starts to become extensive. 
\end{enumerate} 

\begin{figure}[]
\includegraphics[height=3cm]{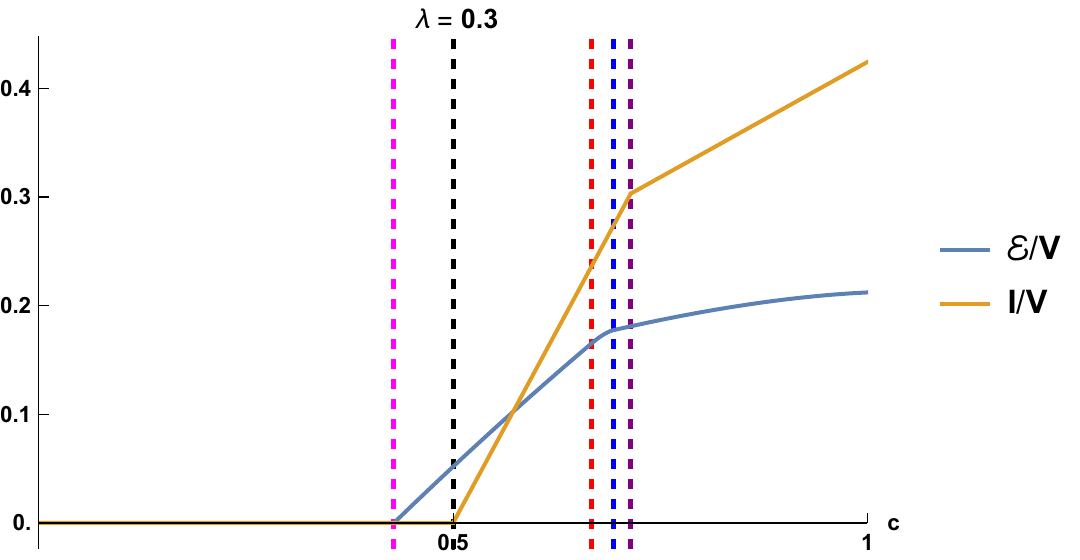}~\includegraphics[height=3cm]{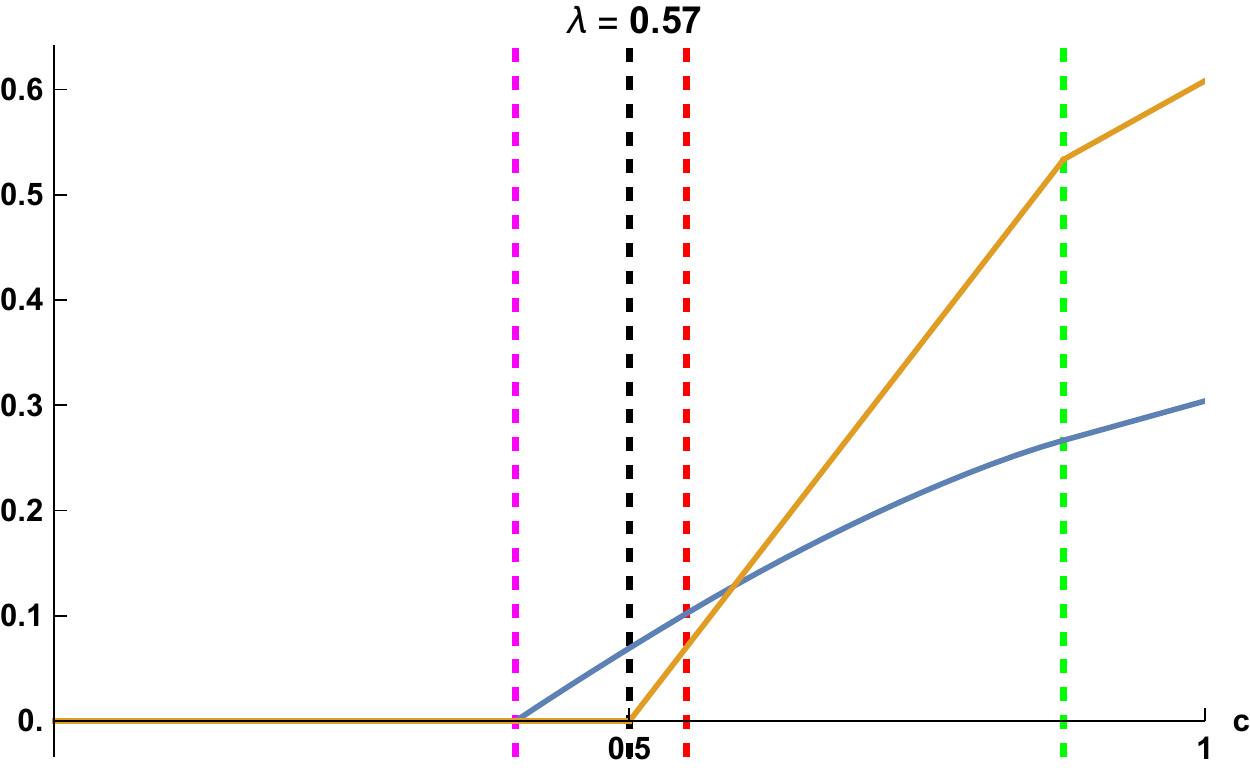}\includegraphics[height=3cm]{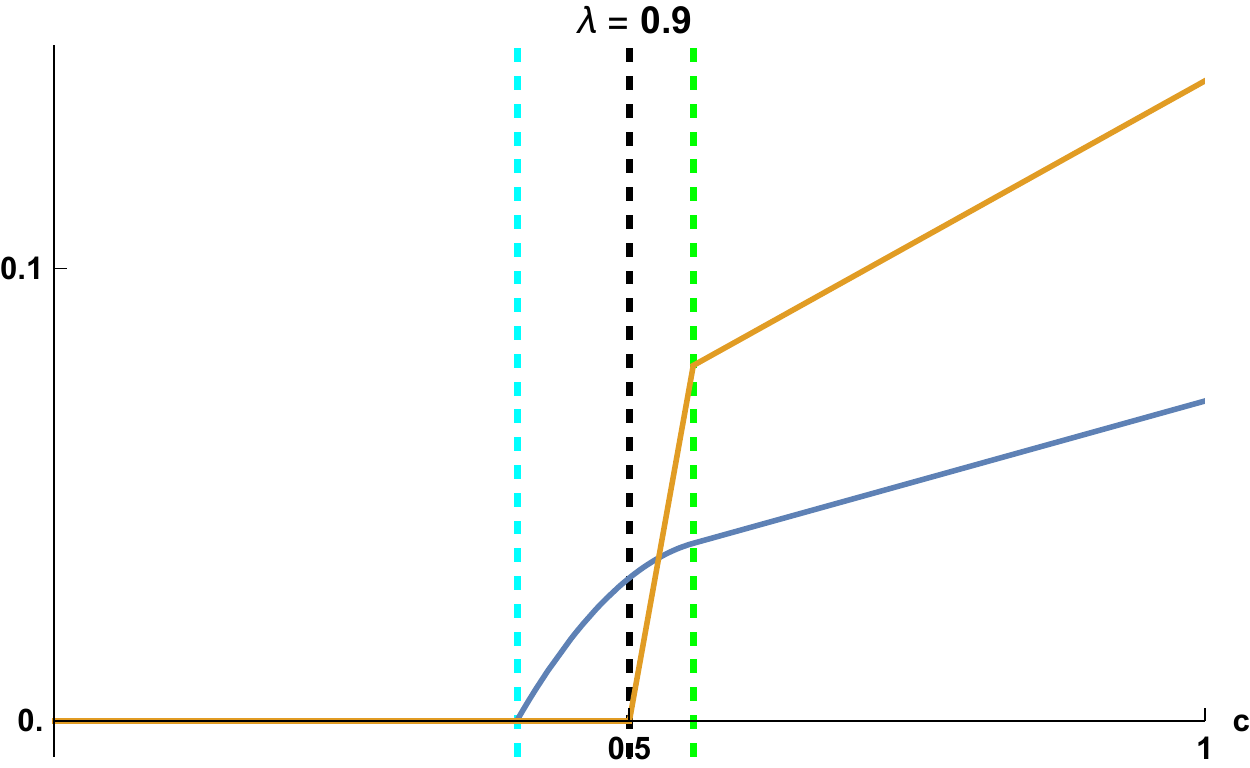}
\vspace{0.3cm}
 \includegraphics[height=3.3cm]{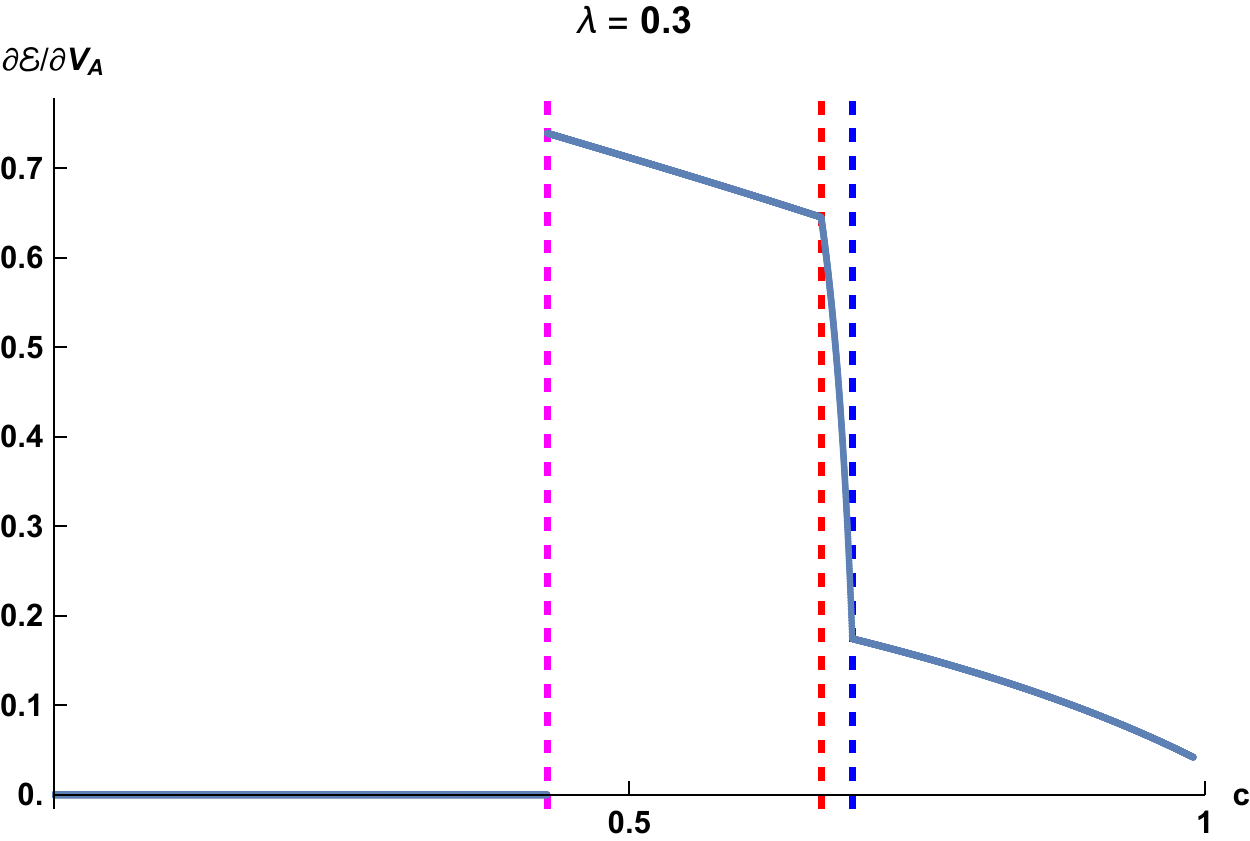}\hspace{0.4cm}\includegraphics[height=3.3cm]{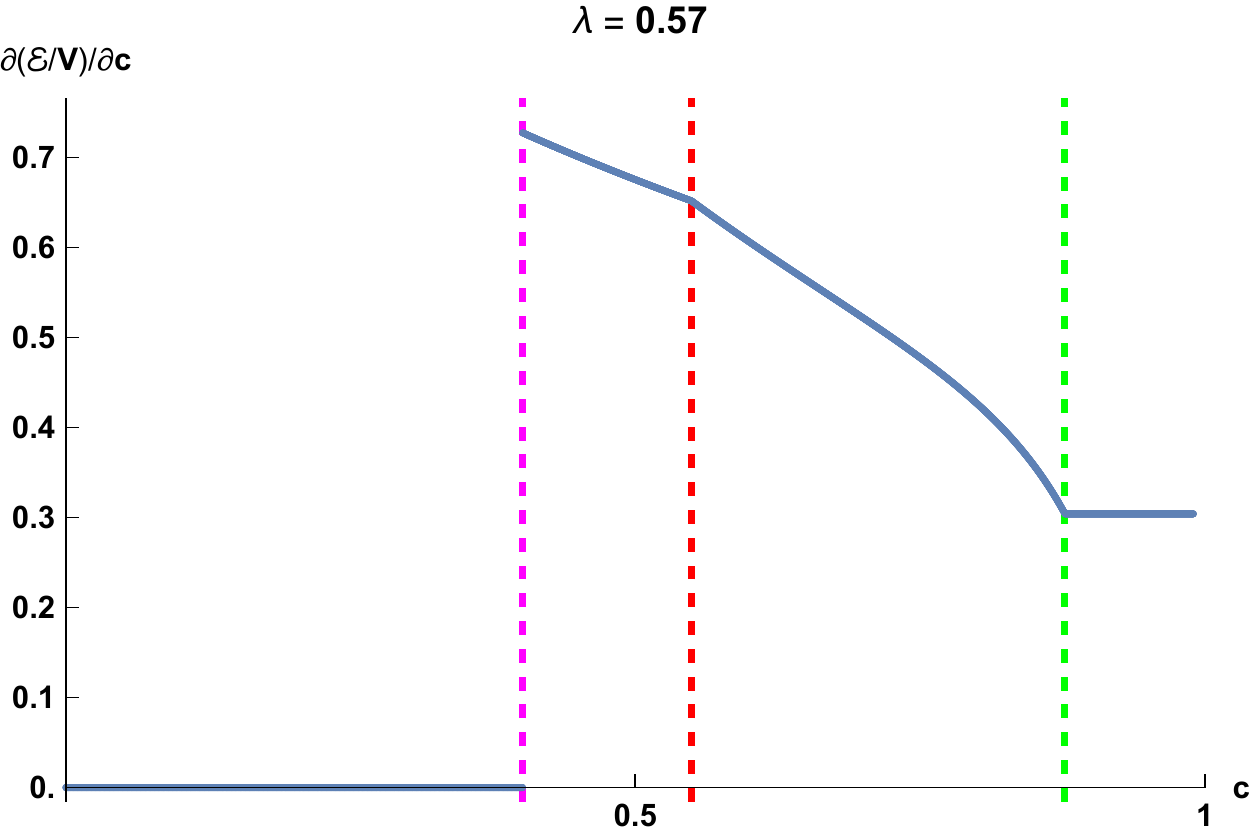}~~\includegraphics[height=3.3cm]{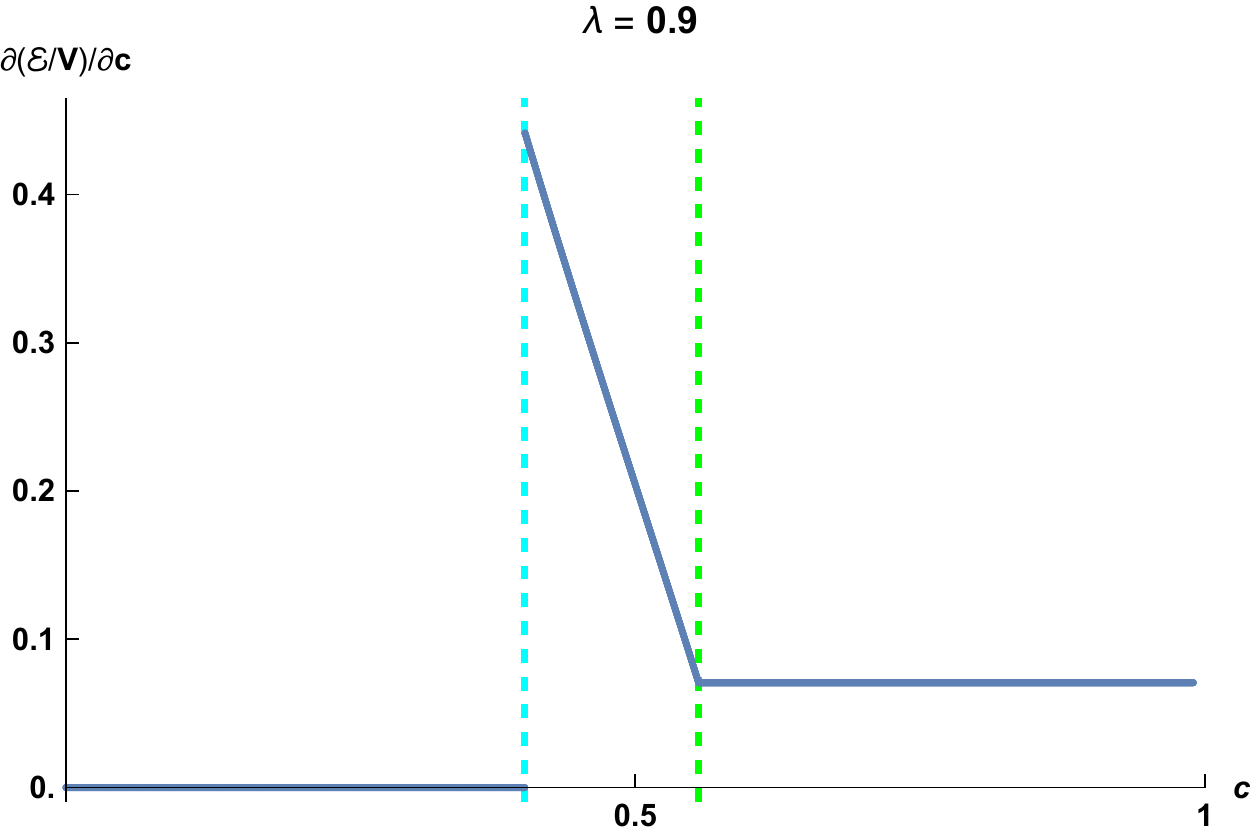}
\caption{Top row: the blue and orange curves respectively show $\sE$ and $I$ as a function of $c$ along three vertical lines in the phase diagram of Fig.~\ref{fig:phase_diagram_a2_inf}(b). Bottom row: $\frac{\partial \sE}{\partial c}$ is shown along the same lines. The relevant phase transitions lines from Fig.~\ref{fig:phase_diagram_a2_inf}(b) are also indicated with dashed vertical lines.}
\label{fig:E_I_plots}
\end{figure}

\subsection{Numerical examples: negativity spectrum in chaotic spin chains}

Throughout this work, we have used an analytic approximation that holds in the thermodynamic limit to calculate various information-theoretic quantities. In this subsection, we present results for the spectrum of the partially transposed density matrix in the canonical ensemble. Here we avoid all approximations and carry out numerical simulations in chaotic spin chains, but are necessarily limited to small sizes. 

We consider a chaotic 1D spin chain with Hamiltonian
\begin{align}
 H = -\sum_{i=1}^N\left(Z_iZ_{i+1}+h_x X_i+h_z Z_i \right).
 \label{H_spin_chain}
\end{align}
This Hamiltonian has been studied frequently as it undergoes a well-understood integrable to chaotic transition \cite{2011PhRvL.106e0405B}.
We find, numerically, that this model has the following partition function
\begin{align}
 g(\beta) = \frac{\log Z(\beta)}{N} = \log \left(2 \cosh \left(f(h_x,h_z) \beta\right)\right)
\end{align}
to extremely precise accuracy. Here, there is only a single fitting parameter, $f$, that depends on the coupling constants chosen. This can be fixed by the low-temperature behavior of the partition function
\begin{align}
 \frac{\log Z(\beta\rightarrow \infty)}{N} = -\frac{\beta E_0}{N} = \beta f(h_x,h_z).
\end{align}
Thus, we have
\begin{align}
 g(\beta) =\log \left(2 \cosh \left(\varepsilon_0 \beta\right)\right),
 \label{g_prediction}
\end{align}
where $\varepsilon_0$ is the ground state energy density. We set $h_x =-1.05$ and $h_z=0.5$ in our calculations below. 

If we want to model the canonical ensemble, it is convenient to study the ``canonical thermal pure quantum states'' of Ref.~\cite{2013PhRvL.111a0401S}
\begin{align}
 \ket{\beta} \propto e^{-\beta H/2}\ket{R}, 
\end{align}
where $\ket{R}$ is a random state vector. Being random, it is approximately an equal weighting of all energy eigenstates
\begin{align}
 \ket{\beta} \simeq \frac{1}{\sqrt{Z(\beta)}} \sum_{E_n }e^{-\beta E_n/2}\ket{E_n}.
\end{align}
The exponential of the Hamiltonian is a dense matrix, so a priori, we will be very limited in the system sizes. This can be improved by approximating the exponential as
\begin{align}
 e^{-\beta H/2} \simeq e^{-\beta (H_A\otimes \mathbbm{1}_B)/2}e^{-\beta (\mathbbm{1}_A\otimes H_B)/2},
\end{align}
where, as in the equilibrium approximation, we have considered the coupling $H_{AB}$ to be small. A Haar random vector can be written as \cite{2009arXiv0910.1768C}
\begin{align}
 \ket{R} = \sum_{i\alpha} X_{i\alpha} \ket{i}_A\otimes\ket{\alpha}_B,
\end{align}
where the $X_{i\alpha}$'s are independent complex Gaussian random variables and $\ket{i}, \ket{\alpha}$ are orthonormal bases on the sub-Hilbert spaces. The (unnormalized) reduced density matrix on subsystem $A$ may then by expressed as 
\begin{align}
 \rho_A(\beta) = e^{-\beta H_A/2}XX^*e^{-\beta H_A/2},
\end{align}
where $X$ is now the $d_A \times d_B$ rectangular matrix with matrix elements $X_{i\alpha}$. This is a great numerical simplification because we avoid ever forming dense, $(d_A d_B) \times (d_A d_B)$ matrices.

In Fig.~\ref{finiteT_neg_Renyi}, we show the exact negativity spectra at both infinite and finite temperature. Certain aspects of the infinite temperature spectra are robust such as the topological transition from the single connected component to two connected components. However, other details are not robust. For example, when all systems are at finite temperature, the spectrum never resembles a semi-circle. 

\begin{figure}[]
 \centering
 \includegraphics[width =.32 \textwidth]{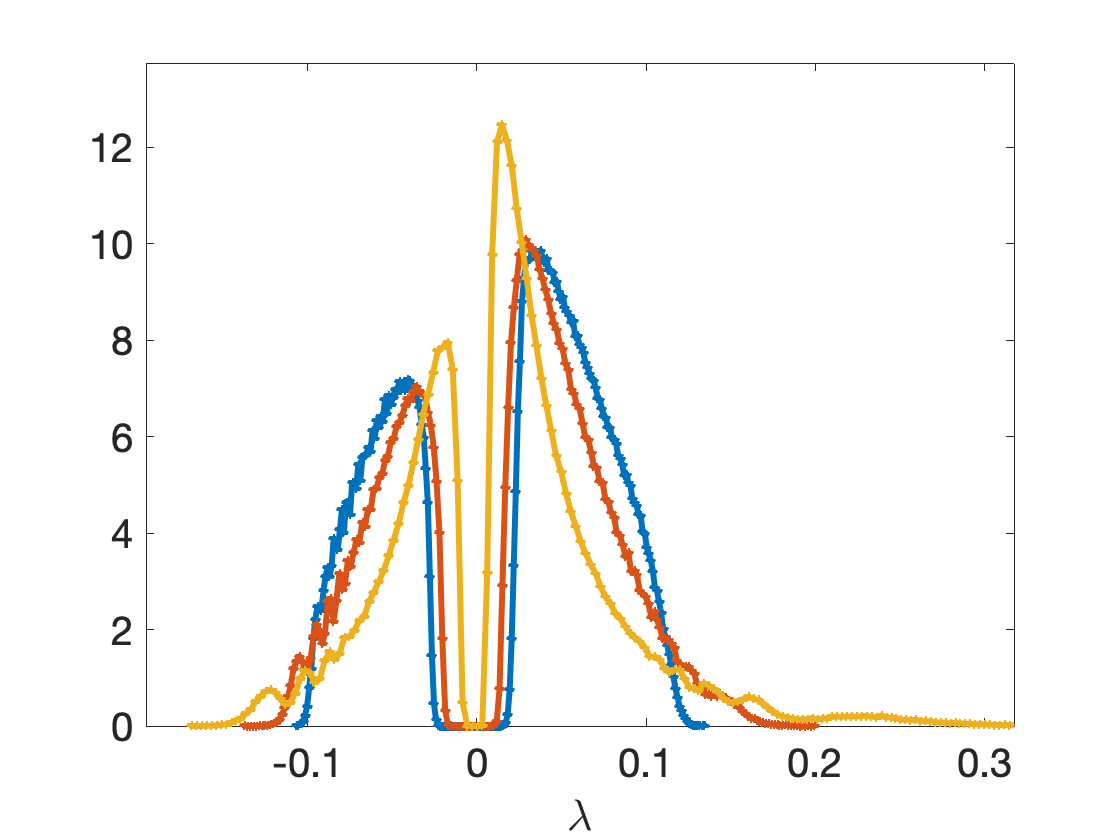}
 \includegraphics[width =.32 \textwidth]{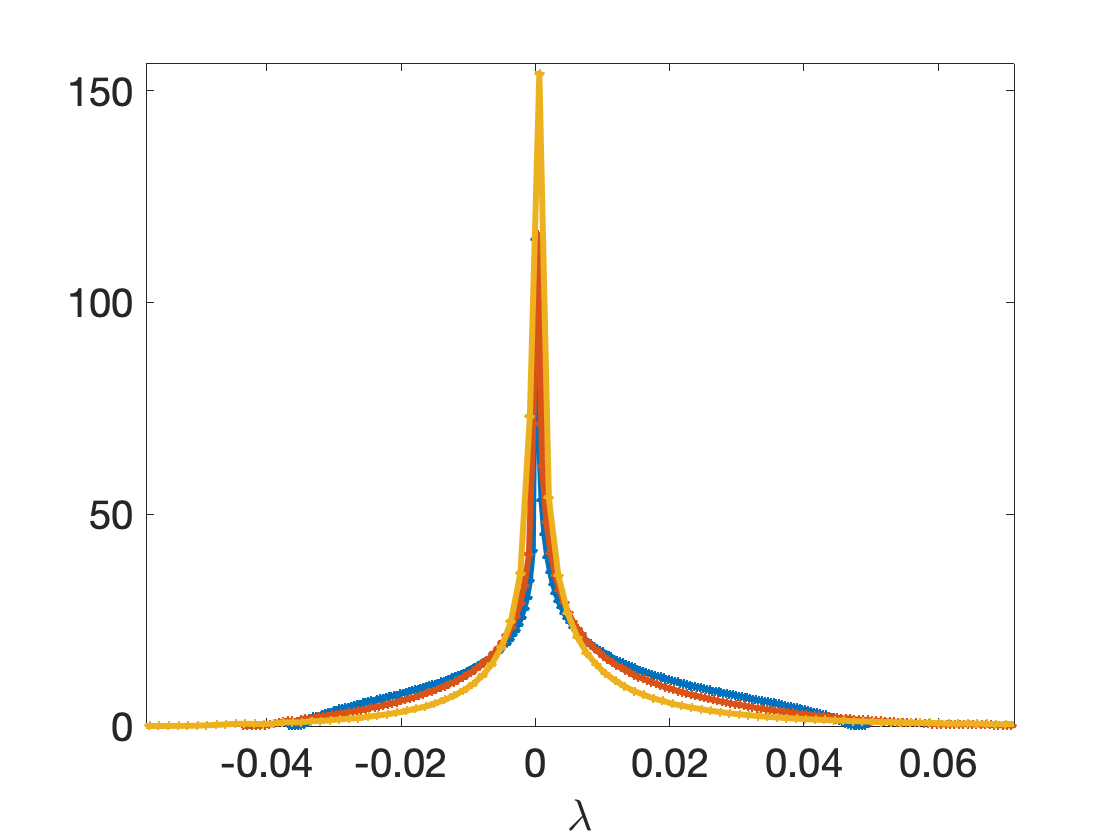}
\includegraphics[width =.32 \textwidth]{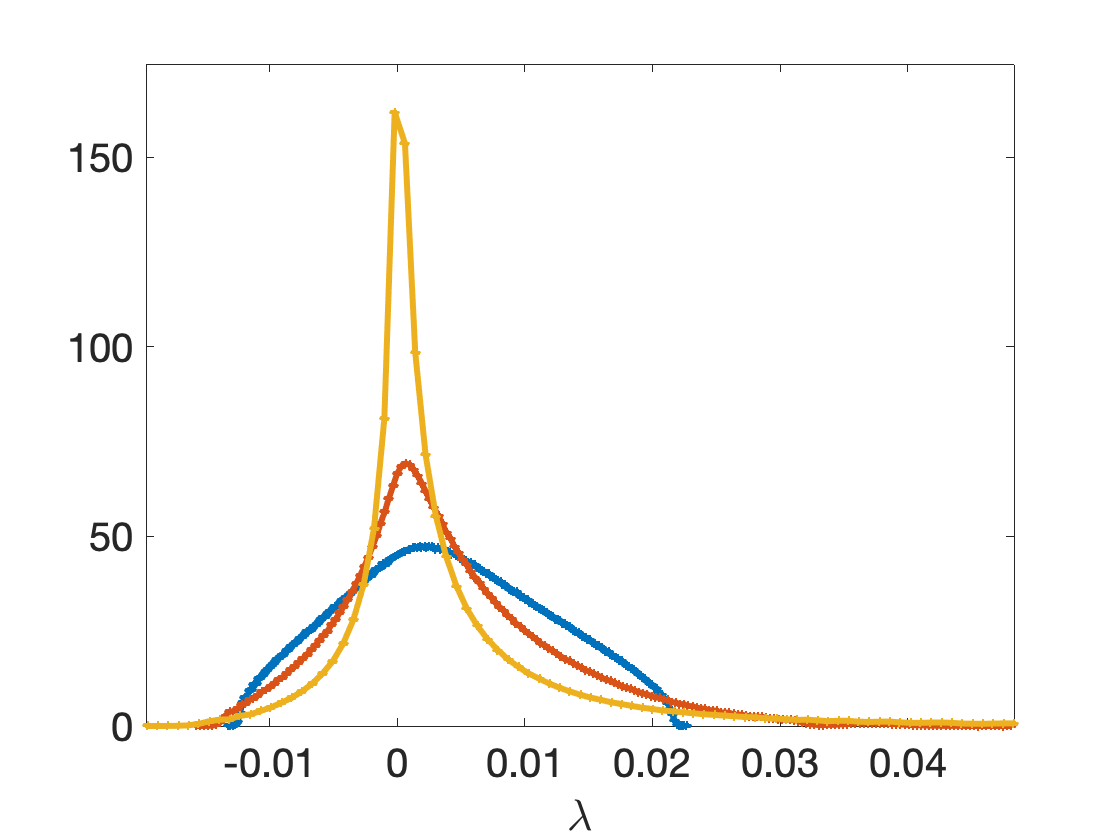}
 \caption{The probability density functions for the negativity spectra at $|\beta\epsilon_0| = 0$ (blue), $\frac{1}{4}$ (orange), and $\frac{1}{2}$ (yellow). From left to right, we have $N_B = 2$, $4$, and $6$. The topological transition is robust to finite temperature. We always take $N_{A_1} = 2$ and $N_{A_2} = 6$.}
 \label{finiteT_neg_Renyi}
\end{figure}




\section{Implications for mixed-state entanglement and black hole radiation} \label{sec:mixen}

We now discuss the operational implications of the above results for $\sE$ and $I$. 
As reviewed in Sec.~\ref{sec:revneg}, the mutual information gives an upper bound the distillable entanglement 
\be
E_d (A_1, A_2) \leq \ha I (A_1, A_2) \label{m_ed} , 
\ee
while the logarithmic negativity gives a lower bound on the exact PPT entanglement cost 
\be 
 E_c^{\rm (ppt, exact)}(A_1, A_2) \geq \sE(A_1, A_2) \ .
 \label{e_ec}
\ee 
We observed that at a finite temperature, for various choices of $\rho^{(\rm eq)}$, there is a region in the entanglement phase diagram where $\sE$ is volume-like while $I$ is sub-volume, and hence $\sE \gg \frac{1}{2}I$. Our results suggest that such behavior is likely to be generic for all finite temperature chaotic systems. In this part of the phase diagram, we then have 
\be 
E_c^{\rm (exact)} \gg E_d^{\rm (exact)}\, . 
\ee
If we further assume that $E_c^{\rm (ppt)}$ is extensive when $E_c^{\rm (ppt, exact)}$ is extensive, then a stronger statement 
can be made, 
\be 
E_c \gg E_d \, . \label{cd} 
\ee
This implies that the preparation of the state of $A$ from EPR pairs by LOCC operations is highly irreversible, and it has significant bound entanglement.

It is interesting to contrast the results here with those from random states in~\cite{hayden2006aspects}, which can be seen as thermalized states at infinite temperature. 
In such states, a regime was found where $E_c$ is extensive while $I$ (and hence $E_d$) is sub-extensive. However, this effect took place for a range of subsystem dimensions which all correspond to $c=\ha$ in the thermodynamic limit, as opposed to an $O(1)$ range of $c$ between some $c_0$ and $\ha$, as we find in the finite temperature examples here. The conclusion about $E_c$ in \cite{hayden2006aspects} was drawn using entanglement of formation as opposed to logarithmic negativity. On calculating $\sE$ in the infinite temperature case in the same regime at $c=\ha$, we do not find that it becomes extensive while $I$ is still sub-extensive. 


We can ask whether the preparation of the state $\rho_A$ by PPT operations, as opposed to LOCC, is also irreversible in this regime. This question cannot be conclusively answered from our results. With the assumption that $E_c^{\rm (ppt)}$ is extensive when $E_c^{\rm (ppt, exact)}$ is extensive, there are two logical possibilities, both with interesting physical consequences:
\begin{enumerate}

\item If $E_d^{\rm (ppt)}$ is comparable to $\sE$, i.e.~volume-like, as in some known systems~\cite{audenaert2003entanglement}, then although LOCC distillable entanglement is small, PPT distillable entanglement is large, implying that finite temperature greatly enhances distillable entanglement if we use PPT operations. 

 \item If $E_d^{\rm (ppt)}$ is comparable to $I$, i.e.~sub-volume-like, then both the PPT and LOCC distillable entanglements are small, and the state is also highly PPT-irreversible. 
 
\end{enumerate}

The results of the previous section apply to generic chaotic many-body systems, and hence also have implications for an evaporating black hole, which is believed to be highly chaotic. 

Consider a black hole formed from the collapse of a star initially in a pure state. The black hole subsequently evaporates by emitting Hawking radiation. 
The evaporation process is very slow in microscopic scales, and we can treat the remaining black hole as well as the radiation as being in macroscopic equilibrium.~\footnote{The two systems can be in separate macroscopic equilibria, because in general the radiation is separated from the black hole after it is emitted.} The full system is still in a pure state, and the reduced density operator of each subsystem can be far from the density operators of thermal ensembles. The results of the previous section can be used to predict 
the entanglement structure {\it within} the radiation subsystem, or entanglement between a part of radiation and the remaining black hole, at both infinite and finite temperatures. For definiteness, we will take $A$ to be the radiation and $B$ to be the black hole.

 In~\cite{1993PhRvL..71.1291P}, from averaging over random pure states, Page found that the entanglement entropy of the radiation undergoes a transition from increasing to decreasing behavior at some time scale $t= t_p$ where $ \log d_A = \log d_B$. Page's calculation was in the infinite temperature case, where from our discussion in section \ref{sec:inft}, we know that $t_p$ is also the time scale at which entanglement within the radiation, as quantified by either $I$ or $\sE$, starts to become extensive. 
In other words, before the Page time $t_p$ there is no entanglement within the radiation.

The natural finite-temperature generalization of the Page time is when $S_A^{(\rm eq)} = S_B^{(\rm eq)}$, corresponding to $c = \ha$, which was already used in~\cite{page2013time}. Our results from the equilibrium approximation and resolvent calculations confirm that this is indeed the time at which the entanglement entropy transitions from increasing to decreasing behavior. 
Interestingly, our results for logarithmic negativity give a new prediction for the quantum-informational properties of the radiation at finite temperature: there are significant entanglement correlations within the radiation long before the Page time. This suggests the 
existence of another time scale $t_b$ when quantum entanglement within the radiation starts becoming extensive. 
For time scales $t \in (t_b, t_p)$, the entanglement correlations within the radiation cannot be distilled using LOCC, i.e.~they are an example of 
``bound entanglement.'' It is an interesting open question whether this entanglement could be distillable using more general operations such as PPT operations.

Our results from the previous section can also be used to derive replica wormhole contributions to the calculation of the negativity between parts of the radiation in Euclidean gravity. We show these replica wormholes explicitly for the model of \cite{2019arXiv191111977P} in Appendix \ref{pssy}. (See Fig.~\ref{fig:gravity_figs}.) Due to specific features of the density of states in JT gravity, the difference between $t_b$ and $t_p$ is suppressed in the large quantity $S_0$ in this model. However, this difference should be macroscopically large for higher-dimensional black holes.

To make predictions for the mixed-state entanglement between the black hole and some part of the radiation, we can take $A_1$ and $B$ to be parts of the radiation and $A_2$ to be the black hole. 

To explore further the nature of entanglement among various parts of the full black hole system, 
it is instructive to consider the Hayden-Preskill thought experiment at a finite temperature: we throw a diary into the black hole, and see when the information of the diary is recoverable from the radiation. We turn to this problem next. We will find that both time scales $t_p$ and $t_b$ can be given operational interpretations in the context of this question.

\section{Implications for information transfer from black hole to radiation} 
\label{transfer_sec}


In this section, we investigate how information is transferred from the black hole to the radiation during the evaporation process using 
the Hayden-Preskill thought experiment~\cite{2007JHEP...09..120H}. 
Suppose we throw a secret diary $D$ into a black hole $B$ at an early stage in the evaporation process. As the black hole evaporates, we collect all of the radiation and refer to it as $R$, and refer to the remaining black hole as $B'$. Suppose we know the initial state of $B$, and have access to a universal quantum computer with which we can act on the radiation $R$. We then ask how much of the radiation we need in order to learn the initial state of the diary, $\rho_D$. 

We can make this question more explicit in the following way. The relevant decompositions of the Hilbert space at the initial and final times respectively are $\sH= \sH_D \otimes \sH_B$ and $\sH= \sH_{B'} \otimes \sH_R$. We can consider the following quantum channel from $\sH_D$ to $\sH_R$, 
\begin{align} \label{n_channel}
 \mathcal{N}(
 \rho_D) = \Tr_{B'}\left[ U \left(\rho_D \otimes \rho_B \right)U^{\dagger} \right],
\end{align}
where $U$ is the time-evolution operator for the black hole and radiation, together with the diary, and $\rho_B = \ket{\psi_0}\bra{\psi_0}_B$ is some fixed initial pure state of the black hole. Asking whether we can learn the state of the diary from the radiation is equivalent to asking whether there exists a universal recovery channel $\sR$ such that $\sR \circ \sN (\rho_D) = \rho_D$ for all $\rho_D$.

Note that while we will mostly discuss the above question in terms of evaporating black holes in this section, it can also be seen as a more general question about thermalized states in quantum many-body systems. Suppose we put some information in a small subsystem $D$, and then let $D$ evolve together with the rest of the system, $B$. Then given some subsystem $R$ in the thermalized state, can we learn the initial state of $D$? 

We will take three different approaches to this question, each of which reveals different aspects of the transfer of information from the black hole to the radiation. We will use the equilibrium approximation in each of these approaches, which allows us to understand the transfer of information both at infinite and finite temperature. Note that all statements at finite temperature below are based on the canonical ensemble universality class, where we take $\sI_{\al} = e^{\beta_1 H_{B'}} \otimes e^{\beta_2 H_{R}}$. We will assume that the Hilbert space of the diary has a finite dimension $d_D$. We will in general consider the case where the diary can be large, so that $d_D$ can be $O(e^{1/G_N})$.
\begin{enumerate} 
\item We first introduce a reference system $Q$ with the same Hilbert space dimension as $D$, 
and consider an initial state in which $D$ is maximally entangled with $Q$. The time-evolution on $\sH_Q$ is trivial, while $\sH_B \otimes \sH_D$ is evolved with $U$. We then look at the mutual information $I(Q, R)$ under time-evolution, which can be seen as a way of quantifying the extent to which the radiation contains information about the diary. The behavior of this quantity motivates us to define two natural time-scales: 
\begin{itemize} 
\item $t_{p_1}$, the time at which $S^{\rm (eq)}_R = S^{\rm (eq)}_{B'} - \log d_D$. $I(Q, R)$ starts increasing from its initial value of zero at this time. 
\item $t_{p_2}$, at which $S^{\rm (eq)}_R = S^{\rm (eq)}_{B'} + \log d_D$. $I(Q, R)$ reaches its maximal value of $2\log d_D$ at this time. 
\end{itemize} 
The standard Page time $t_p$, at which we have $S_R^{\rm (eq)} = S_{B'}^{\rm (eq)}$, lies between these two time scales. 

When $I(Q, R)$ reaches its maximal value at $t_{p_2}$, this can be interpreted by saying that all the information that was initially in the diary is now present in the radiation. This statement can be understood operationally in terms of the quantum channel $\sN$ in \eqref{n_channel}. From the results of \cite{2007PhRvA..75f4304N}, $I(Q, R)= 2 \log d_D$ implies the existence of a universal recovery channel $\sR$ for $\sN$, i.e.~$\sR \circ \sN (\rho_D) = \rho_D$ for all $\rho_D$. This is consistent with the fact that from a gravitational perspective, $t_{p_2}$ is the latest time at which an island can form \cite{2020JHEP...09..002P} (see Fig.~\ref{hayden_preskill_cartoon}). 
\item For some choice of reference state $\sigma_D$, and any state $\rho_D$, we compute the difference in the relative entropies before and after applying the channel $\sN$,
\be 
 D(\sN(\rho_D)|| \sN(\sigma_D))- D(\rho_D || \sigma_D), 
 \ee
in order to find a lower bound for the fidelity of some recovery channel $\sR$ using \eqref{petz_fidelity}. The relevant time scales for this quantity turn out in general to depend on the choice of $\sigma_D$ and $\rho_D$, and are given by 
\begin{itemize}
\item $t_p(\rho_D)$, the time at which $S^{\rm (eq)}_R = S^{\rm (eq)}_{B'} + S(\rho_D)$. The lower bound on the fidelity of some recovery channel $\sR$ first starts to increase from an exponentially small value at this time. 
\item $t_{p_2}(\sigma_D, \rho_D)$, at which $S^{\rm (eq)}_R = S^{\rm (eq)}_{B'} + D(\rho_D || \sigma_D) +S(\rho_D) = S^{\rm (eq)}_{B'} - \text{Tr}[\rho_D \log \sigma_D]$. The lower bound on the fidelity reaches its maximal value of 1 at this time-scale. 
\end{itemize} 
Note that if we take $\rho_D$ to be pure, then $t_p(\rho_D) = t_p$, the standard Page time. If we take $\sigma_D = \mathbf{1}/d_D$, then $t_{p_2}(\sigma_D, \rho_D)$ becomes equal to $t_{p_2}$ defined in the previous point independently of $\rho_D$, implying that universal recovery is possible after $t_{p_2}$. Recall that in the previous point, universal recovery at $t_{p_2}$ was deduced from the complementary perspective that the mutual information $I(Q, R)$ becomes maximal at this time. 

The lower bound on the fidelity provides an operational way of seeing the gradual transfer of information from the black hole to the radiation between times $t_p$ and $t_{p_2}$, as the fidelity increases from its minimal value to one in this range of times. However, this quantity does not seem to have a regime which reflects the growth of $I(Q, R)$ from time $t_{p_1}$ to $t_p$ which we observed in the previous point. 
\item We explicitly calculate the fidelity of the Petz recovery map $\sP$ for $\sN$, taking the initial state $\rho_D$ to be pure to simplify calculations. This reveals the following new time scale, which the lower bound in terms of the relative entropy from the previous point is not sensitive to: 
\begin{itemize} 
\item $t_b$, the time at which $S^{\rm (eq)}_{\ha, R} = S^{\rm (eq)}_{2, B'}$. For a large diary, the fidelity of the Petz map starts growing from its initial value of $F(\rho_D||\sigma_D)$ at this time. Recall from our discussion in section \ref{general_canonical}, that this is also the time scale at which the logarithmic negativity between two parts of the radiation starts to become extensive in the canonical ensemble. \footnote{At the end of this section, we briefly consider an example with the microcanonical ensemble, taking $\sI_{\al}$ as in \eqref{mic_inf_temp}. For this setup, we find that the fidelity of a large diary starts increasing at a time $t_r<t_p$, which can be either earlier or later than $t_b$.}
\end{itemize} 
The time at which the fidelity saturates to unity is $t_{p_2}$, consistent with the conclusions of the two other approaches in the previous points.
\end{enumerate} 
Note that at infinite temperature and for small diaries such that $\log d_D \ll S^{\rm (eq)}_{B'}, S^{\rm (eq)}_R$, all the different time scales above coincide, i.e.~$t_{p_1} = t_b = t_p = t_{p_2}$. More generally, allowing for finite temperature and large diaries, we have $t_{p_1} \leq t_p \leq t_{p_2}$ and $t_{b} \leq t_p \leq t_{p_2}$.



\subsection{
From the perspective of 
mutual information} 
\label{mutual_ref}

We first discuss the transfer of information from the perspective of the mutual information of the radiation with a reference system $Q$. This approach was considered at infinite temperature in the original discussion of~\cite{2007JHEP...09..120H} by using random states, and has been recently generalized to generic chaotic systems at infinite temperature using properties of operator growth~\cite{2021JHEP...03..088L}. We now provide a generalization to finite temperature using the equilibrium approximation. 

We take the initial state of the full system to be 
\begin{align}
 \ket{\Psi_0} = \frac{1}{\sqrt{d_D}}\sum_{n= 0}^{d_D-1} \ket{n}_Q \otimes \ket{n}_D \otimes \ket{\psi_0}_B \label{psi0_hp}.
\end{align}
We will take the time evolution in $Q$ to be trivial, so that the $n$-th Renyi entropy of the time-evolved state in subsystem $A$ is given by 
\be 
\sZ_{n, A} = \braket{{\eta}_A \otimes e_{\bar{A}} \, | \, (\mathbf{1} \otimes \mathbf{1})^{n}_Q \, \otimes \, (U \otimes U^{\dagger})_{DB}^n \, | \, {\rho_0} \, , e }, \quad \rho_0 = \ket{\Psi_0}\bra{\Psi_0} \, . \label{u_hp}
\ee 
From the fact that the time-evolution in $Q$ is trivial, we immediately have at all times
\be 
S_{n, Q} = \log d_D \, . 
\ee
With the initial state \eqref{psi0_hp}, \eqref{u_hp} for $A= R$ becomes 
\be 
\sZ_{n, R} = \frac{1}{d_D^n} \braket{e_{B'} \otimes \eta_R | (U \otimes U^{\dagger})^n | (\tilde \rho_0)_B \otimes \frac{\mathbf{1}_D}{d_D}, e} \label{mixed_2}
\ee
where $\tilde \rho_0 = \ket{\psi_0}\bra{\psi_0}$. Note that \eqref{mixed_2} can also be seen as the time-evolution of the entanglement entropy of a mixed state with initial entropy $\log d_D$. We can then use \eqref{ejj} to find
\be 
S_{n, R} = \begin{cases} 
S_{n, R}^{\rm (eq)} & S_{n, R}^{\rm (eq)} < S_{n, B'}^{\rm (eq)} + \log d_D \\ 
S_{n, B'}^{\rm (eq)} + \log d_D & S_{n, R}^{\rm (eq)} > S_{n, B'}^{\rm (eq)} + \log d_D
\end{cases} 
\ee
Since the state on the full system is pure, 
\be 
\sZ_{n, QR} = \sZ_{n, B'} = \braket{\eta_{B'} \otimes e_R | (U \otimes U^{\dagger})^n | (\tilde \rho_0)_B \otimes \frac{\mathbf{1}_D}{d_D}, e} \label{mixed_3}.
\ee
Again using \eqref{ejj}, we have 
\be 
S_{n, QR} = S_{n, B'} = \begin{cases} 
S_{n, R}^{\rm (eq)} + \log d_D & S_{n, R}^{\rm (eq)} + \log d_D < S_{n, B'}^{\rm (eq)} \\ 
S_{n, B'}^{\rm (eq)} & S_{n, R}^{\rm (eq)} + \log d_D > S_{n, B'}^{\rm (eq)} 
\end{cases} .
\ee
The $n$-th Renyi mutual information between $Q$ and $R$ is then given by 
\be 
I_n (Q, R) = \begin{cases} 
0 & S^{\rm (eq)}_{n, R} < S^{\rm (eq)}_{n, B'} - \log d_D \\
\log d_D + S_{n, R}^{\rm (eq)}- S_{n, B'}^{\rm (eq)} & S^{\rm (eq)}_{n, B'} - \log d_D < S^{\rm (eq)}_{n, R} < S^{\rm (eq)}_{n, B'} + \log d_D \\
2 \log d_D & S^{\rm (eq)}_{n, R} > S^{\rm (eq)}_{n, B'} + \log d_D 
\end{cases} \, 
\ee
and by analytic continuation, 
\be 
I(Q, R) = \begin{cases} 
0 & S^{\rm (eq)}_{R} < S^{\rm (eq)}_{B'} - \log d_D \\
\log d_D + S_{R}^{\rm (eq)}- S_{B'}^{\rm (eq)} & S^{\rm (eq)}_{B'} - \log d_D < S^{\rm (eq)}_{R} < S^{\rm (eq)}_{B'} + \log d_D \\
2 \log d_D & S^{\rm (eq)}_{R} > S^{\rm (eq)}_{B'} + \log d_D 
\end{cases} \, . \label{iqr}
\ee
Note that at all times, $I(Q, B') = 2 \log d_D - I(Q, R)$. From \cite{2007PhRvA..75f4304N}, when $I(Q, R) =2 \log d_D$ and 
$I(Q, B') =0$, there exists a recovery channel $\sR$ satisfying $\sR \otimes \sN (\rho_D) = \rho_D$ for all $\rho_D$ under the channel $\sN$ defined in \eqref{n_channel}. Hence, the universal recovery channel exists after a time scale $t_{p_2}$, at which $S^{\rm (eq)}_{R}= S^{\rm (eq)}_{B'} + \log d_D$. 
This result from the equilibrium approximation is consistent with the expectations from~\cite{2019JHEP...12..007H} and~\cite{2020JHEP...09..002P}. 

\begin{figure}[]
 \centering
 \includegraphics[width = .4\textwidth]{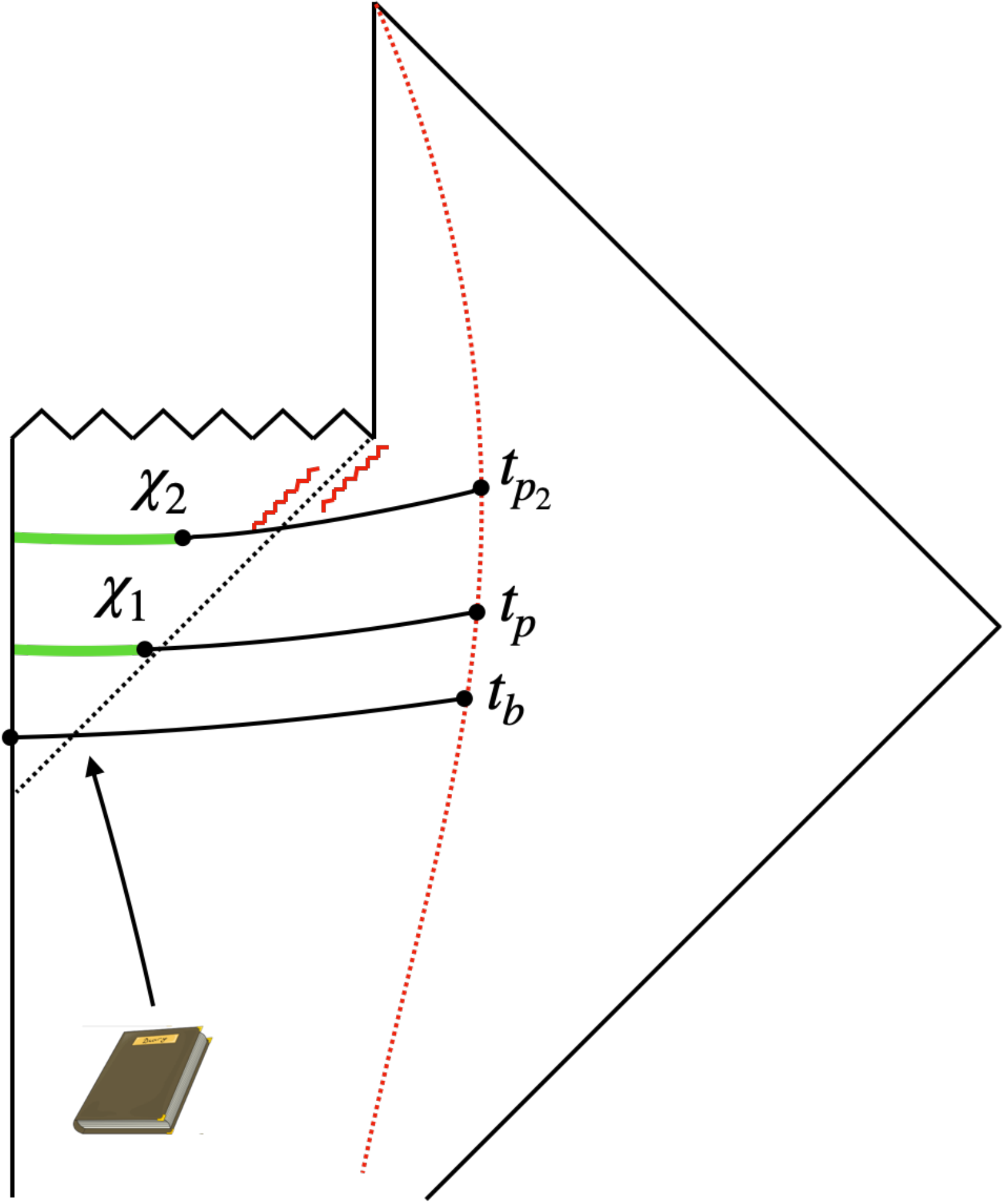}
 \caption{A Penrose diagram is shown for an evaporating black hole with a large diary thrown into it. We ask when the diary can be reconstructed from the radiation (the fields to the right of the dotted red line). At early times, the entanglement wedge of the radiation is trivial, so the diary cannot be recovered. If the diary was initialized in a pure state, an island (green) forms at $t_p$ where the nontrivial extremal surface, $\chi_1$, becomes smaller than the entropy of the radiation. If the diary was initialized in the maximally mixed state, the island will not form until $t_{p_2}$ because the generalized entropy includes the entropy of the diary. While the diary cannot be reconstructed with $O(1)$ fidelity until the island forms, the fidelity exponentially increases from its minimal value starting at the earlier time $t_b$. }
 \label{hayden_preskill_cartoon}
\end{figure}

Note from \eqref{iqr} that while the mutual information between the radiation and the auxiliary system becomes maximal at $t_{p_2}$, it begins to grow from zero at an earlier time $t_{p_1}$, at which $S^{\rm (eq)}_R =S^{\rm (eq)}_B- \log d_D$.

\subsection{Recovery channel: bound from relative entropy} 

We now consider the evolution of the fidelity of recovery channels during the evaporation process. 
In this subsection, we use~\eqref{petz_fidelity} to put a lower bound on the fidelity of some recovery channel $\sR$, by calculating 
 the change of the relative entropy under $\mathcal{N}$ using the equilibrium approximation. In next subsection, we calculate the fidelity of the Petz map explicitly. Note that in both this subsection and the next, we do not include a reference system $Q$, and the input state for $\sN$ is simply $\rho_D$.

\subsubsection{Infinite temperature}

For simplicity, let us first consider the infinite temperature case, and take $\sigma_D$ to be $\mathbf{1}_D/d_D$ and $\rho_D$ to be a pure state. Then for the first term in the expression \eqref{relative_replica} for $ D(\mathcal{N}(\rho_D) || \mathcal{N}(\sigma_D) )$, in the infinite temperature case we can sum over all permutations to find (see for instance \cite{2020arXiv200801089L})
\begin{align}
 \Tr \left[ \mathcal{N}(\rho_D)^n \right] = \frac{1}{( d_Rd_{B'})^n}\sum_{\tau \in \mathcal{S}_n} d_R^{k(\eta^{-1} \tau) } d_{B'}^{k(\tau)} = d_R^{1-n}{}_2F_1\left( 1-n,-n,2,\frac{d_R}{d_{B'}}\right) 
\end{align}
where we consider the regime where $d_R$ and $d_{B'}$ are both large, but with no restrictions on the relative sizes. 
The resulting von Neumann entropy coincides with Page's formula \cite{1993PhRvL..71.1291P}
\begin{align}
 S(\mathcal{N}(\rho_D)) = \lim_{n \rightarrow 1} \frac{1}{1-n}\log \Tr \left[ \mathcal{N}(\rho_D)^n \right] =
 \begin{cases}\log d_R - \frac{d_R}{2 d_{B'}}, & d_R < d_{B'}
 \\
 \log d_{B'} - \frac{d_{B'}}{2 d_{R}}, & d_R > d_{B'}
 \end{cases}.
 \label{page_formula}
\end{align}
Using \eqref{relative_ent_perm_sum}, the second term in \eqref{relative_replica} is given by a similar sum
\begin{align}
 \Tr \left[\mathcal{N}(\rho_D) \mathcal{N}(\sigma_D)^{n-1} \right]= \frac{1}{(d_D d_Rd_{B'})^n}\sum_{\tau \in S_n} d_R^{k(\eta^{-1} \tau) }\left( d_D d_{B'}\right)^{k(\tau)}.
\end{align}
This can be interpreted as the entropy of the radiation if it were coupled to an additional bath with a Hilbert space dimension identical to that of the diary. The $n\rightarrow 1$ limit is given by 
\begin{align}
 \lim_{n \rightarrow 1} \frac{1}{1-n}\log \Tr \left[ \mathcal{N}(\rho_D) \mathcal{N}(\sigma_D)^{n-1} \right]
 =
 \begin{cases}\log d_R - \frac{d_R}{2 d_{B'}d_D}, & d_R < d_{B'}d_D
 \\
 \log d_{B'}d_D - \frac{d_{B'}d_D}{2 d_{R}}, & d_R > d_{B'}d_D
 \end{cases}.
\end{align}
The relative entropy is therefore
\begin{align}
 D(\mathcal{N}(\rho_D)||\mathcal{N}(\sigma_D)) 
 &= \begin{cases}
 \frac{d_R(d_D-1)}{2d_{B'}d_D} , & d_R < d_{B'}
 \\
 \log \frac{d_R}{d_{B'}}+\frac{d_{B'}}{2 d_R} -\frac{d_{R}}{2 d_{B'}d_D} , & d_{B'}<d_R < d_{B'}d_D
 \\
 \log d_D+\frac{d_{B'}(1-d_D)}{2d_R}, & d_R >d_{B'}d_D
 \end{cases}.
 \label{Srel_HP_eq}
\end{align}
Plugging back into \eqref{petz_fidelity}, we obtain a lower bound on the fidelity: 
\begin{align}
 &F(\rho, [\mathcal{R}_{\sigma,\mathcal{N}}\circ \mathcal{N}](\rho)) \geq \begin{cases}
 \frac{1}{d_D}\, {\exp\left({\frac{d_R(d_D-1)}{2d_{B'}d_D}}\right)}, & d_R < d_{B'}
 \\
 \frac{d_R}{d_{B'}d_D}\, \exp\left({\frac{d_{B'}}{2 d_R} -\frac{d_{R}}{2 d_{B'}d_D}}\right) , & d_{B'}<d_R < d_{B'}d_D
 \\
 \exp\left({\frac{d_{B'}(1-d_D)}{2d_R}}\right), & d_R > d_{B'}d_D
 \end{cases}.
 \label{fidelity_inf_temp}
\end{align}
The transitions between the different lines of \eqref{fidelity_inf_temp} occur at the infinite-temperature versions of the times $t_{p}$ and $t_{p_2}$ respectively.
The fidelity is exponentially small prior to $t_p$. After $t_p$, the fidelity is still small but exponentially increases until it reaches a value close to one at $t_{p_2}$.
While the formula for $\exp (D(\mathcal{N}(\rho_D) || 
 \mathcal{N}(\sigma_D) ) - D(\rho_D|| \sigma_D) )$ on the RHS of \eqref{fidelity_inf_temp} is only exact in the limit of large Hilbert space dimensions, we find that it is remarkably accurate even for small system sizes in Fig.~\ref{petz_fidelity_HP}.
\begin{figure}
 \centering
 \includegraphics[width = .6\textwidth]{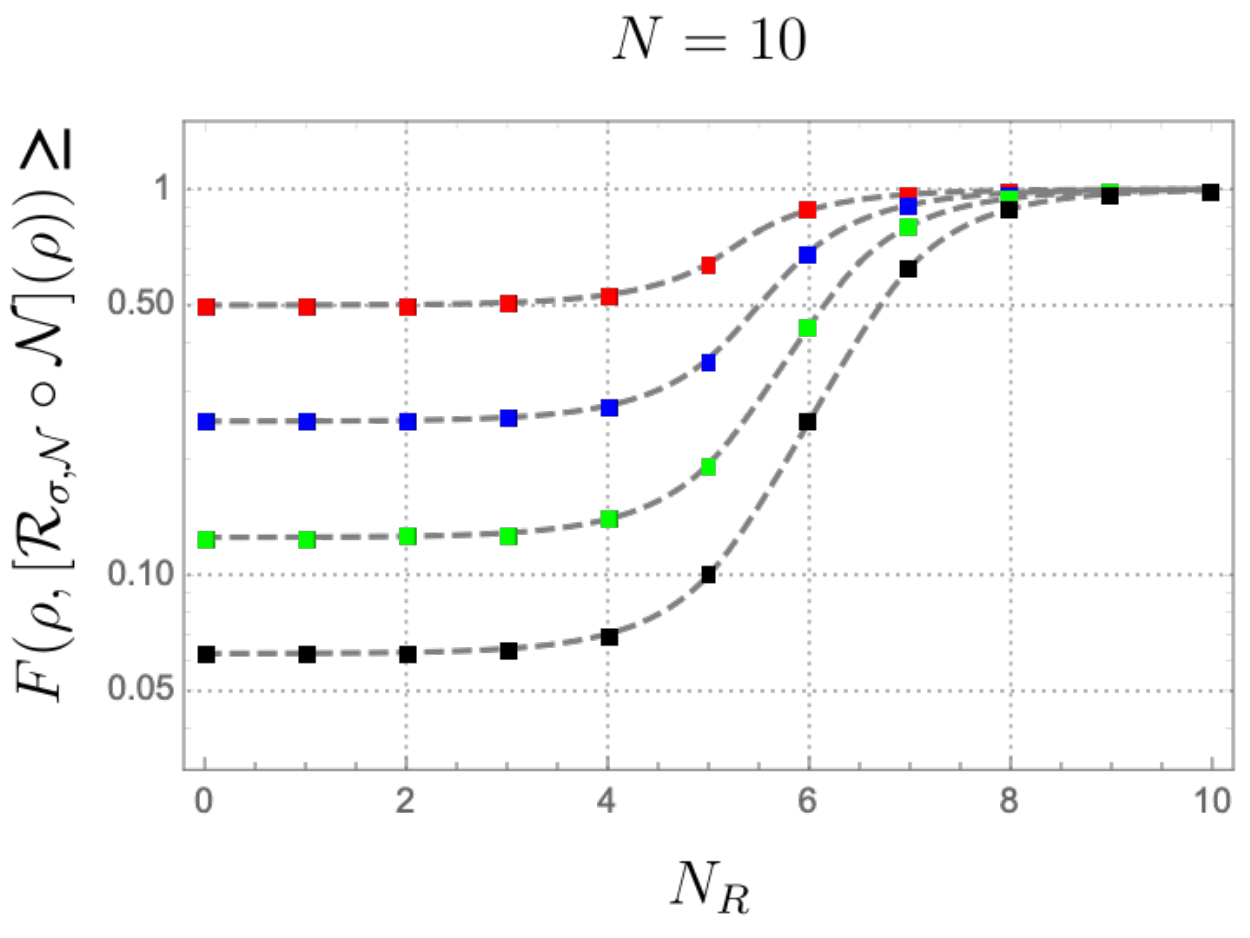}
 \caption{The lower bound on the fidelity of the Petz map to recover the diary thrown into the black hole from the radiation as a function of the number of qubits $N_R$ in the radiation, taking the total number of qubits in the system to be $N=10$. The diary is composed of one (red), two (blue), three (green), or four (black) qubits. The dashed lines are the analytic predictions calculated at large $N$, \eqref{fidelity_inf_temp}, although the agreement is already very precise. In the numerical data, we take a fiducial pure state tensored with the diary state and apply a random unitary matrix, computing the relative entropy before and after the random channel. We take $10^3$ disorder realizations. Error bars are plotted, though barely visible due to small fluctuations.
} 
 \label{petz_fidelity_HP}
\end{figure}

\subsubsection{Finite temperature}


Let us now consider finite temperature, and allow for general choices of the initial state of the diary $\rho_D$ and the reference state $\sigma_D$. Now, in general, we can no longer perform the sum over all permutations appearing in the equilibrium approximation. We can instead use the formula \eqref{relative_entropy_finite_T} based on the dominant permutations in different regimes, to find\footnote{Note that this formula breaks down and becomes infinite in the large $R$ regime if the initial diary states $\rho_D$ and $\sigma_D$ are close to orthogonal. This is guaranteed not to be the case if we choose $\sigma_D$ to be full rank.}
\begin{align}
 D(\mathcal{N}(\rho_D) || 
 \mathcal{N}(\sigma_D) ) \simeq \begin{cases}
 0 ,& S_{R}^{\rm (eq)} <S_{B'}^{\rm (eq)} + S(\rho_D)
 \\
 S_{R}^{\rm (eq)} - S_{B'}^{\rm (eq)} - S(\rho_D), & S_{B'}^{\rm (eq)} + S(\rho_D) < S_{R}^{\rm (eq)} \\
 & <S_{B'}^{\rm (eq)} +D(\rho_D \lvert \rvert \sigma_D) + S(\rho_D)
 \\
 D(\rho_D \lvert \rvert \sigma_D), & S_{R}^{\rm (eq)}> S_{B'}^{\rm (eq)} +D(\rho_D \lvert \rvert \sigma_D)
 \end{cases}.
 \label{relative_entropy_can}
\end{align}
As an aside, note that we can interpret this result in terms of quantum hypothesis testing, from the standard operational interpretation of the relative entropy. Say Alice were to ask Bob a yes or no question and Bob responded to Alice by writing his answer in his diary (i.e.~encoding one of two states $\rho_D$ and $\sigma_D$ in the diary), then throwing it into the black hole, knowing that Alice would collect the radiation from the black hole in order to decode his message using a quantum measurement. If this is repeated over and over, the error rate of Alice misidentifying Bob's answer is given by $e^{-D(\mathcal{N}(\rho_D) || 
 \mathcal{N}(\sigma_D))}$.

\eqref{relative_entropy_can} implies the following bound on the fidelity of recovery on using \eqref{petz_fidelity}: 
\begin{align}
 &F(\rho, [\mathcal{R}_{\sigma,\mathcal{N}}\circ \mathcal{N}](\rho) \geq 
 \begin{cases}
 e^{-D(\rho_D \lvert \rvert \sigma_D)},& S_{R}^{\rm (eq)} <S_{B'}^{\rm (eq)} + S(\rho_D) 
 \\
 e^{S_{R}^{\rm (eq)} - S_{B'}^{\rm (eq)} - S(\rho_D)-D(\rho_D \lvert \rvert \sigma_D)}, & S_{B'}^{\rm (eq)} +S(\rho_D)< S_{R}^{\rm (eq)}
 \\
 & <S_{B'}^{\rm (eq)} +D(\rho_D \lvert \rvert \sigma_D) + S(\rho_D)
 \\
 1, & S_{R}^{\rm (eq)}> S_{B'}^{\rm (eq)} +D(\rho_D \lvert \rvert \sigma_D) + S(\rho_D)
 \end{cases}.
 \label{fidelity_general}
\end{align}
We now see that the lower bound on fidelity first starts increasing at a time $t_p(\rho_D)$, at which $S_{R}^{\rm (eq)} =S_{B'}^{\rm (eq)} + S(\rho_D)$, and reaches its maximal value at 
$t_{p_2}(\sigma_D, \rho_D)$, at which $S^{\rm (eq)}_R = S^{\rm (eq)}_{B'} + D(\rho_D || \sigma_D) +S(\rho_D) = S^{\rm (eq)}_{B'} - \text{Tr}[\rho_D \log \sigma_D]$.

\subsection{Recovery channel: the Petz Map and its fidelity}

We progress to explicitly evaluating the fidelity of the Petz recovery map. The quantum channel that we seek to reverse is
\begin{align}
 \mathcal{N}: \mathcal{L}(\mathcal{H}_{D}) &\rightarrow \mathcal{L}(\mathcal{H}_R)
 \nonumber
 \\
 \rho_D &\mapsto \Tr_{B'}\left[ U^{\dagger} \left(\rho_D \otimes \rho_B \right)U \right],
\end{align}
As a reminder, the Petz map is given by 
\begin{align}
 \mathcal{P}_{\sigma,\mathcal{N}}: \mathcal{L}(\mathcal{H}_{R}) &\rightarrow\mathcal{L}(\mathcal{H}_{D}) 
 \nonumber
 \\
 X &\mapsto \sigma^{\frac{1}{2}} \mathcal{N}^{\dagger}\left(\mathcal{N}(\sigma)^{-\frac{1}{2}} X \mathcal{N}(\sigma)^{-\frac{1}{2}}\right)\sigma^{\frac{1}{2}}.
\end{align}

\subsubsection{Infinite temperature}

We first study Petz recovery at infinite temperature, again taking $\rho_D$ to be pure and $\sigma_D$ to be maximally mixed. Using the equilibrium approximation for this quantity in \eqref{fm_tau} with $\sI_{\al}=\mathbf{1}$, we have $ F(\rho , [\mathcal{P}_{\sigma,\mathcal{N}}\circ \mathcal{N}](\rho))= \lim_{m \rightarrow \ha} F_m$, where 
\begin{align}
 F_m 
 = \frac{1}{d_{D}^{2m+3}(d_R d_{B'})^{2m +2}}\sum_{\tau \in \mathcal{S}_{2m + 2}}d_R^{k(\eta^{-1} \tau)} d_{B'}^{k(\tau)} d_D^{k(\tau)+\zeta(\tau)} 
 \label{fmdef}
\end{align}
where $\zeta(\tau)$ is zero if the first and $(m+2)^{th}$ elements are in different cycles in the permutation $\tau$, and 1 if they are in the same cycle.

At early times, $d_{B'} \gg d_R$ and the identity element will dominate. The identity has $\zeta(e) = 0$, so that 
\begin{align}
 F_m = \frac{1}{d_{D}d_R^{2m +1}} + O\left(d_{B'}^{-1}\right), \quad F(\rho, [\mathcal{P}_{\sigma, \mathcal{N}}\circ \mathcal{N}](\rho)) = \frac{1}{d_D} = F(\rho_D, \sigma_D) \, . \label{early}
\end{align}
At late times,when $d_R \gg d_{B'}d_D$ the cyclic permutation will dominate, which has $\zeta(\eta) = 1$, so that
\begin{align}
 F_m = \frac{1}{(d_{B'}d_D)^{2m +1}} + O\left( d_R^{-1}\right) , \quad F(\rho , [\mathcal{P}_{\sigma,\mathcal{N}}\circ \mathcal{N}](\rho)) =1 \, . \label{late}
\end{align}

To understand how the recovery process improves from \eqref{early} to \eqref{late} as the size of the radiation grows, we can perform the full sum in \eqref{fmdef}, as we explain later in this subsection. Before turning to this detailed calculation, we can try to understand relevant time scales by looking at the leading corrections at both early and late times: 
\begin{enumerate} 
\item To find the leading corrections to \eqref{early}, note that
 the permutations with the next largest value of $k(\tau)$ after the identity are those with a single transposition, e.g.~(12). If it were not for the $d_D^{\zeta(\tau)}$ term, we would need to sum over all such permutations. However, only the permutation that transposes the first and $(m+2)^{th}$ elements has $\zeta(\tau) = 1$. Therefore, this permutation gives the leading correction, and on including it we find
\begin{align}
 F_m = \frac{1}{d_{D}d_R^{2m +1}} + \frac{1}{d_{D}d_{B'}d_R ^{2m }} + O\left(d_D^{-2}d_{B'}^{-1}\right).
\end{align}
so that we find the leading and next-to-leading order contributions to the fidelity at early times to be
\begin{align}
 F(\rho , [\mathcal{P}_{\sigma,\mathcal{N}}\circ \mathcal{N}](\rho)) = \frac{1}{d_D} + \frac{d_R}{d_D d_{B'}}, \label{early_petz}
\end{align}
which grows as more radiation is collected, reflecting the improved recovery. This correction starts to give a contribution comparable to the leading term at time $t_{p}$, which indicates that the fidelity starts to grow from its initial value at $t_p$. Note also that this correction has the same scaling as the correction in the lower bound \eqref{fidelity_inf_temp}, but is twice as large. 

When the diary is small, the terms with $\zeta(\tau) = 0$ can also be important. There are $\binom{2m+2}{2}-1$ leading terms, so that 
\begin{align}
 F_m = \frac{1}{d_{D}d_R^{2m +1}} + \frac{1}{d_{D}d_{B'}d_R ^{2m }} + \frac{m(3+2m)}{d_{D}^2d_{B'}d_R ^{2m }} +O(d_{B'}^{-2}d_D^{-2}).
\end{align}
The replica limit $m \rightarrow -\ha$ gives
\begin{align}
 F(\rho , [\mathcal{P}_{\sigma,\mathcal{N}}\circ \mathcal{N}](\rho)) = \frac{1}{d_D}\left(1 + \frac{d_R}{ d_{B'}}- \frac{d_R}{d_{D}d_{B'}}\right),
\end{align}

\item 
To understand the time scale at which the late-time value \eqref{late} is reached, let us now evaluate the leading correction to \eqref{late}. There are now $m(m+1)$ permutations satisfying both $k(\eta^{-1} \tau) = 2m+1$ and $\zeta(\tau) = 1$, leading to
\begin{align}
 F_m = \frac{1}{(d_{B'}d_D)^{2m +1}} + \frac{ m(m+1) }{d_R ( d_{B'}d_D)^{2m }} + O\left(d_R^{-1}d_D^{-2m-1} \right).
\end{align}
The leading and next-to-leading contributions to the fidelity at late times are therefore
\begin{align}
 F(\rho, [\mathcal{P}_{\sigma,\mathcal{N}}\circ \mathcal{N}](\rho)) = 1 - \frac{d_{B'}d_D }{4d_R} + O\left(d_R^{-1} d_D^0 \right). \label{late_petz}
\end{align}
Note that the correction had to be negative because the fidelity is bounded above by unity. The correction becomes comparable to the leading term at times earlier than $t_{p_2}$. Note also that this correction has the same scaling as, but is half the size of, the lower bound correction in \eqref{fidelity_inf_temp}.

For small diaries, there are $(m+1)^2$ other terms which are also important in the correction to the late-time value, and we find 
\begin{align}
 F(\rho, [\mathcal{P}_{\sigma,\mathcal{N}}\circ \mathcal{N}](\rho)) = 1 - \frac{d_{B'}d_D }{4d_R} + \frac{d_{B'} }{4d_R} + O\left(d_R^{-2} d_D^{-1} \right).
\end{align}
\end{enumerate} 

Hence, for the infinite temperature case, \eqref{early_petz} and \eqref{late_petz} together give the same predictions for the time scales at which the fidelity of $\sP$ first increases and the time scale at which it saturates, as the predictions based on the lower bound for some recovery channel in \eqref{fidelity_inf_temp}. 

Finally, we can also examine the crossover regime of \eqref{fmdef}, where the fidelity becomes $O(1)$. At the time $t_{p_2}$, $d_R = d_{B'} d_D := d$, so the sum simplifies to 
\begin{align}
 F_m \simeq \frac{1}{d^{4m+4}} \sum_{\tau \in \mathcal{S}_{2m +2}} d^{k(\eta^{-1} \tau) + k(\tau) }d_D^{\zeta(\tau)-1}.
\end{align}
The exponent is maximized at $2m+4$ when $\tau$ is a non-crossing permutation that has the first and $(m+2)^{th}$ factors in the same cycle. Out of the $C_{2m + 2}$ total non-crossing permutations (where $C_n$ is the $n-$th Catalan number), only $C_{m + 1}^2$ have $\zeta(\tau) = 1$, a statement we will soon prove. Therefore,
\begin{align}
 F_m \simeq \frac{C_{m+1}^2}{d^{2m+1}}. \label{633}
\end{align}
Taking the $m\rightarrow -\frac{1}{2}$ limit, we find the fidelity at $t_{p_2}$ to have an $O(1)$ value independent of the dimensions, 
\begin{align}
 F(\rho , [\mathcal{P}_{\sigma,\mathcal{N}}\circ \mathcal{N}](\rho)) = \frac{64}{9\pi^2} \simeq 0.72.
\end{align}
This fidelity is markedly larger than the lower bound from \eqref{fidelity_inf_temp}, which is $e^{-\frac{1}{2}} \simeq0.61$.
Again, for small diaries, we include the subleading non-crossing permutation with $\zeta(\tau) = 0$ to find
\begin{align}
 F(\rho , [\mathcal{P}_{\sigma,\mathcal{N}}\circ \mathcal{N}](\rho)) = \frac{64}{9\pi^2}+ \frac{1-\frac{64}{9\pi^2}}{d_D}.
\end{align}

We now turn to the full calculation of  the fidelity in the planar limit, i.e., we carry out the full sum in \eqref{fmdef} analytically. 
Recall that the planar limit corresponds to non-crossing permutations $\tau$ where $k(\eta^{-1}\tau)+k(\tau) = 2m +3$. We claim that the subset of these permutations with $\zeta(\tau) = 1$ and $k(\eta^{-1}\tau) = p$ is given by a product of Narayana numbers
\begin{align}
	NC_{2m+2, \zeta = 1,p} &= \sum_{p_1 + p _2= p} N_{m+1,p_1}N_{m+1,p_2} = \sum_{p_1= 1}^{p} N_{m+1,p_1}N_{m+1,p-p_1}
	\nonumber
	\\
	&=\frac{\binom{m+1}{p-2} \binom{m+1}{p-1} \,
  _4F_3(-m-1,-m,1-p,2-p;2,m-p+3,m-p+4;1)}{m+1} .
  \label{nara_prod}
\end{align}
We note that the previous statement regarding the square of Catalan numbers above \eqref{633} is the special case where we sum the above equation from $p = 2$ to $ p = 2m+2$.

\eqref{nara_prod} can be proven by considering the elements of the permutation group as a circular lattice of $2m + 2$ points then moving to the dual lattice (see Fig.~\ref{complement_proof}). In the dual lattice, we count the number of non-crossing permutations that factorize into non-crossing permutations of $m + 1$ and $m+ 1$ elements. There is a unique non-crossing permutation of the original lattice corresponding to each one of these permutations, commonly known as the ``Kreweras complement.'' This is the maximally extended permutation in the dual graph that does not cross the permutation. It is clear that for each factorized permutation in the dual lattice, the Kreweras complement in the original lattice has $\zeta(\tau) = 1$. Indeed, the number of non-crossing permutations in the dual graph with $k(\tau) = p$ is the same as the number of non-crossing permutations in the original graph with $k(\eta^{-1}\tau) = p$. Using this insight and the prior knowledge of counting the number of non-crossing permutations with a given $k(\tau)$, we arrive at \eqref{nara_prod}.
\begin{figure}
\centering
\includegraphics[height = 7cm]{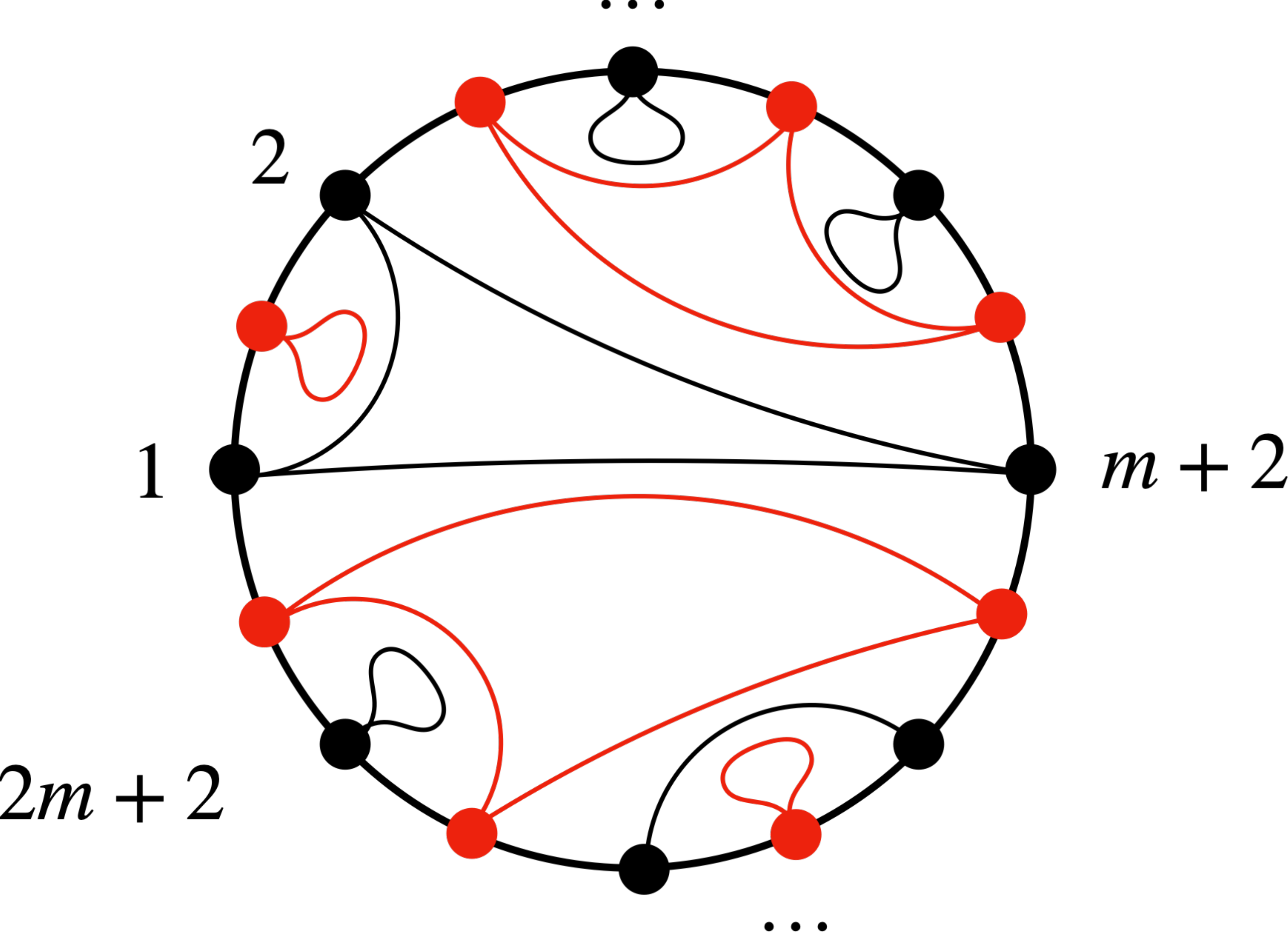}
\caption{A circular lattice of $2m + 2$ elements is shown as black dots. The dual lattice is the red dots. We show an example of a non-crossing permutation on the red lattice that factorizes across the first and $m + 2$nd elements of the original lattice. Clearly, the Kreweras complement (shown in black lines) connects the first and $m + 2$nd elements such that $\zeta(\tau) = 1$. }
\label{complement_proof}
\end{figure}

We can now find the sum over all permutations with $\zeta(\tau) = 1$ in \eqref{fmdef}, which we call $\tilde{F}_m$: 
\begin{align}
	\tilde{F}_m 
 &= \frac{1}{(d_R d_{B'} d_D)^{2m +2}}\sum_{p = 2}^{2m+2}NC_{2m+2, \zeta = 1,p} d_R^{p} (d_Dd_{B'})^{2m+3-p}
 \nonumber
 \\
 &= \frac{1}{d_{B'} d_D}\begin{cases}
 \left(d_R^{-m} {}_2F_1\left(-m,-m-1,2,\frac{d_R}{d_D d_{B'}}\right) \right)^2, & d_R < d_D d_{B'}
\\
 \left( (d_D d_{B'})^{-m} {}_2F_1\left(-m,-m-1,2,\frac{d_D d_{B'}}{d_R}\right) \right)^2, & d_R > d_D d_{B'}
 \end{cases}.
 \end{align}
To enumerate the terms with $\zeta(\tau)=0$ and $k(\eta^{-1}\tau) =p$, we note that this must simply be the remaining non-crossing permutations of which there are $N_{2m+2,p}-NC_{2m+2, \zeta = 1,p} $. We refer to this second contribution to $F_m$ as $\bar{F}_m$, and immediately find
\begin{align}
	\bar{F}_m =\begin{cases} \frac{1}{d_D d_R^{2m+1}}\, {}_2 F_1\left(-2m-1,-2m-2,2 ,\frac{d_R}{d_D d_{B'}}\right) - \frac{\tilde{F}_m}{d_D}, & d_R < d_D d_{B'}
\\
\frac{1}{d_D (d_D d_{B'})^{2m+1}}\, {}_2 F_1\left(-2m-1,-2m-2,2 ,\frac{d_D d_{B'}}{d_R}\right) - \frac{\tilde{F}_m}{d_D}, & d_R > d_D d_{B'}
	\end{cases} 
\end{align}
This expression simplifies in the $m\rightarrow -\frac{1}{2} $ limit to 
\begin{align}
	\bar{F}_{-\frac{1}{2}} =\frac{1}{d_D} - \frac{\tilde{F}_{-\frac{1}{2}}}{d_D}
\end{align}
Therefore, the fidelity is
\begin{align}
	F(\rho , [\mathcal{P}_{\sigma,\mathcal{N}}\circ \mathcal{N}](\rho)) 
	= \begin{cases} 
(1- \frac{1}{d_D})\, \frac{d_R}{d_{B'}d_D} \, {}_2F_1\left(\ha,-\ha,2,\frac{d_R}{d_D d_{B'}}\right)^2 + \frac{1}{d_D}	& d_R < d_D d_{B'} \\
	 (1- \frac{1}{d_D}) \, {}_2F_1\left(\ha,-\ha,2,\frac{d_D d_{B'}}{d_R}\right)^2 + \frac{1}{d_D} & d_R > d_D d_{B'} 
	\end{cases} 
	\label{fidelity_inf}
\end{align}
This expression is plotted in Fig.~\ref{petz_recovery_fidelity}, where it matches perfectly with numerical tests for finite system sizes.

\begin{figure}
 \centering
 \includegraphics[width =.6\textwidth]{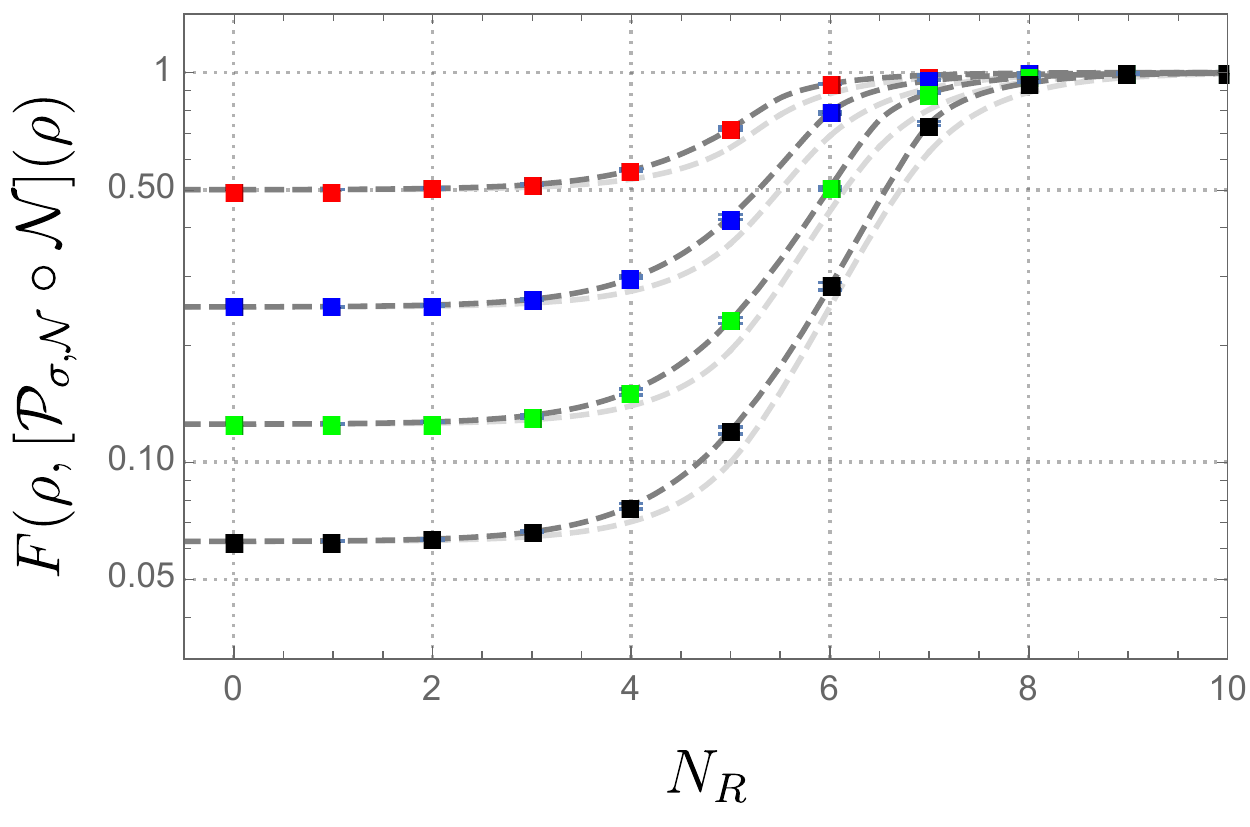}
 \caption{The fidelity of the Petz map for the Hayden-Preskill protocol at infinite temperature, as a function of the number of qubits $N_R$ in the radiation. The total number of qubits in the black hole and the radiation is 10 in all cases, and the different curves correspond to different sizes of the diary, ranging from 1 to 4 qubits from top to bottom. The gray dashed curve is the analytic result given by \eqref{fidelity_inf}. The light gray curves are the lower bounds set by the change in relative entropy. In the numerical data, we take a fiducial pure state tensored with the diary state and apply a random unitary matrix and partial trace, followed by the Petz recovery map. We numerically evaluate the between this recovered state and the initial state. Again, we show the average over $10^3$ disorder realizations and the corresponding error bars.}
 \label{petz_recovery_fidelity}
\end{figure}

\subsubsection{Finite temperature}

At finite temperature, we have far less control over the contributions from subleading permutations, and hence we are able to find the value of the fidelity corresponding to the sum over all permutations in the equilibrium approximation only in some special cases. First, note that from the general discussion in Section \ref{sec:petzA} based on leading permutations, we expect again that at early times when the identity permutation dominates, the fidelity is given by 
\be 
 F(\rho , [\mathcal{P}_{\sigma,\mathcal{N}}\circ \mathcal{N}](\rho)) = F(\rho_D , \sigma_D) .
\ee
where in the last expression we take $\sigma_D$ to be an equilibrium density matrix on $D$, and $\rho_D$ to be some pure state. Similarly, at sufficiently late times, the cyclic permutation should dominate, so that we have 
\be 
 F(\rho , [\mathcal{P}_{\sigma,\mathcal{N}}\circ \mathcal{N}](\rho)) =1 \, . 
\ee 

At intermediate times, the fidelity will interpolate between these values. To probe this regime and find the relevant time scales, let us first look at the leading corrections to the dominant permutations. The first correction to early-time contribution from the identity permutation comes from the permutation that swaps the first and $(m+2)^{th}$ copies,
\begin{align}
 F_m &= \frac{Z_{2m+2,R}}{Z_{1,R}^{2m + 2}}F(\rho_D, \sigma_D)+ \frac{Z_{2,B'}Z_{m+1,R}^2}{Z_{1,B'}^2Z_{1,R}^{2m + 2}} \Tr \left[ \sigma_D^{\frac{1}{2}}\rho_D \sigma_D^{\frac{1}{2}}\rho_D\right].
\end{align}
In the replica limit, this can be rewritten in terms of Renyi entropies as
\begin{align}
 F(\rho, [\mathcal{P}_{\sigma,\mathcal{N}}\circ \mathcal{N}](\rho)) = F(\rho_D,\sigma_D) + e^{S^{\rm (eq)}_{\frac{1}{2}, R}-S^{\rm (eq)}_{2,B'}}\Tr \left[ \sigma_D^{\frac{1}{2}}\rho_D \sigma_D^{\frac{1}{2}}\rho_D\right]. \label{finite_corr}
\end{align} 
We can assume $\Tr \left[ \sigma_D^{\frac{1}{2}}\rho_D \sigma_D^{\frac{1}{2}}\rho_D\right]$ is of roughly the same magnitude as $F(\rho_D, \sigma_D)$ (note that if we take $\sigma_D$ to be that maximally mixed state and $\rho_D$ to be pure, then both are equal to $\frac{1}{d_D}$). This suggests that the fidelity begins to grow significantly from its initial value at the time scale $t_b$, at which $S^{\rm (eq)}_{\frac{1}{2}, R} = S^{\rm (eq)}_{2,B'}$. 

However, in order to reliably calculate the fidelity past $t_b$, we must sum over the contributions from all permutations corresponding to planar diagrams, a calculation we explain in Appendix \ref{petz_app_can} using a generating functional method. For this calculation, we assume that $\sigma_D$ is maximally mixed, $\rho_D$ is pure, the radiation is at infinite temperature, and the black hole is at inverse temperature $\beta$, with the density of states as in $AdS_3$, $\rho(E) = e^{c V \sqrt{E/V}}$. Now if we define 
\be
x \equiv {\log d_R \ov V} , \quad y \equiv {\log d_D \ov V}, 
\ee
where $V$ is the volume of $B'$ and $d_R$ is the dimension of the radiation, 
then the various relevant time scales are given by:
\be
t_b: \quad x = {3 c^2 \ov 8 \b}, \quad t_p: \quad x = {c^2 \ov 2 \b} , \quad t_{p_2}: \quad x = {c^2 \ov 2 \b} + y \ .
\ee
We find that for small diaries (i.e. $y$ is $O(V^{-1})$), the fidelity grows rapidly at $t_p$ and the time scale $t_b$ does not turn out to be relevant,
\be 
F(\rho , [\mathcal{P}_{\sigma,\mathcal{N}}\circ \mathcal{N}](\rho)) \approx \begin{cases} 
 \frac{1}{d_D} & x < \frac{c^2}{2\beta} \\
 1 & x > \frac{c^2}{2\beta} 
\end{cases} 
\label{small_fid}
\ee
with the transition occurring in a region of size $O((V\beta)^{-\frac{1}{2}})$ in $x$. On the other hand, for a sufficiently large diary such that
there is a regime where $\frac{3c^2}{8\beta} < x < \frac{5c^2}{16\beta} +y$, we have 
\be 
F(\rho , [\mathcal{P}_{\sigma,\mathcal{N}}\circ \mathcal{N}](\rho))
= \begin{cases} 
\frac{1}{d_D} & x< \frac{3c^2}{8\beta} \\
 e^{V(-\frac{3c^2}{8\beta} + x-y)} =e^{\log d_R-S_{2, B'}-\log d_D} & \frac{3c^2}{8\beta} < x < \frac{5c^2}{16\beta} +y \\ 
 e^{V (c \sqrt{-\frac{c^2}{4\beta^2} + \frac{x-y}{\beta}} -x+y )} & \frac{5c^2}{16\beta} +y < x < \frac{c^2}{2\beta} +y\\
 1 & x> \frac{c^2}{2\beta} +y
\end{cases} 
\label{large_fid_1}
\ee
So for a sufficiently large diary, the fidelity starts increasing exponentially from its initial value of $\frac{1}{d_D}$ at time $t_b$, and the initial increase is precisely as predicted by \eqref{finite_corr}. Note that the lower bound on the fidelity from the change in relative entropy from \eqref{fidelity_general} is not sensitive to the time scale $t_b$, and only starts growing at the Page time $t_p$. 

When $y$ is $O(1)$ but the diary is not sufficiently large such that such that 
there is a regime where $\frac{3c^2}{8\beta} < x < \frac{5c^2}{16\beta} +y$, the fidelity starts increasing from $\frac{1}{d_D}$ at a time $t_r$ between $t_b$ and $t_p$ defined by 
\be 
t_r : \quad x = \frac{c^2}{2\beta} + 2y - c \sqrt{\frac{y}{\beta}}
\ee
so that we have 
\be 
F(\rho , [\mathcal{P}_{\sigma,\mathcal{N}}\circ \mathcal{N}](\rho))
= \begin{cases} 
\frac{1}{d_D} & x< \frac{c^2}{2\beta} + 2y - c \sqrt{\frac{y}{\beta}} \\
 e^{V (c \sqrt{-\frac{c^2}{4\beta^2} + \frac{x-y}{\beta}} -x+y )} & \frac{c^2}{2\beta} + 2y - c \sqrt{\frac{y}{\beta}} < x < \frac{c^2}{2\beta} +y\\
 1 & x> \frac{c^2}{2\beta} +y
\end{cases} 
\label{int_fid}
\ee
$t_r$ is increasingly earlier for larger $\log d_D$, and eventually becomes equal to $t_b$. 

 \eqref{small_fid},\eqref{large_fid_1} and \eqref{int_fid} are derived with several approximations for the thermodynamic limit. We plot a more exact expression resulting from the calculations of Appendix \ref{petz_app_can} for finite but large volumes in Fig.~\ref{petz_recovery_fidelity_finite.pdf}. Note in particular that for the case where $d_D$ is large, we see the initial exponential growth of fidelity $e^{S^{\rm (eq)}_{\ha, R} - S^{\rm (eq)}_{2, B'} - \log d_D}$ predicted by \eqref{finite_corr}. For a small diary, the curves obtained from increasing system sizes with fixed $\log d_D$ gradually approach an increasingly sharp transition, as expected from \eqref{small_fid}.

The discussion in this section so far has entirely been for the canonical ensemble case. In Appendix \ref{petz_app_mic}, we also consider the example of $\sI_{\al}$ as in the microcanonical example of \eqref{mic_inf_temp}, corresponding to a case where the radiation is divided into two parts $R_1$ and $R_2$ such that there is energy conservation between $R_1$ and $B'$ while $R_2$ is at infinite temperature. For a small diary, we again find that the fidelity grows from $\frac{1}{d_D}$ to $1$ rapidly at $t_p$. For a large diary, we find that the fidelity reaches 1 at $t_{p_2}$, and that there is a time scale $t_r< t_p$ when the fidelity starts to grow above $\frac{1}{d_D}$. But for this example, $t_r$ does not seem to be related to the time $t_b$ at which $\sE(R_1, R_2)$ starts to grow in any regime. 


\begin{figure}[!h]
	\centering
	\includegraphics[width =.48\textwidth]{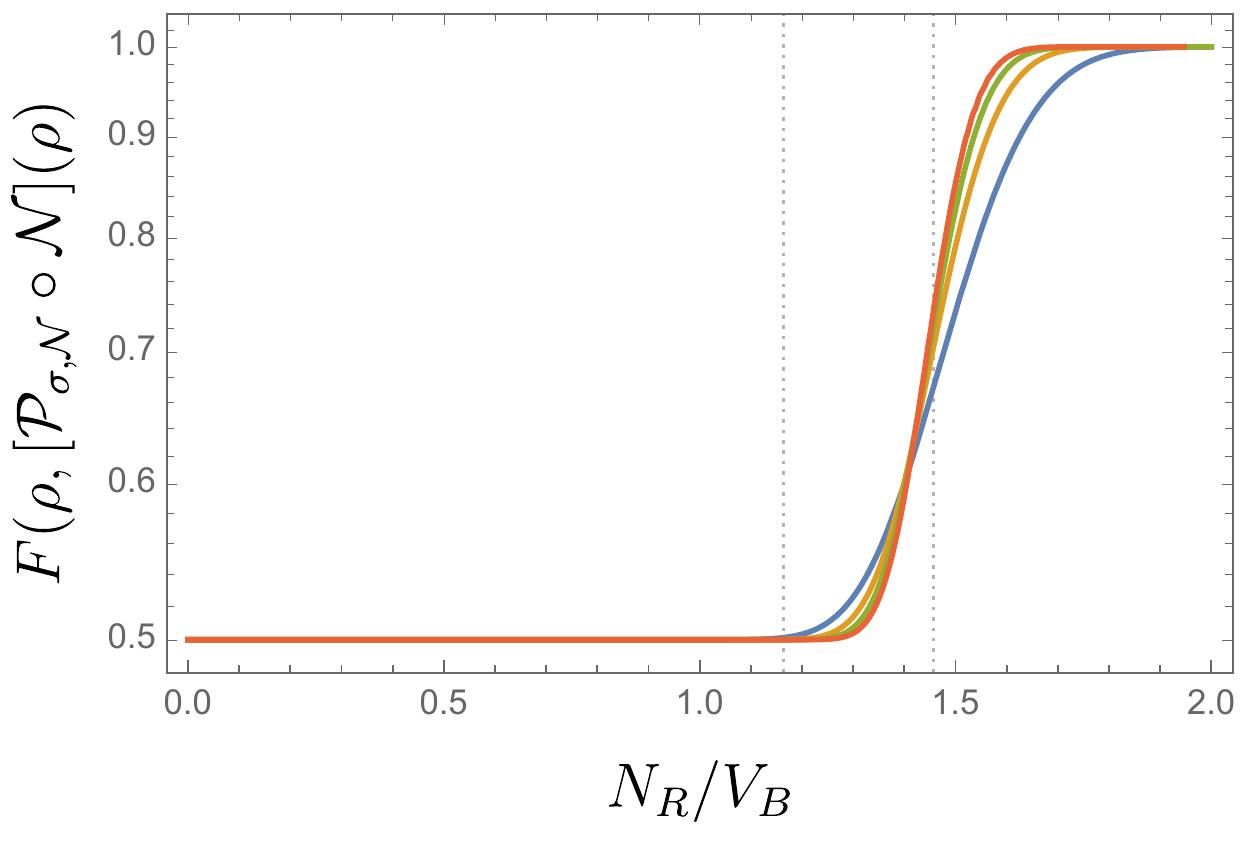}
	\includegraphics[width =.48\textwidth]{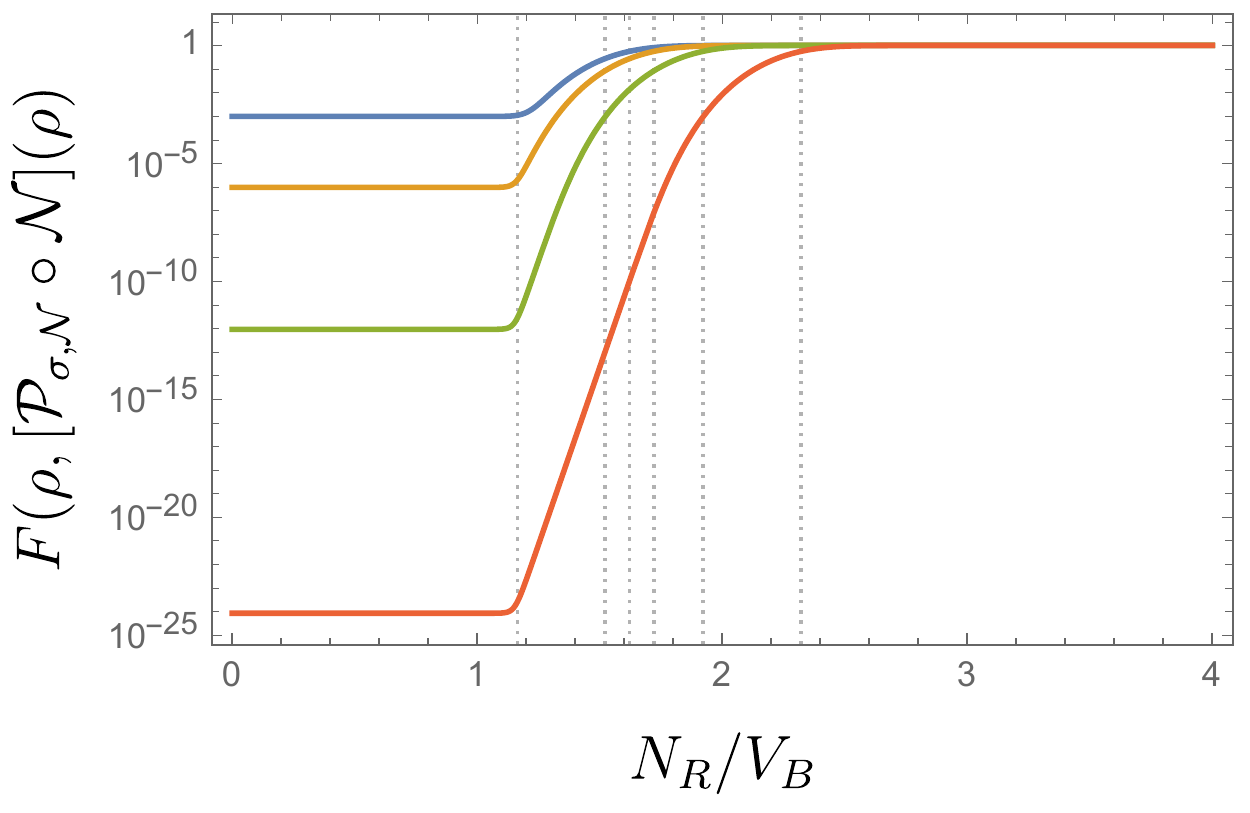}
	\caption{Left: The fidelity of the Petz recovery map is shown for a diary consisting of a single qubit. The different plots are for different $V = 100$, $200$, $300$, and $400$ (from blue to red). The transition approaches the second vertical line, which corresponds to $t_p$ in the thermodynamic limit. The first vertical line is $t_b$. Right: The fidelity of the Petz recovery map for large diaries of sizes 10, 20, 40, and 80 qubits (top to bottom). The first two vertical lines denote $t_b$ and $t_p$ which are separated even in the thermodynamic limit. We also display the $t_{p_2}$'s for the respective diary sizes, always with $V = 100$. We take $R$ to be at infinite temperature and $B'$ to be at finite temperature with Cardy-like density of states $\rho(E) = e^{V\sqrt{\frac{E}{V}}}$ in both plots, taking $\beta = 1/2$. Notably, the fidelity increases significantly between $t_b$ and $t_p$ and does not saturate to one until after $t_{p_2}$.}
	\label{petz_recovery_fidelity_finite.pdf}
\end{figure}

\section{Conclusions and discussions} 
\label{conclusion}

In this paper, we generalized the equilibrium approximation introduced in~\cite{2020arXiv200801089L} to various quantum-informational measures beyond the entanglement entropy, such as Renyi and logarithmic negativities, relative entropy, Petz map fidelity, and reflected entropy. 
We then studied in detail the entanglement correlations within equilibrated pure states at a {\it finite} temperature. We found rich entanglement structures exhibited by the phase diagrams of the logarithmic negativity, mutual information,  
Renyi negativities, and Renyi mutual informations. In addition to their implications for general quantum chaotic systems,
our results also give predictions for the entanglement structure of an evaporating black hole. The verification of these predictions from gravity calculations would help in further probing the quantum nature of black holes as well as informational aspects of quantum gravity. In particular, we predicted the existence of a new time scale $t_b$ before the Page time, when entanglement correlations within the Hawking radiation start to emerge. 

We also studied in detail the process of information recovery from a subsystem, using the Petz map. This issue is central to understanding how and when quantum information is transferred from a black hole to its radiation. We showed that even at finite temperature, the fidelity of the recovery map as well as an upper bound on it provided by relative entropy can 
be expressed in terms of natural quantum-informational measures in thermal density matrices. Intriguingly, we found evidence that $t_b$ is the relevant time scale at which the fidelity of recovery starts to increase for a large diary, which suggests that correlations that are detected by negativity but not detected by mutual information play an important role in information recovery.

There are many future directions to explore. Our formulations of the equilibrium approximation for relative entropy and reflected entropy should find other applications. For example, it would be interesting to explore in detail the universal behavior of reflected entropy for finite temperature systems, which we only briefly touched upon. Such studies should also 
provide further understanding of multipartite entanglement in a quantum many-body system, especially 
among different parts of the radiation and the black hole. 

It would be highly desirable to understand better the nature of the entanglement in the region of the phase diagram where $\sE \gg I$. The question is of particular interest for black hole physics. In this regard, it would be interesting to understand whether it is possible to obtain a Lorentzian derivation of the non-zero negativity before the Page time, and see whether there is any semi-classical or geometric description of bound entanglement. Another interesting question is about how new phases which might not correspond to analytic continuation, similar to the ones we found in Fig.~\ref{fig:phase_diagram_a2_inf}, would be manifested in a gravity calculation.

From a purely quantum information-theoretic perspective, it would be interesting to investigate if there are deeper direct connections between logarithmic negativity and recoverability of quantum information.  In our calculations for the Hayden-Preskill protocol, the known lower bound on the fidelity of recovery from the relative entropy turned out not to be sensitive to the $t_b$ time scale at which both the Petz map fidelity and the logarithmic negativity start to grow. Our results thus hint that there could be strictly stronger lower bounds on fidelity of recovery if one considered notions related to the partial transpose.

\acknowledgments

We thank Chris Akers, Tom Faulkner, Simon Lin, and Pratik Rath for sharing their manuscripts with us. This work is supported by the Office of High Energy Physics of U.S. Department of Energy under grant Contract Numbers DE-SC0012567 and DE-SC0020360 (MIT contract \# 578218).

\begin{appendix} 
\begin{section}{Properties of the permutation group}
\label{app:perm}

We can define a notion of distance between elements $\sigma, \tau$ of the permutation group $\sS_n$ with the Cayley distance, 
\be 
d(\sigma, \tau) = n - k(\sigma \tau^{-1}). \label{cayley}
\ee
This definition of the distance is consistent with the triangle inequality, 
\be 
d(\sigma, \tau) + d(\tau, \rho)\geq d(\sigma, \rho). \label{triangle}
\ee
\eqref{cayley} and \eqref{triangle} can be used to show that
\begin{align} 
k(\eta \tau^{-1})+ k(\tau)&\leq n+1 \\ k(\eta^{-1} \tau^{-1})+ k(\tau)&\leq n+1 \\ k(\eta^{-1}\tau) + k(\eta \tau) & \leq \begin{cases} n+2 & n \text{ even} \\ n+1 & n \text{ odd}
\end{cases} 
\label{twosums}
\end{align} 
where we use the fact $k(\eta)=1$ for all $n$, $k(\eta^2)=2$ for even $n$, and $k(\eta^2)=1$ for odd $n$. 
We then also have 
\be 
\begin{gathered} 
k(\eta^{-1} \tau)+ k(\eta \tau) + k(\tau) \\= \frac{k(\eta \tau^{-1}) + k(\tau)}{2}+ \frac{k(\eta^{-1} \tau^{-1}) + k(\tau)}{2} + \frac{k(\eta\tau^{-1}) +k(\eta\tau)}{2} \leq \begin{cases} \frac{3n}{2} +2 & n \text{ even}\\
\frac{3n+3}{2}& n\text{ odd}
\end{cases} \label{3sum}
\end{gathered}
\ee
where the upper bound follows from the inequalities for the individual terms in the second line, and we have used (for instance) that $k(\eta\tau) = k(\tau^{-1}\eta^{-1}) = k(\tau^{-1}(\eta^{-1} \tau^{-1})\tau) = k(\eta^{-1}\tau^{-1})$. We will now find the set of permutations that saturates \eqref{3sum}, which is the same as the set that saturates the inequalities for each of the three terms on the LHS of the second line. 

First note that the permutations that saturate the inequality for both the first and the second term are in one-to-one correspondence with the non-crossing partitions of $n$ elements. Given a non-crossing partition $\{a^1_1, ..., a^1_{m_1} \}, \{ a^2_1, ..., a^2_{m_2} \}, ..., \{a^k, ..., a^k_{m_k}\}$, where the elements of each set are listed in ascending order, the corresponding permutations are given by 
\begin{align} 
\tau_1 = (a^1_1~a^1_2 ... ~ a^1_{m_1} ) ( a^2_1~ ... ~a^2_{m_2} ) ... (a_1^k ~ ... ~ a^k_{m_k}) :\quad k(\eta^{-1} \tau_1^{-1})+ k(\tau_1)= n+1\\
\tau_2= (a^1_{m_1} ~ ... ~a^1_2 ~ a^1_1 ) ~ ( a^2_{m_2} ~... ~a^2_1 ) ... (a^k_{m_k} ~ ... ~ a_1^k) :\quad k(\eta \tau_2^{-1})+ k(\tau_2)= n+1
\end{align} 
The permutations that maximize the sum of the first two terms in the second line of \eqref{3sum} are those that lie in the intersection of the above values of $\tau_1$ and $\tau_2$. The difference in the ordering of the elements within the cycles in $\tau_1$ and $\tau_2$ does not change the permutation if and only if the permutation consists entirely of one-cycles and two-cycles. Hence, the permutations that simultaneously saturate the inequalities for the first and second terms are in one-to-one correspondence with the non-crossing partitions of $n$ elements that have blocks only of size 1 and 2. 

Next, note that for any permutation $\tau$ that saturates the inequalities for both the first and the second terms, we have
\be 
k(\eta^{-1}\tau^{-1}) + k(\eta \tau^{-1}) = 2n+2- 2 k(\tau) 
\ee 
To maximize the RHS, we need to minimize $k(\tau)$ among permutations that only have one-cycles and two-cycles. 

For even $n$, $k(\tau)$ is minimized for permutations that only have two cycles, where it is given by $n/2$. Hence, the permutations saturating \eqref{3sum} for even $n$ are the ones corresponding to non-crossing partitions of $n$ elements that have blocks only of size 2. We refer to this set of permutations as $\{\tau\}^{\ast}$. For any such permutation, $k(\eta \tau^{-1})= k(\eta \tau)= \frac{n}{2}+1$. The number of such permutations is equal to $C_{n/2}$, where $C_{n/2}$ is the Catalan number. 

Similarly, for odd $n$, the permutations saturating \eqref{3sum} are the ones corresponding to non-crossing partitions of $n$ elements that have $\frac{n-1}{2}$ blocks of size 2 and 1 block of size 1. We refer to this set of permutations as $\{\tau\}^{\ast}_{\rm odd}$. For any such permutation, $k(\tau)= k(\eta \tau^{-1})= k(\eta \tau)= \frac{n+1}{2}$. The number of such permutations is equal to $n~ C_{\frac{n-1}{2}}$. 
\end{section} 

\begin{figure}[]
\includegraphics[width=5cm]{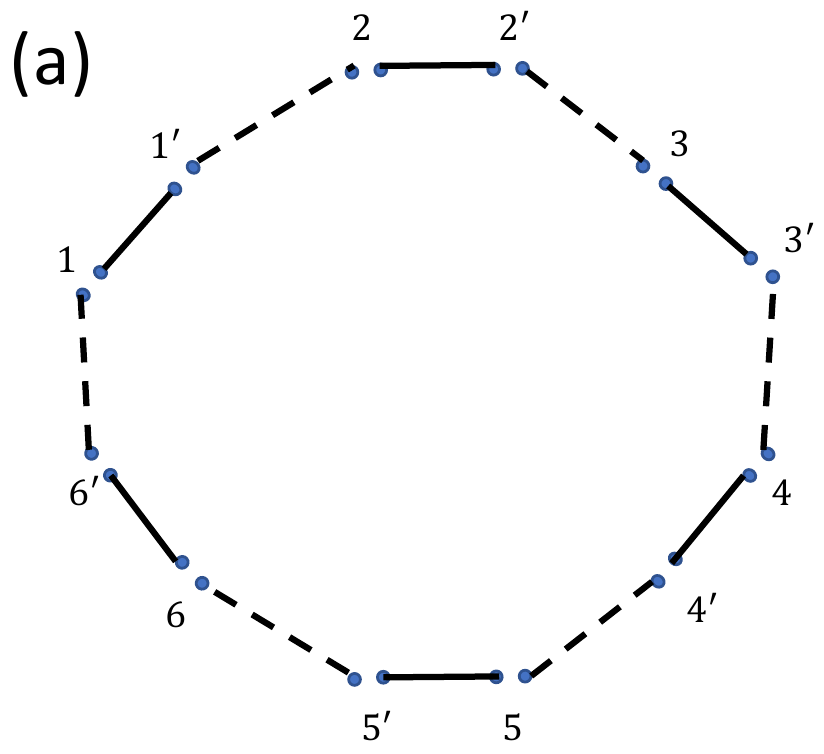} \includegraphics[width=5cm]{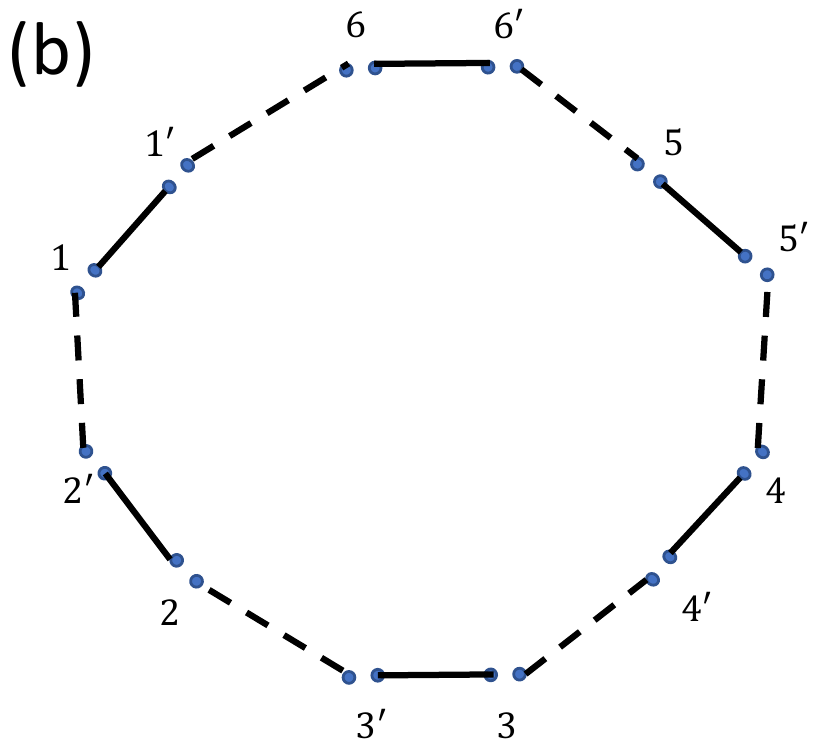}
\includegraphics[width=15cm]{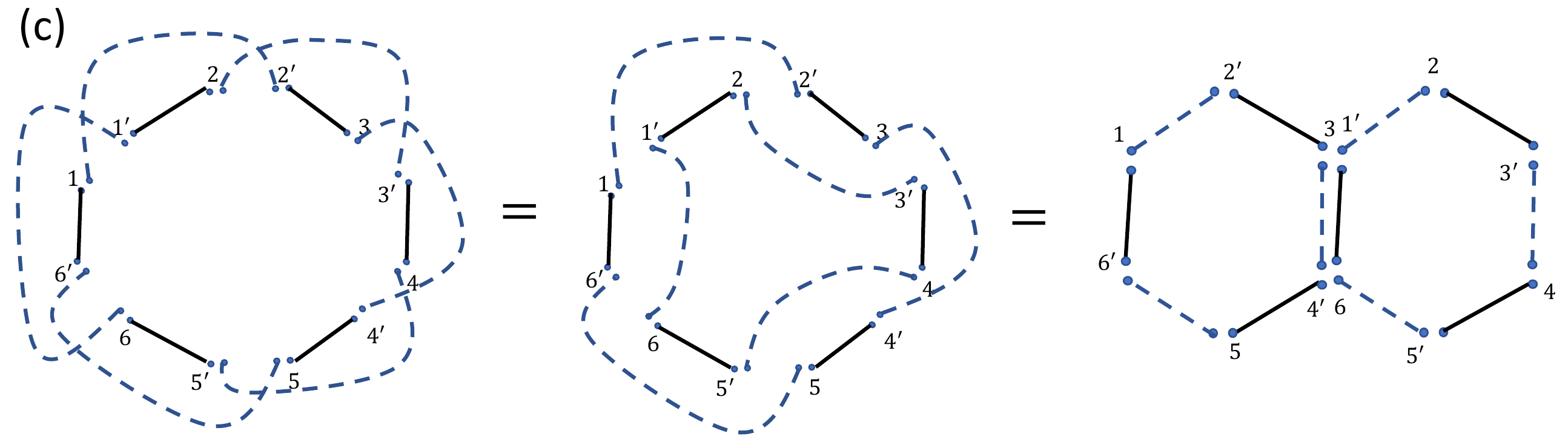}
\caption{Boundary conditions for the diagrams corresponding to the three terms in the second line of \eqref{3sum}.}
\label{3fig}
\end{figure} 
An alternative way to understand the inequality \eqref{3sum} and the permutations that saturate it is in terms of 'tHooft-like double-line diagrams. The first line of \eqref{3sum} is equal to the total number of loops in diagrams of the kind shown in Fig.~\ref{fig:negativity_figs}, which involve sets of three lines, so we cannot directly modify 'tHooft's method to that expression. It is useful to separately consider the three terms in the second line of \eqref{3sum}, each of which can be equated with the total number of loops in a diagram with a different boundary condition. The boundary conditions for the first two terms are as shown in Fig.~\ref{3fig}(a) and (b), and the ones for the third term are explained in \ref{3fig}(c), where it will be convenient to use the third form to apply 'tHooft's method. 

By adding an extra loop around each of the diagrams, they can be seen as double-line diagrams, which can then be mapped to polygons by replacing the double lines with single lines. An example of the mapping to polygons for a particular $\tau$ that saturates \eqref{3sum}, with the boundary conditions of Figs.~\ref{3fig}(a) and (c), is shown in Fig.~\ref{fig:thooft}. The total number of loops in the double-line diagram is equal to the total number of faces of the corresponding polygon. If the polygon can be flattened on a surface of minimum genus $h$, then 
\be 
F = E-V +2 - 2h. 
\ee
\begin{figure}[]
\centering
\includegraphics[height=2.5cm]{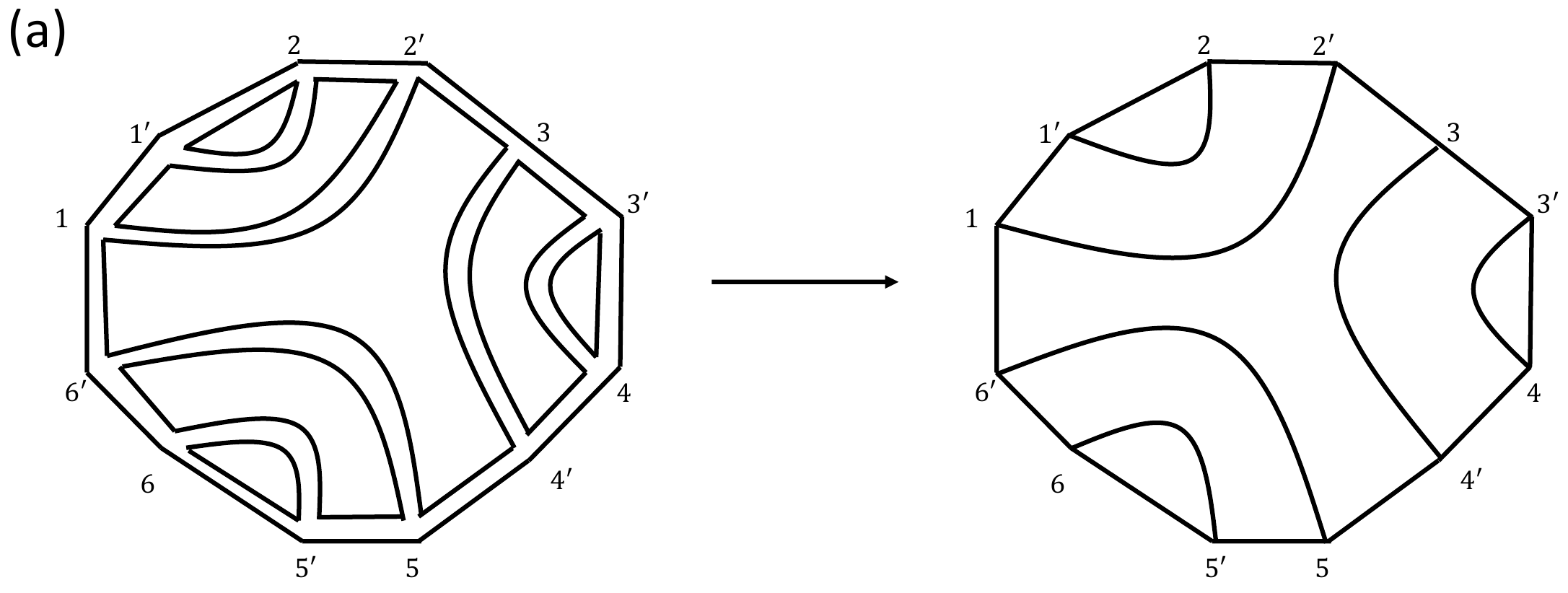}
\includegraphics[height=2.5cm]{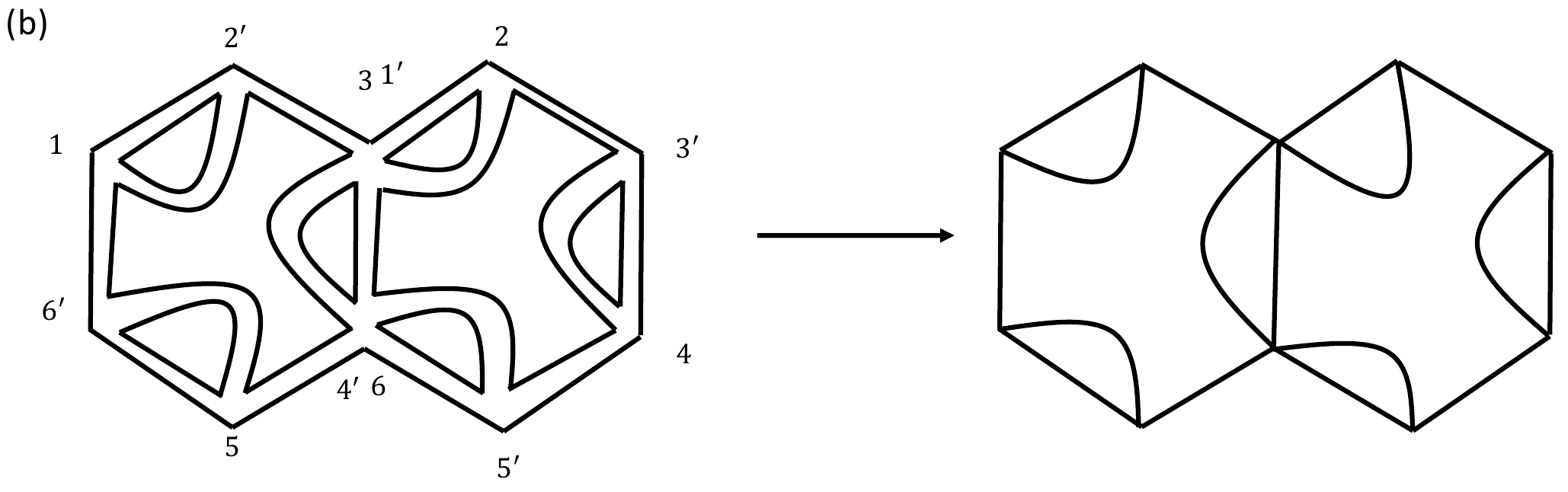}
\caption{For $n=6$ and $\tau=(12)(34)(56)$, we show the planar double-line diagrams and the corresponding polygons that we get for the first and third terms in the second line of \eqref{3sum}, using the boundary conditions from \eqref{3fig} (a) and (c).}
\label{fig:thooft}
\end{figure}
Now for any given $\tau$, in the polygons corresponding to the boundary conditions in Fig.~\ref{3fig} (a) and (b), we always have $E = 3n$ and $V = 2n$, so if for instance $F_{(a)}$ is the number of faces in the diagram for $\tau$ with the boundary conditions of Fig.~\ref{3fig}(a) and $h_{(a)}$ is the corresponding genus, we have 
\be 
\begin{gathered} 
k(\tau) + k(\tau^{-1} \eta) = F_{(a)}-1 = n+ 1- 2h_{(a)} \leq n+1, \\
 k(\tau) + k(\tau^{-1} \eta^{-1}) = F_{(b)}-1 = n+ 1- 2h_{(b)} \leq n+1
\end{gathered} 
\ee
where we subtract 1 from the total number of faces to remove the contribution from the extra line we added around the diagrams.
In the polygons corresponding to the boundary conditions in the last figure in \ref{3fig} (c), we have $E = 3n -1$ and $V = 2n -2$, so 
\be 
k(\eta \tau^{-1}) + k(\tau \eta) = F_{(c)}-1 = n +2 - 2h_{(c)} \leq n+2
\ee
Hence, the set of permutations that saturate \eqref{3sum} are those that give planar diagrams for each of the Figs.~\ref{3fig} (a), (b) and (c). 

\section{Estimate of fluctuations in the equilibrium approximation for the negativity}
\label{app:fluctuations}

In this appendix, we show that the fluctuations in $\sZ_n^{\rm(PT)}$ around the equilibrium value, as measured self-consistently by the quantity $\Delta_N$ defined in \eqref{Delta}, are suppressed relative to the equilibrium value. For this, we need to find the equilibrium approximation for the quantity $(\sZ_n^{\rm (PT)})^2$, and compare it to $([\sZ_n^{\rm (PT)}]_{\text{eq approx}})^2$. Let $d_X$ be the effective Hilbert space dimension of subsystem $X$, and $\alpha_2 := \eta \otimes \eta$. Then 
\be 
\begin{gathered} 
\left((\sZ_n^{\rm (PT)})^2\right)_{~\text{eq approx}} = \sum_{\tau_2 \in \sS_{2n}}\bra{e_B \otimes \eta_{A_1}\otimes \eta^{-1}_{A_2}} \bra{e_B \otimes \eta_{A_1}\otimes \eta^{-1}_{A_2}} \ket{\sI_{\al},\tau_2} \\
\sim \sum_{\tau_2 \in \sS_{2n}} d_{B}^{k(\tau_2)}d_{A_1}^{k(\tau_2 \alpha_2^{-1})} d_{A_2}^{k(\tau_2 \alpha_2)}
\end{gathered} 
\ee
Note that 
\be 
([\sZ_n^{\rm(PT)}]_{\text{eq approx}})^2 = \sum_{\tau, \sigma \in \sS_n} \bra{e_B \otimes \eta_{A_1}\otimes \eta^{-1}_{A_2}} \bra{e_B \otimes \eta_{A_1}\otimes \eta^{-1}_{A_2}} \ket{\sI_{\al},\tau \otimes \sigma} 
\ee
So 
\be 
\begin{gathered} 
\Delta_N^2 = \sum_{\tau_2 \in \sS_{2n}, \tau_2 \neq \tau\otimes \sigma} \bra{e_B \otimes \eta_{A_1}\otimes \eta^{-1}_{A_2}} \bra{e_B \otimes \eta_{A_1}\otimes \eta^{-1}_{A_2}} \ket{\sI_{\al},\tau_2} \\
\sim \sum_{\tau_2 \in \sS_{2n}, \tau_2 \neq \tau \otimes \sigma} d_{B}^{k(\tau_2)}d_{A_1}^{k(\tau_2 \alpha_2^{-1})} d_{A_2}^{k(\tau_2 \alpha_2)}
\end{gathered} 
\ee

Let us first consider the entanglement saturation phase, where $d_{A_1} \sim d_{A_2} \sim d_B\sim d$. Then 
\be 
[(\sZ_n^{\rm (PT)})^2]_{\text{eq approx}} \sim \sum_{\tau_2 \in \sS_{2n}} d^{k(\tau_2)+ k(\tau_2 \alpha_2^{-1}) + k(\tau_2 \alpha_2)}
\label{es_squared}
\ee
Using the Cayley distance and the triangle inequality for three permutations $\sigma_2, \tau_2, \rho_2 \in \sS_{2n}$, we have 
\be 
k(\sigma_2 \tau_2^{-1}) + k(\tau_2 \rho_2^{-1}) \leq 2n + k(\sigma_2 \rho_2^{-1})
\ee
which gives the following upper bound for the exponent in \eqref{es_squared}: 
\be 
\begin{gathered} 
k(\tau_2)+ k(\tau_2 \alpha_2^{-1}) + k(\tau_2 \alpha_2)\\ = \frac{k(\tau_2^{-1})+ k(\tau_2 \alpha_2^{-1})}{2} + \frac{k( \alpha_2\tau_2^{-1}) + k(\tau_2 \alpha_2) }{2}+ \frac{k(\tau_2^{-1}) + k(\tau_2 \alpha_2)}{2} \leq \begin{cases} 3n + 4 & n \text{ even} \\ 3n + 3 & n \text{ odd}
\end{cases} 
\end{gathered}
\label{three_2}
\ee
We can then check that if $\tau, \sigma \in \sS_n$ both saturate \eqref{3sum}, then $\tau\otimes \sigma \in \sS_{2n}$ saturates \eqref{three_2}. We will now show using 'tHooft-like line diagrams that for any $\tau_2 \neq \tau \otimes \sigma$, \eqref{three_2} is not saturated. The three terms on the second line of \eqref{three_2} can be equated with the total number of loops respectively in the diagrams with boundary conditions shown in Fig.~\ref{fig:dev} (a), (b) and (c). The leading contribution comes from the permutations that give planar diagrams for each of the three boundary conditions. In particular, any non-factorized permutation gives a non-planar diagram for each of (a), (b) and (c), corresponding to genuses that can be labelled $h_1, h_2, h_3 \geq 1$. With manipulations of the diagrams to the ones used in appendix \ref{app:perm}, we have 
\be 
\begin{gathered} 
k(\tau_2^{-1})+ k(\tau_2 \alpha_2^{-1}) = 2n +2 -2 h_1, \\ k(\tau_2^{-1}) + k(\tau_2 \alpha_2)= 2n +2 -2 h_2, \\
 k( \alpha_2\tau_2^{-1}) + k(\tau_2 \alpha_2) = 2n +4 - 2h_3. 
 \end{gathered} 
\ee
Hence, all terms contributing to $\Delta_N^2$ are of the form 
\be 
d^{(2n+2-2h_1)/2+ (2n+2-2h_2)/2 + (2n+4-2h_3)/2} \leq d^{3n/2+2-3} 
\ee
We therefore find that in the entanglement saturation phase, the fluctuations are suppressed as 
\be 
\frac{\Delta_N}{[\sZ_n^{\text{(PT)}}]_{\text{eq approx}}} \sim d^{-3/2} \sim Z_1^{-1/2}
\ee
where $Z_1$ is the effect dimension of the full Hilbert space. 
\begin{figure}[]
\includegraphics[width=8cm]{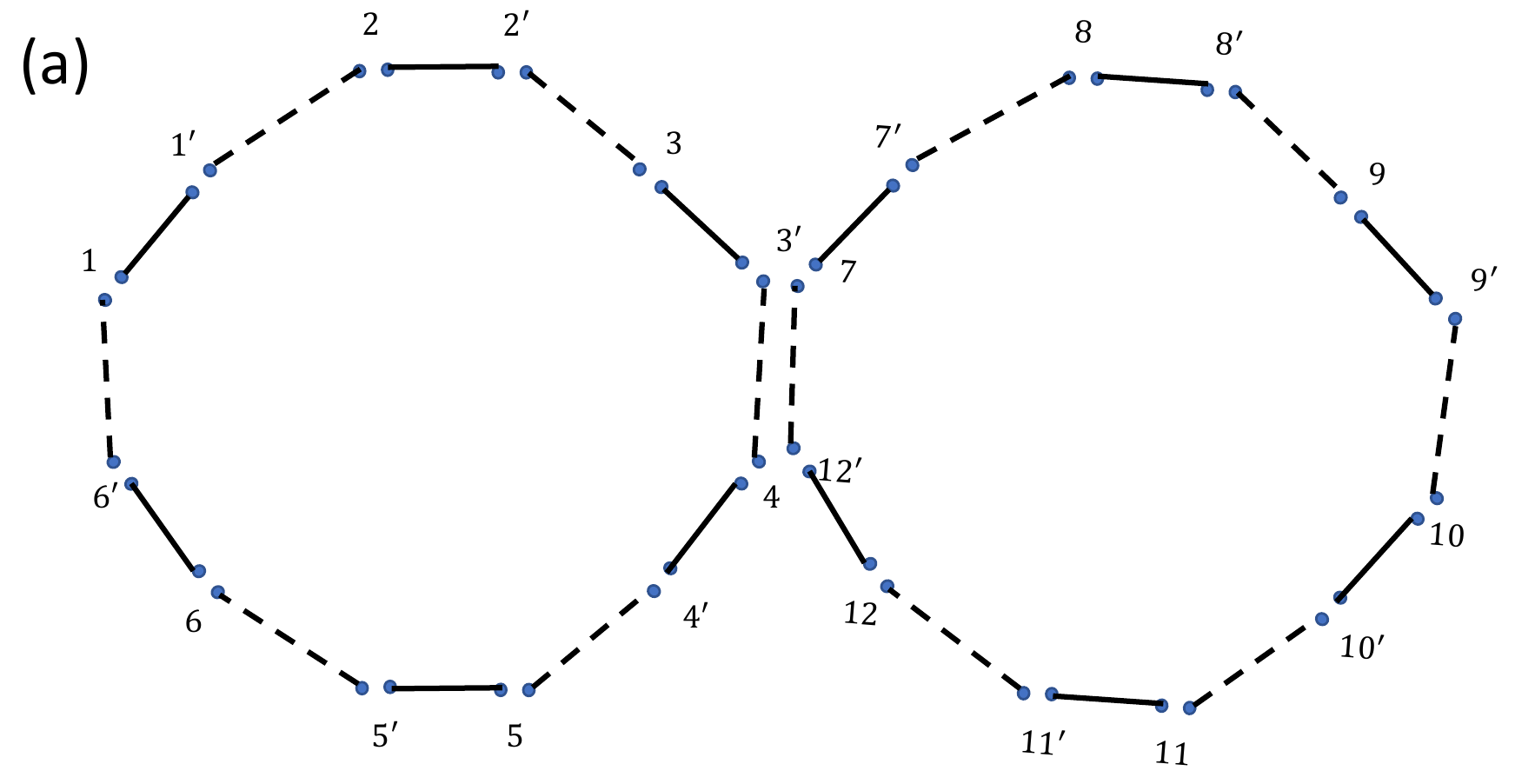}
\includegraphics[width=8cm]{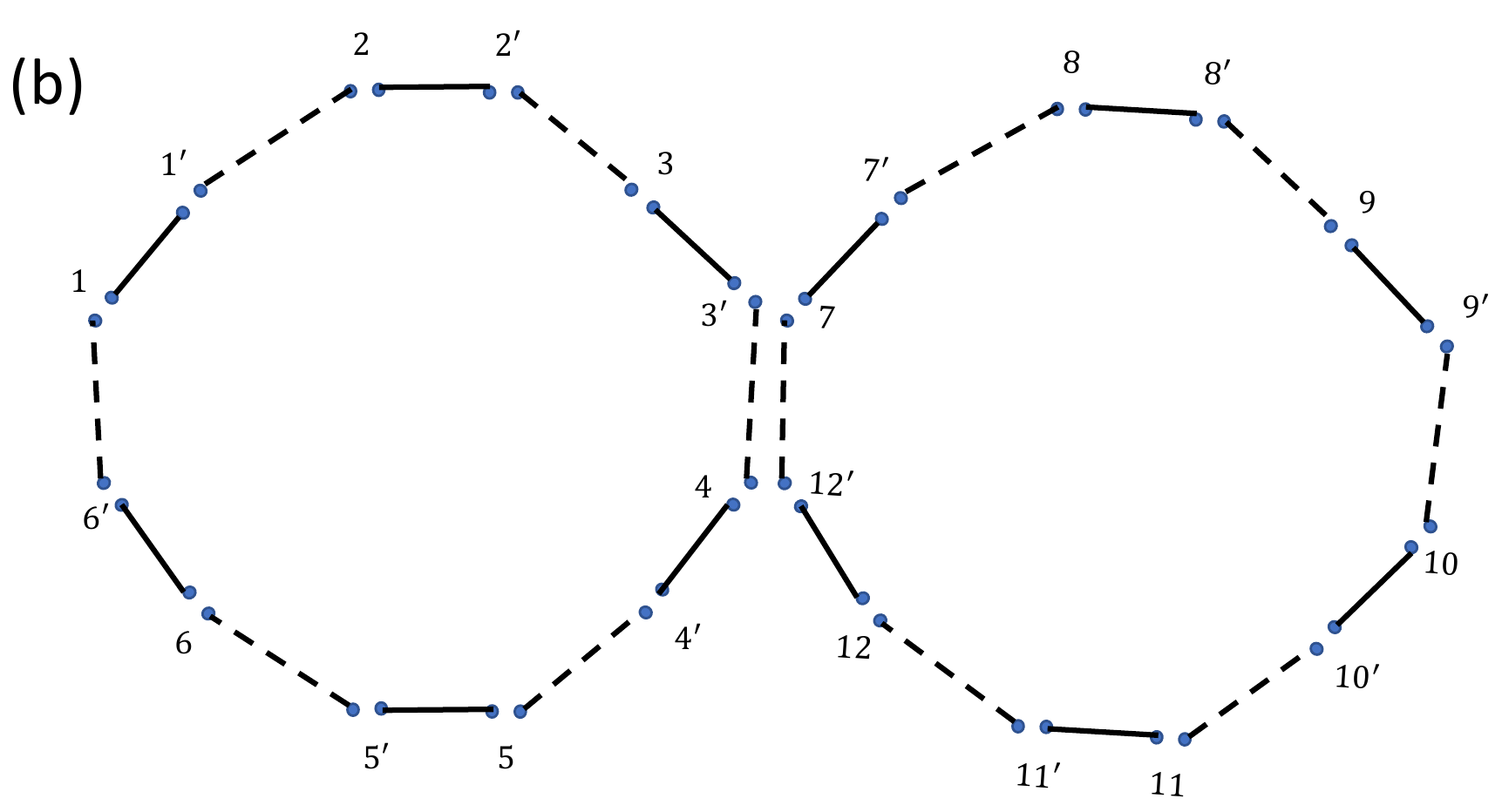}
\includegraphics[height=3cm]{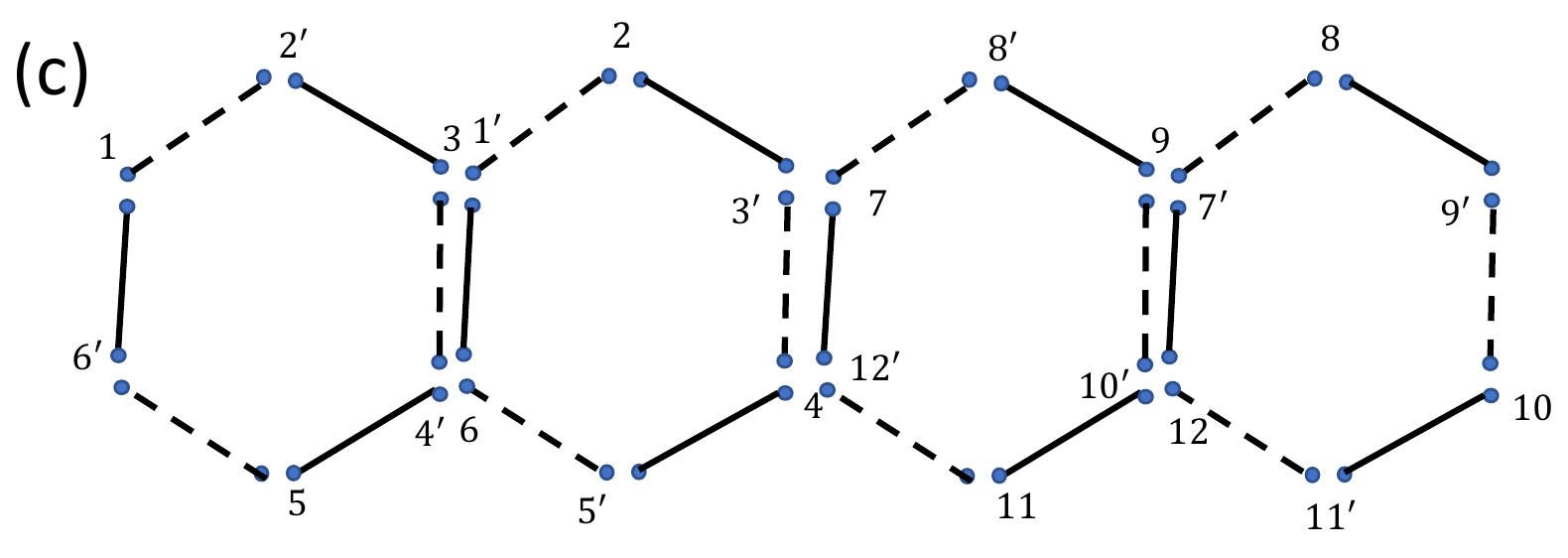}
\caption{Boundary conditions for the diagrams corresponding to the three terms in the second line of \eqref{three_2} are shown in order in (a), (b) and (c).}
\label{fig:dev}
\end{figure} 

Next, consider the maximally entangled phase. In this case WLOG, $d_{A_1}\gg d_{A_2}, d_{B}$, so the dominant contribution to $[\sZ_n^{\rm (PT)}]_{\text{eq approx}}$ comes from $\tau= \eta$, and the dominant contribution to $[(\sZ_n^{\rm PT})^2]_{\text{eq approx}}$ comes from $\tau_2= \alpha = \eta \otimes \eta$. From any $\tau_2 \neq \eta \otimes \eta$, and in particular any non-factorized $\tau_2$, the contribution is suppressed relative to the leading contribution by a factor of at least $d_{A_1}^{-1} \sim D^{-1}$, so 
\be 
\frac{\Delta_N}{[\sZ_n^{\text{(PT)}}]_{\text{eq approx}}}\sim D^{-1/2}
\ee

Finally, consider the PPT phase. In this case $d_B \gg d_{A_1}, d_B$, so the dominant contribution to $[\sZ_n^{\rm (PT)}]_{\text{eq approx}}$ comes from $\tau= e$, and the dominant contribution to $[(\sZ_n^{\rm (PT)})^2]_{\text{eq approx}}$ comes from $\tau_2= \alpha = e \otimes e$. Again we have 
\be 
\frac{\Delta_N}{[\sZ_n^{\text{(PT)}}]_{\text{eq approx}}}\sim D^{-1/2}. 
\ee

\section{Resolvent calculations of logarithmic negativity in various cases} \label{app:resol}

\subsection{General diagrammatic approach}

In order to find the the logarithmic negativity, it is useful to first find the equilibrium approximation for the resolvent 
\be 
R_N(\lambda) = \frac{1}{\lambda}\sum_{n=0}^{\infty} \frac{1}{\lambda^n} \text{Tr}[(\rho_A^{T_2})^n]\, . \label{Rn_example}
\ee
It is useful to see $R_N$ as the trace of a matrix 
\be 
R_{pq} = \frac{1}{\lambda}\sum_{n=0}^{\infty} \frac{1}{\lambda^n} {(\rho_A^{T_2})^n}_{pq} , \quad \ket{p} = \ket{p_1}_{A_1} \ket{p_2}_{A_2} , \quad \ket{q} = \ket{q_1}_{A_1} \ket{q_2}_{A_2} 
\ee
We can apply the equilibrium approximation to each ${(\rho_A^{T_2})^n}_{pq}$. The common lower half of the diagrams for all permutations in this case can be deduced from Fig.~\ref{fig:negativity_figs} (a) for $\text{Tr}[(\rho_A^{T_2})^n]$ by erasing the dashed line connecting $i'_{a_{n}}$ and $i_{a_1}$, and the dotted line connecting $i'_{\bar{a}_{1}}$ and $i_{\bar{a}_n}$, and instead taking the inner product of $\ket{i_{a_1}}$, $\ket{i'_{\bar{a}_{1}}}$, $\ket{i_{a_n}}$, $\ket{i'_{\bar{a}_{n}}}$ with $\ket{p_1}$, $\ket{p_2}$, $\ket{q_1}$, $\ket{q_2}$ respectively. The resulting lines are shown in Fig.~\ref{fig:rij_ext}, which also explains how the factors of $\frac{1}{\lambda^{n+1}}$ and $\frac{1}{Z_1^n}$ that are common to all terms with index $n$ in \eqref{Rn_example} (the second factor comes from the equilibrium approximation) are incorporated into these lines. 
\begin{figure}[] 
\centering
\includegraphics[width=12cm]{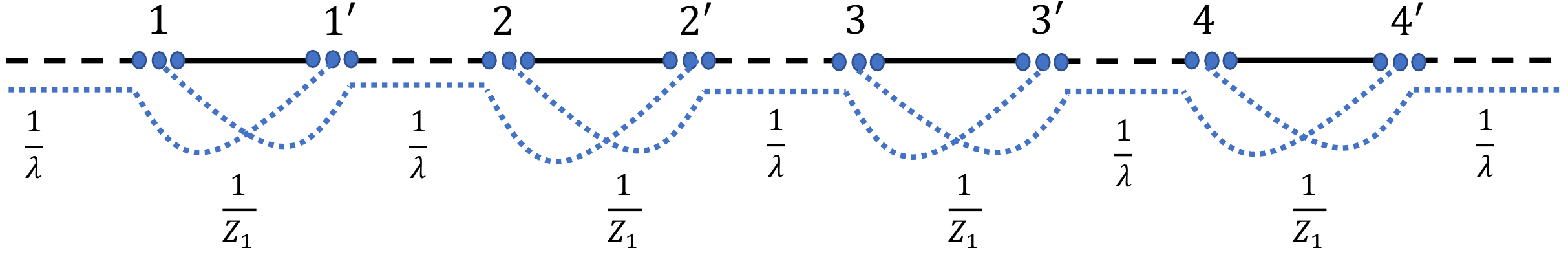}
\caption{``Boundary" lines for the equilibrium approximation for $R_{pq}$ with $n=4$.}
\label{fig:rij_ext}
\end{figure}
In the limit where the effective dimension of $A_1$ is much larger than that of $A_2$, it is sufficient to consider contributions from planar diagrams for all $n$ to $R_{pq}$. We can write $R_{pq}$ in terms of a self-energy $\Sigma_{pq}$ as shown in Fig.~\ref{fig:rpq}(a). $\Sigma_{pq}$ is a sum of diagrams without any disconnected parts connected by $\frac{1}{\lambda}\delta_{pq}$, to which the first few contributions are shown in Fig.~\ref{fig:rpq}(b). We take $\ket{p_1}, \ket{q_1}$ and $\ket{p_2}, \ket{q_2}$ to be elements of the energy eigenbasis in $A_1$ and $A_2$ respectively, so that we approximately have that 
\be 
 \bra{p_1}\bra{p_2}\sI_{\al} \ket{q_1} \ket{q_2} \propto \delta_{p_1 \, q_1}\, \delta_{p_2 \, q_2}
\ee
for both the canonical and microcanonical ensembles, and hence from the diagrams contributing to $\Sigma_{pq}$ we can see that both $\Sigma_{pq}$ and $R_{pq}$ are proportional to $\delta_{pq}$. 
\begin{figure}[] 
\centering
\includegraphics[width=12cm]{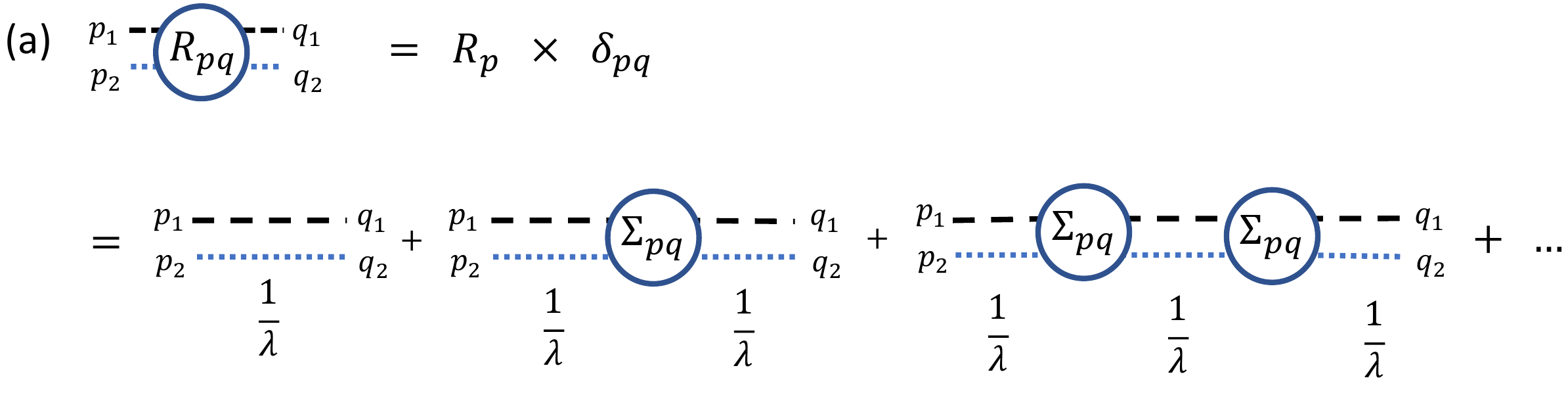}
\includegraphics[width=12cm]{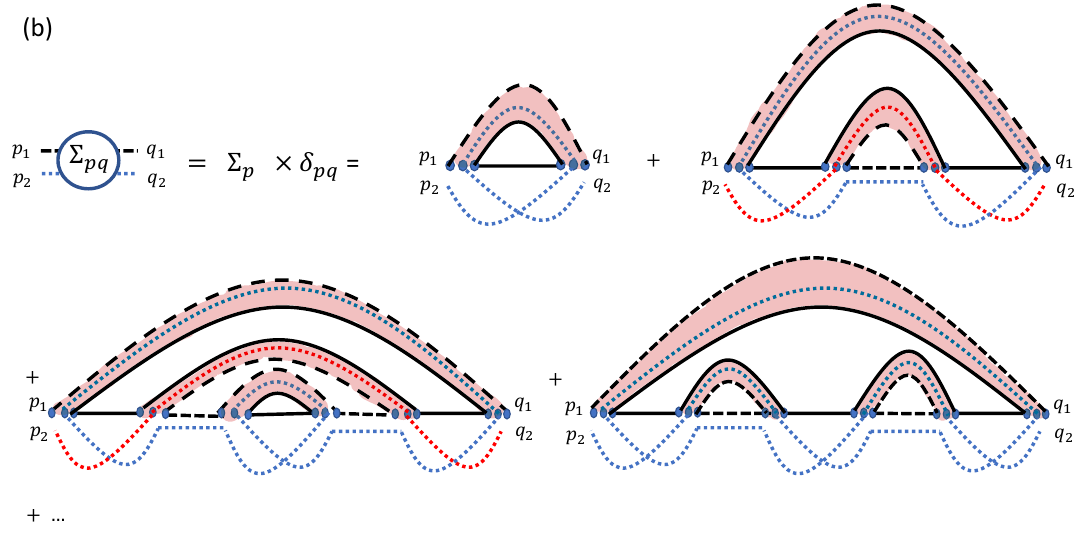}
\caption{Diagrammatic representation of $R_{pq}$ and the first few diagrams contributing to $\Sigma_{pq}$ in the case where $A_1$ is larger than $A_2$. }
\label{fig:rpq}
\end{figure}

We can immediately see by summing the geometric series on the RHS of Fig.~\ref{fig:rpq}(a) that 
\be 
R_N = \sum_p R_p , \quad R_{p} = \frac{1}{\lambda - \Sigma_{p}} \, . 
\ee
We can systematically include all planar contributions to $\Sigma_{p}$ using the Schwinger-Dyson equation shown diagrammatically in Fig.~\ref{fig:rpq_sd}. For general choices of $\sI_{\al}$, this Schwinger-Dyson equation in general leads to a complicated set of equations relating $R_p$ for all different $p$ to each other. Below we consider a few choices of $\sI_{\al}$ for which the Schwinger-Dyson equation simplifies. 
\begin{figure}[] 
\centering
\includegraphics[width=15cm]{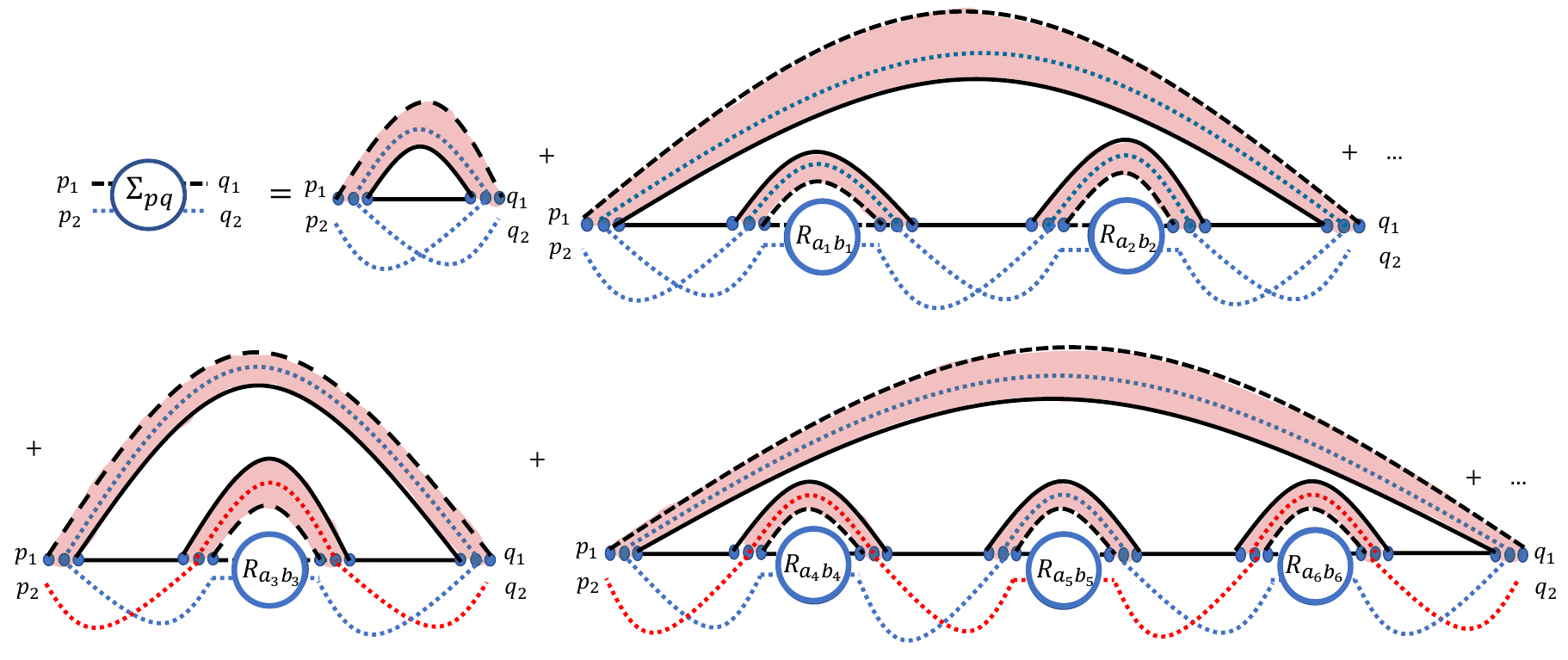}
\caption{Rewriting of the RHS of Fig.~\ref{fig:rpq}(b) as a Schwinger-Dyson equation. All $a_i, b_i$ are independently summed over (unless they have delta functions among each other or with $p, q$ according to the diagram).}
\label{fig:rpq_sd}
\end{figure}

\subsection{Infinite temperature}
\label{app:mic_inf}
We first consider the case where the effective identity operator is simply the identity, 
\be 
\sI_{\al} = \mathbf{1}_{A_1} \otimes \mathbf{1}_{A_2} \otimes \mathbf{1}_{B}\, . 
\ee
Here the Schwinger-Dyson equation for $R_p$ becomes independent of the index $p$ in $A$, and we have 
\be 
R_N^{\infty} = d_A R_1, \quad R_1 = \frac{1}{\lambda - \Sigma_1}\, . \label{r1_simple}
\ee
Each line on the RHS of Fig.~\ref{fig:rpq_sd} now simplifies to a geometric series, and hence
\be
\begin{gathered} 
\Sigma_1 = \left(\frac{1}{d_A} + \alpha R_1 \right)\, \left(1 + R_1^2 \beta + (R_1^2 \beta)^2 + ...\right) 
= \left(\frac{1}{d_A} + \alpha R_1\right)\, \frac{1}{1- R_1^2 \beta}, \\
\alpha = \frac{1}{d_A d_B}, \quad \beta = \frac{1}{(d_{A_2}d_B)^2}
\end{gathered} 
\ee
Substituting this into \eqref{r1_simple}, we get a cubic equation for $R_1$, 
\be 
\beta \lambda R_1^3 + (\alpha-\beta) R_1^2 + \left(\frac{1}{d_A}-\lambda\right) R_1+ 1= 0 \, . \label{rinf_cubic}
\ee
It is difficult to obtain a general analytic expression for $\sZ^{\rm (PT)}:= \sum_i |\lambda_i| = e^{\sE}$ by finding the analytic form of $D_N(\lambda)$ from the solutions to \eqref{rinf_cubic}. However, the density and the logarithmic negativity can be found numerically from \eqref{rinf_cubic}, and turn out to agree with the analytic continuation in \eqref{legn}, \eqref{zeroinf}, \eqref{neg_inf_es}, as discussed in \cite{2021PRXQ....2c0347S}. That is, we have 
\be 
\sZ^{\rm (PT)}_{\infty}(d_{A_1}, d_{A_2}, d_B) = \begin{cases} 
1 & \log d_{A} < \log d_B \\
d_{A_1} & \log{(d_{A_1} d_B)} <\log d_{A_2} \\
d_{A_2} & \log(d_{A_2} d_B) < \log d_{A_1} \\
\left(\frac{d_{A}}{d_B}\right)^{1/2} & \log d_{A}> \log d_B, ~~ |\log{d_{A_1}}-\log d_{A_2}|< \log d_B 
\end{cases} \, . 
\label{inf_zpt}
\ee
Now consider the regime where $\frac{d_{A_1}}{d_{A_2} d_B}\ll 1$. This corresponds to being outside the ME phase. Then since $\alpha \gg \beta$ in this regime, $R_1$ is $O(1)$, and $\lambda$ is $O(1/d_A)$, \eqref{rinf_cubic} simplifies to a quadratic equation for $R_1$, 
\be 
 \alpha \, R_1^2 + \left(\frac{1}{d_A}-\lambda\right) R_1+ 1= 0 . \label{r1_quad}
\ee
The same quadratic equation can be obtained diagrammatically by ignoring all contributions to the Schwinger-Dyson equation in Fig.~\ref{fig:rpq_sd} except the first term in each line on the RHS. This corresponds to including contributions from permutations with only one-cycles and two-cycles to $\sZ_n^{\rm (PT)}$ for all $n$. We can now solve \eqref{r1_quad} to get a simple semicircle form of $D_N(\lambda)$ from $R_N(\lambda)$, 
\be 
D_{\infty}^{\rm sc}(\lambda, d_A, d_B) = \frac{d_A^2 d_B}{2 \pi} \sqrt{ \frac{4}{d_A d_B} - \left(\lambda - \frac{1}{d_A}\right)^2 } , \label{sc_inf}
\ee
 which can be integrated analytically to get $\sZ^{\rm (PT)}$, 
 \be 
 \sZ_{\infty}^{\rm (PT),\, sc} (d_A, d_B) = \begin{cases} 
1 & \log d_A < \log d_B \\
\left(\frac{d_{A}}{d_B}\right)^{1/2} &\log d_A > \log d_B
\end{cases} \, . 
 \ee
As expected, this approximation misses the ME phase of \eqref{inf_zpt}. 

\subsection{Canonical ensemble with $A$ at infinite temperature}
\label{app:mic_can}

Next, let us consider the case where $A$ is at infinite temperature and $B$ is at finite temperature in the canonical ensemble, 
that is, 
\be 
\sI_{\al} = \mathbf{1}_{A_1} \otimes \mathbf{1}_{A_2} \otimes e^{-\beta H_B} \, . \label{ainfcan} 
\ee
Then the RHS of Fig.~\ref{fig:rpq_sd} implies that $R_p$, $\Sigma_p$ again become independent of the index $p$, but in this case the RHS does not simplify to a geometric series, so that we have 
\be 
R_N = d_A\, R_1, \quad R_1 = \frac{1}{\lambda - \Sigma_1}, 
\ee
where 
\be 
\Sigma_1 = \frac{1}{d_A} +\sum_{n=1}^{\infty} \frac{Z_{2n+1, B}}{(d_A Z_{1,B})^{2n+1}} \, (d_{A_1} \, R_1)^{2n} + \sum_{n=1}^{\infty} \frac{Z_{2n, B}\, }{(d_A Z_{1,B})^{2n}} \, d_{A_2} \, (d_{A_1} \, R_1)^{2n-1} 
\ee
where $Z_{n, B} = \text{Tr}[e^{-n \beta H_B }]$. 
The resulting equation for $R_N$ can be written as 
\begin{align}
 \lambda R_N = d_A + d_{A_2}\int dE \rho(E) \sum_{k = 1}^{\infty} \left[\left(\frac{R_Ne^{-\beta E}}{d_A d_{A_2}Z_{1,B}} \right)^{2k-1}+ d_{A_2} \left(\frac{R_Ne^{-\beta E}}{d_A d_{A_2}Z_{1,B}} \right)^{2k}\right].
\end{align}
Completing the geometric sums, we find
\begin{align}
 \lambda R_N = d_A + \int dE \rho(E)\frac{d_{A_2}^2 R_N \left(d_{A} Z_{1,B} e^{\beta E}+R_N\right)}{d_{A}^2
 d_{A_2}^2 Z_{1,B}^2 e^{2 \beta E}-R_N^2}.
 \label{R_negativity}
\end{align}
where $\rho(E) = e^{V s(E/V)}$ is the density of states for $B$. On specifying the density of states, this equation can be can be solved numerically for $R_N(\lambda)$ and used to obtain $\sE$. The solution with a gaussian entropy density is shown in Fig.~\ref{fig:resolvent_checks} as a function of $c$ at $\lambda = \ha$. The result agrees with the expressions for $\sE$ in the NE and ES phases in \eqref{ksn} and \eqref{ksn2} and the naive phase transition line between them in \eqref{n_es_e_can}. 

\begin{figure}[!h] 
\centering 
\includegraphics[width=.7\textwidth]{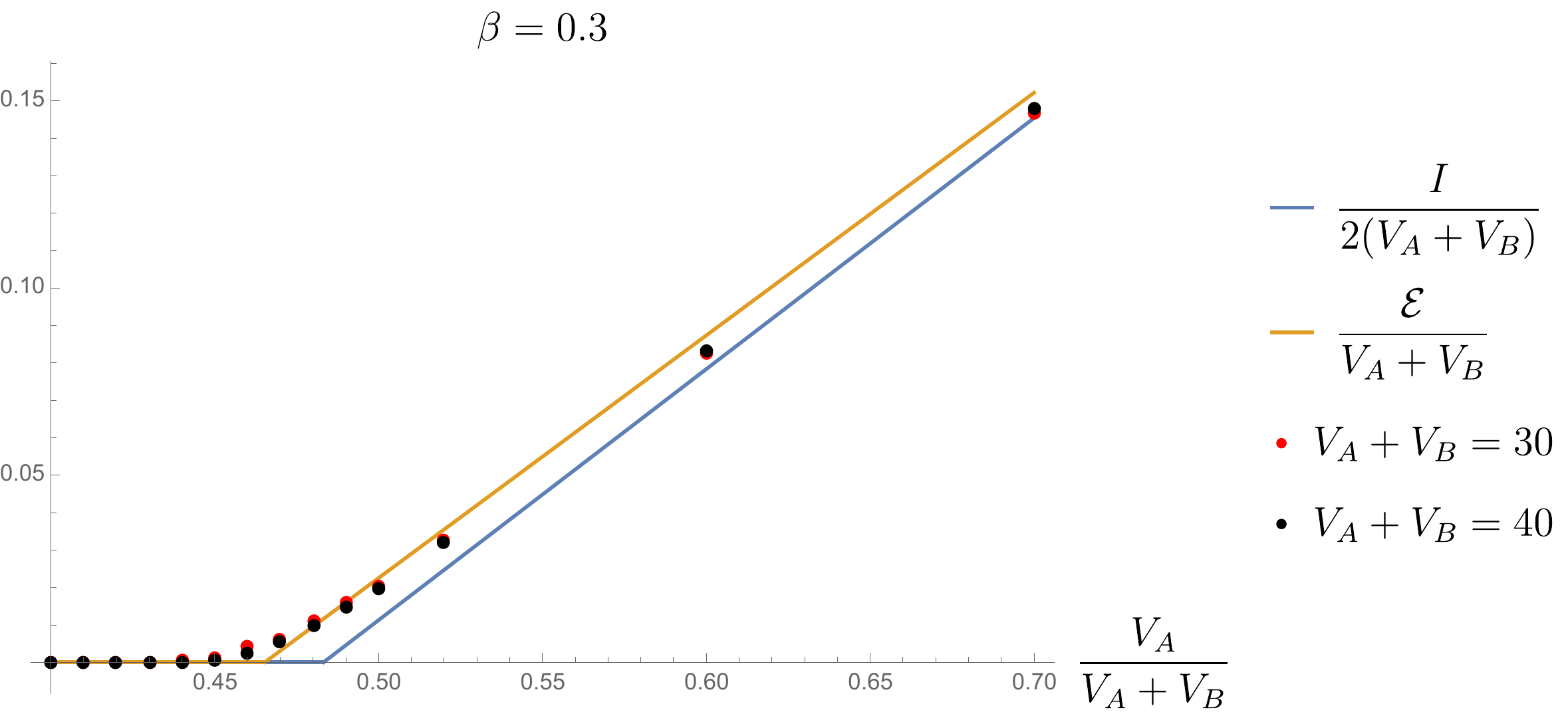}
\caption{Left: For $\sI_{\al}$ given by \eqref{ainfcan}, the logarithmic negativity is computed numerically using the resolvent \eqref{R_negativity} and compared to the naive analytic continuation in \eqref{ksn}, \eqref{ksn2} and \eqref{n_es_e_can}, with excellent agreement. 
}
\label{fig:resolvent_checks}
\end{figure}

Similar to the infinite temperature case, we can consider the regime where the effective dimensions $d_{A_1}, d_{A_2}, d_{B}$ of the different subsystems are such that $\frac{d_{A_1}}{d_{A_2} d_B}\ll 1$. Assuming that the same set of permutations contributes in this regime at finite temperature as the one that contributed at infinite temperature, we can then include only the first diagram from each line on the RHS of Fig.~\ref{fig:rpq_sd}. We then have
\be 
\Sigma_1 = \frac{1}{d_A} + \frac{Z_{2, B}}{d_A\, (Z_{1,B})^{2}}\, R_1
\ee 
which gives a quadratic equation 
\be 
 \alpha' \, R_1^2 + \left(\frac{1}{d_A}-\lambda\right) R_1+ 1= 0 , \quad \alpha' = \frac{Z_{2, B}}{(Z_{1, B})^2 d_A} \label{r1_quad_2}
\ee
Similar to \eqref{r1_quad}, this gives a semicircle form of the density of eigenvalues,
which for this case gives 
 \be 
 \sZ^{\rm (PT)}_{\infty, {\rm sc}} = \begin{cases} 
1 & \log d_A < S^{\rm (eq)}_{2,B} \\
e^{\frac{1}{2}(\log d_A - S^{\rm (eq)}_{2, B})}&\log d_A > S^{\rm (eq)}_{2,B}
\end{cases} \, . \label{zsc_inf}
 \ee
 This gives the value of the logarithmic negativity we would expect in the ES phase from naive analytic continuation, and also the transition between the NE and ES phases we would expect from naive analytic continuation.

\subsubsection{Toy model for black hole evaporation in JT gravity}
\label{pssy}

As discussed in \cite{2020arXiv200801089L}, the rules for calculating the entanglement entropy of the radiation in the holographic model of black hole evaporation in \cite{2019arXiv191111977P} can be derived from the equilibrium approximation, taking $A$ to be the radiation $R$, $B$ to be the black hole, and 
\be 
\sI_{\al} = \mathbf{1}_R \otimes f(H_B) e^{-\beta H_B} \label{jt}
\ee
which is a special case of \eqref{ainfcan}, with an additional factor $f(H_B)$ which captures the effect of end-of-the-world (EOW) branes, and with the density of states as in JT gravity, so that 
\begin{align}
 \mathcal{Z}_{m,B} = e^{S_0} z_m(\beta) \label{zm_jt}
\end{align}
where $S_0$ is the large ground state entropy and $z_m(\beta)$ is $O(1)$. The rules for calculating $\sZ_n^{\rm (PT)}$ in this model can similarly be derived from the equilibrium approximation. The negativity within the radiation in this model was discussed in Refs.~\cite{2021arXiv210902649K, 2021arXiv211011947D}.

Like in other cases, the dominant permutation in $\sZ_n^{\rm (PT)}$ for this choice of $\sI_{\al}$ is always one out of $e$, $\eta$ or $\eta^{-1}$, and $\tau_{ES}$. We can read off the ``boundary'' expressions for $\sZ_n^{\rm (PT)}(\tau)$ in this holographic model, in terms of $d_{R_1}, d_{R_2}, Z_{n, B}$, from the diagrams in Fig. \ref{fig:negativity_figs} (where $Z_{n, B}= \text{Tr}[(e^{-\beta H_B} f(H_B))^n]$). Using the relation between bulk and boundary partition functions in holography, we then find geometries like the ones shown in Fig. \ref{fig:gravity_figs} for evaluating $\sZ_n^{\rm (PT)}$ in the ES and ME phases. Both involve ``replica wormholes,'' which are connected Euclidean gravity path integrals between multiple asymptotic boundaries. Note that more generally, \eqref{mastereq} can be used to systematically derive replica wormhole prescriptions for calculating $\sZ_n^{\rm (PT)}$ in other gravity setups, including the model of \cite{2020JHEP...05..013A}. 
\begin{figure}[!h]
 \centering
 \includegraphics[width=12cm]{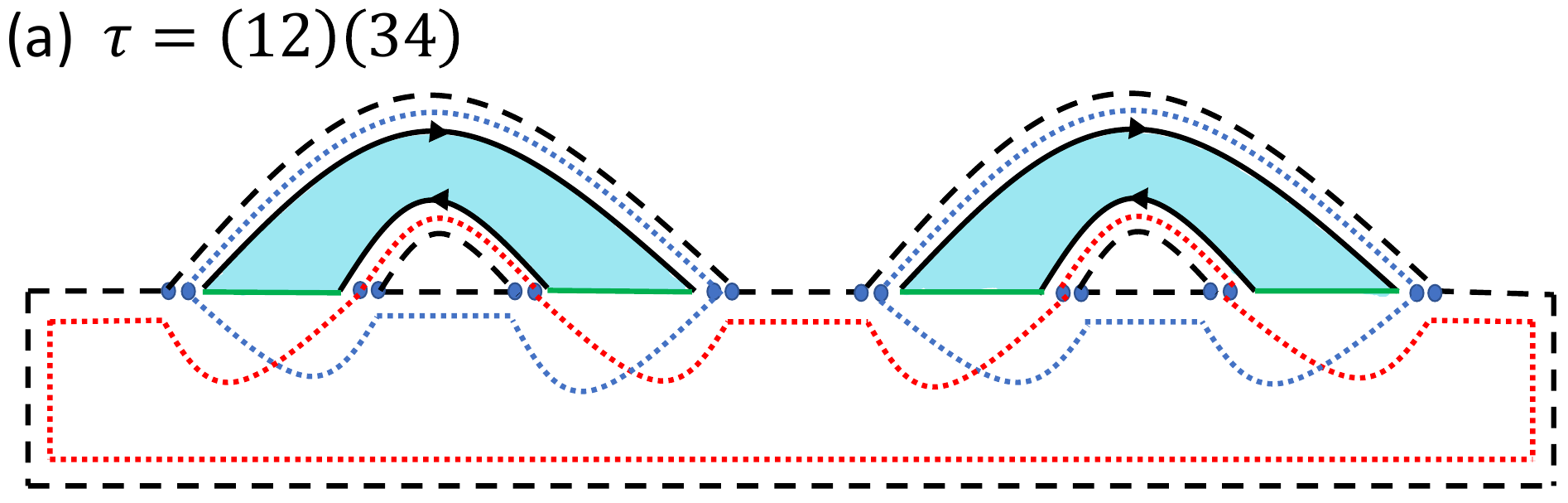}
 \includegraphics[width=12cm]{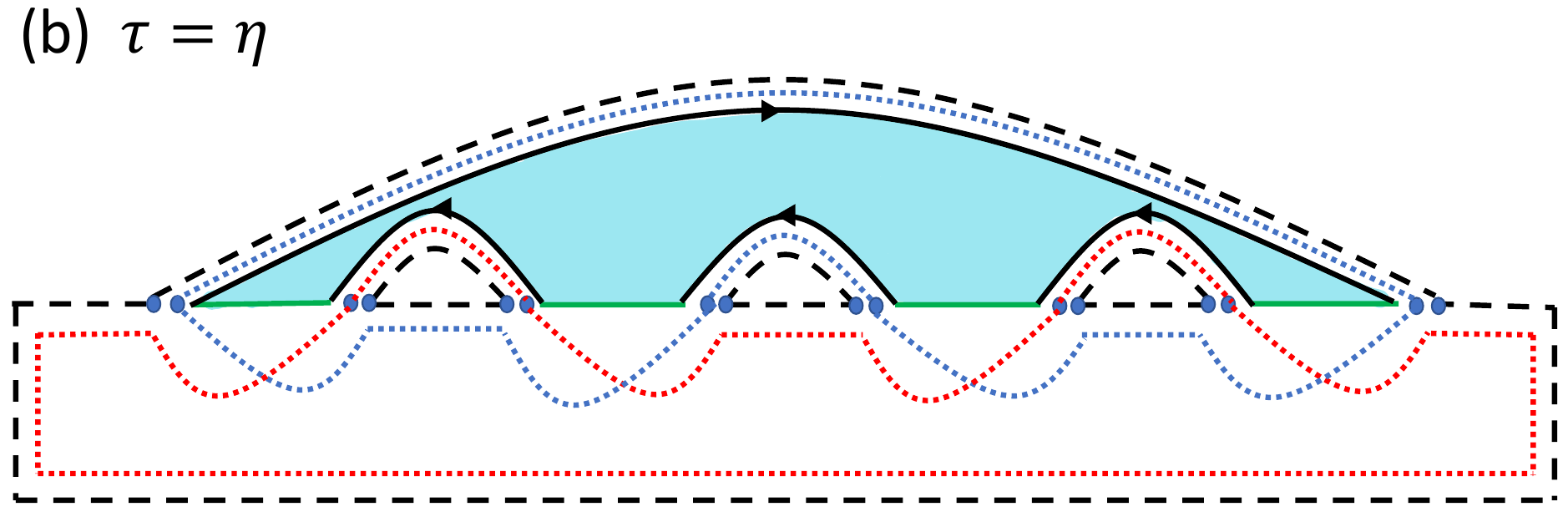}
 \caption{Euclidean gravity path integrals with replica wormholes for calculating $\sZ_n^{\rm (PT)}$ in the model of \cite{2019arXiv191111977P}, for $n=4$. The dominant contribution in the ES phase, shown in (a), involves connected partition functions $Z_{2, B}$ between two asymptotic boundaries, while the ME phase contribution in (b) involves $Z_{n, B}$. The black lines with arrows are asymptotic boundaries in JT gravity (each with length $\beta$, although the lengths appear to differ in the figure), and the green lines are EOW branes. The dashed and dotted loops simply give factors of $d_{A_1}$ and $d_{A_2}$ respectively.}
 \label{fig:gravity_figs}
\end{figure}

Note that \eqref{zm_jt} implies in particular that 
\bea 
& S_{2, B}^{(\rm eq)} = S_0 + s_2 (\b) , \quad S_{B}^{(\rm eq)} = S_0 + s_1 (\b) , \nonumber \\ & s_1 (\b) > s_2 (\b) \sim O(1) \label{jt_s}
\eea
From analytic continuation of the expressions for $\sZ_n^{\rm (PT)}$ for even $n$ to $n\rightarrow 1$ in this model, the expressions for the logarithmic negativity in \eqref{ksn}-\eqref{ksn2} become 
\begin{align}
& \sE_{NE}(R_1, R_2) = 0 \\
 & \sE_{ME}(R_1, R_2) = \log d_{R_1} \quad \text{ or } \quad \sE_{ME} = \log d_{R_2}\\
 & \sE_{ES}(R_1, R_2) = \frac{1}{2}\left(\log d_{R} - S_0 - s_2(\beta) + \log \frac{8}{3\pi}\right)
\end{align} 
Note that we have explicitly included a term $\log \frac{8}{3\pi}$ coming from the degeneracy of the permutations in the ES phase in this case (which we have ignored for other choices of $\sI_{\al}$ in the paper), as it contributes at the same order as $\log \frac{z_2(\beta)}{z_1(\beta)^2}$. Recall that the Page time $t_p$ corresponds to $c= \ha$. The transition from the NE phase, where the logarithmic negativity within the radiation is sub-extensive, to the ES phase, where it becomes extensive, takes place at 
\be 
c = c_0\equiv \frac{1}{2}\left(1- \frac{s_1(\beta)- s_2(\beta) + \log 8\pi/3}{S_0} \right)
\ee
For this model, the difference between $c= \frac{1}{2}$ and $c= c_0$ is suppressed due to the special structure of the density of states in JT gravity compared to other black hole models.

It is also interesting to consider the negativity between the black hole and a subsystem of the radiation, $R_1$. In this case, the permutation that dominates in $\sZ_n^{\rm (PT)}$ at early times when $e^{S_0} \gg d_{R_1} d_{R_2}$ is 
$\tau = \eta^{-1}$, leading on analytic continuation to 
\begin{align}
 \mathcal{E}(R_1, B) = \log d_{R_1}.
\end{align}
irrespective of the relative sizes of $R_1$ and $R_2$. This means that each Hawking quantum is maximally entangled with the black hole at early times, as expected. After the Page time, if $d_{R_1} \gg d_{R_2}e^{S_0}$, $\tau = \eta$ dominates, and by analytic continuation the 
negativity is then
\begin{align}
 \mathcal{E}(R_1, B) = S_0 + \log\ \frac{z_{\frac{1}{2}}(\beta)^2}{z_1(\beta)} = S^{\rm (eq)}_{\frac{1}{2}, B}\, .
\end{align}
Most interestingly, we can also consider the regime after the Page time where $d_{R_2} \gg d_{R_1}e^{S_0}$ i.e.~how small amounts of the radiation are entangled with the black hole. The usual story of black hole evaporation says that these quanta are maximally entangled with the black hole, more specifically, their pair behind the horizon. This entanglement played a starring role in the firewall paradox \cite{2013JHEP...02..062A}. But the dominant term in $\sZ_n^{\rm (PT)}$ is the identity, so that on analytic continuation, 
\begin{align}
 \mathcal{E}(R_1, B) = 0.
\end{align}
This is the statement that Hawking quanta after the Page time are not actually entangled with the black hole, but only entangled with the early radiation, avoiding paradoxes with entanglement monogamy. 

Note that we can numerically compute the resolvent and spectrum in this model using the general equation \eqref{R_negativity}, now with 
\be 
\rho (E) = e^{S_0} \, \frac{\sqrt{2E}}{2\pi^2} \, \sinh (2 \pi \sqrt{2E})
\ee
This problem is studied systematically in Ref.~\cite{2021arXiv211011947D}. For the special case of the ``microcanonical ensemble'' in \cite{2019arXiv191111977P}, the equation for the resolvent reduces to the infinite temperature case in \eqref{rinf_cubic}, with the dimension of the black hole given by $e^{\bf S}$, where {\bf S} is the microcanonical entropy of the black hole.



\subsection{Microcanonical ensemble with $B$ at infinite temperature}
\label{app:mic_b}
We now consider $\sI_{\al}$ given by \eqref{mic_case1}, which we repeat here\footnote{Similar resolvent calculations are done in Ref.~\cite{hassan_forthcoming}.}: 
\be \label{mic_case1_repeat}
\sI_{\al} = \sum_{E^{A_1}_{a_1} + E^{A_2}_{a_2} \in I_{E, \Delta}} \ket{a_1}\bra{a_1} \otimes \ket{a_2}\bra{a_2} \otimes \mathbf{1}_B\, . 
\ee
\textbf{(i) Contributions to $\sZ_n^{\rm (PT)}$ from different permutations}\\
Let us first understand the contributions to $\sZ_n^{\rm (PT)}$ from different permutations in this case. Recall that the contribution from $\tau$ can be expanded as 
\bea \label{diag_exp}
\begin{split}
&\vev{\eta_{A_1} \otimes \eta^{-1}_{A_2} \otimes e_{B} | \sI_\al , \tau} = \sum_{i_1, i'_1, ... i_n, i'_n} \braket{\eta_{A_1} \otimes \eta^{-1}_{A_2} \otimes e_{B}|i_1 \bar{i'}_1 ... i_n \bar{i'}_n} \\
& \quad \quad \quad \times \braket{ i_1|\sI_{\al}|i'_{\tau(1)}} \, \braket{ i_2|\sI_{\al}|i'_{\tau(2)}} \, ... \, \braket{ i_n|\sI_{\al}|i'_{\tau(n)}}
\end{split}
\eea
where 
\be 
 \ket{i_m} = \ket{i_{m_a}}_{A_1} \ket{i_{m_{\bar{a}}}}_{A_2} \ket{i_{m_b}}_B, \quad \ket{\bar{i'}_m} = \ket{\bar{i'}_{m_a}}_{A_1} \ket{\bar{i'}_{m_{\bar{a}}}}_{A_2} \ket{\bar{i'}_{m_b}}_B \, . 
\ee
It is useful to take the states $\ket{i_{m_a}}_{A_1}$, $\ket{i_{m_{\bar{a}}}}_{A_2}$, $\ket{i_{m_b}}_B$ to be energy eigenstates in $A_1, A_2, B$. Then note that for \eqref{mic_case1}, each factor in the second line of \eqref{diag_exp} simultaneously gives delta functions between the indices on $A_1$, $A_2$ and $B$ in the bra and the ket, and enforces the condition that the energies in $A_1$ and $A_2$ add up to $E$. For instance, 
\be 
\braket{ i_{a_1} \, i_{\bar{a}_1}\, i_{b_1}\, |\sI_{\al}| i'_{a_1} \, i'_{\bar{a}_1}\, i'_{b_1}\,} = \delta_{i_{a_1} i'_{a_1}}\, \delta_{i_{\bar{a}_1} i'_{\bar{a}_1}}\, \delta_{i_{b_1} i'_{b_1}} \delta_{E^{A_1}_{a_1}+ E^{A_2}_{\bar{a}_1}, \, E}
\ee
In Fig.~\ref{fig:mic_infb_taus}, we explain the resulting contributions from different $\tau$ with the help of diagrams, showing how the energies of different index loops are related. 
\begin{figure}[] 
\centering
\includegraphics[width=10cm]{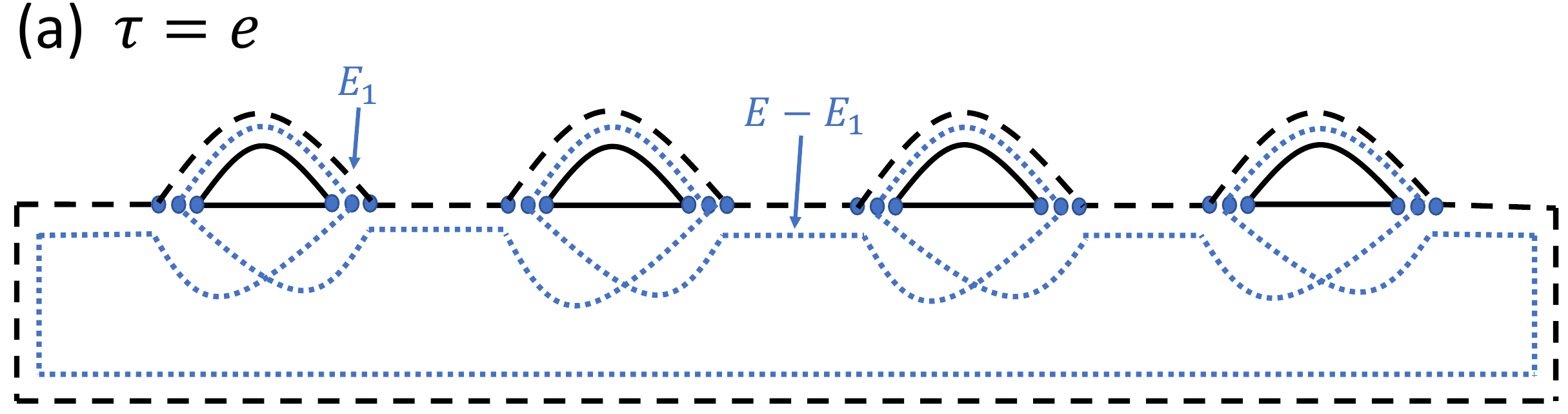}

\vspace{0.2cm}

\includegraphics[width=10cm]{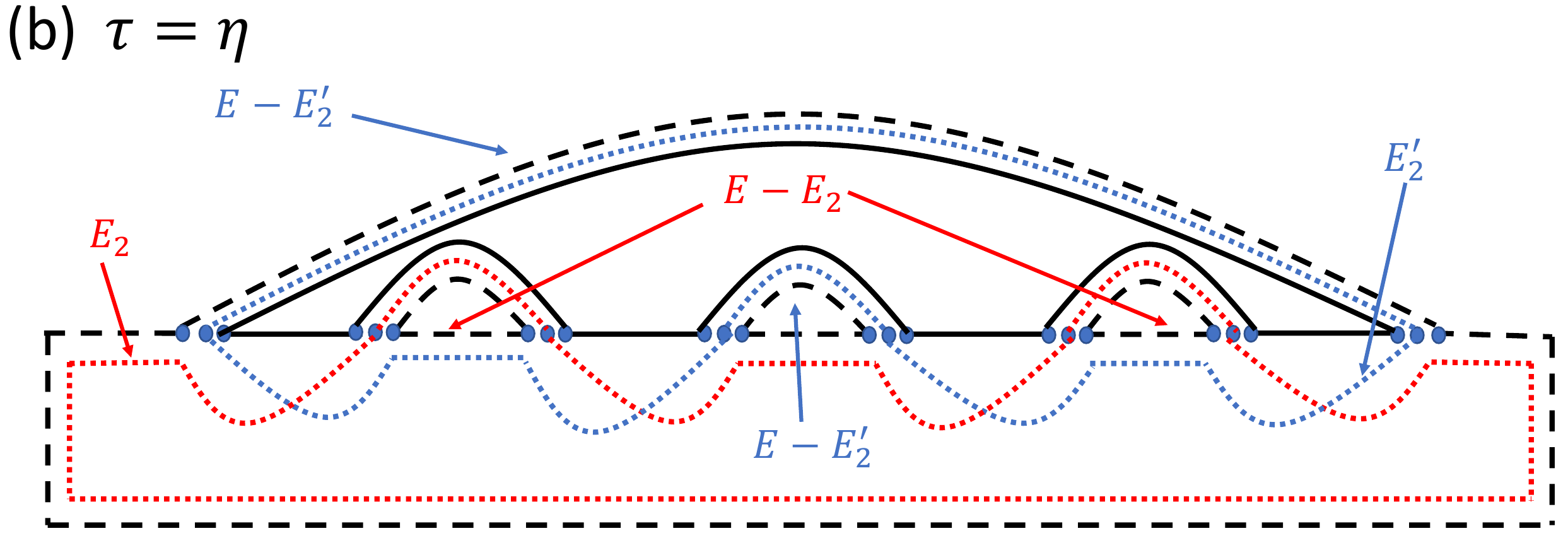} 

\vspace{0.2cm}

\includegraphics[width=10cm]{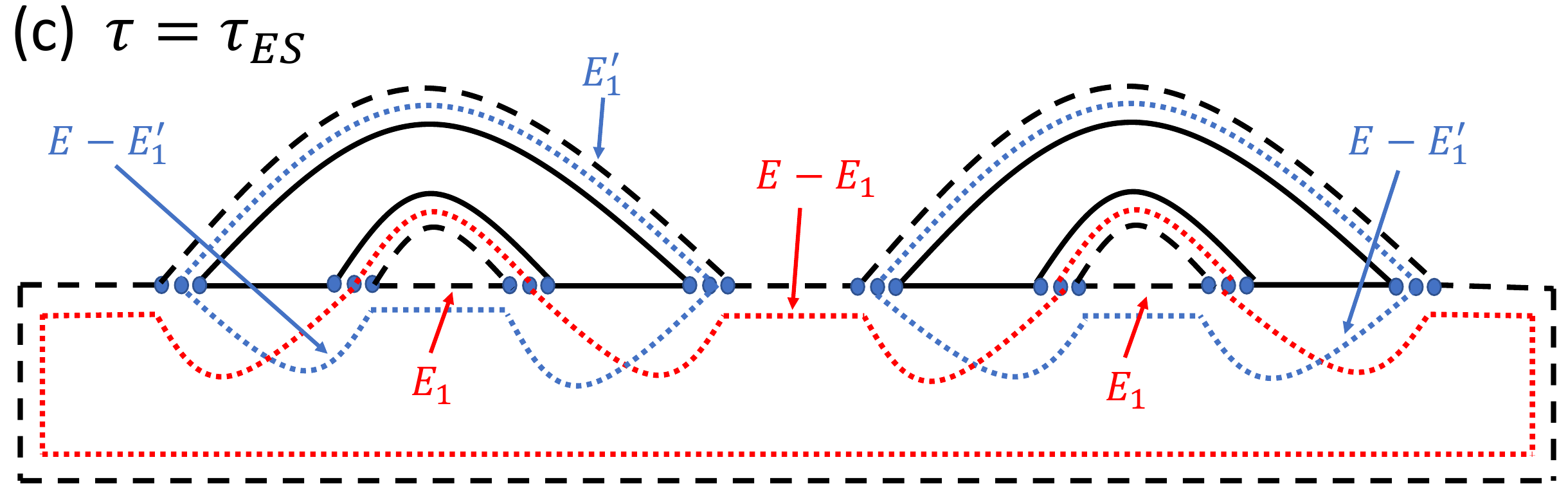}
\caption{Diagrams for different possible dominant contributions to $\sZ_n^{\rm (PT)}$ in the microcanonical ensemble with $B$ at infinite temperature. Recall that dashed, dotted and solid lines respectively represent $A_1$, $A_2$, $B$. $E_s, E_s'$ are energies in $A_s$ for $s=1,2$, and each of these energies is independently summed over in the expression for $\sZ_n^{\rm (PT)}(\tau)$.}
\label{fig:mic_infb_taus}
\end{figure}

Then if the density of states in subsystem $P$ at energy $E_P$ is $d^P_{E_P}$, we find 
\bea
 &\sZ_n^{\rm (PT)}(e) = \frac{1}{N_E^n} d_B^n \sum_{E_1} d^{A_1}_{E_1} d^{A_2}_{E- E_1} , \label{dim1} \\ 
 &\sZ_n^{\rm (PT)}(\tau_{ES}) = \frac{1}{N_E^n} d_B^{n/2} \left( \sum_{E_1} (d^{A_1}_{E_1})^{n/2} d^{A_2}_{E- E_1} \right) \left( \sum_{{E_1'}} d^{A_1}_{E'_1} (d^{A_2}_{E- E'_1})^{n/2} \right) , \label{dim2} \\
&\sZ_n^{\rm (PT)}(\eta) = \frac{1}{N_E^n} d_B \left(\sum_{E_2} d^{A_2}_{E_2} (d^{A_1}_{E- E_2})^{n/2}\right) \, \left(\sum_{E'_2} d^{A_2}_{E'_2} (d^{A_1}_{E- E'_2})^{n/2}\right) \label{dim3}
\eea
where 
\be
N_E= \text{Tr}[\sI_{\al}] = d_B \sum_{E_1} d^{A_1}_{E_1} d^{A_2}_{E-E_1}.
\ee
 On using the saddle point approximation for each of these terms, we find that the saddle point values of the energies in the resulting expressions can be identified as the saddle points that appear in expressions for Renyi entropies of equilibrium density matrices for various subsystems and indices. As a result, we can further simplify \eqref{dim1}-\eqref{dim3} to get the expressions in terms of equilibrium Renyi entropies in \eqref{er_mic}-\eqref{fin_es_mic}. \\

\noindent \textbf{(ii) Resolvent and logarithmic negativity}\\ 

Let us now understand what the Schwinger-Dyson equations in Fig.~\ref{fig:rpq}(a) and \ref{fig:rpq_sd} give for this choice of $\sI_{\al}$. Recall that we take take $\ket{p_s}$, $\ket{q_s}$ to be energy eigenstates in $A_s$. Then we can see that $R_p$ depends only on the energies $E_1$ and $E_2$ of $\ket{p_1}_{A_1}$ and $\ket{p_2}_{A_2}$. So we can express the resolvent as a sum 
\be \label{resolvent_mic}
R_N(\lambda) = \sum_{E_1, E_2} d^{A_1}_{E_1} d^{A_2}_{E_2} \, \tilde{R}(E_1, E_2), \quad \tilde{R}(E_1, E_2) = \begin{cases} 
R_1(E_1) & E_1 + E_2 = E\\ 
R_2(E_1, E_2) & E_1 + E_2 \neq E
\end{cases} 
\ee
The set of diagrams contributing to $\tilde{R}(E_1, E_2)$ differs depending on whether or not the external energies $E_1$ and $E_2$ add up to $E$, giving the two cases $R_1$ and $R_2$ in \eqref{resolvent_mic}. For example, out of the diagrams explicitly shown on the RHS of Fig.~\ref{fig:rpq}(b), in the case where $E_1 + E_2=E$, all diagrams on the RHS of give a non-zero contribution, whereas in the case where $E_1 + E_2\neq E$, only the second diagram gives a non-zero contribution, as the remaining diagrams enforce $E_1 + E_2=E$. Then we have 
\be 
R_1(E_1) = \frac{1}{\lambda - \Sigma_1(E_1)}, \quad R_2(E_1, E_2) = \frac{1}{\lambda - \Sigma_2(E_1, E_2)} \label{geom}
\ee 
where $\Sigma_1, \Sigma_2$ are expressed in terms of $R_1, R_2$ through the Schwinger-Dyson equations in Fig.~\ref{fig:r1_sd} and \ref{fig:r2_sd}. 
\begin{figure}[]
\centering 
\includegraphics[width=15cm]{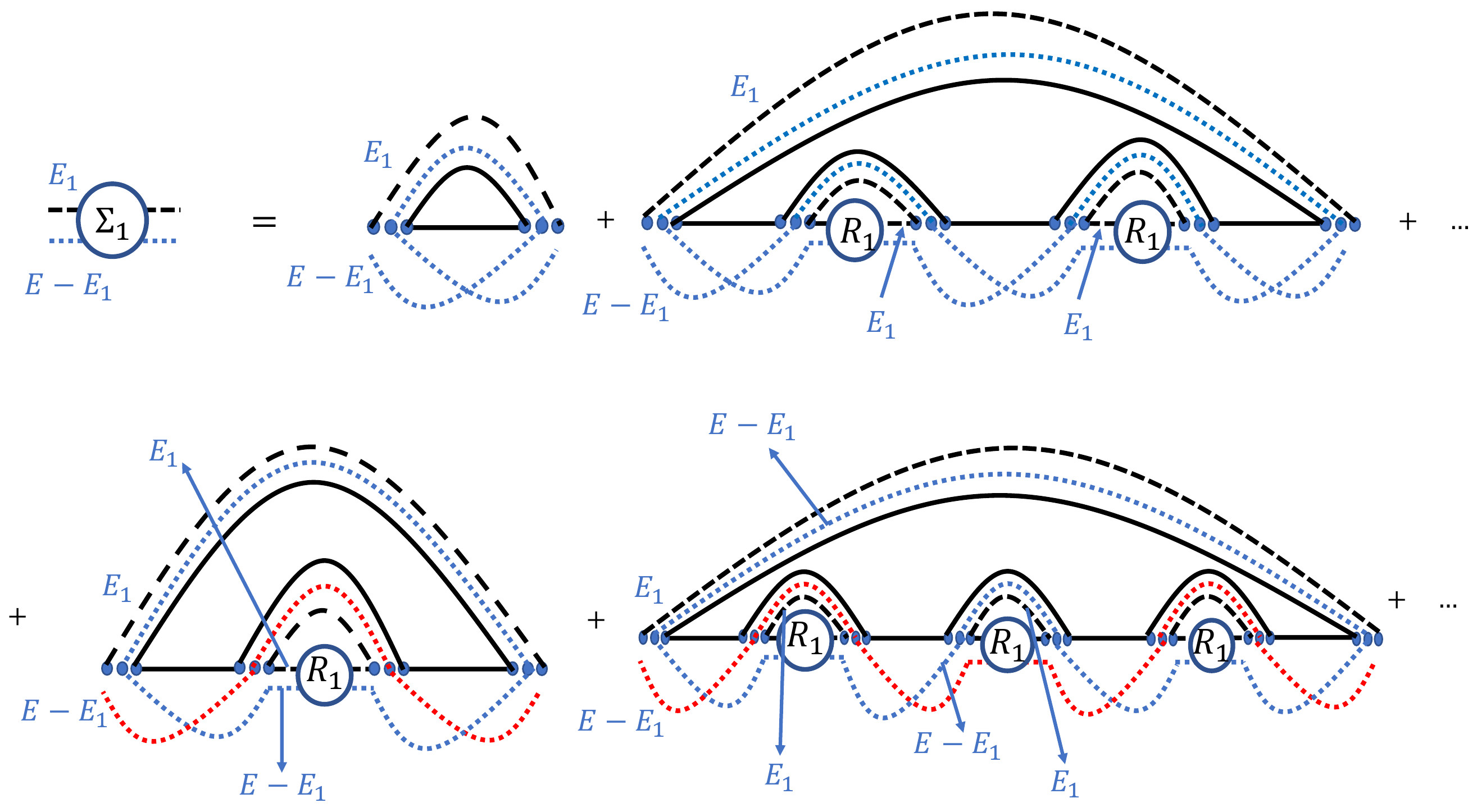}
\caption{Schwinger-Dyson equation for different contributions to $\Sigma_1(E_1)$.}
\label{fig:r1_sd}
\end{figure} 
\begin{figure}[]
\centering 
\includegraphics[width=2.5cm]{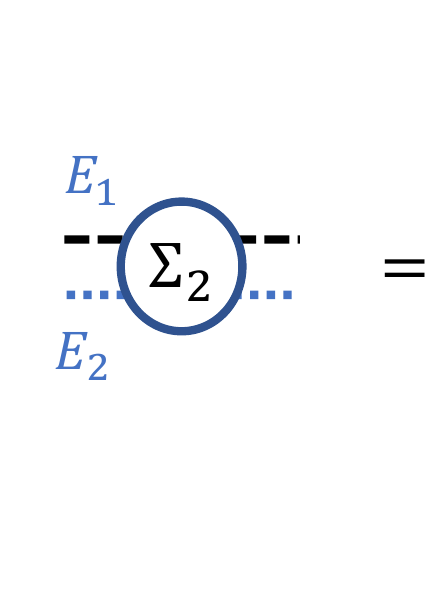}\includegraphics[width=12.5cm]{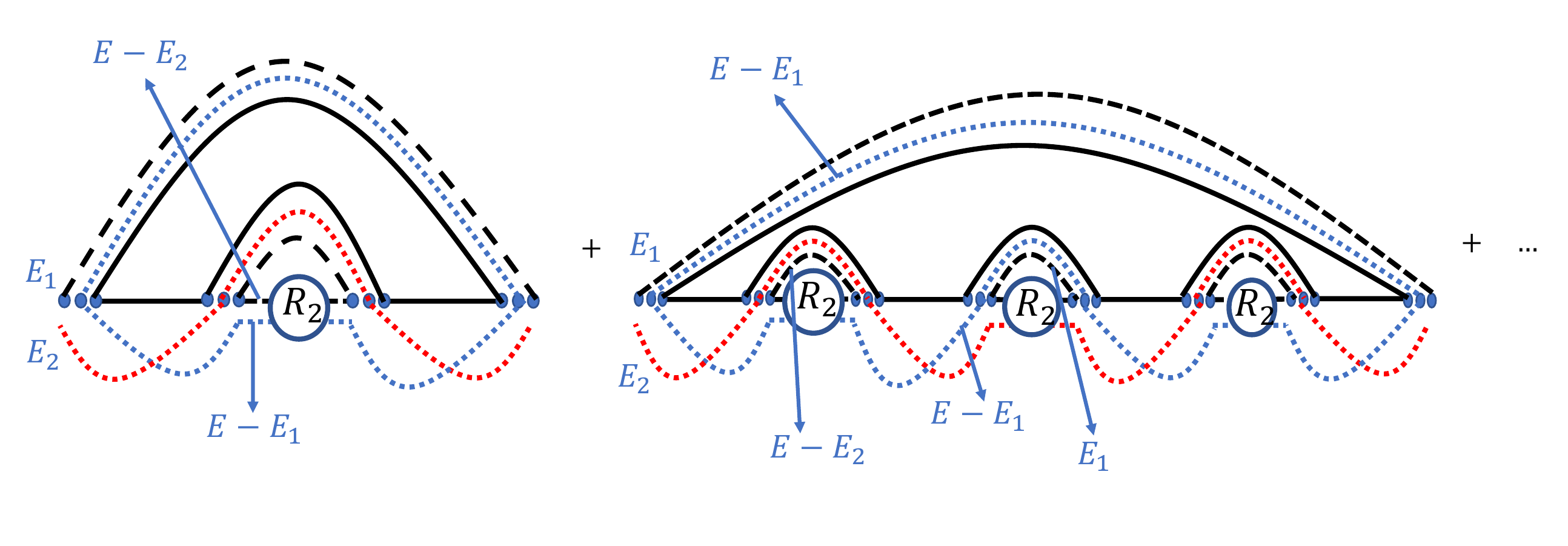}
\caption{Schwinger-Dyson equation for different contributions to $\Sigma_2(E_1, E_2)$.}
\label{fig:r2_sd}
\end{figure}

\noindent \textit{Contribution from permutations with only one-cycles and two-cycles}

To evaluate $\Sigma_1$ and $\Sigma_2$, we can first consider the contributions from all diagrams which have only one-cycles and two-cycles, based again on the expectation that this should capture the regime where the effective dimensions are such that $\frac{d_{A_1}}{d_B d_{A_2}}\ll 1$. We can restrict to such contributions by keeping only the first term in each line on the RHS of Fig.~\ref{fig:r1_sd}, and only the first term on the RHS of Fig.~\ref{fig:r2_sd}. 

We then find
\be 
\Sigma_1(E_1) = \frac{d_B}{N_E} + \frac{d^{A_2}_{E-E_1} d^{A_1}_{E_1} d_B}{N_E^2} R_1(E_1)
\ee
which together with \eqref{geom} gives a quadratic equation for $R_1$, 
\be 
 \alpha_E \, R_1^2 + \left(\frac{d_B}{N_E}-\lambda\right) R_1+ 1= 0 , \quad \alpha_E = \frac{d^{A_1}_{E_1} d^{A_2}_{E- E_1} d_B}{N_E^2} \label{r1_quad_E}
\ee
which has the solution 
\be 
R_1(E_1, E- E_1) = \frac{\lambda - \frac{d_B}{N_E}}{2\alpha_E} - \frac{1}{2\alpha_E} \sqrt{\left(\lambda - \frac{d_B}{N_E}\right)^2 - 4 \alpha_E} \, . 
\ee
From the first diagram on the RHS of \eqref{fig:r2_sd}, we find 
\be 
\Sigma_2(E_1, E_2) = \frac{d_B d^{A_1}_{E-E_2} d^{A_2}_{E-E_1}}{N_E^2} R_2(E-E_2, E-E_1)
\ee
Using the analogous equation for $\Sigma_2(E-E_2,E- E_1)$ together with \eqref{geom}, we get a quadratic equation for $R_2(E_1, E_2)$, 
\be 
 \lambda \alpha_1 R_2^2 -(\lambda^2 +\alpha_1 -\alpha_2) R_2 + \lambda = 0, \quad \alpha_1 = \frac{d^{A_1}_{E_1} d^{A_2}_{E_2} d_B}{N_E^2}, ~~ \alpha_2 = \frac{d^{A_1}_{E-E_2} d^{A_2}_{E-E_1} d_B}{ N_E^2} \, \label{a12}
\ee
which has the solution 
\be 
R_2(E_1, E_2) = \frac{\lambda^2 + \alpha_1 -\alpha_2 - \sqrt{(\lambda- a_1) (\lambda- a_2) (\lambda +a_2)(\lambda + a_1)}}{2 \lambda \alpha_1}
\label{r2_quad}
\ee
where 
\be 
\quad a_1 = \sqrt{\alpha_1}+ \sqrt{\alpha_2}, \quad a_2 = |\sqrt{\alpha_1}- \sqrt{\alpha_2}| \, . 
\ee
By substituting $R_1$ and $R_2$ back into \eqref{resolvent_mic} and finding the imaginary part, we can find the density of eigenvalues of $\rho_A^{T_2}$: 
\be 
D_N(\lambda) = \sum_{E_1}\frac{1}{p_{E_1}} D_{\infty}^{\rm sc}(\tilde{\lambda}, d^{A_1}_{E_1} d^{A_2}_{E-E_1}, d_B ) + \sum_{E_1\neq E_2} D_{E_1, E_2}(\lambda) 
\label{dn_sum}
\ee
where the first term is expressed in terms of the infinite temperature value of $D_N(\lambda)$ in the semi-circle regime from \eqref{sc_inf}, and we have introduced 
\be 
p_{E_1} = \frac{d^{A_1}_{E_1} d^{A_2}_{E-E_1} d_B}{N_E} , \quad \tilde{\lambda} = \frac{\lambda}{p_{E_1}} \, . 
\ee
The second term of \eqref{dn_sum} is found by noting that the second term in \eqref{r2_quad} is imaginary between $-a_1$ and $-a_2$, and between $a_2$ and $a_1$, 
\be
\begin{gathered} 
D_{E_1, E_2}(\lambda) = \frac{1}{2} (d^{A_1}_{E_1}d^{A_2}_{E_2} -d^{A_1}_{E- E_2}d^{A_2}_{E-E_1}) \delta(\lambda) \\ + \frac{N_E^2}{2\pi |\lambda| d_B}\, \sqrt{a_1^2 -\lambda^2} \sqrt{\lambda^2 - a_2^2} \, \theta(a_1^2 -\lambda^2) \theta(\lambda^2- a_2^2)
\end{gathered} 
\ee
So the exponential of the logarithmic negativity, $\sum_i |\lambda_i|$, is given by 
\be 
\sZ^{\rm(PT)}= \sum_{E_1} p_{E_1} \sZ^{\rm (PT)}_{\infty,\,{\rm sc}}(d^{A_1}_{E_1}d^{A_2}_{E-E_1}, d_B) + \sum_{E_1, E_2} \frac{N_E^2 }{\pi d_B} \int_{a_2}^{a_1} d\lambda \sqrt{a_1^2 -\lambda^2} \sqrt{\lambda^2 - a_2^2} 
\label{ln_two}
\ee
where $\sZ^{\rm (PT)}_{\infty,{\rm sc}}$ is given by \eqref{zsc_inf}. 

The integral in \eqref{ln_two} evaluates to 
\be 
 \int_{a_2}^{a_1} d\lambda \sqrt{a_1^2 -\lambda^2} \sqrt{\lambda^2 - a_2^2} = \frac{2}{3} a_1^3 \, E\left(1-\frac{a_2^2}{a_1^2}\right) + \frac{2}{3} a_1 a_2^2 \, E\left(1-\frac{a_2^2}{a_1^2}\right) - \frac{4}{3} a_1 a_2^2\, K\left(1-\frac{a_2^2}{a_1^2}\right)
\ee
where $K$ and $E$ respectively refer to complete elliptic integrals of the first and second kind. 
Now since $a_2/a_1 \approx 1 + e^{-c V}+...$ for some positive coefficient $c$, to find the leading contribution in volume we can approximate both of the elliptic functions appearing above with the leading orders in their Taylor expansions as their argument goes to 0, which are both given by $\pi/2$. We can then approximate 
\be 
\int_{a_2}^{a_1} d\lambda \sqrt{a_1^2 -\lambda^2} \sqrt{\lambda^2 - a_2^2} \approx \frac{\pi}{3} a_1 \left(a_1^2- a_2^2 \right) = \frac{4\pi}{3} (\sqrt{\alpha_1} + \sqrt{\alpha_2})\sqrt{\alpha_1\, \alpha_2}
\ee

So we find 
\be 
\begin{gathered} 
\sZ^{\rm (PT)} = \frac{d_B}{N_E}\sum_{E_1,\, d^{A_1}_{E_1}d^{A_2}_{E-E_1}< d_B} d^{A_1}_{E_1} d^{A_2}_{E-E_1} + \frac{\sqrt{d_B}}{N_E}\sum_{E_1, \, d^{A_1}_{E_1}d^{A_2}_{E-E_1}> d_B} (d^{A_1}_{E_1})^{3/2} (d^{A_2}_{E-E_1})^{3/2} \\
+ \frac{8}{3}\frac{\sqrt{d_B}}{N_E} \, \sum_{E_1} d^{A_1}_{E_1} \left( d^{A_2}_{E-E_1}\right)^{\frac{1}{2}} \, \times \, \sum_{E_2}d^{A_2}_{E_2} \left(d^{A_1}_{E-E_2}\right)^{\frac{1}{2}} \label{all_three}
\end{gathered} 
\ee
The saddle point energy density for the first two terms in the last line is the same, and is equal to $\epsilon = E/V_{A}$. When $c = \frac{S^{\rm (eq)}_{A}}{S^{\rm (eq)}_{A}+S^{\rm (eq)}_{B}}<\frac{1}{2}$, this saddle point lies in the regime of the first term, which then contributes 1. The contribution from the second term is given by its boundary value in this regime, which is always less than the saddle-point value of the same expression, given by 
$e^{\frac{1}{2}(S^{\rm (eq)}_{A_1} + S^{\rm (eq)}_{A_2} - S^{\rm (eq)}_{B})}$. 
For $c>\frac{1}{2}$, the second term contributes $e^{\frac{1}{2}(S^{\rm (eq)}_{A_1} + S^{\rm (eq)}_{A_2} - S^{\rm (eq)}_{B})}$ and dominates over the first. 
The last term can always be approximated with its saddle point value, which is given by 
$e^{{\frac{1}{2} (S^{\rm (eq)}_{\frac{1}{2}, A_1} + S^{\rm (eq)}_{\frac{1}{2}, A_2}- S^{\rm (eq)}_{B})}}$ (where we have ignored an overall factor that gives an $O(1)$ contribution to negativity). For $S^{\rm (eq)}_{\ha, A_1} + S^{\rm (eq)}_{\ha, A_2}> S^{\rm (eq)}_{B}$, 
this contribution dominates over the other two terms, giving 
\be 
\sE = \frac{1}{2} \left( S^{\rm (eq)}_{\frac{1}{2}, A_1} + S^{\rm (eq)}_{\frac{1}{2}, A_2}- S^{\rm (eq)}_{B} \right) \label{val_mic}
\ee
Both the ES phase value of the negativity \eqref{val_mic} and the transition line $S^{\rm (eq)}_{\ha, A_1} + S^{\rm (eq)}_{\ha, A_2} = S^{\rm (eq)}_{B}$ in this approximation thus agree with the naive analytic continuation in \eqref{mksn2} and \eqref{inf_b_an}. \\

\noindent \textit{Contribution from all planar permutations}

Let us now sum over all contributions on the RHS of Figs. \ref{fig:r1_sd} and \ref{fig:r2_sd}. While in this case we are not able to find the logarithmic negativity in all regimes from this calculation either analytically or numerically, we can use the expression for the full sum to systematically understand the regime in which the above result from all permutations with only one- and two- cycles should be valid.

On the RHS of Fig.~\ref{fig:r1_sd}(b), each of the two lines can be obtained by summing over a geometric series, so that we get 
\be
\Sigma_1 = \left(\frac{d_B}{N_E} + \alpha_E \, R_1\right) \frac{1}{1-\beta_E\, R_1^2} , \quad \beta_E = \frac{(d^{A_1}_{E_1})^2}{N_E^2}
\ee
Here we have suppressed the arguments of both $\Sigma_1$ and $R_1$, which are the same in all expressions. This gives a cubic equation for $R_1(E_1)$: 
\be 
\beta_E \lambda R_1^3 + (\alpha_E-\beta_E) R_1^2 + \left(\frac{d_B}{N_E}-\lambda\right) R_1+ 1= 0 \, . 
\ee
We see that this reduces to the contribution from only one-cycles and two-cycles in \eqref{r1_quad_E} when $\alpha_E \gg \beta_E$ for all $E_1$, i.e.~when
\be 
 d^{A_2}_{E-E_1} d_B \gg d^{A_1}_{E_1}, \quad \text{for all $E_1$} \, . \label{first_c}
\ee

Comparing this to the cubic equation for the infinite temperature case \eqref{rinf_cubic}, we find that the first term in \eqref{resolvent_mic} is given by 
\be 
 \sum_{E_1} d^{A_1}_{E_1} d^{A_2}_{E- E_1} R_1(E_1, E- E_1) = \sum_{E_1} \frac{1}{p_{E_1}} R_N^{\infty}(d^{A_1}_{E_1},d^{A_2}_{E-E_1}, d_B) 
\ee
where $R_N^{\infty}$ is the infinite temperature resolvent. The contribution to $\sZ^{\rm (PT)}$ from this term is therefore 
\be 
\sZ^{\rm (PT)}_1 = \sum_{E_1} p_{E_1} \sZ^{\rm (PT)}_{\infty}(d^{A_1}_{E_1}, d^{A_2}_{E-E_1}, d_B)\, . 
\ee

Now for evaluating $\Sigma_2$, the RHS of \eqref{fig:r2_sd} also gives a geometric series (suppressing the $\lambda$ argument in all $\Sigma_2$, $R_2$ below): 
\be \begin{gathered} 
\Sigma_2(E_1, E_2) = \frac{d^{A_1}_{E-E_2}d^{A_2}_{E-E_1} d_B}{N_E^2} R_2(E-E_2, E-E_1) (1+ \tilde{\Delta} + \tilde{\Delta}^2 + ...), \\
\tilde{\Delta} = \frac{d^{A_1}_{E_1}d^{A_1}_{E-E_2}}{N_E^2} R(E_1, E_2) R(E-E_2, E-E_1)
\end{gathered} 
\label{sigma2}
\ee
To simplify the notation, let us fix some $E_1$ and $E_2$, and denote 
\be 
\Sigma_2 := \Sigma_2(E_1, E_2), \quad \Sigma_2' := \Sigma_2(E-E_2, E-E_1), \quad R_2 := \Sigma_2(E_1, E_2), \quad R_2' := R_2(E-E_2, E-E_1)
\ee
Then from \eqref{sigma2} and \eqref{geom}, we have 
\begin{align} 
&\lambda - \frac{1}{R_2}= \Sigma_2 = \frac{d^{A_1}_{E-E_2} d^{A_2}_{E-E_1} d_B}{N_E^2}\, R'_2 \, \frac{1}{1-\frac{d^{A_1}_{E_1} d^{A_1}_{E-E_2}}{N_E^2} R_2 R_2'}\label{r_eq_1} \\
&\lambda - \frac{1}{R'_2}= \Sigma'_2 = \frac{d^{A_1}_{E_1} d^{A_2}_{E_2} d_B}{N_E^2} R_2 \, \frac{1}{1-\frac{d^{A_1}_{E-E_2} d^{A_1}_{E_1}}{N_E^2} R_2 R_2'} \label{r_eq_2}
\end{align} 
From \eqref{r_eq_1} and \eqref{r_eq_2}, we get the following cubic equation for $R_2$: 
\be 
\alpha_3 \lambda^2 R_2^3 + \lambda (\alpha_4+\alpha_1-2\alpha_3) R_2^2 -(\lambda^2 + \alpha_4- \alpha_3 -\alpha_2+\alpha_1) R_2 + \lambda = 0 \,. 
\ee
where 
\be 
\alpha_3= \frac{(d^{A_1}_{E_1})^2 d^{A_2}_{E_2}}{d^{A_2}_{E-E_1} N_E^2}, \quad \alpha_4= \frac{d^{A_1}_{E_1} d^{A_1}_{E-E_2}}{N_E^2}, 
\ee
and $\alpha_1, \alpha_2$ were defined in \eqref{a12}. This reduces to the contribution from one-cycles and two-cycles in \eqref{a12} if $\alpha_1\gg \alpha_3, ~~\alpha_1 \gg \alpha_4$, i.e.~if we have \eqref{first_c} and the additional condition 
\be 
d^{A_2}_{E_1} \, d_B \gg d^{A_2}_{E-E_1} \quad {\text{for all } E_1. }
\ee

The solutions to this equation are 
\be 
R_2 =
\frac{2a-b-c}{3 a \lambda} + \frac{e^{- q i 2 \pi/3}}{6 a \lambda^2} 2^{1/3} \frac{Z}{(X+ \sqrt{Y})^{1/3}} + \frac{e^{i q 2 \pi/3}}{6 a \lambda^2} 2^{2/3} (X+ \sqrt{Y})^{1/3}
\ee
with $q = 0, 1, -1$, and 
\be 
\begin{gathered} 
X = \lambda^3 (- 9 a (a+b+c)\lambda^2 -(2a -b -c)(a^2 - 2(b+c)^2 -a(b+c-9d))) \\
Z = \lambda^2 (6 a \lambda^2 + 2(a^2 +(b+c)^2 - a(b+c+3d)))\\
Y = X^2 - \frac{Z^3}{2} \\
\end{gathered} 
\ee
In the limit $\lambda\rightarrow \infty$, 
\be
\begin{gathered} 
Z \rightarrow 6 a \lambda^4 (1 + O(\frac{1}{\lambda^2})), \\
(X+ \sqrt{Y})^{1/3} = e^{i \pi/6} 2^{1/3} \sqrt{3} a^{1/2} \lambda^2 (1+ i \frac{a(a+b+c)}{2\sqrt{3} a^{3/2}\lambda})
\end{gathered} 
\ee
and hence 
\be 
R_2 = \frac{2a-b-c}{3 a \lambda} + \frac{e^{-i(\frac{2 \pi q}{3} + \frac{\pi}{6})}}{\sqrt{3}\sqrt{a}} (1 - i \frac{a+b+c}{2\sqrt{3} a^{3/2}\lambda}) + \frac{e^{i(\frac{2 \pi q}{3} + \frac{\pi}{6})}}{\sqrt{3}\sqrt{a}} (1 + i \frac{a+b+c}{2\sqrt{3} a^{3/2}\lambda}) + O\left(\frac{1}{\lambda^2}\right)
\ee
For the solutions with $q=0, 1$, the leading order in the $\lambda \rightarrow \infty$ limit is $O(\lambda^0)$, which implies that the leading behaviour of $R$ coming from these solutions is also $O(\lambda^0)$. Since $R$ should approach $\frac{d_A}{\lambda}$, these solutions are not allowed. For $q=-1$, we find in this limit
\be 
R_2(\lambda) = \frac{1}{\lambda} + O(\frac{1}{\lambda^2})
\ee
$R_1(E_1;\lambda)$ also approaches $\frac{1}{\lambda}$. We therefore find 
\be 
R_N(\lambda) = \sum_{E_{1}, E_{2}} \frac{d^{A_1}_{E_1} d^{A_2}_{E_2}}{\lambda} + O(\frac{1}{\lambda^2})= \frac{d_A}{\lambda} + O(\frac{1}{\lambda^2})
\ee
which is the expected behaviour. 

When $\lambda$ is sufficiently large, $Y<0$, and $R_2$ solution is real. For $Y>0$, we find 
\be 
\text{Im} R_2 = \frac{1}{4\sqrt{3} a \lambda^2} \mid \frac{Z}{(|X+ \sqrt{Y}|)^{1/3}} 2^{1/3} - 2^{2/3} (|X+ \sqrt{Y}|)^{1/3} \mid
\ee
This expression can in principle be used to find $D(\lambda)$ and $\sZ^{\rm (PT)}$ for this case, but finding the support of $\text{Im} R_2$ and performing the integral to find the contribution to $\sZ^{\rm (PT)}$ is difficult to do analytically.

\subsection{Microcanonical ensemble with $A_2$ at infinite temperature}
\label{app:mic_a2}
\textbf{(i) Contributions to $\sZ_n^{\rm (PT)}$ from different permutations}\\
Unlike in the earlier example \eqref{mic_case1}, when $\sI_{\al}$ is given by \eqref{mic_inf_temp}, the contributions to $\sZ_n^{(PT)}$ from different $\tau$ can be expressed in a simple way in terms of the cycle numbers of $\tau$, $\eta \tau$ and $\eta \tau^{-1}$. In this case, in any diagram, if we take the energy in any index loop of $A_1$ to be $E_1$, the energy conservation constraint between $A_1$ and $B$ immediately sets all energies in index loops of $A_1$ to be $E_1$, and the energies in index loops of $B$ to $E-E_1$. Some examples are shown in Fig.~\ref{fig:mic2_infb_taus}. 
\begin{figure}[] 
\centering
\includegraphics[width=10cm]{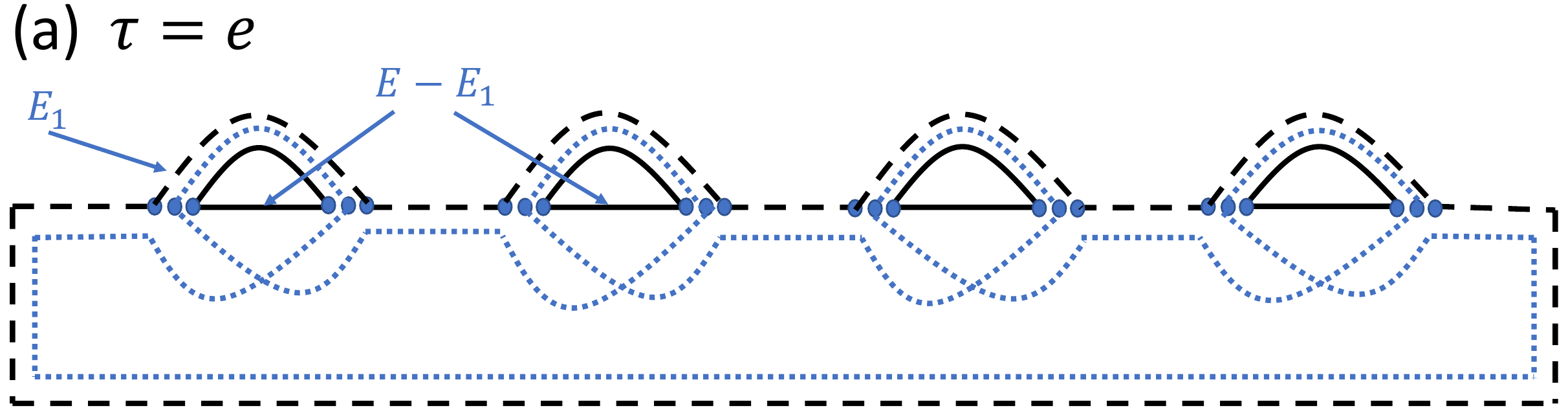}

\vspace{0.2cm}

\includegraphics[width=10cm]{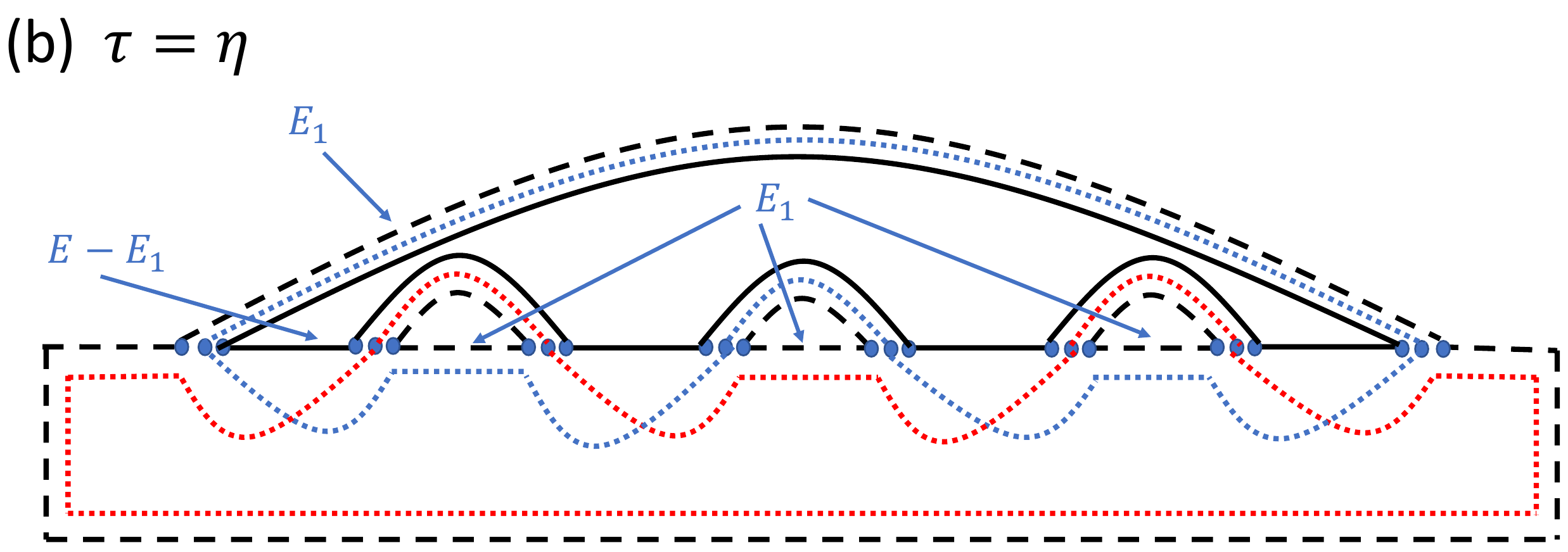} 

\vspace{0.2cm}

\includegraphics[width=10cm]{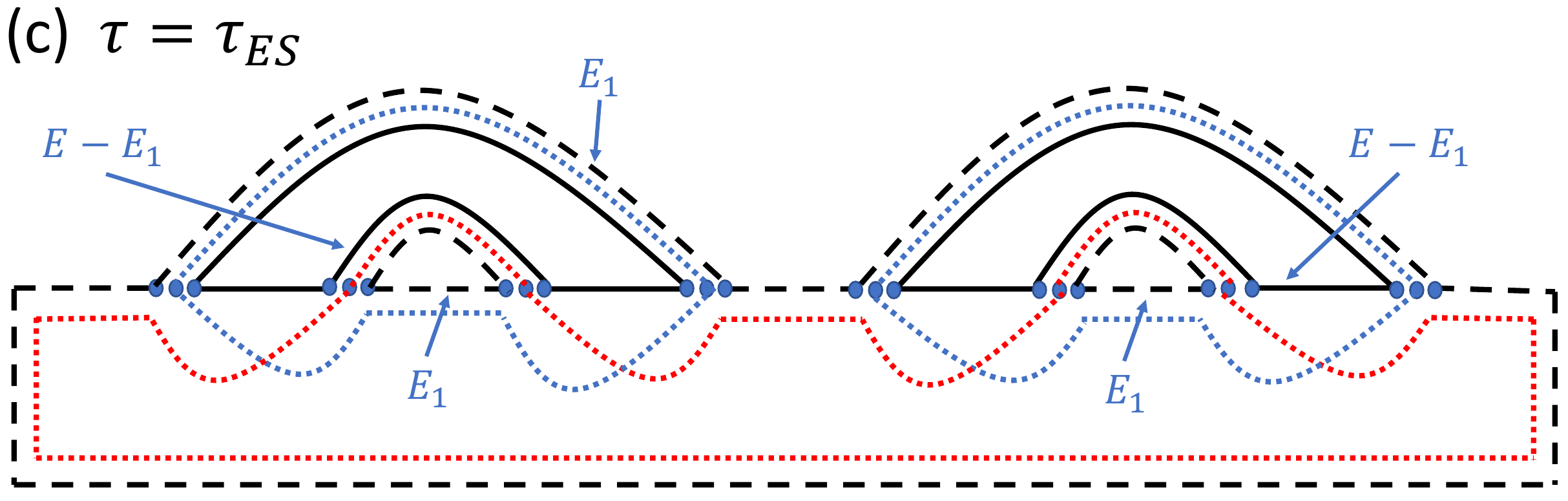}
\caption{Diagrams for different possible dominant contributions to $\sZ_n^{\rm (PT)}$ in the microcanonical ensemble with $A_2$ at infinite temperature. Recall that dashed, dotted and solid lines respectively represent $A_1$, $A_2$, $B$. In this example, all energies in $A_1$ get set to some $E_1$ which is summed over, and each of the energies in $B$ gets set to $E-E_1$.}
\label{fig:mic2_infb_taus}
\end{figure}

We therefore find in this case that 
\be 
\sZ_n^{\rm (PT)} = \sum_{\tau} \sZ_n^{\rm (PT)}(\tau) , \quad \sZ_n^{\rm (PT)}(\tau) = \frac{1}{N_E^n}({d_{A_2}})^{k(\eta\tau)} ~ \sum_{E_1}(d^{A_1}_{E_{1}})^{k(\eta \tau^{-1})}~ (d^{B}_{E- E_1})^{k(\tau)} \label{zn_mic2}
\ee
where $N_E = d_{A_2} \sum_{E_1} d^{A_1}_{E_1}d^{B}_{E-E_1}$.
Then on using the saddle-point approximation in these expressions and comparing to the saddle points appearing in various equilibrium Renyi entropies, we get the expressions for different possible dominant permutations in \eqref{mic2_e}-\eqref{mic2_es}. \\ 

\noindent\textbf{(ii) Logarithmic negativity via resolvent} 

In this example, due to the simplicity of the expressions for $\sZ_n^{\rm (PT)}$, it is more useful to write down an expression for the resolvent by directly summing over $Z_n^{\rm(PT)}/\lambda^n$ as in the definition \eqref{Rndef}, instead of using the Schwinger-Dyson equation like in previous examples. 
Recall that in the infinite temperature case
\be 
\sZ_{n, \infty}^{\rm (PT)} = \frac{1}{(d_A d_B)^n}\sum_{\tau \in \sS_n}({d_{A_2}})^{k(\eta\tau)} ~ \sum_{E_1}(d_{A_1})^{k(\eta \tau^{-1})}~ (d_{B})^{k(\tau)} \label{zn_inf_new}
\ee
Then from \eqref{zn_mic2}, \eqref{zn_inf_new} and \eqref{Rndef}, we find 
\be 
R_N(\lambda) = \sum_{E_1} \frac{1}{p_{E_1}} R_N^{\infty}(\tilde{\lambda}, d^{A_1}_{E_1}, d_{A_2}, d^B_{E - E_2}), 
\label{r_mic}
\ee
where 
\be 
p_{E_1} = \frac{d^{A_1}_{E_1}\, d_{A_2} \, d^{B}_{E-E_1}}{N_E}, \quad \tilde{\lambda} = \lambda/p_{E_1}. 
\ee
Using \eqref{Dndef}, we then find that 
 \be 
 D_N(\lambda) = \sum_{E_1} \frac{1}{p_{E_1}} D_N^{\infty}(\tilde{\lambda}, d^{A_1}_{E_1},\, d_{A_2},\, d^B_{E - E_1})
 \ee
where $D_N^{\infty}$ is the spectral density of the partially transposed density matrix at infinite temperature, and hence 
\be 
\sZ^{\rm (PT)} := \sum_i |\lambda_i| = \sum_{E_1} p_{E_1}\, \sZ^{\rm (PT)}_{\infty}(d_{A_1}^{E_1}\, , d_{A_2}\, , d_B^{E- E_1})
\label{inf}
\ee
where $\sZ^{\rm (PT)}_{\infty}$ is given by \eqref{inf_zpt}. We therefore have
\be 
\begin{gathered} 
\sZ^{\rm (PT)} \approx \sum_{E_1, ~d^{A_1}_{E_1}d_{A_2}<d^B_{E- E_1}} p_{E_1} \quad + d_{A_2} \times \sum_{E_1, ~d_{A_2} d^{B}_{E-E_1}< d^{A_1}_{E_1}} p_{E_1} + \sum_{E_1, ~ d^{A_1}_{E_1} d^B_{E-E_1}< d_{A_2}} p_{E_1} d^{A_1}_{E_1} \\[10pt] + ~{d_{A_2}}^{1/2} \times \sum_{\substack{E_1,~~ d^{A_1}_{E_1}d_{A_2}>d^B_{E-E_1} ~,\\ ~ \frac{1}{d^B_{E-E_1}}<\frac{d^{A_1}_{E_1}}{d_{A_2}}< d^B_{E-E_1}}} p_{E_1} (d^{A_1}_{E_1})^{1/2} (d^{B}_{E-E_1})^{-1/2}
\label{4_cases}
\end{gathered} 
\ee 
In the limit where $V$ is large, each term in \eqref{4_cases} can be evaluated at the energy that maximizes its exponent, and $\sZ^{\rm(PT)}$ is approximately equal to the maximum among the four terms. For each of the terms, there are some values of the volumes $V_{A_1}, V_{A_2}, V_B$ such that the saddle-point energy lies within the range of $E_1$ specified for that term, and other values where it does not. In the latter case, we must use the boundary value rather than the saddle-point value for that term. 

Before further analyzing \eqref{4_cases}, it will be useful to introduce some definitions. Recall that as we vary the sizes of various subsystems, we keep the average energy density $\frac{E}{V_{A_1}+V_B}$ fixed to $\epsilon$, and the infinite temperature entropy density in $B$ fixed to $s_0$. If $\epsilon_1 = \frac{E_1}{V_{A_1}}$ is an energy density in $A$, we define the corresponding energy density in $B$ due to energy conservation as $\bar{\epsilon}_1$, 
\be 
\bar{\epsilon}_1 = \frac{\epsilon(V_{A_1}+V_B)- \epsilon_1 V_{A_1}}{V_B} \, . 
\ee
Also, we define $\epsilon_1^{(\al)}$ as the solution to the equation 
\be 
s'(\epsilon_1^{(\al)}) = \alpha s'\left( \overline{\epsilon_1^{(\al)}} \right) \, . 
\ee
For example, $\epsilon_1^{(1)}$ is the saddle point for evaluating $N_E = \text{Tr}[\sI_{\al}]$, and is equal to $\epsilon$ for any form of $s(\epsilon)$. With $s(\epsilon)$ as in \eqref{s_half}, 
\be 
\epsilon_1^{(\alpha)} = \frac{\epsilon \, (V_{A_1} + V_B)}{V_{A_1}+ V_B\, \alpha^2} \, . 
\ee
Also, note that the equilibrium entropies of the full system and different subsystems are given by 
\be 
\begin{gathered} 
S^{\rm (eq)}= V_{A_2} s_0 + (V_{A_1} + V_B) \, s(\epsilon), \quad S^{\rm (eq)}_{A_1} = V_{A_1} s(\epsilon), \quad S^{\rm (eq)}_{B} = V_{B} s(\epsilon), \\
S^{\rm (eq)}_{A_2} = S^{\rm (eq)}_{n, A_2}= V_{A_2} s_0 \, , \\
S_{\alpha, \, A_1}^{\rm (eq)} = V_{A_1} \frac{\alpha s(\epsilon) - s(\epsilon_1^{(\alpha)})}{\alpha-1} + V_B \frac{\alpha s(\epsilon)-\alpha s(\overline{\epsilon_1^{(\alpha)}})}{\alpha-1} , \quad \alpha \neq 1 \\
S_{\alpha, \, B}^{\rm (eq)} = V_{B} \frac{\alpha s(\epsilon) - s(\overline{\epsilon_1^{(1/\alpha)})}}{\alpha-1} + V_{A_1} \frac{\alpha s(\epsilon)-\alpha s(\epsilon_1^{(1/\alpha)})}{\alpha-1}, \quad \alpha\neq 1 \, .\\
\end{gathered} 
\ee
Considering when the dominant contributions to different terms in \eqref{4_cases} are given by their saddle-point or boundary values, we then find that 
\be 
\sE(A_1, A_2) = \log \sZ^{\rm (PT)} \approx \left[ \text{max}_{i=1, ..., 4}\, f_i \, -\, S^{\rm (eq)}\right] 
\ee
 where 
 \be 
 \begin{gathered} 
 f_1 = \begin{cases} 
S^{\rm (eq)} & S^{\rm (eq)}_A < S^{\rm (eq)}_B \\ 
S^{\rm (eq)}_{A_2}+ V_{A_1} s(\theta_1) + V_B s\left(\bar{\theta}_1\right) & S^{\rm (eq)}_A > S^{\rm (eq)}_B\end{cases} 
 \end{gathered} 
 \label{f1}
\ee
\be 
 \begin{gathered} 
 f_2= \begin{cases} 
 S^{\rm (eq)}_{A_2} + S^{\rm (eq)}& S^{\rm (eq)}_{A_2} + S^{\rm (eq)}_{B}< S^{\rm (eq)}_{A_1}\\ 
 2 S^{\rm (eq)}_{A_2}+ V_{A_1}\, s(\theta_2) + V_{A_2}\, s(\bar{\theta}_2) & S^{\rm (eq)}_{A_2} + S^{\rm (eq)}_{B}> S^{\rm (eq)}_{A_1} 
 \end{cases} 
 \end{gathered} 
 \label{f2}
 \ee
 \be 
 \begin{gathered} 
 f_3= \begin{cases} 
S^{\rm (eq)}_{\frac{1}{2}, \, A_1} + S^{\rm (eq)} \quad \quad & V_{A_1}\,s(\epsilon_1^{(1/2)}) + V_B \, s\left(\overline{\epsilon_1^{(1/2)}} \right) < V_{A_2} s_0 \\
S^{\rm (eq)}_{A_2} + 2 V_{A_1} s(\theta_3) + V_B\, s \left(\bar{\theta}_3 \right) \quad \quad & V_{A_1} s(\epsilon_1^{(1/2)}) + V_B \, s\left(\overline{\epsilon_1^{(1/2)}} \right) > V_{A_2} s_0 
 \end{cases} 
 \end{gathered} 
 \label{f3}
 \ee
 \be 
 \begin{gathered} 
 f_4(c, \lambda) = \begin{cases} 
\frac{3}{2}S^{\rm (eq)}_{A_2} + S^{\rm (eq)}_{A_1, \frac{1}{3}} \quad & V_{A_1} s(\epsilon_1^{(1/3)}) + V_{A_2}s_0 > V_B\, s\left( \overline{\epsilon_1^{(1/3)}}\right), \text{ and }\\ + \frac{1}{2}(S^{\rm (eq)}_{A_1}+ S^{\rm (eq)}_{B}) \quad & V_{A_2}\, s_0 < V_{A_1} s(\epsilon_1^{(1/3)}) + V_B\, s\left(\overline{\epsilon_1^{(1/3)}}\right) , \text{ and } \\
 & V_{A_1}s(\epsilon_1^{(1/3)})< V_{A_2} s_0 + V_B\, s\left(\overline{\epsilon_1^{(1/3)}}\right)\\ 
\text{max}_{i=1, 2, 3}\, [\frac{3}{2} S^{\rm (eq)}_{A_2}+ \frac{3}{2} V_{A_1}s(\theta_i) \\
+ \frac{1}{2}V_B s \left(\bar{\theta}_{i} \right) ] & \text{otherwise} 
 \end{cases} 
 \end{gathered}
 \label{f4} 
 \ee
In the above expressions, $\theta_i$ are the values of the energy density lying at the boundaries of the regimes corresponding to the different terms in \eqref{4_cases}, and are defined implicitly as solutions to the equations 
\be
\begin{gathered} 
V_{A_1} s(\theta_1) + V_{A_2} s_0 = V_B s\left(\bar{\theta}_1 \right), \\
V_{A_2} s_0 + V_B s\left(\bar{\theta}_2\right) = V_{A_1} s(\theta_2), \\
V_{A_1} s(\theta_3) + V_B s\left( \bar{\theta}_3\right) = V_{A_2} s_0, 
\end{gathered} 
\ee
which we can solve numerically for the entropy density in \eqref{s_half}. For certain values of the subsystem volumes, we will find that solutions for some of the $\theta_i$ do not exist; this corresponds to cases where the range of energies corresponding to some of the terms in \eqref{4_cases} do not exist.

 \begin{figure}[]
\centering
\includegraphics[width=7cm]{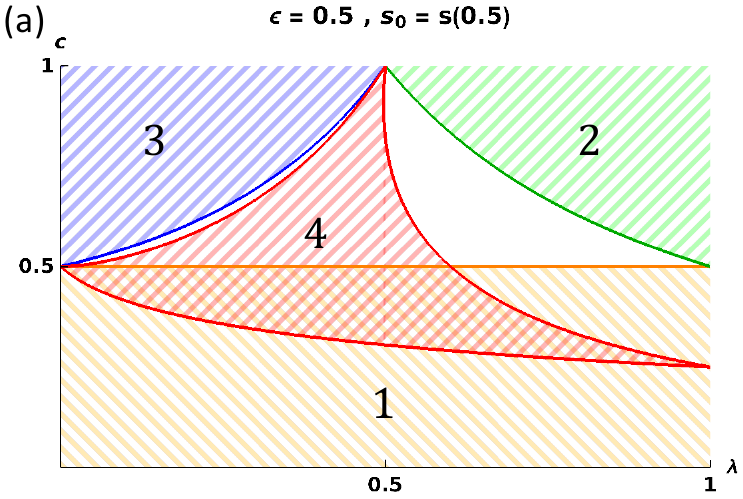}
\includegraphics[width=7cm]{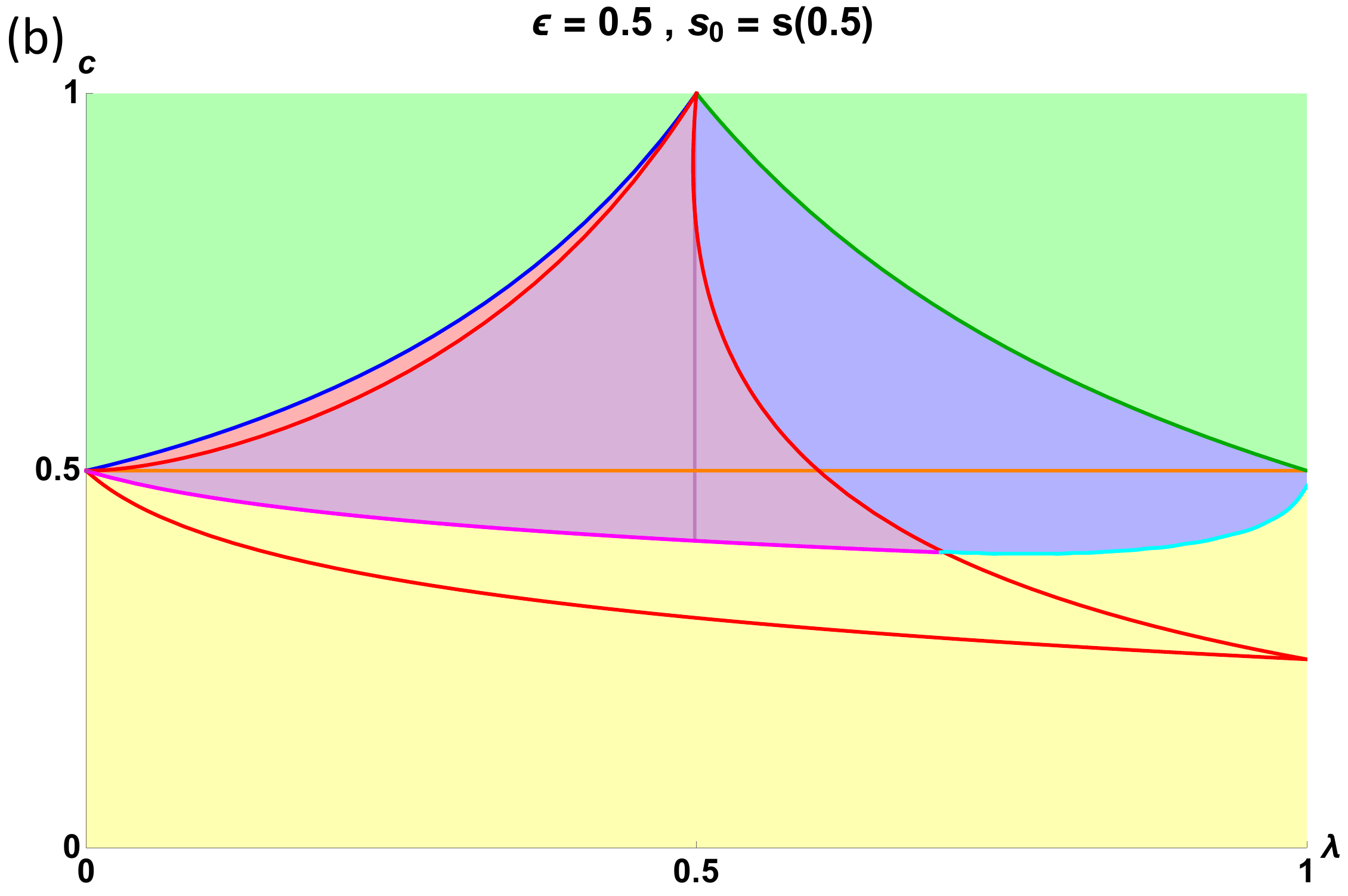}
\caption{Regions 1-4 in (a) show the parameter ranges for which $f_1$ through $f_4$ are given by the saddle-point values in the first lines of \eqref{f1}-\eqref{f4}. The resulting phase diagram for the logarithmic negativity is shown in (b). Different shaded regions correspond to distinct phases. All lines from (a), including those that do not correspond to phase transition lines of the logarithmic negativity, are also shown in (b) to make it easier to compare regions in the two figures.}
\label{fig:regions}
\end{figure} 
Let us try to understand the phase diagram for the logarithmic negativity in terms of the parameters $c = \frac{S^{\rm (eq)}_A}{S^{\rm (eq)}_A+S^{\rm (eq)}_B}$ and $\lambda = \frac{S^{\rm (eq)}_{A_1}}{S^{\rm (eq)}_{A}}$. We first find the ranges of $c, \lambda$ for which the different terms $f_1$ through $f_4$ are given by their saddle-point values (the first lines of equations \eqref{f1}-\eqref{f4}) in Fig.~\ref{fig:regions}(a). 

The parameter ranges where $f_2$ and $f_3$ take their saddle-point values are shown respectively by the green and blue regions in Fig.~\ref{fig:regions}(a). In these regions, we find that $f_2$ and $f_3$ respectively are also larger than the other $f_i$. Hence, in these regions the logarithmic negativity is given by 
\be \label{ln_me}
\sE = \begin{cases} S^{\rm (eq)}_{1/2, \, A_1} & \lambda <\frac{1}{2} \\ 
S^{\rm (eq)}_{A_2} & \lambda > \frac{1}{2} \, 
\end{cases} 
\ee
There is some intersection between the orange and red shaded regions in Fig.~\ref{fig:regions}(a) where $f_1$ and $f_4$ respectively take their saddle point values. 
On comparing the first lines of \eqref{f1} and \eqref{f4}, we find that the latter dominates for 
\be 
\frac{1}{2}S_{A_2}^{\rm eq} + S_{1/3, A_1}^{\rm eq} - \frac{1}{2} S_{A_1}^{\rm eq} - \frac{1}{2} S^{\rm eq}_{B}>0 \, . \label{n_1_again} 
\ee
The condition \eqref{n_1_again} is shown with the pink line in Fig.~\ref{fig:regions}(b). In the purple shaded region of Fig.~\ref{fig:regions}(b), we find that $f_4$ is also larger than $f_2$ and $f_3$, and hence in this region the logarithmic negativity is given by 
\be 
\sE = \frac{1}{2}S_{A_2}^{\rm eq} + S_{1/3, A_1}^{\rm eq} - \frac{1}{2} S_{A_1}^{\rm eq} - \frac{1}{2} S^{\rm eq}_{B} \, . \label{ln_es} 
\ee

In the unshaded regions of Fig.~\ref{fig:regions} (a), none of the terms of \eqref{4_cases} take their saddle-point values. In the left unshaded region of Fig.~\ref{fig:regions} (a), which is the same as the region shaded in red in Fig.~\ref{fig:regions} (b), we find that the dominant contribution is given by the second lines of $f_3$ and $f_4$, which coincide with each other. In this region, the logarithmic negativity is given by 
\be 
\sE = S^{\rm (eq)}_{A_2} + V_{A_1} (s(\theta_3)- s(\epsilon)) - V_B s(\epsilon) \,. 
\label{ln_4}
\ee
In the blue shaded region of Fig.~\ref{fig:regions}(b), which includes both the right unshaded region of Fig.~\ref{fig:regions}(a) and part of the orange shaded region there, the dominant contribution is given by the second lines of $f_2$ and $f_4$, which coincide. The logarithmic negativity here is given by 
\be 
\sE = V_{A_1} \, (2 s(\theta_2)- s(\epsilon)) - V_B \, s(\epsilon) \, . 
\label{ln_5}
\ee
Finally, in the yellow shaded region of Fig.~\ref{fig:regions}(b), we have 
\be 
\sE = 0 \, . \label{ln_zero}
\ee
Each of the expressions \eqref{ln_me}, \eqref{ln_es}, \eqref{ln_4}, \eqref{ln_5}, \eqref{ln_zero} corresponds to a distinct phase, and there is a discontinuity in either the first or the second derivative of $\sE$ across each of the phase transition lines, as shown in Fig.~\ref{fig:E_I_plots}.

\begin{section}{Maxima of functions of permutations at finite temperature} 
\label{finitetemp_perms}
In this appendix, we use general properties of the function $f(\beta)$ in any system to show that $G(\beta, \tau)$ (as defined in \eqref{func_G}) is maximized by $\tau=e$, while $G(\beta, \tau) + G(\beta, \eta^{-1}\tau)$ is maximized by both $\tau=e$ and $\tau=\eta$. 

Recall 
\be 
G(\beta, \tau) = \sum_{i=1}^{k(\tau)} f(c_i \beta) \label{onesum}
\ee
and 
\be 
G(\beta, \tau) + G(\beta, \eta^{-1}\tau) = \sum_{i=1}^{k(\tau)} f(c_i \beta) + \sum_{j=1}^{k(\eta^{-1} \tau)} f(a_j \beta) \label{bothsum}
\ee
Both \eqref{onesum} and \eqref{bothsum} are examples of sums of the form
\be 
P(S, m, \{c_i\}):= \sum_{i=1}^m f(c_i \beta) , \quad \sum_{i=1}^m c_i = S, \quad 1 \leq c_i \leq n \label{ci_cond}
\ee 
where the $m$ numbers $c_i$ are all positive integers, and $S$ is either $n$ or $2n$. In fact, note that the above constraints together further restrict the range of the $c_i$, to
\be 
1 \leq c_i \leq S - (m-1) \, . \label{new_const}
\ee
To simplify the problem of finding $\tau$ that maximize \eqref{onesum} and \eqref{bothsum}, we first ask: with the constraint in \eqref{new_const}, what are the values of $m, c_i$ for which the function $P(S, m, \{c_i\})$ is maximized? 

We first fix $m$, and see what set of $m$ numbers $c_i$ satisfying \eqref{new_const} maximizes $P(S, m, \{c_i\})$. Note that since each $c_i$ lies between 1 and $S - (m-1)$, it can be written as 
\be 
c_i = \lambda + (1-\lambda) (S-(m-1))
\ee
where $\lambda =\frac{S-m+1-c_i}{S-m}$ lies between 0 and 1. Then using \eqref{in_1} for each term in $P(m, S, \{c_i\})$, we find 
\be 
\begin{gathered} 
P(S, m, \{ c_i \}) \leq \sum_{i=1}^m \left( \frac{S-m+1-c_i}{S-m} f( \beta) + \frac{c_i-1}{S-m} f\left((S-m+1)\beta\right) \right)\\
 \leq Q(m, S):= (m-1)f(\beta) + f\left((S-m +1)\beta \right) 
\end{gathered}
\label{P_eq}
\ee

Note further that $Q(m, S)$ is an increasing function of $m$ for both choices of $S$, since \eqref{in_5} implies in particular that 
\be 
f(\beta) - \beta f'((2n-m+1)\beta)\geq 0, \quad f(\beta) - \beta f'((n-m+1)\beta) \geq 0. 
\ee
Hence, the following statements hold: 
\begin{enumerate} 
\item 
\be 
G(\tau) = \sum_{i=1}^{k(\tau)} f(c_i \beta) \leq n f(\beta) \label{Gtau_max}
\ee
This is because on comparing the definition of $G(\tau)$ to that of $P(S, m, \{c_i\})$, we see that $S=n$, and the maximum value of $m$ that can appear in $G(\tau)$ is $n$. But the RHS of \eqref{Gtau_max} is the value of the LHS for $\tau=e$, so $G(\tau)= \sum_{i=1}^{k(\tau)} f(c_i \beta)$ is maximized by $\tau=e$. 
\item 
\be 
G(\tau) + G(\eta\tau^{-1}) =\sum_{i=1}^{k(\tau)} f(c_i \beta) + \sum_{j=1}^{k(\eta \tau^{-1})} f(a_j \beta) \leq n f(\beta) +f(n\beta)
\ee
This follows since the maximum number of terms in this full expression is $n+1$ since $k(\tau) + k(\eta \tau^{-1})\leq n+1$, and here $S=2n$. The RHS is the value of the LHS for $\tau=e$ and $\tau=\eta$, so $G(\tau)+ G(\eta \tau^{-1})$ is maximized by both.

\end{enumerate}

\end{section}

\section{$\sZ_n^{\rm (PT)}$ in the microcanonical ensemble with energy conservation in $AB$} \label{app:mic_hom} 

Let us now consider the equilibrium approximation with $\sI_{\al}$ as in \eqref{full_mic}. Unlike in the simpler example in the previous subsection, we are not able to compute the logarithmic negativity through the resolvent in this example. However, we can find the phase diagram for 
$\sZ_n^{\rm (PT)}$ and the Renyi negativities, which have the same qualitative features as those for the canonical example and the simple example of the previous subsection. 
 
Let us assume that like in other universality classes, the dominant contribution is always given by one out of $e$, $\eta$, $\eta^{-1}$, and $\tau_{ES}$. Let us try to understand the contributions that we get from each of these permutations. In each case, we can use diagrams similar to Fig.~\ref{fig:mic_infb_taus}, now taking into account energy conservation between all three subsystems $A_1$, $A_2$, and $B$. We then find: 

\begin{enumerate} 
\item 
 \be 
 \begin{gathered} 
 \sZ_n^{\rm (PT)}(e) = \frac{1}{N_E^n} \sum_{E_{1}, E_{2}} d^{A_1}_{E_1} d^{A_2}_{E_2}(d^B_{E-E_{1}- E_{2}})^n
 \end{gathered} 
 \ee
Expressing each density of states in terms of the corresponding entropy density (e.g.~$d^{A_1}_{E_1} = \exp({V_{A_1} s(\frac{E_1}{V_{A_1}})})$~), we find that the saddle-point equations for the energy densities $\epsilon_1 = \frac{E_1}{V_{A_1}}$ and $\epsilon_1 = \frac{E_2}{V_{A_2}}$ are 
 \begin{align} 
 &s'(\epsilon_1) = n s'\left( \frac{V\epsilon - V_{A_1} \epsilon_1 - V_{A_2}\epsilon_2}{V_B}\right) \label{1_saddle} \\
 &s'(\epsilon_2)= n s'\left( \frac{V\epsilon - V_{A_1} \epsilon_1 - V_{A_2}\epsilon_2}{V_B}\right)
 \label{2_saddle}
 \end{align} 
 where $\epsilon = \frac{E}{V}$. 
Since the RHS of \eqref{1_saddle} and \eqref{2_saddle} is the same, we can assume that $\epsilon_1 = \epsilon_2 = \epsilon_e$, and solve
 \be 
 s'(\epsilon_e) = n s'\left( \frac{ V\epsilon - V_A \epsilon_e}{V_B}\right) \label{fullmic_e}
 \ee
 We therefore find 
\be 
\sZ_n^{\rm(PT)}(e) \approx e^{ V_{A} s(\epsilon_e) + n V_B s\left( \frac{V \epsilon -V_A \epsilon_e}{V_B}\right) -V n s(\epsilon)} \, . \label{fullmic_e2} 
\ee
 Taking the $n\rightarrow 1$ limit of \eqref{fullmic_e} and \eqref{fullmic_e2} gives 
 \be 
 \sE = 0 \, . 
 \ee
 
 \item 
 \be 
 \begin{gathered}
 \sZ_n^{\rm (PT)}(\eta) = \frac{1}{N_E^n} \sum_{E_{B}, E_{2}, E'_{2}} d^B_{E_B} \, d^{A_2}_{E_{2}} \, d^{A_2}_{E'_2} \, (d^{A_1}_{E-E_{2}-E_B})^{\frac{n}{2}} \, (d^{A_1}_{E-E'_{2}-E_B})^{\frac{n}{2}}\\
 \end{gathered} 
 \ee
 The saddle-point equations for the energy densities $\epsilon_B = \frac{E_B}{V_B}$, $\epsilon_2= \frac{E_2}{V_{A_2}}$, and $\epsilon'_2= \frac{E'_2}{V_{A_2}}$ are 
 \begin{align} 
& s'(\epsilon_B) - \frac{n}{2} s'\left(\frac{\epsilon V - V_{A_2} \epsilon_{2}- V_B \epsilon_B}{V_{A_1}} \right) - \frac{n}{2} s'\left(\frac{\epsilon V - V_{A_2} \epsilon'_2- V_B \epsilon_B}{V_{A_1}} \right) = 0 \\
& s'(\epsilon_2) - \frac{n}{2} s'\left(\frac{\epsilon V - V_{A_2}\epsilon_{2}- V_B\epsilon_B}{V_{A_1}} \right)= 0 \label{e2_fullmic}\\
& s'(\epsilon'_{2}) - \frac{n}{2} s'\left(\frac{\epsilon V - V_{A_2}\epsilon'_{2}- V_B \epsilon_B}{V_{A_1}} \right)= 0 \label{e2p_fullmic}
 \end{align} 
 Since the system of equations is unchanged on exchanging $\epsilon_2$ with $\epsilon'_2$ everywhere, the solutions for both variables will be the same. Hence, we can simplify the equations by setting $\epsilon_2 = \epsilon'_2 = \epsilon_2^{\eta}$. Then we must solve the following two equations for $\epsilon_2^{\eta}$ and $\epsilon_B^{\eta}$: 
 \begin{align} 
 &s'(\epsilon_B^{\eta}) - n s'\left(\frac{V \epsilon - V_{A_2} \epsilon_{2}^{\eta}- V_B\epsilon_B^{\eta}}{V_{A_1}} \right) = 0 \label{miceta_1} \\
&s'(\epsilon_{2}^{\eta}) - \frac{n}{2} s'\left(\frac{V \epsilon - V_{A_2} \epsilon_{2}^{\eta}- V_B\epsilon_B^{\eta}}{V_{A_1}} \right) = 0 \label{miceta_2}
\end{align} 
to find 
 \be 
 \sZ_n^{\rm (PT)}(\eta) \approx \exp(V_B s(\epsilon_B^{\eta}) + 2 V_{A_2} s(\epsilon_2^{\eta}) + n V_{A_1} s\left(\frac{V - \epsilon V_{A_2}\epsilon_2^{\eta} -V_B\epsilon_B^{\eta}}{V_{A_1}} \right)- n V s(\epsilon)) \label{eta_final}
 \ee
For general forms of $s(\epsilon)$, there is no relation between \eqref{eta_final} for general $n$ and the Renyi entropies of $\rho^{\rm (eq)}$. But for $n=1$, \eqref{miceta_1}, \eqref{miceta_2} and \eqref{eta_final} simplify such that we have 
\be
\sE = S^{\rm (eq)}_{\frac{1}{2}, A_2} 
\ee

Similarly, taking the $n\rightarrow 1$ limit of $\sZ_n^{\rm (PT)}(\eta^{-1})$, we find 
\be
\sE = S^{\rm (eq)}_{\frac{1}{2}, A_1} \, . 
\ee
 
\item Now consider the contribution from $\tau_{ES}$:
 \be 
 \begin{gathered} 
 \sZ_n^{\rm (PT)}(\tau_{ES}) = \frac{1}{N_E^n}\sum_{\substack{E_{1}, E'_{1}, E_{B}^1,\\ E_{B}^2,..., E_{B}^{n/2}}}\left( \prod_{j=1}^{n/2} d^B_{E_B^j}\right)
~ d^{A_2}_{E- E_{1}- E_{B}^1} \left( \prod_{j=1}^{n/2} d^{A_2}_{E- E'_{1}-E_{B}^j}\right)\\
~ \times d^{A_1}_{E'_{1}} d^{A_1}_{E_{1}} \left(\prod_{j=2}^{n/2} d^{A_1}_{E_{1} + E_{B}^1-E_{B}^j}\right) 
 \end{gathered} 
 \ee
 The $n+2$ saddle point equations are 
 \begin{align}
 &s'(\epsilon_{B}^1) -s'\left(\frac{V\epsilon- V_{A_1} \epsilon_{1}- V_B\epsilon_{B}^1}{V_{A_2}} \right) - s'\left(\frac{V \epsilon - V_{A_1} \epsilon'_{1} -V_B\epsilon_{B}^1}{V_{A_2}} \right) \nonumber \\ 
 &+ \sum_{j=2}^{n/2} s'\left( \epsilon_{1} + \frac{V_B}{V_{A_1}}(\epsilon_{B}^1- \epsilon_{B}^j)\right) = 0\\
 &s'(\epsilon_{1}) -s'\left(\frac{\epsilon - V_{A_1} \epsilon_{1}- V_B\epsilon_{B}^1}{V_{A_2}} \right) + \sum_{j=2}^{n/2} s'\left(\epsilon_{1}+\frac{V_B}{V_{A_2}}(\epsilon_{B}^1-\epsilon_{B}^j)\right) = 0\\
 &s'(\epsilon'_{1}) - \sum_{j=1}^{n/2} s'\left(\frac{\epsilon V - V_{A_1} \epsilon'_{1} - V_B\epsilon_{B}^j}{V_{A_2}} \right)=0 \\
 &s'(\epsilon_{B}^j) - s'\left(\frac{V\epsilon-V_{A_1} \epsilon'_{1}-V_B\epsilon_{B}^j}{V_{A_2}} \right) - s'\left(\epsilon_{1}+\frac{V_B}{V_{A_1}}(\epsilon_{B}^1-\epsilon_{B}^j)\right) = 0 , \quad j= 2, ..., n 
\end{align} 
 Note that this set of equations is invariant under permuting the $\epsilon_{B}^j$ from $j=2$ to $n$. We can therefore assume that these variables have the same solution, so that we have a simpler set of four equations for any $n$: 
 \begin{align} 
 &s'(\epsilon_{B}^1) -s'\left(\frac{V\epsilon- V_{A_1} \epsilon_{1}- V_B\epsilon_{B}^1}{V_{A_2}} \right) - s'\left(\frac{V\epsilon -V_{A_1} \epsilon'_{1} - V_B\epsilon_{B}^1}{V_{A_2}} \right) \nonumber \\ &+ (\frac{n}{2}-1) s'\left( \epsilon_{1} + \frac{V_B}{V_{A_1}}(\epsilon_{B}^1- \epsilon_{B}^j)\right) = 0 \label{mices-1}\\
 &s'(\epsilon_{1}) -s'\left(\frac{V\epsilon - V_{A_1} \epsilon_{1}-V_B\epsilon_{B}^1}{V_{A_2}} \right) + (\frac{n}{2}-1)s'\left(\epsilon_{1}+\frac{V_B}{V_{A_1}}(\epsilon_{B}^1-\epsilon_{B}^j)\right) = 0\\
 &s'(\epsilon'_{1}) - s'\left(\frac{V\epsilon - V_{A_1} \epsilon'_{1} - V_B\epsilon_{B}^1}{V_{A_2}} \right)- (\frac{n}{2}-1)s'\left(\frac{V\epsilon - V_{A_1} \epsilon'_{1} - V_B\epsilon_{B}^j}{V_{A_2}} \right)=0 \\
 &s'(\epsilon_{B}^j) - s'\left(\frac{V\epsilon-V_{A_1} \epsilon'_{1}-V_B\epsilon_{B}^j}{V_{A_2}} \right) - s'\left(\epsilon_{1}+\frac{V_B}{V_{A_1}}(\epsilon_{B}^1-\epsilon_{B}^j)\right) = 0 , \quad j= 2, ..., n \label{mices-4}
 \end{align} 
 In terms of the solution $\{\bar\epsilon_{1}, \bar \epsilon'_{1}, \bar \epsilon_{B}^j, \bar \epsilon_{B}^1\}$ to these equations, we can write 
 \be
 \begin{gathered}
 \sZ_n^{\rm (PT)}(\tau_{\rm ES}) \approx \exp( V_B(\frac{n}{2}-1) s(\bar{\epsilon}_{B}^j) + V_B s(\bar{\epsilon}_{B}^1) + V_{A_2}s(\frac{V\epsilon- V_{A_1} \bar{\epsilon}_{1} - V_B \bar{\epsilon}_{B}^1}{V_{A_2}}) \\ + V_{A_2} (\frac{n}{2}-1) s\left(\frac{V\epsilon - V_{A_1} \bar{\epsilon}'_{1} - V_B\bar{\epsilon}_{B}^j}{V_{A_2}} \right)+V_{A_2} s\left(\frac{V\epsilon - V_{A_1} \bar{\epsilon}'_{1} - V_B\bar{\epsilon}_B^1}{V_{A_2}} \right)\\+V_{A_1} s(\bar{\epsilon}'_{1}) +V_{A_1} s(\bar{\epsilon}_{1}) + V_{A_1} (\frac{n}{2}-1) s\left(\bar{\epsilon}_{1}+\frac{V_B}{V_{A_1}}(\bar{\epsilon}_{B}^1-\bar{\epsilon}_{B}^j)\right)-V n s(\epsilon))
 \end{gathered} 
 \ee
In this case, it is not clear how to take the $n\rightarrow 1$ limit, so we do not get simple expressions for $\sE$ like in the other phases. 
 \end{enumerate} 
 
\eqref{mices-1}-\eqref{mices-4} are difficult to solve for general forms of $s(\epsilon)$, so that it is difficult to find even the phase diagram for $\sZ_n^{\rm (PT)}$ for general $s(\epsilon)$. As one simple example, let us consider the Gaussian form of the entropy density
 \be 
 s(\epsilon) = \log 2 - \frac{1}{2} \epsilon^2 \label{gaussian}
 \ee
This form is an approximation for the entropy density of discrete systems, such as those with $g(\beta)$ given by \eqref{example_disc}, close to $\epsilon = 0$. The phase diagrams for $\sZ_n^{\rm (PT)}$ with this form of $s(\epsilon)$ for the $n=4$ and $n=6$ are shown in Fig.~\ref{fig:gaussian_mc}. 
 
 \begin{figure}[]
\centering 
\includegraphics[height=4cm]{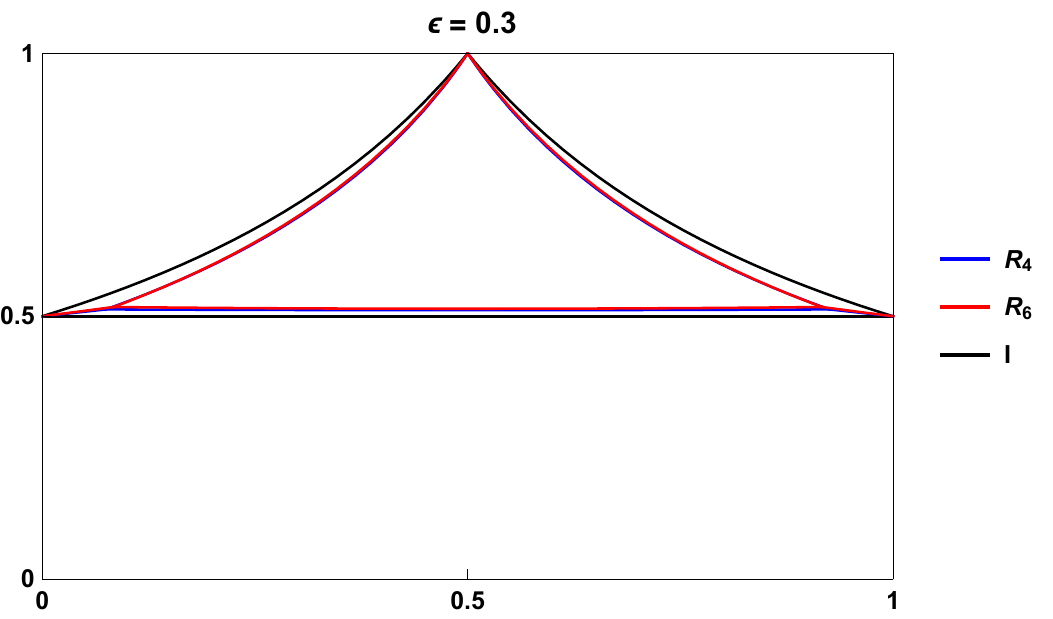} 
\includegraphics[height=4cm]{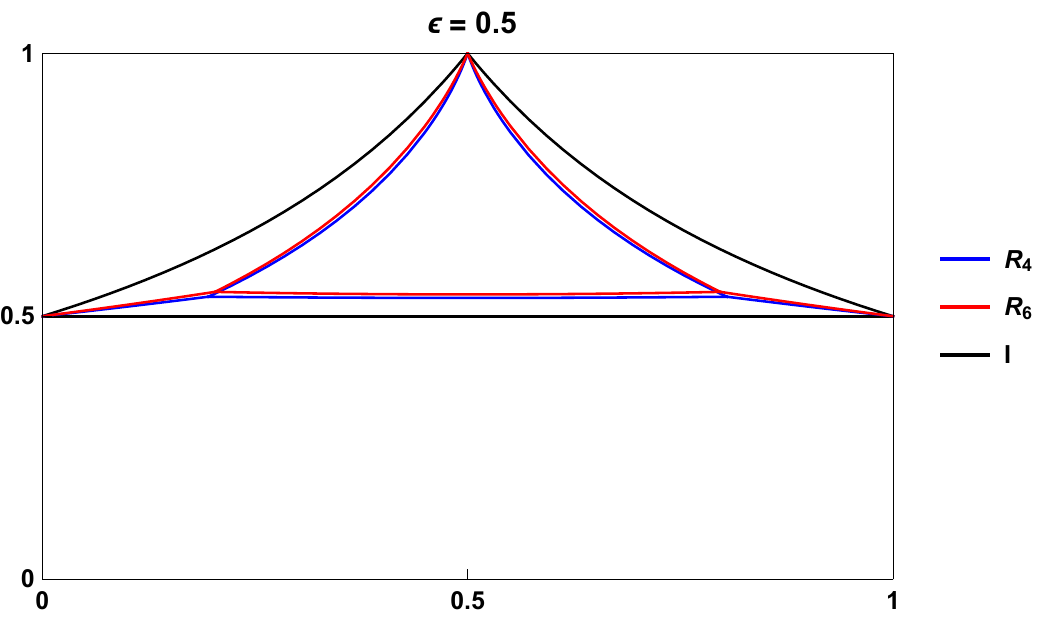} 
\caption{Phase transition lines for the $n=4, 6$ Renyi negativities in the homogeneous microcanonical ensemble case, with $s(\epsilon)$ given by \eqref{gaussian}.}
\label{fig:gaussian_mc}
\end{figure}

\begin{section}{Finite temperature Page curve in inhomogeneous systems}
\label{app:page_finite}

In this appendix, we show that the expression \eqref{endm} for the von Neumann entropy $S_A$ in an equilibrated pure state holds in two examples where $AB$ is inhomogeneous. In such examples, the Page transition in different Renyi entropies $S_{n,A}$ takes place at different values of $c = \frac{S^{\rm(eq)}_A}{S^{\rm(eq)}_A+S^{\rm(eq)}_{B}}$. Hence, we cannot a priori deduce the form of the von Neumann entropy by analytic continuation, and instead we must use the resolvent for the reduced density matrix $\rho_A$, defined in \eqref{Rdef}. 

\subsection{Canonical ensemble with $A$ at infinite temperature and $\bar A$ at finite temperature}

Consider the effective identity operator 
\be 
\sI_{\al} = \mathbf{1}_A \otimes e^{-\beta H_{B}}, \label{ainf_n}
\ee
which we used previously to compute the negativity below \eqref{ainfcan}. For this choice of $\sI_{\al}$, 
\be 
S_{n, A} = \text{min}(\log d_A, S^{\rm (eq)}_{n, B}) \label{Renyi_inf}
\ee
Since $S^{\rm (eq)}_{n, B}$ is $n$-dependent for $\beta>0$, it is clear that the transition from increasing to decreasing behavior as a function of the volume fraction $\frac{V_A}{V}$ depends on $n$. 

To find the von Neumann entropy, we can use a Schwinger-Keldysh equation to evaluate the resolvent. The calculation is exactly analogous to the derivation of \eqref{R_negativity} using the Schwinger-Keldysh equation, and we find an integral expression 
\begin{align}
 \lambda \sR &= d_{A} + \int dE \rho (E)\sum_{k = 1}^{\infty}\left( \frac{\sR e^{-\beta E}}{d_{A}Z_{1,B}}\right)^k
 \quad = d_{A} + \int dE \rho (E)\frac{\sR}{d_A Z_{1,B} e^{\beta E}-\sR}.
 \label{R_entropy}
\end{align}
The von Neumann entropy can be evaluated numerically using this expression, confirming \eqref{endm}, as shown in Fig.~\ref{fig:resolvent_checks_b}.

\begin{figure}[] 
\centering 
\includegraphics[width=10cm]{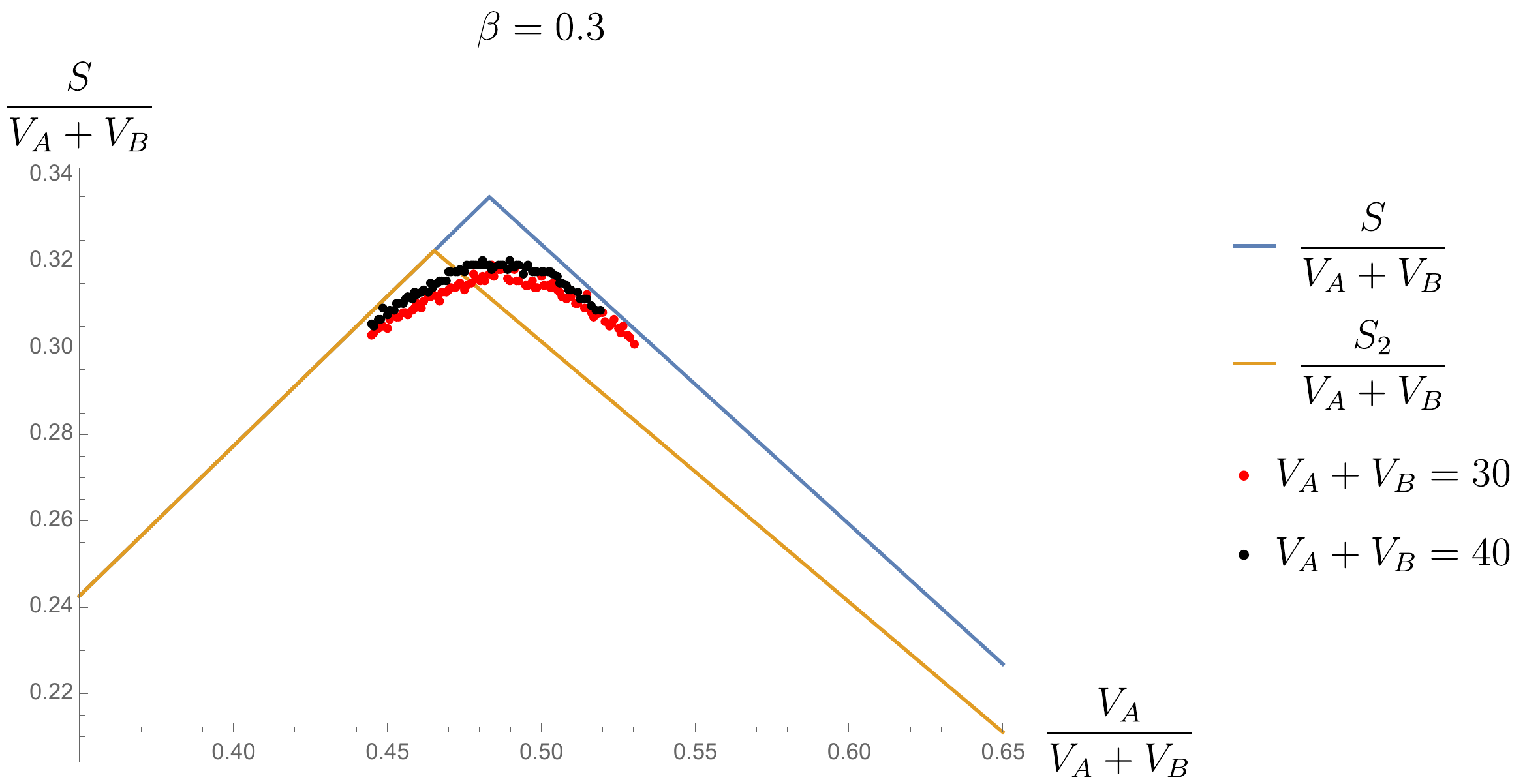}
\caption{Numerical evaluation of the von Neumann entropy of $A$ with $\sI_{\al}$ as in \eqref{ainf_n} using the resolvent \eqref{R_entropy}. We can see that the Page transition occurs at the point predicted by \eqref{endm} and not at, for instance, the transition for the second Renyi entropy from \eqref{Renyi_inf}. For simplicity, we have used a Gaussian density of states $\rho(E) \propto \exp\left[-\frac{E^2}{2N_B}\right]$. }
\label{fig:resolvent_checks_b}
\end{figure}

\subsection{Microcanonical ensemble with different entropy densities in $A$ and $B$}
If we take the effective identity operator to be 
\be 
\sI_{\al} = \sum_{p,\, q, \, E^A_p + E^{\bar A}_q = E} \ket{p}\bra{p}_{A} \otimes \ket{q}\bra{q}_{B}, 
\ee
then the equilibrium approximation for $\sZ_n^{(A)}$ in \eqref{fen} gives
\be 
\sZ_n^{(A)} \approx \sum_{\tau} {(d_{\sE}^A)^{k(\tau\eta^{-1})} (d^{B}_{E-\sE})^{k(\tau)} \ov (N_E)^n}, \quad N_E = \sum_{\sE} d_{\sE}^A \, d^{B}_{E-\sE} 
\ee
Let us consider the resolvent 
\ie
R(\lam) & = {1 \ov \lam} \sum_{n=0}^\infty \lam^{-n} \sZ_n^{(A)}
= \sum_{\sE} {1 \ov \lam} \sum_{n=0}^\infty \lam^{-n} \frac{(d_{\sE}^A d^{B}_{E-\sE})^n}{N_E^n} \sum_{\tau} {(d_{\sE}^A)^{k(\tau\eta^{-1})} (d^{B}_{E-\sE})^{k(\tau)} \ov (d_{\sE}^A d^{B}_{E-\sE})^n} \cr
& = \sum_{\sE} {N_E \ov d_{\sE}^A d^{B}_{E-\sE}} R_\infty (\tilde \lam; d_{\sE}^A, d^{B}_{E-\sE})
= \sum_\sE {1 \ov p_\sE} R_\infty (\tilde \lam; d_{\sE}^A, d^{B}_{E-\sE})
\label{egs}
\fe
where
 \begin{align} 
R_\infty (\lambda; d_A, d_{B}) &= \frac{d_A - d_{B}}{2\lambda} + \frac{d_Ad_{B}}{2} - \frac{d_A d_{B}}{2\lambda}\sqrt{\left(\lambda- \lam_+ \right)\left(\lambda-\lam_- \right)}, \label{resolvent}
\end{align}
is the resolvent for $\rho_A$ in the infinite temperature case, and 
\be 
\tilde \lam = {\lam N_E \ov d_{\sE}^A d^{B}_{E-\sE}}, \qquad p_\sE 
= {d_{\sE}^A d^{B}_{E-\sE} \ov N_E } \ .
\ee
\eqref{egs} can be used to express the density of eigenvalues $D(\lambda)$ of $\rho_A$, and its von Neumann and Renyi entropies, all in terms of their infinite temperature values, 
\bega 
D (\lam) = \sum_{\sE} 
{1 \ov p_\sE} D_\infty (\tilde \lam; d_{\sE}^A, d^{B}_{E-\sE}) \\
\label{entropy}
S_A = - \int D \lam \, \lam \log \lam = \sum_\sE p_\sE S_A^{\infty} (\sE) - \sum_\sE p_\sE \log p_{\sE}, \quad 
\\
\label{zn_se}
\sZ_{n,A} = \int d \lam \, D(\lam) \lam^n = \sum_\sE (p_\sE)^n \, \sZ_{n, A}^{\infty} (d_{\sE}^A ; d^{B}_{E-\sE}) \ .
\end{gather} 

Let us first consider $S_n^{ (A)}$ for $n\neq 1$. Since we know that 
\be 
\sZ_{n,A}^{\infty}(d_A, d_{B}) = \begin{cases} 
d_A^{1-n} & d_A< d_{B} \\
d_{B}^{1-n} & d_{B}< d_{A}
\end{cases} 
\ee
we have 
\be 
\sZ_{n,A} = \sum_{\sE, \, d^A_{\sE}< d^{B}_{E-\sE}} \frac{(d^{B}_{E-\sE})^n d^A_{\sE} }{N_E^n} + \sum_{\sE, \, d^A_{\sE}> d^{B}_{E-\sE}} \frac{d^{B}_{E-\sE} (d^A_{\sE})^n}{N_E^n} 
\ee
Assuming that the saddle points for each of the above two integrals lie within their respective ranges, we then have 
\be 
\sZ_{n,A} \approx e^{-(n-1)S_{n, A}^{\rm (eq)}} + e^{-(n-1)S_{n, B}^{\rm (eq)}} \quad \Rightarrow \quad
S_{n,A} = \text{min}(S_{n, A}^{\rm (eq)}, \, S_{n, B}^{\rm (eq)}) \label{n_page}
\ee
If we take the entropy densities in $A$ and $B$ to have distinct forms, for instance, 
\be 
d^A_{E} = e^{V_A \, s_A(\frac{E}{V_A})}, \quad d^{B}_{E} = e^{V_{B} \, s_{B}(\frac{E}{V_{B}})}, \quad s_A(\epsilon) = g \, \epsilon^{1/2}, \quad s_{B}(\epsilon) = g \, \epsilon^{3/4} \, , \label{ahalf}
\ee 
then as shown in Fig.~\ref{fig:n_mc}, \eqref{n_page} implies that $S_n$ for different $n$ transition from increasing to decreasing behaviour as a function of the volume fraction $\frac{V_A}{V}$ at different points. 

 \begin{figure}[] 
 \begin{center} 
 \includegraphics[width=10cm]{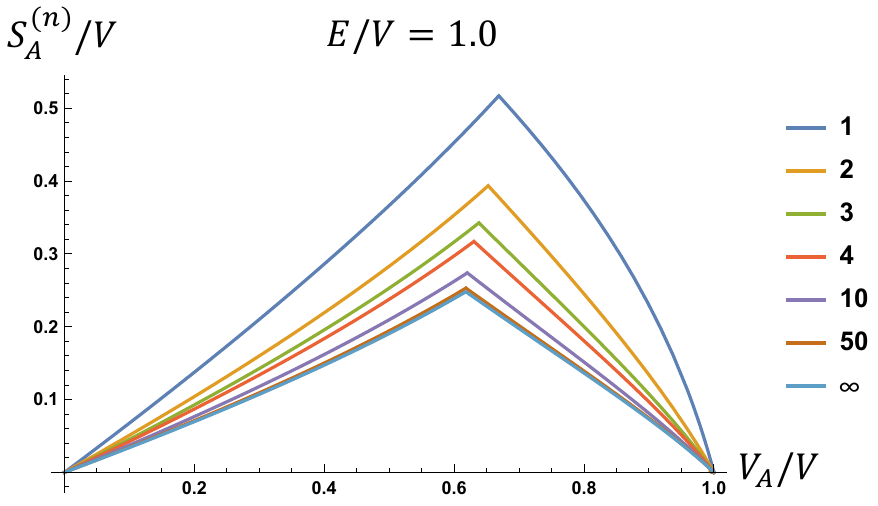}
 \end{center}
 \caption{Renyi and von Neumann entropies as a function of volume fraction for the microcanonical ensemble with inhomogeneous entropy densities in $A$ and $B$.}
 \label{fig:n_mc}
 \end{figure} 
\end{section}

Let us now consider the von Neumann entropy in \eqref{entropy}. We have 
\be 
\begin{gathered} 
S_A = \sum_{\sE} \frac{d^A_{\sE} d^{B}_{E-\sE}}{N_E} \text{min}\left( \log d^A_{\sE}, \log d^{B}_{E-\sE}\right)- \sum_{\sE} \frac{d^{A}_{\sE} d^{B}_{E-\sE}}{N_E} \log \left( \frac{d^{A}_{\sE} d^{B}_{E-\sE}}{N_E} \right)
\end{gathered} 
\label{sa}
\ee
Let us evaluate each of the terms on the right-hand side in the saddle-point approximation. First note that 
\be 
N_E = \sum_{\sE} d^{A}_{\sE} d^{B}_{E-\sE} = V_A\int d \epsilon ~ e^{V_A s_A(\epsilon) + V_{B}s_{B}\left(\frac{E - \epsilon V_A }{V_{B}}\right)} \approx \sqrt{\frac{2\pi}{B}} \, e^{S^{\rm (eq)}_A+ S^{\rm (eq)}_{B}} 
\ee
where 
\be 
B = -\frac{1}{V_A} s''_A(\bar{\epsilon}_A) - \frac{1}{V_{B}} s''_{B}(\bar{\epsilon}_{B}) \, . 
\ee
Note that the von Neumann entropies $S^{\rm (eq)}_A$ and $S^{\rm (eq)}_{B}$ appear here as they are respectively the values of $V_A s_A(\epsilon)$ and $V_A s_A(\frac{E- V_A\epsilon}{V_{\bar A}})$ at the saddle-point value $\bar{\epsilon}$ of $\epsilon$, which is the solution to 
\be 
s'_A(\bar \epsilon) = s'_{B}\left(\frac{E- V_A \bar \epsilon}{V_{B}}\right) \, . \label{saddle_1}
\ee
We can use the saddle-point approximation for the second term in \eqref{sa} to see that it gives a contribution at $O(\log V)$. The saddle-point approximation for the first term is 
\be 
\begin{gathered} 
S_A \approx V_A \int d\epsilon \frac{e^{-\frac{1}{2}B \, V_A^2 \, (\epsilon- \bar{\epsilon})^2}}{\sqrt{2\pi/B}} \text{min} (S^{\rm (eq)}_A+ V_A \beta(\epsilon - \bar{\epsilon})+... \, , \, S^{\rm (eq)}_{B} - V_A \beta(\epsilon - \bar{\epsilon})+...\, )
\end{gathered} 
\label{sa_int}
\ee
Assume without loss of generality that $S^{\rm (eq)}_A< S^{\rm (eq)}_{B}$. We can write 
\be
\begin{gathered} 
S_A \approx S^{\rm (eq)}_A \, \sqrt{\frac{B}{2\pi}} V_A \int_{2 V_A \beta (\epsilon - \bar \epsilon)\leq S_{B}^{\rm (eq)}-S_A^{\rm (eq)}} d \epsilon \, e^{-\frac{1}{2}B \, V_A^2 \, (\epsilon- \bar{\epsilon})^2}\, \\
+ \sqrt{\frac{B}{2\pi}} V_A \int_{2 V_A \beta (\epsilon - \bar \epsilon)\leq S_{B}^{\rm (eq)}-S_A^{\rm (eq)}} d \epsilon \, e^{-\frac{1}{2}B \, V_A^2 \, (\epsilon- \bar{\epsilon})^2}\, V_A\beta(\epsilon - \bar \epsilon) \\ + \sqrt{\frac{B}{2\pi}} V_A \int_{2 V_A \beta (\epsilon - \bar \epsilon)> S_{B}^{\rm (eq)}-S_A^{\rm (eq)}} \, d \epsilon \, e^{-\frac{1}{2}B \, V_A^2 \, (\epsilon- \bar{\epsilon})^2}\, (S^{\rm (eq)}_{B}- V_A\beta(\epsilon - \bar \epsilon) )
\end{gathered} 
\label{approx_3terms}
\ee
Let us consider the above expression in different regimes of the difference $S^{\rm (eq)}_{B}- S^{\rm (eq)}_{A}$. 
\begin{enumerate} 
\item If $S^{\rm (eq)}_{B} -S_A^{\rm (eq)} \gg \frac{1}{\sqrt{B}}$, then the first term in \eqref{approx_3terms} is approximately $S_A$, while the remaining terms are approximately zero. 
\item If $S^{\rm (eq)}_A = S^{\rm (eq)}_{B}$, then \eqref{approx_3terms} becomes 
\begin{align} 
S_A &= S^{\rm (eq)}_A \, \sqrt{\frac{B}{2\pi}} V_A \int_{-\infty}^{\infty} d \epsilon \, e^{-\frac{1}{2}B \, V_A^2 \, (\epsilon- \bar{\epsilon})^2} - \sqrt{\frac{B}{2\pi}} V_A \int_{-\infty}^{\infty} d \epsilon \, e^{-\frac{1}{2}B \, V_A^2 \, (\epsilon- \bar{\epsilon})^2}\, V_A \beta |\epsilon - \bar \epsilon| \nonumber
\\ &=S^{\rm (eq)}_A - \sqrt{\frac{2 \beta^2}{\pi B}}
\end{align}
 \item If $S_{B}^{\rm (eq)} - S_{A}^{\rm (eq)} \sim \frac{1}{\sqrt{B}}$, then the three terms in \eqref{approx_3terms} should be evaluated more carefully, and we get corrections to the value of $S^{\rm (eq)}_A$ of order $\sqrt{V}$. This gives a smoothed out transition between the increasing and decreasing parts of the Page curve as a function of $V_A$. However, since $S_A^{\rm (eq)}$ and $S_{B}^{\rm (eq)}$ scale roughly linearly with $V_A$ and $V_{B}$, the smoothing out of the transition is over a range of values of $V_A$ of order $\sqrt{V}$. So as a function of the parameter $c:= \frac{V_A}{V}$, the transition between the increasing and decreasing parts of the Page curve is still sharp in the thermodynamic limit.
\end{enumerate} 
The observations about enhanced corrections of order $\sqrt{V}$ close to the Page transition in points 2 and 3 were made earlier in \cite{murthy} and \cite{dong_enhanced}. Similar effects were observed in the canonical ensemble in the model of \cite{2019arXiv191111977P}. Since we are interested in the sharp transition as a function of $c$ in this discussion, we can ignore the effects of such corrections, and we then find that at leading order \eqref{sa_int} implies 
\be 
S_A = \text{min}(S_A^{\rm (eq)}, S_{B}^{\rm (eq)}), 
\label{S_suff}
\ee
confirming \eqref{endm} for this example. As shown in Fig.~\ref{fig:n_mc}, when we take the inhomogeneous entropy densities in \eqref{ahalf}, the Page transition of the von Neumann entropy coming from \eqref{S_suff} takes place at a distinct value of $\frac{V_A}{V}$ from the transitions of each of Renyi entropies from \eqref{n_page}.

\section{Evolution of Petz map fidelity at finite temperature}
\label{petz_appendix}

\subsection{Canonical ensemble} \label{petz_app_can}
To evaluate the fidelity of the Petz map reconstruction $F(\rho , [\mathcal{P}_{\sigma,\mathcal{N}}\circ \mathcal{N}](\rho))$ systematically at finite temperature for the canonical ensemble, we use a method similar to Section 3 of \cite{2019arXiv191111977P}. We take the reference state $\sigma_D$ to be maximally mixed and $\rho_D$ to be pure. For simplicity, we take $R$ to be at infinite temperature, so that the effective identity operator is
\be 
\sI_{\al} = e^{-\beta H_{B'}} \otimes \mathbf{1}_R \, . 
\ee
 It is useful to define
\be
F_{n_1, n_2} \equiv F^c_{n_1, n_2} + F^d_{n_1, n_2} 
\ee
for two non-negative integers $n_1, n_2$, where 
\begin{align} 
F^{c}_{n_1, n_2} \equiv \frac{1}{Z_1^{n_1+n_2 +2}} \frac{1}{d_D^{n_1 + n_2 +2}}\sum_{\tau\in P_c} \braket{\eta_R \otimes e_{B'}| \sI_{\al}, \tau} d_D^{k(\tau)} \\ 
 F^{d}_{n_1, n_2} \equiv \frac{1}{Z_1^{n_1+n_2 +2}} \frac{1}{d_D^{n_1 + n_2 +3}}\sum_{\tau\in P_d} \braket{\eta_R \otimes e_{B'}| \sI_{\al}, \tau} d_D^{k(\tau)}
\end{align} 
where $P_c$ refers to the set of permutations in $S_{n_1 + n_2 +2}$ such that the first and $n_1 + 2$-th element are in the same cycle, and $P_d$ refers to the rest. The equilibrium approximation for the fidelity is then given by \footnote{Note that here we use the alternative analytic continuation from \eqref{ren_alt}.} 
\be 
F(\rho , [\mathcal{P}_{\sigma,\mathcal{N}}\circ \mathcal{N}](\rho))= F^c + F^d, \quad F^{c, d} = \lim_{n_1\rightarrow -\ha, \, n_2\rightarrow -\ha}F^{c, d}_{n_1, n_2}
\ee

Note that 
\begin{align} 
F^c_{n_1, n_2} + d_D F^d_{n_1, n_2} &= \frac{1}{(Z_1 d_D)^{n_1+ n_2+2}} \sum_{\tau \in S_{n_1+n_2 +2}} \braket{\eta_R \otimes e_{B'} \otimes e_{D'} | (\sI_{\al})_{B'R} \otimes \mathbf{1}_{D'}, \tau} \\
&= \sZ_{n_1+ n_2 +2, R}
\end{align}
where we get the last expression by identifying the second-to-last expression to be the equilibrium approximation for $\sZ_{n_1+ n_2 +2, R}$ for an equilibrated pure state $\ket{\Psi}$ in a Hilbert space $\sH_{B'R} \otimes \sH_{D'}$, with the effective identity operator $(\sI_{\al})_{B'R} \otimes \mathbf{1}_{D'}$, where $D'$ is an auxiliary system with Hilbert space dimension $d_D$. This implies that 
\be 
F^c + d_D F^d = \sZ_{1, R} = 1\, , \label{cd_rel}
\ee
so that it is sufficient to calculate $F^c$ and express the fidelity as 
\be 
F(\rho , [\mathcal{P}_{\sigma,\mathcal{N}}\circ \mathcal{N}](\rho)) = \left( 1- \frac{1}{d_D}\right) F^c + \frac{1}{d_D}\, . 
\ee

In order to systematically obtain $F^c$ using $F^c_{n_1, n_2}$, we can define the generating functional
\be 
F^{c}(\lambda_1, \lambda_2) = \sum_{n_1, n_2=0}^{\infty} \frac{F^{c}_{n_1, n_2}}{\lambda_1^{n_1} \lambda_2^{n_2}} \, . \label{f_lam}
\ee
Then for non-negative integers $n_1, n_2$, 
\be 
F^{c}_{n_1, n_2} = \oint_{\infty} \frac{d\lambda_1}{2\pi i}\frac{d\lambda_2}{2\pi i} \, \lambda_1^{n_1-1} \lambda_2^{n_2-1} \, F^{c}(\lambda_1, \lambda_2) \label{fc_n1_n2}
\ee
where the contour is taken to be around the point at infinity. 
Now in all permutations $\tau$ contributing to \eqref{f_lam}, the first and $(n_1+2)^{th}$
 element are in a common cycle. Suppose this cycle also includes $m_1$ elements out of the $n_1$ elements between the first and $(n_1 + 2)^{th}$ element, and $m_2$ elements after the $(n_1 + 2)^{th}$ element. We can then consider the total contribution from a fixed $m_1$, $m_2$ to all $n_1, n_2$, as shown in Figure \ref{fig:petz_sum}. Note that in the process we have introduced factors of 
$R(\lambda)$, the resolvent for $\rho_R \equiv \text{Tr}_{B'D'}[\ket{\Psi} \bra{\Psi}]$, with $\ket{\Psi}$ as defined above \eqref{cd_rel}. 
\begin{figure}[!h]
\centering 
\includegraphics[width=7cm]{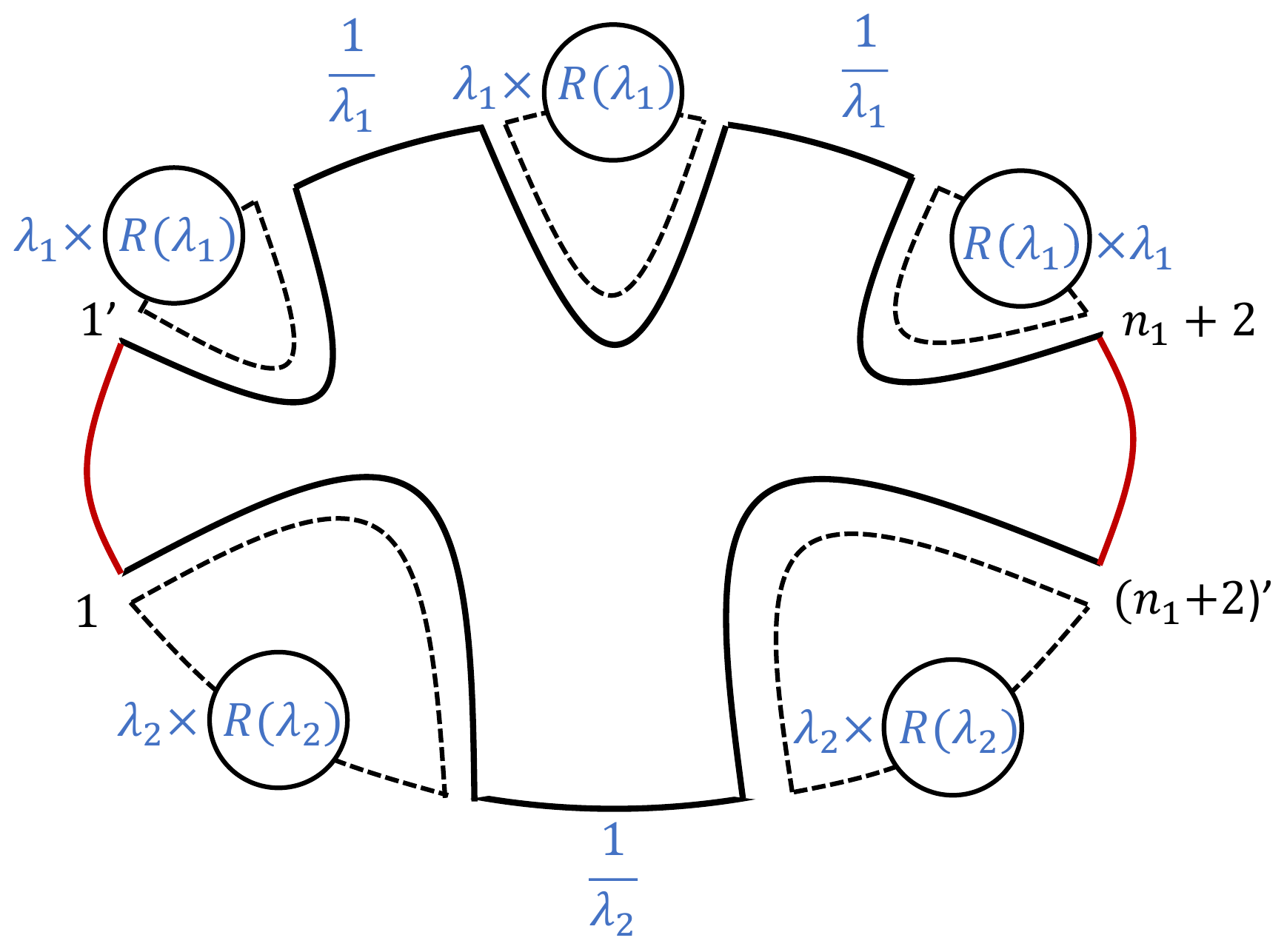}
\caption{Contribution from a fixed $m_1 = 2$ and $m_2=1$ to all $n_1$ and $n_2$ in the expression for $F^{c}(\lambda_1, \lambda_2) $ in \eqref{f_lam}. The solid loop gives a factor of $\frac{d_DZ_{m_1 + m_2 + 2, B'}}{(d_D d_R Z_{1, B'} )^{m_1+m_2+2} }$, and we get $m_1$, $m_1+1$, $m_2$ and $m_2+1$ factors respectively of $\frac{1}{\lambda_1}$, $\lambda_1 R(\lambda_1)$, $\frac{1}{\lambda_2}$ and $\lambda_2 R(\lambda_2)$, leading to \eqref{fcm12}.}
\label{fig:petz_sum}
\end{figure} 

 We then find 
\be 
\label{fcm12}
F^c(\lambda_1, \lambda_2) = \sum_{m_1, m_2=0}^{\infty} \frac{d_D Z_{m_1 + m_2+2, B'}}{(Z_{1,B'} \, d_R \, d_D)^{m_1 + m_2 +2}} \, R(\lambda_1)^{m_1 +1} R(\lambda_2)^{m_2 +1} \, \lambda_1 \lambda_2 \, .  
\ee
and using \eqref{fc_n1_n2}, 
\be 
\begin{gathered} 
F^c_{n_1, n_2} = d_D \int_0^{\infty} dE\, \rho(E) \, \left(\oint_{\infty} \frac{d\lambda}{2\pi i} \lambda^{n_1}R(\lambda)\frac{e^{-\beta E}}{Z_{1, B'} d_D} \frac{1}{d_R- \frac{e^{-\beta E} R(\lambda)}{Z_{1, B'} d_D}}\right) \\ \times \, \left(\oint_{\infty} \frac{d\lambda}{2\pi i} \lambda^{n_2}R(\lambda)\frac{e^{-\beta E}}{Z_{1, B'} d_D} \frac{1}{d_R- \frac{e^{-\beta E} R(\lambda)}{Z_{1, B'} d_D}}\right) 
\end{gathered} 
\label{fc_full_def}
\ee
Note that $\lambda^{-\ha}$ has a branch point at infinity, so the above expression cannot be analytically continued as written. However, we can first deform the contour for integer $n_1, n_2$ to surround the branch cut on the real axis coming from $R(\lambda)$, and then analytically continue to $n_1= n_2 =-\ha$, to get 
\be 
\begin{gathered} 
F^c = d_D \int_0^{\infty} dE\, \rho(E) \, \left(\oint_{\sC} \frac{d\lambda}{2\pi i} \lambda^{-\ha}R(\lambda)\frac{e^{-\beta E}}{Z_{1, B'} d_D} \frac{1}{d_R- \frac{e^{-\beta E} R(\lambda)}{Z_{1, B'} d_D}}\right)^2 
\end{gathered} 
\label{fc_full}
\ee
where $\sC$ is the contour that surrounds the branch cut on the real axis. In order to further simplify this expression, we use the following approximation for the resolvent: 
\be 
R(\lambda) \approx \frac{d_R}{\lambda - \lambda_0} \, \label{R_approx_petz}
\ee
where $\lambda_0$ is the smallest eigenvalue of $\rho_R$, which turns out to be 
\be 
\lambda_0 =  \frac{1}{d_R}g, \quad g = \frac{1}{Z_{1, B'}}\int_{E_0}^{\infty} dE \, \rho(E)\, e^{-\beta E} \label{lambda0}
\ee
with $E_0$ defined through 
\be 
\int_0^{E_0} dE\, \rho(E) = \frac{d_R}{d_D} \label{e0}\, . 
\ee 
In deriving \eqref{R_approx_petz}-\eqref{e0}, we use similar methods to \cite{2019arXiv191111977P}, which we review for this context at the end of this subsection. Putting \eqref{R_approx_petz} into \eqref{fc_full}, we find 
\be 
F^c \approx \frac{1}{Z_{1, B'}} \int_0^{\infty} dE\, \rho(E)\, \frac{e^{-2\beta E}}{d_D Z_{1, B'} \lambda_0 +{e^{-\beta E}}} \, . \label{fc_e}
\ee
Now to further analyse \eqref{fc_e}, let us take the density of states to be as in $AdS_3$, $\rho(E) = e^{c V\sqrt{E/V}}$, where $V$ is the volume of $B'$.  
Then $Z_{1,B'}$ can be evaluated using the saddle point approximation and has the behavior 
\be 
Z_{1,B'} = c' V^\ha e^{- \b E_c + c V \sqrt{\ep_c}} = c' V^\ha e^{{c^2 V \ov 4 \b}} , \quad E_c = V \ep_c , \quad \sqrt{\ep_c} = {c \ov 2 \b} , \quad S =V {c^2 \ov 2 \b} 
\ee
where $S$ is the thermal entropy and $c'$ is some $O(1)$ constant. The second equilibrium Renyi entropy can also be evaluated from the saddle point 
\be
e^{- S_{2,B'}} = {Z_{1,B'} (2 \b) \ov Z_{1,B'}^2 (\b)} \sim V^{-\ha} e^{-{3 c^2 V \ov 8 \b}} , \quad S_{2, B'} = {3 c^2 V \ov 8 \b} \ .
\ee
Now if we define 
\be
x \equiv {\log d_R \ov V} , \quad y \equiv {\log d_D \ov V}, 
\ee
then the various relevant time scales are given by:
\be
t_b: \quad x = {3 c^2 \ov 8 \b}, \quad t_p: \quad x = {c^2 \ov 2 \b} , \quad t_{p_2}: \quad x = {c^2 \ov 2 \b} + y \ .
\ee
We will always take $x \sim O(1)$ while take $V \to \infty$, but $y$ can be $O(V^{-1})$ or $O(1)$ depending on whether the diary is ``small'' or ``large.'' 

Now let us examine the behavior of $g$ defined in~\eqref{lambda0}. Note that for $\epsilon_0 = E_0/V$, 
\bega 
V \int_0^{\ep_0} d\ep \, e^{c V \sqrt{\ep}} = {2 \ov c V} \sqrt{\ep_0} e^{c V \sqrt{\ep_0}} = {d_R \ov d_D} \\
\quad \Rightarrow \quad \sqrt{\ep_0} \approx {x - y \ov c} + O(V^{-1} \log V ), \quad  \ .
 \end{gather} 
 so that $g$ has the form 
 \bega 
 g = \ha \le[1 +{\rm erf} \le(\sqrt{V \b} (\sqrt{\ep_c} - \sqrt{\ep_0}) \ri) \ri] + 
\sqrt{\b \ov\pi c^2 V} e^{- V \b (\sqrt{\ep_c} - \sqrt{\ep_0})^2}  \\
 = \bca 
 1 + O(e^{-V}) & x-y < {c^2 \ov 2 \b} - O(V^{-\ha}) \cr
 {\sqrt{\ep_0 \ov \ep_c} e^{- V \b (\sqrt{\ep_c} - \sqrt{\ep_0})^2} \ov 2 \sqrt{\pi V \b} \le(\sqrt{\ep_0} - \sqrt{\ep_c}) \ri)} & x-y > {c^2 \ov 2 \b} + O(V^{-\ha}) \cr
 \eca \ .
 \label{23}
 \end{gather} 
 
During the evaporation process, $\log d_R$ increases while $V$ decreases. We would like to find the evolution of $F^c$ as a function of $x$ to see how the information recovery improves during the evaporation process.

Note first that from \eqref{23}, for $x-y > {c^2 \ov 2 \b} + O(V^{-\ha})$, the first term in the denominator of the integrand of \eqref{fc_e} can always be ignored relative to the second, implying that the $F^c\approx 1$ and hence $F\approx 1$ at $t_{p_2}$ and later times.
For $t< t_{p_2}$, $\lam_0 \approx {1 \ov d_R}$, and we can simplify \eqref{fc_e} in this regime to 
\be 
F^c \approx \sqrt{V} e^{-V \frac{c^2}{4\beta}} \int_0^{\infty} d\epsilon \, e^{c V \sqrt{\epsilon}} \frac{e^{-2\beta V \epsilon}}{e^{V(\frac{c^2}{4\beta}-x +y)}+ e^{-\beta V \epsilon}} \, . 
\label{115}
\ee
The second term in the denominator dominates for 
\be
\epsilon < \epsilon_u \equiv -\frac{c^2}{4\beta^2} + \frac{x-y}{\beta} \label{ep_small}
\ee
and the first term dominates otherwise. (Note that the regime \eqref{ep_small} exists only for $x-y \geq \frac{c^2}{4\beta}$.) So we can write (ignoring sub-exponential prefactors)
\be 
F^c \sim e^{-V c^2/(4\beta)} \int_0^{\epsilon_u} e^{c V \sqrt{\epsilon}} e^{- \beta V \epsilon} + e^{-V c^2/(4\beta)} e^{\beta V \epsilon_u} \int_{\epsilon_u}^{\infty} e^{c V \sqrt{\epsilon}} e^{-2 \beta V \epsilon}
\label{fc_twoterms} 
\ee
Recall that the saddle point for the first integral is at $\epsilon_c = \frac{c^2}{4\beta^2}$, and the saddle point for the second integral is at $\epsilon^{(2)}_c =\frac{c^2}{16\beta^2}$. 
We consider three regimes: 
\begin{enumerate} 
\item $x -y < \frac{c^2}{4\beta}$ : We only have the second term in \eqref{fc_twoterms}, which can be approximated with its saddle point, so 
\be 
F^c \sim e^{V(-\frac{3c^2}{8\beta} + x-y)} , \quad x -y < \frac{c^2}{4\beta} \label{29}
\ee
\item $ \frac{c^2}{4\beta} < x - y < \frac{5c^2}{16\beta}$: Here the first term should be approximated with the boundary value at $\epsilon_u$, and the second term with the saddle point at $\epsilon^{(2)}_c$. So we find 
\be 
F^c \sim e^{V (c \sqrt{-\frac{c^2}{4\beta^2} + \frac{x-y}{\beta}} -x+y )} + e^{V(-\frac{3c^2}{8\beta} + x-y)}
\ee
The second term is always greater than the first in this regime, so we have 
\be 
F^c \approx e^{V(-\frac{3c^2}{8\beta} + x-y)} , \quad \frac{c^2}{4\beta} < x-y< \frac{5c^2}{16\beta} \, . 
\ee
\item $ \frac{5c^2}{16\beta} < x - y < \frac{c^2}{2\beta}$: In this regime, both terms are approximated with their boundary values, which turn out to be the same, so we have 
\be 
F^c \approx 2 e^{V (c \sqrt{-\frac{c^2}{4\beta^2} + \frac{x-y}{\beta}} -x+y )}, \quad  \frac{5c^2}{16\beta} < x-y < \frac{c^2}{2\beta} \label{32}
\ee
\end{enumerate} 
For a small diary, we can neglect $y$ in \eqref{29}-\eqref{32}, and we therefore find that $(1- \frac{1}{d_D})F^c$ is exponentially suppressed in volume before $t_p$, while $\frac{1}{d_D}$ is $O(1)$. We therefore have 
\be 
F(\rho , [\mathcal{P}_{\sigma,\mathcal{N}}\circ \mathcal{N}](\rho)) \approx \begin{cases} 
 \frac{1}{d_D} & x < \frac{c^2}{2\beta} \\
 1 & x > \frac{c^2}{2\beta}
\end{cases} 
\label{smalldiary_c}
\ee
So the fidelity for the recovery of a small diary improves rapidly from $\frac{1}{d_D}$ to 1 at the Page time.

For a diary sufficiently large such that 
there is a regime where $\frac{3c^2}{8\beta} < x < \frac{5c^2}{16\beta} +y$, we have 
\be 
F(\rho , [\mathcal{P}_{\sigma,\mathcal{N}}\circ \mathcal{N}](\rho))
= \begin{cases} 
\frac{1}{d_D} & x< \frac{3c^2}{8\beta} \\
 e^{V(-\frac{3c^2}{8\beta} + x-y)} =e^{\log d_R-S_{2, B'}-\log d_D} & \frac{3c^2}{8\beta} < x < \frac{5c^2}{16\beta} +y \\ 
 e^{V (c \sqrt{-\frac{c^2}{4\beta^2} + \frac{x-y}{\beta}} -x+y )} & \frac{5c^2}{16\beta} +y < x < \frac{c^2}{2\beta} +y\\
 1 & x> \frac{c^2}{2\beta} +y
\end{cases} 
\label{large_fid}
\ee
So in this case, the fidelity starts increasing from its initial value of $\frac{1}{d_D}$ at time $t_b$ (independently of $\log d_D$ as long as there is a non-trivial regime corresponding to the second line), and the initial improvement is exponential in $\log d_R- S_{2, B'}$, precisely as predicted by the leading correction in \eqref{finite_corr}. 

For a diary that is $O(1)$ but not sufficiently large such that there is a regime where $\frac{3c^2}{8\beta} < x < \frac{5c^2}{16\beta} +y$, we have 
\be 
F(\rho , [\mathcal{P}_{\sigma,\mathcal{N}}\circ \mathcal{N}](\rho))
= \begin{cases} 
\frac{1}{d_D} & x< \frac{c^2}{2\beta} + 2y - c \sqrt{\frac{y}{\beta}} \\
 e^{V (c \sqrt{-\frac{c^2}{4\beta^2} + \frac{x-y}{\beta}} -x+y )} & \frac{c^2}{2\beta} + 2y - c \sqrt{\frac{y}{\beta}} < x < \frac{c^2}{2\beta} +y\\
 1 & x> \frac{c^2}{2\beta} +y
\end{cases} 
\label{int_fid_2}
\ee
So in this case, the fidelity starts to increase from $\frac{1}{d_D}$ at a time 
\be 
t_r : \quad x = \frac{c^2}{2\beta} + 2y - c \sqrt{\frac{y}{\beta}}
\ee
which is macroscopically earlier than $t_{p}$ when $y$ is $O(1)$, and becomes increasingly earlier as the diary becomes larger.

Further, note that in \eqref{fc_twoterms}, we can evaluate both integrals exactly using error functions. The first integral is approximately
\begin{align}
F^c \approx \frac{\sqrt{\pi } c \left(1-\text{erf}\left(\frac{\sqrt{V} \left(c-\sqrt{4 \beta 
  (x-y)-c^2}\right)}{2 \sqrt{\beta }}\right)\right)}{2 \beta ^{3/2}}.
\end{align}
Near $t_{p_2}$, this term dominates over the second integral. This shows that the transition regime is $O((V\beta)^{-\frac{1}{2}})$ in $x-y$, as this is the regime where the error function becomes $O(1)$.

We can also use \eqref{e0} and \eqref{lambda0} to get an exact expression for $\lambda_0$ in terms of $d_R, d_D, V$, which can then be used to evaluate the integral \eqref{fc_e} for large values of $V$ numerically. This leads to the evolution of the fidelity shown in Fig. \ref{petz_recovery_fidelity_finite.pdf}.

Let us now justify the approximation we used for $R(\lambda)$ in \eqref{R_approx_petz}. We use similar steps to Section 2 of \cite{2019arXiv191111977P}. The Schwinger-Dyson equation for $R(\lambda)$ is 
\begin{align} 
&\lambda R = d_R + \sum_{n=1}^{\infty} d_D Z_{n, B'} \frac{R^n}{d_R^n Z_{1, B'}^n d_D^n} \\
& \Rightarrow \lambda R = d_R + d_D R \int_0^{\infty} dE\, \rho(E)\, \frac{e^{-\beta E} }{d_R d_D Z_{1, B'} - e^{-\beta E} R} \label{r_gen}
\end{align}
 It is useful to first find the smallest eigenvalue $\lambda_0$ of $\rho_R$. For $\lambda< \lambda_0$, we can see from the definition of $R$ that $R$ is real and negative, and we can write 
\be 
\lambda \approx \frac{d_R}{R} - \frac{d_D}{R} \int_0^{E'_0} dE\, \rho(E) + d_D \int_{E'_0}^{\infty} dE\, \rho(E)\, \frac{e^{-\beta E} }{d_R d_D Z_{1, B'} } \label{l_near_l0}
\ee
where $E'_0$ for a given set of $d_R, d_D,V_{B'}$ and $\lambda$ is defined implicitly by 
\be 
- e^{-\beta E'_0} = \frac{d_R d_D Z_{1, B'}}{R(\lambda)} \, . 
\ee
In the two terms in \eqref{l_near_l0}, we have assumed that different terms dominate in the denominator of the integrand of \eqref{r_gen}. 
Since $\lambda =\lambda_0$ is a branch point of $R(\lambda)$, we have $\frac{d\lambda}{dR}=0$ at this point, which implies 
\be 
\int_0^{E_0} dE\, \rho(E) \approx \frac{d_R}{d_D} \label{e0_2}
\ee 
where we use $E_0$ to refer to the value of $E'_0$ at $\lambda_0$. We can treat \eqref{e0} as the definition of $E_0$. In terms of $E_0$, $\lambda_0$ is given by 
\be 
\lambda_0 =  \frac{1}{d_R Z_{1, B'}}\int_{E_0}^{\infty} dE \, \rho(E)\, e^{-\beta E} \label{lambda0_2}
\ee

Let us now return to \eqref{r_gen} to obtain an approximation for the resolvent. We can divide the integral into two parts at $E_0$: 
\be
\begin{gathered} 
\lambda R = d_R + d_D R \int_0^{E_0} dE\, \rho(E)\, \frac{e^{-\beta E} }{d_R d_D Z_{1, B'} - e^{-\beta E} R} \\+ d_D R \int_{E_0}^{\infty} dE\, \rho(E)\,e^{-\beta E} \frac{e^{-\beta E} }{d_R d_D Z_{1, B'} - e^{-\beta E} R}
\end{gathered} 
\ee
Note that $e^{-\beta E_0}R(\lambda_0) = - d_R d_D Z_{1, B'}$, and $e^{-\beta E}|R(\lambda)|< e^{-\beta E_0}|R(\lambda_0)|$ for $E> E_0$ and $\lambda> \lambda_0$, using the definition of $R$. Hence, the denominator in the second term of the integrand can be approximated as $d_R d_D Z_{1, B'}$, and we have
\be \begin{gathered}
\lambda R \approx d_R + d_D R \int_0^{E_0} dE\, \rho(E)\, \frac{e^{-\beta E} }{d_R d_D Z_{1, B'} - e^{-\beta E} R} \\+ \frac{1}{d_R Z_{1, B'}}R \int_{E_0}^{\infty} dE\, \rho(E)\,e^{-\beta E} \\
\Rightarrow \lambda R \approx d_R + d_D R \int_0^{E_0} dE\, \rho(E)\, \frac{e^{-\beta E} }{d_R d_D Z_{1, B'} - e^{-\beta E} R} + R \lambda_0 \\
\Rightarrow R = \frac{d_R}{\lambda - \lambda_0} + \frac{d_D}{\lambda- \lambda_0} R \int_0^{E_0} dE\, \rho(E)\, \frac{e^{-\beta E} }{d_R d_D Z_{1, B'} - e^{-\beta E} R} 
\end{gathered} 
\ee
Treating the second term as a perturbation, we find to first order 
\be 
R \approx \frac{d_R}{\lambda - \lambda_0} + \frac{d_D}{\lambda-\lambda_0} \int_0^{E_0} dE \rho(E)\, \frac{e^{-\beta E}}{Z_{1,B'} d_D} \frac{1}{\lambda -\lambda_0 - e^{-\beta E}/(Z_{1, B'} d_D)} \label{final_approx}
\ee
Let us now understand whether our approximation in treating the second term as a perturbation is self-consistent. Suppose we require that for some small $\epsilon$, 
\be 
\lambda - \lambda_0 - \frac{e^{-\beta \epsilon}}{Z_{1, B'} d_D}>0 \, . \label{ep}
\ee
The integral in the second term can be divided into two parts, 
\be 
\begin{gathered} 
I = \frac{d_D}{\lambda-\lambda_0} \int_0^{\epsilon} dE \rho(E)\, \frac{e^{-\beta E}}{Z_{1,B'} d_D} \frac{1}{\lambda -\lambda_0 - e^{-\beta E}/(Z_{1, B'} d_D)} \\ + d_D \int_{\epsilon}^{E_0} dE \rho(E)\, \frac{e^{-\beta E}/{(Z_{1,B'} d_D)}}{\lambda- \lambda_0} \frac{1}{\lambda -\lambda_0 - e^{-\beta E}/(Z_{1, B'} d_D)} 
\end{gathered} 
\label{corr}
\ee
If $\epsilon$ is small, the first term is small. The integrand in the second term is positive and $<\rho(E)$ due to \eqref{ep}, so from \eqref{e0}, the second term is $<d_R$, and hence smaller than the leading term in \eqref{final_approx}. So the approximation \eqref{final_approx} is valid as long as \eqref{ep} is satisfied for small $\epsilon$, and to a first approximation, we can ignore the second term in \eqref{final_approx} and write $R(\lambda)$ as in \eqref{R_approx_petz}.

\subsection{Microcanonical ensemble}
\label{petz_app_mic}
Let us now consider the question of Petz map fidelity taking the effective identity operator to be as in \eqref{mic_inf_temp}, with $A_1= R_1$ and $A_2=R_2$ corresponding to two parts of the radiation, and $B$ to the black hole. We again take $\rho_D$ to be pure and $\sigma_D$ to be maximally mixed. $R_2$ is taken to be at infinite temperature while there is energy conservation between $R_1$ and $B$. We take the average energy density $\ep$ in $R_1 B$ to be 
\be 
\ep = {E \ov V_{R_1} + V_B} 
\ee
and view the evaporation process as a process where $S_R^{(\rm eq)} = V_{R_1} s (\ep) + \log d_{R_2}$ increases
while $S_B^{(\rm eq)} = V_B s (\ep)$ decreases. $t_p$ is the time when 
\be 
t_p : \quad V_{R_1} s ( \ep) + \log d_{R_2} = V_B s ( \ep) \ \label{tp_mic}
\ee
and 
 $t_{p_2}$ is defined as the time when 
\be 
t_{p_2}: \quad V_{R_1} s (\ep) + \log d_{R_2} = V_B s (\ep) + \log d_D \ .
\ee
Recall that in this setup, the time $t_b$ at which the logarithmic negativity starts to grow is given for sufficiently small $\lambda \equiv \frac{\log d_{R_2}}{V_{R_1} s(\epsilon) + \log d_{R_2}}$ by 
\be \label{uhs}
t_b: \quad \log d_{R_2} + 2 S_{{1 \ov 3}, R_1}^{(\rm eq)} - V_{R_1} s ( \ep) = V_B s (\ep) \ . 
\ee

The equilibrium approximation for the Petz map fidelity is given by 
\be 
F(\rho , [\mathcal{P}_{\sigma,\mathcal{N}}\circ \mathcal{N}](\rho)) = \lim_{m\rightarrow \ha} F^E_m, \label{f_e_12}
\ee 
where 
\be 
\begin{gathered} 
F_m^{E} \equiv \sum_{E_1} \frac{1}{d_D^{2m+3} (N_E d_{R_2})^{2m+2}} \sum_{\tau \in \sS_{2m+2}} (d^{R_1}_{E_1} d_{R_2})^{k(\eta^{-1}\tau)} (d^{B'}_{E-E_1})^{k(\tau)} d_D^{k(\tau)+\zeta(\tau)}\\
= \sum_{E_1}\left( \frac{d^{R_1}_{E_1}d^{B'}_{E- E_1}}{N_E}\right)^{2m+2}F_m^{\infty}(d_D, \, d^{R_1}_{E_1} d_{R_2}, \, d^{B'}_{E- E_1}) 
\end{gathered} 
\ee
with 
\be 
N_E \equiv \sum_{E_1}d^{R_1}_{E_1}d^{B'}_{E- E_1} \approx c_1 e^{(V_1 + V_B) s ( \ep) } 
\ee
where $c_1$ and all $c_i$ we will introduce below are $O(1)$ constants, 
and 
\be 
F_m^{\infty}(d_D, d_R, d_{B'}) \equiv \frac{1}{d_D^{2m+3} (d_R d_B)^{2m+2}} \sum_{\tau \in \sS_{2m+2}} d_R^{k(\eta^{-1}\tau)}d_{B'}^{k(\tau)} d_D^{k(\tau)+\zeta(\tau)}
\ee
can be identified to be the infinite-temperature value of $F_m$ in \eqref{fmdef}. 
Then using \eqref{f_e_12} and \eqref{fidelity_inf}, 
\be 
F(\rho , [\mathcal{P}_{\sigma,\mathcal{N}}\circ \mathcal{N}](\rho)) = \frac{1}{d_D} +\le(1-\frac{1}{d_D} \ri) (F_1 + F_2 ) 
\ee
where 
\bega
F_1 = \frac{1}{d_D} \frac{d_{R_2}}{N_E} \sum_{E_1 < V_{R_1} \ep_u} (d^{R_1}_{E_1})^2 \,  \, {}_2F_1\left(\ha, -\ha, 2, \frac{d_{R_2} d^{R_1}_{E_1}}{d^{B'}_{E- E_1}d_D} \right)^2 , \label{f1def} \\
F_2 =  \sum_{E_1 > V_{R_1} \ep_u} \frac{d^{R_1}_{E_1}d^{B'}_{E- E_1}}{N_E} \, {}_2F_1\left(\ha, -\ha, 2, \frac{d^{B'}_{E- E_1}d_D}{d_{R_2} d^{R_1}_{E_1}} \right)^2  \label{f2def}
\end{gather} 
and $\ep_u$ is defined as the solution to  
\be\label{ff0}
V_{R_1} s (\ep_u) + \log d_{R_2} = \log d_D + V_B s (\bar \ep_u) , \quad \bar \ep_u = {E - V_{R_1} \ep_u \ov V_B} \ .
\ee
We note that $F_1$ is always dominated by the value at upper limit of the sum in \eqref{f1def}, 
\bega \label{ff1}
F_1 = c_2 e^{\Lam_1} , \quad \Lam_1 = 2 V_{R_1} s (\ep_u) - (V_{R_1} + V_B) s ( \ep) + \log d_{R_2} - \log d_D
= \Lam_2 , \\
 \Lam_2 = V_{R_1} (s (\ep_u) - s (\ep)) + V_B (s (\bar \ep_u) - s ( \ep))
\end{gather} 
and $F_2$ is dominated by either its saddle point or its lower limit, i.e. 
\be \label{ff2}
F_2 = \bca 1 & \ep_u < \ep \cr
 c_3 e^{\Lam_2}  & \ep_u > \ep 
\eca \ .
\ee
Note that by definition $\Lam_2 \leq 0$, and thus $F_1$ is always exponentially suppressed except at $\ep_u = \ep$. 

From~\eqref{ff0} when we decrease the values of $V_1$ and $\log d_{R_2}$ (i.e. going to earlier times), the value of $\ep_u$ should increase. We can assume that $\ep_u$ is a smooth and monotonically decreasing function of $t$. 
At $t_{p2}$ we have $\ep_u = \ep$ while for $t < t_{p2}$ ($t > t_{p2}$) we have $\ep_u > \ep$ ($\ep_u < \ep$). 
From~\eqref{ff1}--\eqref{ff2}, we then conclude that for any $d_D$, $F_1 + F_2$ is exponentially small in volume for $t < t_{p_2}$ and becomes $1$ at $t_{p_2}$ over a very short range of time (volume suppressed). 

Then for a small diary, precisely as in \eqref{smalldiary_c} for the canonical example, we have 
\be 
F(\rho , [\mathcal{P}_{\sigma,\mathcal{N}}\circ \mathcal{N}](\rho)) = \begin{cases} 
\frac{1}{d_D} & t< t_p \\
1 & t > t_p 
\end{cases} \, . 
\ee 
For a large diary, we have 
\be 
F(\rho , [\mathcal{P}_{\sigma,\mathcal{N}}\circ \mathcal{N}](\rho)) = \begin{cases} 
\frac{1}{d_D} & t< t_r \\
\frac{1}{d_D} e^{V_{R_1} (2s (\ep_u) -s ( \ep)) + \log d_{R_2} -V_B s ( \ep)} & \\=e^{V_{R_1} (s (\ep_u) - s (\ep)) + V_B (s (\bar \ep_u) - s ( \ep))} & t_r < t < t_{p_2} \\
1 & t> t_{p_2}
\end{cases} \, . 
\ee
The growth of the fidelity above $1/d_D$ for a large diary starts at the time scale 
 \be 
 t_r : \quad V_{R_1}(2 s(\epsilon_u) - s(\epsilon))+ V_{R_2} s_0 = V_B s(\epsilon) 
 \ee
 Comparing this to $t_p$ defined in \eqref{tp_mic} and noting that $\epsilon_u> \epsilon$ for $t< t_{p_2}$, we can see that $t_r< t_p$. Note that unlike in the canonical ensemble example in the previous subsection, $t_r$ does not converge to $t_b$ defined by \eqref{uhs} for sufficiently large diaries. For a given $\lambda$, depending on $\log d_D$, $t_r$ can be either earlier or later than $t_b$.

\end{appendix}

\bibliographystyle{JHEP}
\bibliography{main}

\providecommand{\href}[2]{#2}\begingroup\raggedright\begin{thebibliography}{10}

\bibitem{2020arXiv200801089L}
H.~{Liu} and S.~{Vardhan}, {\it {Entanglement entropies of equilibrated pure
  states in quantum many-body systems and gravity}},  {\em arXiv e-prints}
  (Aug., 2020) arXiv:2008.01089, [\href{http://arxiv.org/abs/2008.01089}{{\tt
  arXiv:2008.01089}}].

\bibitem{2020JHEP...09..002P}
G.~{Penington}, {\it {Entanglement wedge reconstruction and the information
  paradox}},  {\em Journal of High Energy Physics} {\bf 2020} (Sept., 2020) 2,
  [\href{http://arxiv.org/abs/1905.08255}{{\tt arXiv:1905.08255}}].

\bibitem{2019JHEP...12..063A}
A.~{Almheiri}, N.~{Engelhardt}, D.~{Marolf}, and H.~{Maxfield}, {\it {The
  entropy of bulk quantum fields and the entanglement wedge of an evaporating
  black hole}},  {\em Journal of High Energy Physics} {\bf 2019} (Dec., 2019)
  63, [\href{http://arxiv.org/abs/1905.08762}{{\tt arXiv:1905.08762}}].

\bibitem{2019arXiv191111977P}
G.~{Penington}, S.~H. {Shenker}, D.~{Stanford}, and Z.~{Yang}, {\it {Replica
  wormholes and the black hole interior}},  {\em arXiv e-prints} (Nov., 2019)
  arXiv:1911.11977, [\href{http://arxiv.org/abs/1911.11977}{{\tt
  arXiv:1911.11977}}].

\bibitem{2020JHEP...05..013A}
A.~{Almheiri}, T.~{Hartman}, J.~{Maldacena}, E.~{Shaghoulian}, and
  A.~{Tajdini}, {\it {Replica wormholes and the entropy of Hawking radiation}},
   {\em Journal of High Energy Physics} {\bf 2020} (May, 2020) 13,
  [\href{http://arxiv.org/abs/1911.12333}{{\tt arXiv:1911.12333}}].

\bibitem{2021arXiv211002959V}
S.~{Vardhan}, J.~{Kudler-Flam}, H.~{Shapourian}, and H.~{Liu}, {\it {Bound
  entanglement in thermalized states and black hole radiation}},  {\em arXiv
  e-prints} (Oct., 2021) arXiv:2110.02959,
  [\href{http://arxiv.org/abs/2110.02959}{{\tt arXiv:2110.02959}}].

\bibitem{audenaert2003entanglement}
K.~{Audenaert}, M.~B. {Plenio}, and J.~{Eisert}, {\it {Entanglement Cost under
  Positive-Partial-Transpose-Preserving Operations}},  {\em \prl} {\bf 90}
  (Jan., 2003) 027901, [\href{http://arxiv.org/abs/quant-ph/0207146}{{\tt
  quant-ph/0207146}}].

\bibitem{2004JMP....45..829C}
M.~{Christandl} and A.~{Winter}, {\it {``Squashed entanglement'': An additive
  entanglement measure}},  {\em Journal of Mathematical Physics} {\bf 45}
  (Mar., 2004) 829--840, [\href{http://arxiv.org/abs/quant-ph/0308088}{{\tt
  quant-ph/0308088}}].

\bibitem{2021PRXQ....2c0347S}
H.~{Shapourian}, S.~{Liu}, J.~{Kudler-Flam}, and A.~{Vishwanath}, {\it
  {Entanglement Negativity Spectrum of Random Mixed States: A Diagrammatic
  Approach}},  {\em PRX Quantum} {\bf 2} (Sept., 2021) 030347,
  [\href{http://arxiv.org/abs/2011.01277}{{\tt arXiv:2011.01277}}].

\bibitem{2007JHEP...09..120H}
P.~{Hayden} and J.~{Preskill}, {\it {Black holes as mirrors: quantum
  information in random subsystems}},  {\em Journal of High Energy Physics}
  {\bf 2007} (Sept., 2007) 120, [\href{http://arxiv.org/abs/0708.4025}{{\tt
  arXiv:0708.4025}}].

\bibitem{2019JHEP...12..007H}
P.~{Hayden} and G.~{Penington}, {\it {Learning the Alpha-bits of black holes}},
   {\em Journal of High Energy Physics} {\bf 2019} (Dec., 2019) 7,
  [\href{http://arxiv.org/abs/1807.06041}{{\tt arXiv:1807.06041}}].

\bibitem{werner1989quantum}
R.~F. {Werner}, {\it {Quantum states with Einstein-Podolsky-Rosen correlations
  admitting a hidden-variable model}},  {\em \pra} {\bf 40} (Oct., 1989)
  4277--4281.

\bibitem{gurvits2004classical}
L.~Gurvits, {\it Classical complexity and quantum entanglement},  {\em Journal
  of Computer and System Sciences} {\bf 69} (2004), no.~3 448--484.

\bibitem{1996PhLA..223....1H}
M.~{Horodecki}, P.~{Horodecki}, and R.~{Horodecki}, {\it {Separability of mixed
  states: necessary and sufficient conditions}},  {\em Physics Letters A} {\bf
  223} (Feb., 1996) 1--8, [\href{http://arxiv.org/abs/quant-ph/9605038}{{\tt
  quant-ph/9605038}}].

\bibitem{1998PhRvA..58..883Z}
K.~{{\.Z}yczkowski}, P.~{Horodecki}, A.~{Sanpera}, and M.~{Lewenstein}, {\it
  {Volume of the set of separable states}},  {\em \pra} {\bf 58} (Aug., 1998)
  883--892, [\href{http://arxiv.org/abs/quant-ph/9804024}{{\tt
  quant-ph/9804024}}].

\bibitem{1996PhRvL..77.1413P}
A.~{Peres}, {\it {Separability Criterion for Density Matrices}},  {\em \prl}
  {\bf 77} (Aug., 1996) 1413--1415,
  [\href{http://arxiv.org/abs/quant-ph/9604005}{{\tt quant-ph/9604005}}].

\bibitem{eisert1999comparison}
J.~{Eisert} and M.~B. {Plenio}, {\it {A comparison of entanglement measures}},
  {\em Journal of Modern Optics} {\bf 46} (Jan., 1999) 145--154,
  [\href{http://arxiv.org/abs/quant-ph/9807034}{{\tt quant-ph/9807034}}].

\bibitem{2000PhRvL..84.2726S}
R.~{Simon}, {\it {Peres-Horodecki Separability Criterion for Continuous
  Variable Systems}},  {\em \prl} {\bf 84} (Mar., 2000) 2726--2729,
  [\href{http://arxiv.org/abs/quant-ph/9909044}{{\tt quant-ph/9909044}}].

\bibitem{vidal2002computable}
G.~{Vidal} and R.~F. {Werner}, {\it {Computable measure of entanglement}},
  {\em \pra} {\bf 65} (Mar., 2002) 032314,
  [\href{http://arxiv.org/abs/quant-ph/0102117}{{\tt quant-ph/0102117}}].

\bibitem{plenio2005logarithmic}
M.~B. {Plenio}, {\it {Logarithmic Negativity: A Full Entanglement Monotone That
  is not Convex}},  {\em \prl} {\bf 95} (Aug., 2005) 090503,
  [\href{http://arxiv.org/abs/quant-ph/0505071}{{\tt quant-ph/0505071}}].

\bibitem{bennett1996mixed}
C.~H. {Bennett}, D.~P. {Divincenzo}, J.~A. {Smolin}, and W.~K. {Wootters}, {\it
  {Mixed-state entanglement and quantum error correction}},  {\em Physical
  Review A} {\bf 54} (Nov., 1996) 3824--3851,
  [\href{http://arxiv.org/abs/quant-ph/9604024}{{\tt quant-ph/9604024}}].

\bibitem{hayden2001asymptotic}
P.~M. {Hayden}, M.~{Horodecki}, and B.~M. {Terhal}, {\it {The asymptotic
  entanglement cost of preparing a quantum state}},  {\em Journal of Physics A
  Mathematical General} {\bf 34} (Sept., 2001) 6891--6898,
  [\href{http://arxiv.org/abs/quant-ph/0008134}{{\tt quant-ph/0008134}}].

\bibitem{1996PhRvA..53.2046B}
C.~H. {Bennett}, H.~J. {Bernstein}, S.~{Popescu}, and B.~{Schumacher}, {\it
  {Concentrating partial entanglement by local operations}},  {\em \pra} {\bf
  53} (Apr., 1996) 2046--2052,
  [\href{http://arxiv.org/abs/quant-ph/9511030}{{\tt quant-ph/9511030}}].

\bibitem{horodecki1998mixed}
M.~{Horodecki}, P.~{Horodecki}, and R.~{Horodecki}, {\it {Mixed-State
  Entanglement and Distillation: Is there a ``Bound'' Entanglement in
  Nature?}},  {\em \prl} {\bf 80} (June, 1998) 5239--5242,
  [\href{http://arxiv.org/abs/quant-ph/9801069}{{\tt quant-ph/9801069}}].

\bibitem{2001PhRvL..87y7902E}
T.~{Eggeling}, K.~G. {Vollbrecht}, R.~F. {Werner}, and M.~M. {Wolf}, {\it
  {Distillability via Protocols Respecting the Positivity of Partial
  Transpose}},  {\em \prl} {\bf 87} (Dec., 2001) 257902,
  [\href{http://arxiv.org/abs/quant-ph/0104095}{{\tt quant-ph/0104095}}].

\bibitem{2017PhRvL.119r0506W}
X.~{Wang} and R.~{Duan}, {\it {Irreversibility of Asymptotic Entanglement
  Manipulation Under Quantum Operations Completely Preserving Positivity of
  Partial Transpose}},  {\em \prl} {\bf 119} (Nov., 2017) 180506,
  [\href{http://arxiv.org/abs/1606.09421}{{\tt arXiv:1606.09421}}].

\bibitem{elben2020mixed}
A.~Elben, R.~Kueng, H.-Y.~R. Huang, R.~van Bijnen, C.~Kokail, M.~Dalmonte,
  P.~Calabrese, B.~Kraus, J.~Preskill, P.~Zoller, et~al., {\it Mixed-state
  entanglement from local randomized measurements},  {\em Physical Review
  Letters} {\bf 125} (2020), no.~20 200501.

\bibitem{2014PhRvB..90f4401C}
C.-M. {Chung}, V.~{Alba}, L.~{Bonnes}, P.~{Chen}, and A.~M. {L{\"a}uchli}, {\it
  {Entanglement negativity via the replica trick: A quantum Monte Carlo
  approach}},  {\em \prb} {\bf 90} (Aug., 2014) 064401,
  [\href{http://arxiv.org/abs/1312.1168}{{\tt arXiv:1312.1168}}].

\bibitem{2018PhRvL.121o0503G}
J.~{Gray}, L.~{Banchi}, A.~{Bayat}, and S.~{Bose}, {\it
  {Machine-Learning-Assisted Many-Body Entanglement Measurement}},  {\em \prl}
  {\bf 121} (Oct., 2018) 150503, [\href{http://arxiv.org/abs/1709.04923}{{\tt
  arXiv:1709.04923}}].

\bibitem{2020PhRvL.125n0603W}
K.-H. {Wu}, T.-C. {Lu}, C.-M. {Chung}, Y.-J. {Kao}, and T.~{Grover}, {\it
  {Entanglement Renyi Negativity across a Finite Temperature Transition: A
  Monte Carlo study}},  {\em \prl} {\bf 125} (Oct., 2020) 140603,
  [\href{http://arxiv.org/abs/1912.03313}{{\tt arXiv:1912.03313}}].

\bibitem{2020PhRvL.125k6801L}
T.-C. {Lu}, T.~H. {Hsieh}, and T.~{Grover}, {\it {Detecting Topological Order
  at Finite Temperature Using Entanglement Negativity}},  {\em \prl} {\bf 125}
  (Sept., 2020) 116801, [\href{http://arxiv.org/abs/1912.04293}{{\tt
  arXiv:1912.04293}}].

\bibitem{2020PhRvB.102f4304W}
E.~{Wybo}, M.~{Knap}, and F.~{Pollmann}, {\it {Entanglement dynamics of a
  many-body localized system coupled to a bath}},  {\em \prb} {\bf 102} (Aug.,
  2020) 064304, [\href{http://arxiv.org/abs/2004.13072}{{\tt
  arXiv:2004.13072}}].

\bibitem{2020arXiv200706305E}
A.~{Elben}, R.~{Kueng}, H.-Y. {Huang}, R.~{van Bijnen}, C.~{Kokail},
  M.~{Dalmonte}, P.~{Calabrese}, B.~{Kraus}, J.~{Preskill}, P.~{Zoller}, and
  B.~{Vermersch}, {\it {Mixed-state entanglement from local randomized
  measurements}},  {\em arXiv e-prints} (July, 2020) arXiv:2007.06305,
  [\href{http://arxiv.org/abs/2007.06305}{{\tt arXiv:2007.06305}}].

\bibitem{2020arXiv200811727L}
T.-C. {Lu} and T.~{Grover}, {\it {Entanglement transitions as a probe of
  qausiparticles and quantum thermalization}},  {\em arXiv e-prints} (Aug.,
  2020) arXiv:2008.11727, [\href{http://arxiv.org/abs/2008.11727}{{\tt
  arXiv:2008.11727}}].

\bibitem{2021arXiv211007384W}
E.~{Wybo}, M.~{Knap}, and F.~{Pollmann}, {\it {Dynamics of Negativity of a
  Wannier-Stark Many-Body Localized System Coupled to a Bath}},  {\em arXiv
  e-prints} (Oct., 2021) arXiv:2110.07384,
  [\href{http://arxiv.org/abs/2110.07384}{{\tt arXiv:2110.07384}}].

\bibitem{cmp/1104115260}
D.~Petz, {\it {Sufficient subalgebras and the relative entropy of states of a
  von Neumann algebra}},  {\em Communications in Mathematical Physics} {\bf
  105} (1986), no.~1 123 -- 131.

\bibitem{10.1093/qmath/39.1.97}
D.~PETZ, {\it {SUFFICIENCY OF CHANNELS OVER VON NEUMANN ALGEBRAS}},  {\em The
  Quarterly Journal of Mathematics} {\bf 39} (03, 1988) 97--108,
  [\href{http://arxiv.org/abs/https://academic.oup.com/qjmath/article-pdf/39/1/97/4559225/39-1-97.pdf}{{\tt
  https://academic.oup.com/qjmath/article-pdf/39/1/97/4559225/39-1-97.pdf}}].

\bibitem{2003RvMaP..15...79P}
D.~{Petz}, {\it {Monotonicity of Quantum Relative Entropy Revisited}},  {\em
  Reviews in Mathematical Physics} {\bf 15} (Jan., 2003) 79--91,
  [\href{http://arxiv.org/abs/quant-ph/0209053}{{\tt quant-ph/0209053}}].

\bibitem{2015arXiv150907127J}
M.~{Junge}, R.~{Renner}, D.~{Sutter}, M.~M. {Wilde}, and A.~{Winter}, {\it
  {Universal recovery maps and approximate sufficiency of quantum relative
  entropy}},  {\em arXiv e-prints} (Sept., 2015) arXiv:1509.07127,
  [\href{http://arxiv.org/abs/1509.07127}{{\tt arXiv:1509.07127}}].

\bibitem{2020arXiv200608002F}
T.~{Faulkner}, S.~{Hollands}, B.~{Swingle}, and Y.~{Wang}, {\it {Approximate
  recovery and relative entropy I. general von Neumann subalgebras}},  {\em
  arXiv e-prints} (June, 2020) arXiv:2006.08002,
  [\href{http://arxiv.org/abs/2006.08002}{{\tt arXiv:2006.08002}}].

\bibitem{2019arXiv190500577D}
S.~{Dutta} and T.~{Faulkner}, {\it {A canonical purification for the
  entanglement wedge cross-section}},  {\em arXiv e-prints} (May, 2019)
  arXiv:1905.00577, [\href{http://arxiv.org/abs/1905.00577}{{\tt
  arXiv:1905.00577}}].

\bibitem{SR_RTN1}
C.~{Akers}, T.~{Faulkner}, S.~{Lin}, and P.~{Rath}, {\it Reflected entropy in
  random tensor networks},  {\em to appear}.

\bibitem{SR_RTN2}
C.~{Akers}, T.~{Faulkner}, S.~{Lin}, and P.~{Rath}, {\it The page curve for
  reflected entropy},  {\em to appear}.

\bibitem{2019CMaPh.376..609C}
S.~X. {Cui}, P.~{Hayden}, T.~{He}, M.~{Headrick}, B.~{Stoica}, and M.~{Walter},
  {\it {Bit Threads and Holographic Monogamy}},  {\em Communications in
  Mathematical Physics} {\bf 376} (July, 2019) 609--648,
  [\href{http://arxiv.org/abs/1808.05234}{{\tt arXiv:1808.05234}}].

\bibitem{2020JHEP...04..208A}
C.~{Akers} and P.~{Rath}, {\it {Entanglement wedge cross sections require
  tripartite entanglement}},  {\em Journal of High Energy Physics} {\bf 2020}
  (Apr., 2020) 208, [\href{http://arxiv.org/abs/1911.07852}{{\tt
  arXiv:1911.07852}}].

\bibitem{2021arXiv210111029D}
X.~{Dong}, X.-L. {Qi}, and M.~{Walter}, {\it {Holographic entanglement
  negativity and replica symmetry breaking}},  {\em arXiv e-prints} (Jan.,
  2021) arXiv:2101.11029, [\href{http://arxiv.org/abs/2101.11029}{{\tt
  arXiv:2101.11029}}].

\bibitem{2011PhRvL.106e0405B}
M.~C. {Ba{\~n}uls}, J.~I. {Cirac}, and M.~B. {Hastings}, {\it {Strong and Weak
  Thermalization of Infinite Nonintegrable Quantum Systems}},  {\em \prl} {\bf
  106} (Feb., 2011) 050405, [\href{http://arxiv.org/abs/1007.3957}{{\tt
  arXiv:1007.3957}}].

\bibitem{2013PhRvL.111a0401S}
S.~{Sugiura} and A.~{Shimizu}, {\it {Canonical Thermal Pure Quantum State}},
  {\em \prl} {\bf 111} (July, 2013) 010401,
  [\href{http://arxiv.org/abs/1302.3138}{{\tt arXiv:1302.3138}}].

\bibitem{2009arXiv0910.1768C}
B.~{Collins} and I.~{Nechita}, {\it {Gaussianization and eigenvalue statistics
  for random quantum channels (III)}},  {\em arXiv e-prints} (Oct., 2009)
  arXiv:0910.1768, [\href{http://arxiv.org/abs/0910.1768}{{\tt
  arXiv:0910.1768}}].

\bibitem{hayden2006aspects}
P.~{Hayden}, D.~W. {Leung}, and A.~{Winter}, {\it {Aspects of Generic
  Entanglement}},  {\em Communications in Mathematical Physics} {\bf 265}
  (July, 2006) 95--117, [\href{http://arxiv.org/abs/quant-ph/0407049}{{\tt
  quant-ph/0407049}}].

\bibitem{1993PhRvL..71.1291P}
D.~N. {Page}, {\it {Average entropy of a subsystem}},  {\em \prl} {\bf 71}
  (Aug., 1993) 1291--1294, [\href{http://arxiv.org/abs/gr-qc/9305007}{{\tt
  gr-qc/9305007}}].

\bibitem{page2013time}
D.~N. {Page}, {\it {Time dependence of Hawking radiation entropy}},  {\em
  Journal of Cosmology and Astroparticle Physics} {\bf 2013} (Sept., 2013) 028,
  [\href{http://arxiv.org/abs/1301.4995}{{\tt arXiv:1301.4995}}].

\bibitem{2007PhRvA..75f4304N}
M.~A. {Nielsen} and D.~{Poulin}, {\it {Algebraic and information-theoretic
  conditions for operator quantum error correction}},  {\em \pra} {\bf 75}
  (June, 2007) 064304, [\href{http://arxiv.org/abs/quant-ph/0506069}{{\tt
  quant-ph/0506069}}].

\bibitem{2021JHEP...03..088L}
H.~{Liu} and S.~{Vardhan}, {\it {A dynamical mechanism for the Page curve from
  quantum chaos}},  {\em Journal of High Energy Physics} {\bf 2021} (Mar.,
  2021) 88, [\href{http://arxiv.org/abs/2002.05734}{{\tt arXiv:2002.05734}}].

\bibitem{2021arXiv210902649K}
J.~{Kudler-Flam}, V.~{Narovlansky}, and S.~{Ryu}, {\it {Negativity Spectra in
  Random Tensor Networks and Holography}},  {\em arXiv e-prints} (Sept., 2021)
  arXiv:2109.02649, [\href{http://arxiv.org/abs/2109.02649}{{\tt
  arXiv:2109.02649}}].

\bibitem{2021arXiv211011947D}
X.~{Dong}, S.~{McBride}, and W.~W. {Weng}, {\it {Replica Wormholes and
  Holographic Entanglement Negativity}},  {\em arXiv e-prints} (Oct., 2021)
  arXiv:2110.11947, [\href{http://arxiv.org/abs/2110.11947}{{\tt
  arXiv:2110.11947}}].

\bibitem{2013JHEP...02..062A}
A.~{Almheiri}, D.~{Marolf}, J.~{Polchinski}, and J.~{Sully}, {\it {Black holes:
  complementarity or firewalls?}},  {\em Journal of High Energy Physics} {\bf
  2013} (Feb., 2013) 62, [\href{http://arxiv.org/abs/1207.3123}{{\tt
  arXiv:1207.3123}}].

\bibitem{hassan_forthcoming}
K.~Hejazi and H.~Shapourian, {\it {Symmetry protected entanglement in random
  mixed states}},  {\em to appear} (2021).

\bibitem{murthy}
C.~{Murthy} and M.~{Srednicki}, {\it {Structure of chaotic eigenstates and
  their entanglement entropy}},  {\em Phys. Rev. E} {\bf 100} (Aug., 2019)
  022131, [\href{http://arxiv.org/abs/1906.04295}{{\tt arXiv:1906.04295}}].

\bibitem{dong_enhanced}
X.~{Dong} and H.~{Wang}, {\it {Enhanced corrections near holographic
  entanglement transitions: a chaotic case study}},  {\em Journal of High
  Energy Physics} {\bf 2020} (Nov., 2020) 7,
  [\href{http://arxiv.org/abs/2006.10051}{{\tt arXiv:2006.10051}}].

\end{thebibliography}\endgroup
 
\end{document}